%% file: master.tex
\documentclass[a4paper,twoside,11pt]{report}
\usepackage{amsmath}
\usepackage{amsfonts}
\usepackage{amssymb}
\usepackage{amstext}
\usepackage{euscript}
\usepackage{multirow}
\usepackage{slashed}
\usepackage{graphicx}
\usepackage[dvips,a4paper,twoside,inner=30mm,outer=20mm,bmargin=25mm,tmargin=30mm,headheight=10mm,headsep=5mm,footnotesep=5mm]{geometry}
\usepackage{fancyhdr}
\usepackage[dvips]{feynmp}
\usepackage{dsfont}

\usepackage{floatflt}

\input{./Auxilliary/commands}

\title{}
\author{Johannes Eiglsperger}

\begin{document}
\input{./0-Titlepage/0-Titlepage}
\cleardoublepage

\tableofcontents
\begin{fmffile}{feynman}

\chapter{Introduction}
\input{./1-Introduction/1-Introduction}

\chapter{Basic Concepts of Potential Models}
\input{./2-Basics/2-Basics}\label{Chapt:BasicConcepts}
\section{Motivation}
\input{./2-Basics/2-Motivation}
\section{Schr\"odinger Equation}
\input{./2-Basics/2-Schrodinger}
\section{The Potential Model}
\input{./2-Basics/2-Potential-model}
\subsection{The Coulomb-plus-Linear Potential}
\input{./2-Basics/2-Coulomb-plus-Linear}
\subsection{Relativistic Corrections}
\input{./2-Basics/2-Relativistic-corrections}
\section{Beyond the Basic Model}
\input{./2-Basics/2-Beyond-Basic-Model}

\chapter{The Breit Interaction}\label{Chapt:CpL-Breit}
\input{./3-Breit/3-Intro}
\section{Perturbative Corrections}
\input{./3-Breit/3-Breit-Quark-Antiquark}
\section{Potential Model}
\input{./3-Breit/3-Potential-Model}
\subsection{Charmonium}
\input{./3-Breit/3-Charmonium}
\subsection{Bottomonium}
\input{./3-Breit/3-Bottomonium}
\section{Mass-Radius Relationship}\label{Chapt:CpL-Breit-Radius-Mass}
\input{./3-Breit/3-Radius-Mass}
\subsection{Charmonium}
\input{./3-Breit/3-Radius-Mass-Charmonium}
\subsection{Bottomonium}
\input{./3-Breit/3-Radius-Mass-Bottomonium}
\section{Summary}
\input{./3-Breit/3-Summary}

\chapter[One-Gluon Exchange without Retardation]{One-Gluon Exchange\\without Retardation}\label{Chapt:CpL-OGE}
\input{./4-OGE/4-Intro}
\section{Perturbative Corrections}
\input{./4-OGE/4-OGE}
\section{Potential Model}
\input{./4-OGE/4-Potential-Model}
\subsection{Charmonium}
\input{./4-OGE/4-Charmonium}
\subsection{Bottomonium}
\input{./4-OGE/4-Bottomonium}
\section{Summary}
\input{./4-OGE/4-Summary}

\chapter{One-Gluon Exchange plus Induced Interaction}\label{Chapt:CpL-induced}
\input{./5-OGE-induced/5-Intro}
\section{Induced Interaction}
\input{./5-OGE-induced/5-Induced-interaction}
\subsection{Pseudoscalar Meson Exchange}
\input{./5-OGE-induced/5-Pseudoscalar-exchange}
\subsection{Vector Meson Exchange}
\input{./5-OGE-induced/5-Vector-exchange}
\subsection{$1S$-State Exchange Induced Interaction}
\input{./5-OGE-induced/5-1S-state-Induced-interaction}
\section{Perturbative Corrections}
\input{./5-OGE-induced/5-OGE-Induced}
\section{Potential Model}
\input{./5-OGE-induced/5-Potential-Model}
\subsection{Charmonium}
\input{./5-OGE-induced/5-Charmonium}
\subsection{Bottomonium}
\input{./5-OGE-induced/5-Bottomonium}
\section{Summary}
\input{./5-OGE-induced/5-Summary}

\chapter{Fourth-order Gluonic Potential}\label{Chapt:Fogtpp}
\input{./6-Fogtpp/6-Intro}
\section{Perturbative Corrections}
\input{./6-Fogtpp/6-Fogtpp}
\section{Potential Model}
\input{./6-Fogtpp/6-Potential-Model}
\subsection{Charmonium}\label{Sec:Fogtpp-Charmonium}
\input{./6-Fogtpp/6-Charmonium}
\subsection{Bottomonium}
\input{./6-Fogtpp/6-Bottomonium}
\section{Additional Spin-Independent Terms of Order $\mathcal{O}(m^{-2})$
}
\input{./6-Fogtpp/6-Titard-Yndurain}
\section{Summary}
\input{./6-Fogtpp/6-Summary}


\chapter{Fourth-order Gluonic Potential plus Induced Interaction}\label{Chapt:FogtppInduced}
\section{Perturbative Corrections}
\input{./7-Fogtpp-induced/7-Perturbative-Corrections}\section{Potential Model}
\input{./7-Fogtpp-induced/7-Potential-Model}
\subsection{Charmonium}
\input{./7-Fogtpp-induced/7-Charmonium}
\subsection{Bottomonium}
\input{./7-Fogtpp-induced/7-Bottomonium}
\subsection{Remarks}
\input{./7-Fogtpp-induced/7-Remarks}
\section{Mass-Radius-Relationship}\label{Chapt:FogtppInducedRadiusMass}
\input{./7-Fogtpp-induced/7-Radius-Mass}
\subsection{Charmonium}\label{Sec:FogtppInducedRadiusMassCharmonium}
\input{./7-Fogtpp-induced/7-Radius-Mass-Charmonium}
\subsection{Bottomonium}\label{Sec:FogtppInducedRadiusMassBottomonium}
\input{./7-Fogtpp-induced/7-Radius-Mass-Bottomonium}
\section{Summary}
\input{./7-Fogtpp-induced/7-Summary}


\chapter{Summary and Conclusion}\label{Chapt:Summary}
\section{Summary}
\input{./9-Summary/9-Discussion}
\section{Conclusion}
\input{./9-Summary/9-Conclusion}
\section{Outlook}
\input{./9-Summary/9-Outlook}


\appendix

\chapter{Conventions and cross sections}
\section{Conventions}
\input{./A-Appendices/A-Conventions}\label{Sec:Conventions}
\section{Cross sections}
\input{./A-Appendices/A-Cross-sections}\label{Sec:Cross-sections}

\chapter{Perturbative QCD Potentials}\label{Chapt:PerturbativeQCDPotentials}
\input{./A-Appendices/B-Pert-QCD-Potentials}
\section{Breit Interaction for Quark-Antiquark Systems}
\input{./A-Appendices/B-Breit}\label{Sec:BreitInteraction}
\section{One-Gluon Exchange without Retardation}
\input{./A-Appendices/B-OGE}\label{Sec:OneGluonExchangeW/ORetardation}
\section{Choice of Kinematics}
\input{./A-Appendices/B-Kinematics}\label{Sec:Kinematics}

\chapter{Computing $\nabla_i\nabla_jW(r)$}
\input{./A-Appendices/C-Computation-nabla-nabla-W}\label{Sec:NablaNablaW}

\chapter{Expectation Values of Momentum Dependent Terms}\label{Chapt:Expect-values-of-MD}
\input{./A-Appendices/D-MD-Expectation-values}
\section{Options for $\langle H^{\textrm{MD}}\rangle$}\label{Sec:MD-interpretations}
\input{./A-Appendices/D-MD-Interpretations}
\section{Momentum Dependent Terms of the Breit Interaction}\label{Sec:MD-Breit-unique}
\input{./A-Appendices/D-MD-Interpretations-and-Retardation}
\end{fmffile}
\nocite{Cohen199, Cohen299, Messiah99, Lucha89, YndurainQG93, YndurainQFT96, PilkuhnRQM}
\bibliographystyle{./Auxilliary/mybibstyle}
\bibliography{./Auxilliary/bibliography}

\chapter*{Acknowledgements\markboth{Acknowledgements}{Acknowledgements}}\input{./0-Titlepage/0-Acknowledgements}
\chapter*{Correction History}\input{./0-Titlepage/0-CorrectionHistory}
\end{document}

%% file: Auxilliary/commands.tex
\makeatletter
\def\cleardoublepage{\clearpage\if@twoside \ifodd\c@page\else%
	\hbox{}%
	\thispagestyle{empty}
	\newpage%
	\if@twocolumn\hbox{}\newpage\fi\fi\fi}
\makeatother

\def\slfrac#1#2{{\mathord{\mathchoice   %
        {\kern.1em\raise.5ex\hbox{$#1$}\kern-.1em / \kern-.15em\lower.25ex\hbox{$#2$}}
        {\kern.1em\raise.5ex\hbox{$#1$}\kern-.1em / \kern-.15em\lower.25ex\hbox{$#2$}}
        {\kern.1em\raise.4ex\hbox{$#1$}\kern-.1em / \kern-.14em\lower.25ex\hbox{$#2$}}
        {\kern.1em\raise.2ex\hbox{$#1$}\kern-.1em / \kern-.1em\lower.25ex\hbox{$#2$}}}}}

\def\TUM#1{%
\textcolor{gray}{
\dimen1=#1\dimen1=.1143\dimen1%
\dimen2=#1\dimen2=.419\dimen2%
\dimen3=#1\dimen3=.0857\dimen3%
\dimen4=\dimen1\advance\dimen4 by\dimen2%
\setbox0=\vbox{\hrule width\dimen3 height\dimen1 depth0pt\vskip\dimen2}%
\setbox1=\vbox{\hrule width\dimen1 height\dimen4 depth0pt}%
\setbox2=\vbox{\hrule width\dimen3 height\dimen1 depth0pt}%
\setbox3=\hbox{\copy0\copy1\copy0\copy1\box2\copy1\copy0\copy1\box0\box1}%
\leavevmode\vbox{\box3}}}
\def\oTUM#1{%
\dimen1=#1\dimen1=.1143\dimen1%
\dimen2=#1\dimen2=.419\dimen2%
\dimen3=#1\dimen3=.0857\dimen3%
\dimen0=#1\dimen0=.018\dimen0%
\dimen4=\dimen1\advance\dimen4 by-\dimen0%
\setbox1=\vbox{\hrule width\dimen0 height\dimen4 depth0pt}%
\advance\dimen4 by\dimen2%
\setbox8=\vbox{\hrule width\dimen0 height\dimen4 depth0pt}%
\advance\dimen4 by-\dimen2\advance\dimen4 by-\dimen0%
\setbox4=\vbox{\hrule width\dimen4 height\dimen0 depth0pt}%
\advance\dimen4 by\dimen1\advance\dimen4 by\dimen3%
\setbox6=\vbox{\hrule width\dimen4 height\dimen0 depth0pt}%
\advance\dimen4 by\dimen3\advance\dimen4 by\dimen0%
\setbox9=\vbox{\hrule width\dimen4 height\dimen0 depth0pt}%
\advance\dimen4 by\dimen1%
\setbox7=\vbox{\hrule width\dimen4 height\dimen0 depth0pt}%
\dimen4=\dimen3%
\setbox5=\vbox{\hrule width\dimen4 height\dimen0 depth0pt}%
\advance\dimen4 by-\dimen0%
\setbox2=\vbox{\hrule width\dimen4 height\dimen0 depth0pt}%
\dimen4=\dimen2\advance\dimen4 by\dimen0%
\setbox3=\vbox{\hrule width\dimen0 height\dimen4 depth0pt}%
\setbox0=\vbox{\hbox{\box9\lower\dimen2\copy3\lower\dimen2\copy5%
\lower\dimen2\copy3\box7}\kern-\dimen2\nointerlineskip%
\hbox{\raise\dimen2\box1\raise\dimen2\box2\copy3\copy4\copy3%
\raise\dimen2\copy5\copy3\box6\copy3\raise\dimen2\copy5\copy3\copy4\copy3%
\raise\dimen2\box5\box3\box4\box8}}%
\leavevmode\box0}

%% file: 0-Titlepage/0-Titlepage.tex
\newlength{\titleheight}
\setlength{\titleheight}{120pt}

\begin{titlepage}
\begin{center}
\begin{minipage}{.84\textwidth}
\begin{center}

{\centering
\rule{\textwidth}{1pt}
}

\vspace{1.5cm}
{\bf \huge \centering 
Quarkonium Spectroscopy:\\
Beyond One-Gluon Exchange\\

}

\vspace{1.1\titleheight}

{\Large 
Diploma Thesis by
}

\vspace{7pt}
 
{\Large 
\textsc{Johannes Eiglsperger}
}

\vspace{2cm}

{\Large 
January, 2007
}

\vspace{4cm}

\newlength{\tumwidth}
\newlength{\restwidth}
\settowidth{\tumwidth}{{\Large Technische Universit\"at M\"unchen}}
\parbox{\tumwidth}{
\Large \centering \vspace{2pt}
Technische Universit\"at M\"unchen

\smallskip

Physik-Department

\smallskip

T39 (Prof.~Dr.~Wolfram Weise)
}

\vspace{3cm}

\rule{\textwidth}{1pt}

\vspace{3pt}


\setlength{\restwidth}{0pt}
\addtolength{\restwidth}{\textwidth}
\addtolength{\restwidth}{-\tumwidth}
\vspace{20pt}
\parbox{.48\restwidth}{\centering \oTUM{.17\textwidth}}\hfill
\parbox{.48\restwidth}{\centering \includegraphics[width=.21\restwidth]{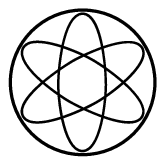}}


\end{center}
\end{minipage}
\end{center}
\end{titlepage}

%
%
%
%
%

%% file: 1-Introduction/1-Introduction.tex
The discovery of the first charmonium state \cite{Aubert:1974js, Augustin:1974xw} in November 1974 led to a revolution in particle physics, especially in hadron spectroscopy. It promoted the quark model and non-Abelian gauge field theories over competing scenarios such as the 'democracy of hadrons' \cite{DeRujula:1975ge}, leading to the prevailing picture of particle physics. The following years of research resulted in the subsequent observation of other charmonium resonances and the discovery of the first bottomonium states in 1977 \cite{Herb:1977ek}. Since that time a lot of theoretical effort has been directed toward understanding hadron spectra and properties within quark models. Quarkonium systems, i.e. bound states of a heavy quark and antiquark, have played a particularly important role in helping to establish the gauge field theory of quarks and gluons and their interactions, Quantum Chromodynamics (QCD).

QCD features two remarkable properties. First, asymptotic freedom implies that at very high energies and momenta, quarks and gluons interact only weakly and act as quasifree particles \cite{Gross:1973id, Politzer:1973fx}. Second, confinement presumably results from the fact that at low energies the force between quarks increases with their distance, so that quarks are always tied into hadrons and cannot be removed individually. 

Confinement, a consequence of the non-Abelian structure of the theory, makes it hard to calculate quantities for bound states within QCD as one cannot apply perturbative QCD. By analogy with positronium, and given the large masses of the charm and bottom quarks, non-relativistic phenomenological potential models have been applied as tools for quarkonium spectroscopy. To accommodate the properties of QCD these models, e.g. \cite{Eichten78, Richardson:1978bt, Buchmuller:1980su}, are based on a short range part motivated by perturbative QCD and a phenomenological long range part accounting for confinement.\medskip\\
After more than a decade of quarkonium physics, the discovery of new resonances above and below the $D\overline{D}$-threshold, often referred to as the Charmonium Renaissance, has revived the interest in this field \cite{Brambilla:2004wf,Swanson:2006st}. 
\medskip\\
From the first potential models for quarkonium in the late 70's to nowadays there have been many new developments in the field of quarkonium spectroscopy. Among others rigorous QCD evaluations \cite{Titard:1993nn, Titard:1994id}, lattice simulations \cite{Okamoto:2001jb, diPierro:2003bu, Juge:2005nr, Koma:2006fw} or Effective Field Theory (EFT) calculations \cite{Barchielli:1988zp, Brambilla:1999xj, Brambilla:1999xf, Brambilla:2000gk, Brambilla:2001fw, Brambilla:2004jw} are applied to gain more fundamental insight. Nevertheless phenomenological potential models still play an important role in heavy meson spectroscopy. Unfortunately some of the more refined models of the mid 80's to nowadays tend to use potentials with a high degree of uncontrolled phenomenological input \cite{Gupta:1982kp, Godfrey:1985xj, Gupta:1989jd, Zeng:1994vj, Ebert:2002pp, Radford:2007vd}.\medskip\\
In this context the task of this diploma thesis is to build up successively a potential model, which yields quantitative quarkonium spectra by using a minimum of parameters and minimal phenomenological input. Consequently the study concentrates on input with fundamental origin, either already used, checking whether essential for the model or not, or so far unaccounted for. A second task of this work is to investigate the relation between radius and mass of the $q\overline{q}$-states.\medskip\\
This thesis is organized as follows: Chapter \ref{Chapt:BasicConcepts} gives an introduction to basic concepts of quarkonium potential models.

In Chapters \ref{Chapt:CpL-Breit} \& \ref{Chapt:CpL-OGE} two versions of the basic Coulomb-plus-Linear model, the first one including retardation corrections and the second one without them, are discussed. In addition the relation between mass and radius of $q\overline{q}$-states within the model is investigated in Chapter \ref{Chapt:CpL-Breit}.

Chapter \ref{Chapt:CpL-induced} introduces the concept of an induced t-channel interaction and discusses a model including this interaction.

A model based on a perturbative QCD potential including processes up to order $\mathcal{O}(\alpha_s^2)$, and a linear confining potential is investigated in Chapter \ref{Chapt:Fogtpp}.

In Chapter \ref{Chapt:FogtppInduced} the model arising from a combination of the induced interaction and a higher order perturbative QCD potential is analyzed and discussed in detail. Further the radius-mass relationship is investigated for this model.

The thesis ends with a short summary in Chapter \ref{Chapt:Summary}. In this chapter the results of the previous chapters are briefly summarized and a conclusion and outlook is given.\bigskip\bigskip\\
The models discussed in this work are denoted with acronyms. The acronym reflects the principle structure of the model:
\begin{itemize}
\item \,\!\![CpL-B]\\
Coulomb-plus-Linear model with perturbative QCD part due to Breit interaction;
\item \,\!\![CpL-3]\\
Coulomb-plus-Linear model with perturbative QCD part due to one-gluon exchange and momentum dependent terms according to option 3 in Appendix \ref{Chapt:Expect-values-of-MD};
\item \,\!\![CpLpI-3]\\
Coulomb-plus-Linear model with induced interaction, perturbative QCD part due to one-gluon exchange and momentum dependent terms according to option 3 in Appendix \ref{Chapt:Expect-values-of-MD};
\item \,\!\![CpL*-2]\\
Coulomb-plus-Linear model with perturbative QCD part to order $\mathcal{O}(\alpha_s^2)$ and momentum dependent terms according to option 2 in Appendix \ref{Chapt:Expect-values-of-MD};
\item \,\!\![CpL*pI-2]\\
Coulomb-plus-Linear model with induced interaction, perturbative QCD part to order $\mathcal{O}(\alpha_s^2)$ and momentum dependent terms according to option 2 in Appendix \ref{Chapt:Expect-values-of-MD}.
\end{itemize}
\bigskip
As common in this field of physics all equations are presented in rationalized units with
\begin{eqnarray*}
\hbar=c=1\:.
\end{eqnarray*}
All numerical computations within this work have been performed with Mathematica$^\text{\textregistered}$ 5.2. 

%% file: 2-Basics/2-Basics.tex
This chapter presents basic concepts of quarkonium potential models, as well as a brief survey of modifications to the basic model introduced in recent years.

%% file: 2-Basics/2-Motivation.tex
Within relativistic quantum field theory the appropriate framework for the description of bound states is the Bethe-Salpeter formalism. In QCD with heavy charm and bottom quarks, the characteristic scale $\Lambda_{QCD}\sim0.2\:\textrm{GeV}$ is small compared to the quark masses, $m_c\sim1.3\:\textrm{GeV}$ and $m_b\sim 4.5\:\textrm{GeV}$. A systematic expansion in powers of $1/m_q$ is possible ("non-relativistic QCD"). Bound state problems can be dealt with non-relativistically. Following these thoughts one treats the bound state problem by solving a Schr\"odinger equation using an appropriate potential.

%% file: 2-Basics/2-Schrodinger.tex
The starting point for calculating the wavefunctions and eigenvalues of quarkonium states is the time-independent Schr\"odinger equation with a central potential and the Hamiltonian
\begin{eqnarray}
H&=&\frac{\vec{p}_1^{\,2}}{2m_1}+\frac{\vec{p}_1^{\,2}}{2m_1}+V_0(|\vec{r}_1-\vec{r}_2|)=\frac{\vec{P}^2}{2M}+\frac{\vec{p}^{\,2}}{2m_\textrm{red}}+V_0(|\vec{r}\,|)=H_{\textrm{cm}}+H_{\textrm{rel}}\:,
\end{eqnarray}
where we have separated the relative motion and the motion of the center of mass with
\vspace{-0.2cm}
\begin{eqnarray}\label{Eq:CM-Trafo}
\begin{array}{l@{\extracolsep{2cm}}l}
H_{\textrm{cm}}=\frac{\vec{P}^{\,2}}{2M}\:, & H_{\textrm{rel}}=\frac{\vec{p}^{\,2}}{2m_\textrm{red}}+V_0(\vec{r})\:, \\ 
\vec{P}=\vec{p}_1+\vec{p}_2\:, & \vec{p}=\frac{m_2\vec{p}_1-m_1\vec{p}_2}{m_1+m_2}\:, \\ 
\vec{R}=\frac{m_1\vec{r}_1+m_2\vec{r}_2}{m_1+m_2}\:, & \vec{r}=\vec{r}_1-\vec{r}_2\:, \\ 
M=m_1+m_2\:, & m_\textrm{red}=\frac{m_1m_2}{m_1+m_2}\:.
\end{array}
\end{eqnarray}
In the center-of-mass frame and substituting
\begin{eqnarray}
\vec{p}&\longrightarrow&-i\vec{\nabla}\:,
\end{eqnarray}
the resulting coordinate space Schr\"odinger equation is:
\begin{eqnarray}\label{Eq:SE-Coordinate-Space}
\left[-\frac{\Delta}{2m_\textrm{red}}+V_0(r)\right]\psi(\vec{r}\,)&=&E\psi(\vec{r}\,)\:.
\end{eqnarray}
For a radially symmetric potential the wave functions written in spherical coordinates are 
\begin{eqnarray}
\psi(r,\theta,\phi)&=&R_{kl}(r)Y_{lm}(\theta,\phi)\:.
\end{eqnarray}
The radial wave functions satisfy the equation
\begin{eqnarray}
\left[-\frac{1}{2m_\textrm{red}}\left(\frac{\partial^2}{\partial r^2}+\frac{2}{r}\frac{\partial}{\partial r}\right)+\frac{l(l+1)}{2m_\textrm{red} r^2}+V_0(r)\right]R_{kl}(r)=E_{kl}R_{kl}(r)\:.
\end{eqnarray}
We introduce as usual the reduced radial wavefunction
\begin{eqnarray}
u_{kl}(r)=rR_{kl}(r)\:,
\end{eqnarray}
and arrive at
\begin{eqnarray}\label{Eq:SE-reduced-radial-wavefunction}
\left[-\frac{1}{2m_\textrm{red}}\frac{\mathrm{d}^2}{\mathrm{d}r^2}+\frac{l(l+1)}{2m_\textrm{red} r^2}+V_0(r)\right]u_{kl}(r)=E_{kl}u_{kl}(r)\:,
\end{eqnarray}
with the normalization condition
\begin{eqnarray}
\int \mathrm{d}^3 r\:|\psi(\vec{r}\,)|^2&=&\int\mathrm{d}\Omega\,\mathrm{d}r\:r^2\big(R_{kl}(r)\big)^2|Y_{lm}(\theta,\phi)|^2=1\:,\nonumber\\
\int\mathrm{d}r\:r^2\big(R_{kl}(r)\big)^2&=&\int\mathrm{d}r\:\big(u_{kl}(r)\big)^2=1\:.
\end{eqnarray}

%% file: 2-Basics/2-Potential-model.tex
Various quarkonium potential models have been used over the years, some of which are purely phenomenological e.g. \cite{Martin:1980jx}, others use perturbative QCD as a guide for the short range part of the potential and a phenomenological long range part to account for confinement e.g. \cite{Eichten78, Richardson:1978bt}. In the region tested by experiments most of these models coincide up to an adjustable constant in the energy. In this section we concentrate on the so called Coulomb-plus-Linear potentials, as these are the most commonly used ones.

%% file: 2-Basics/2-Coulomb-plus-Linear.tex
\begin{center}
\begin{figure}[t!]
\begin{center}
\includegraphics[width=0.9\textwidth]{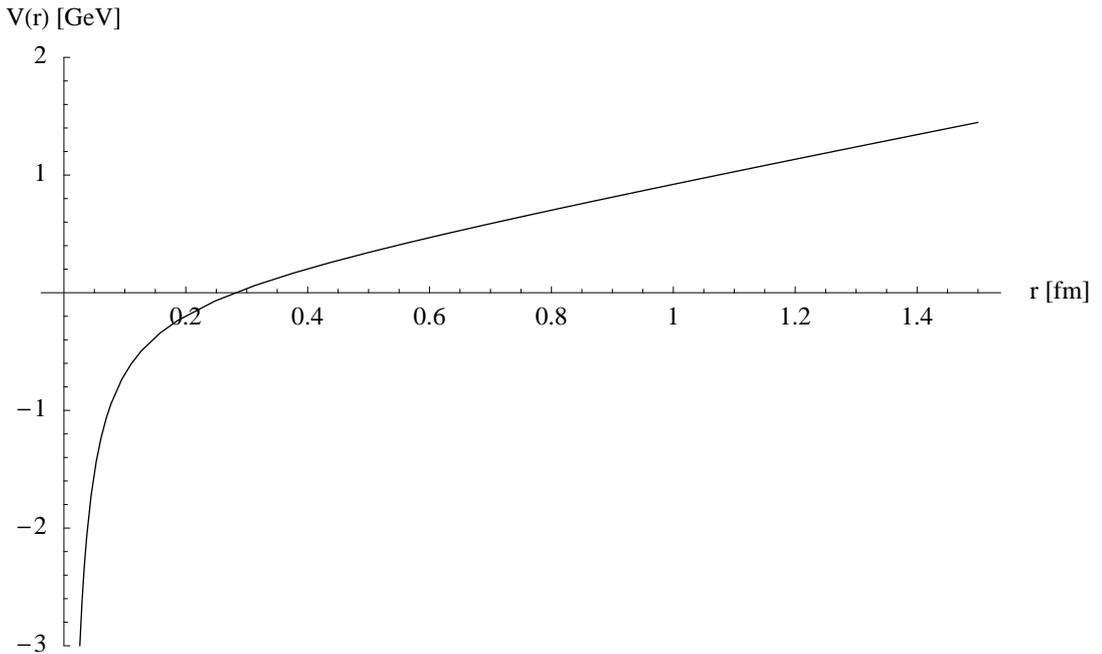}
\end{center}
\caption{\label{Fig:Funnel-Potential} The Coulomb-plus-Linear potential with parameters $\alpha_s=0.3$, $\sigma=1.0\: \textrm{GeV/fm}$.}
\end{figure} 
\end{center}

As the strong interaction is described by an asymptotically free theory, one expects that the short-distance structure of quarkonium is adequately described by perturbative QCD with a small "running" coupling constant. The leading term of the quark-antiquark potential - arising from a perturbative QCD calculation (see Appendix \ref{Chapt:PerturbativeQCDPotentials}) - is essentially Coulomb-like
\begin{eqnarray}
V_{0,\textrm{pert}}(r)&=&-\frac{4\alpha_s}{3r}\:.
\end{eqnarray}
The color charges of the quark and antiquark are subject to confinement. Lattice gauge theories observe, in the limit of static quarks, a linear behaviour of the potential at large separation $r$ (e.g. \cite{Bernard:2000gd}). Consequently,
\begin{eqnarray}
V_{\textrm{conf}}(r)=\sigma r\:,
\end{eqnarray}
is expected to be a reasonable choice for the long range part of the potential. Hence the potential we use in Eq. (\ref{Eq:SE-reduced-radial-wavefunction}) to obtain eigenvalues and wavefunctions for the $q\overline{q}$ bound states is given by the Coulomb-plus-Linear potential
\begin{eqnarray}\label{Eq:Funnel-Potential}
V_0(r)=V_{0,\textrm{pert}}(r)+V_{\textrm{conf}}(r)+C=-\frac{4\alpha_s}{3r}+\sigma r+C\:.
\end{eqnarray}
The Coulomb-plus-linear potential (Fig. \ref{Fig:Funnel-Potential}), sometimes referred to as Cornell potential or funnel potential, has first been proposed by the Cornell group around Eichten \cite{Eichten:1974af, Eichten78, Eichten80} to reproduce the charmonium spectrum.

In contrast to many models of this type we restrict ourselves to $C=0$ in the present work. While there is no principle argument against a non-zero constant $C$, we wish to keep the number of parameters at a minimum and determine the leading order quarkonium masses as
\begin{eqnarray}
M_{0,kl}&=&2m_q+E_{kl}\:,
\end{eqnarray}
where $m_q$ is the quark mass.

%% file: 2-Basics/2-Relativistic-corrections.tex
The Coulomb-plus-linear potential itself is spin-independent and therefore not sufficient to reproduce the full structure of the quarkonium spectra. One has to introduce spin-dependent corrections to this potential to be able to describe the quarkonium spectra in detail.

In the calculation of quarkonium masses, corrections to the potential $V_0$ used in the \newline Schr\"odinger equation,
\begin{eqnarray}
V(\vec{r}\,)&=&V_0(r)+\delta V(\vec{r}\,)\:,
\end{eqnarray}
which arise either from the perturbative QCD part or as corrections to the confining part of the potential, are usually treated perturbatively:
\begin{eqnarray}
M(k^{2S+1}l_j)&=&M_{0,kl}+\delta M_{kjlS}\:,\nonumber\\
\delta M_{kjlS}&=&\int\mathrm{d}^3 r\:\psi^\dagger(\vec{r}\,)\:\delta V(\vec{r}\,)\:\psi(\vec{r}\,)=\langle \delta V(\vec{r}\,)\rangle\:.
\end{eqnarray}
Level splitting are in general due to three interaction types: spin-spin, spin-orbit and tensor interaction.
The spin-dependent potentials  are expressed in terms of $\vec{r}$, $\vec{p}$, $\vec{s}_1$ and $\vec{s}_2$. One introduces the total spin
\begin{eqnarray}
\vec{S}&=&\vec{s}_1+\vec{s}_2\:,
\end{eqnarray}
the angular momentum
\begin{eqnarray}
\vec{L}&=&\vec{r}\times\vec{p}\:,
\end{eqnarray}
and the tensor operator
\begin{eqnarray}
S_{12}&=&12\left[\frac{(\vec{s}_1\cdot\vec{r}\,)(\vec{s}_2\cdot\vec{r}\,)}{r^2}-\frac{1}{3}(\vec{s}_1\cdot\vec{s}_2)\right]\:.
\end{eqnarray}
Experiments provide us with information on the total angular momentum $\vec{J}$, the angular momentum $\vec{L}$ and the total spin $\vec{S}$ of the quarkonium states. Expectation values of $\langle\vec{s}_1\cdot\vec{s}_2\rangle$, $\langle\vec{L}\cdot\vec{S}\rangle$, $\langle S_{12}\rangle$ are then expressed in terms of the eigenvalues $j$, $l$, $S$ of total angular momentum $\vec{J}$, angular momentum $\vec{L}$ and total spin $\vec{S}$.

\paragraph{Spin-spin coefficient\\}
We have
\begin{eqnarray}
\vec{S}^2&=&S(S+1)=(\vec{s}_1+\vec{s}_2)^2=s_1(s_1+1)+s_2(s_2+1)+2\vec{s}_1\cdot\vec{s}_2\:,
\end{eqnarray}
and therefore
\begin{eqnarray}
\vec{s}_1\cdot\vec{s}_2&=&\frac{1}{2}\left[\vec{S}^{\,2}-s_1(s_1+1)-s_2(s_2+1)\right]\:.
\end{eqnarray}
In the case of $s_1=s_2=1/2$ this implies
\begin{eqnarray}
\vec{s}_1\cdot\vec{s}_2&=&\left\lbrace\begin{array}{ll}
-\frac{3}{4}& \textrm{for spin singlet } S=0\\
&\\
+\frac{1}{4}& \textrm{for spin triplet } S=1
\end{array}\right.\:.
\end{eqnarray}

\paragraph{Spin-orbit coefficient\\}
For the calculation of $(\vec{L}\cdot\vec{S})$ we employ
\begin{eqnarray}
\vec{J}^{\,2}&=&j(j+1)=(\vec{L}+\vec{S})^2=\vec{L}^{2}+\vec{S}^{\,2}+2(\vec{L}\cdot\vec{S})=\nonumber\\
&=&l(l+1)+S(S+1)+2(\vec{L}\cdot{S})\:,
\end{eqnarray}
to find
\begin{eqnarray}
\vec{L}\cdot\vec{S}&=&\frac{1}{2}\left[j(j+1)-l(l+1)-S(S+1)\right]\:.
\end{eqnarray}
The spin-orbit term obviously vanishes for $l=0$ or $S=0$ and thus only contributes for $l\neq0$ and $S=1$. Tab. \ref{Tab:LS-Coeff} gives matrix elements of $\vec{L}\cdot\vec{S}$ for states with eigenvalues $j$, $l$, $S=1$.

\begin{table}[ht!]
\begin{center}
\begin{displaymath}
\begin{array}{|c|c|c|c|}
\hline
&&&\\
 j & \quad l+1 \quad&  l &  l-1 \\
&&&\\
\hline
&&&\\
\quad\vec{L}\cdot\vec{S} \quad& l & \quad-1\quad &\quad -(l+1)\quad\\
&&&\\
\hline
\end{array}
\end{displaymath}
\caption{\label{Tab:LS-Coeff} Spin-orbit coupling for $l\neq0$ and $S=1$.}
\end{center}
\end{table}

\paragraph{Tensor coefficient\\}
The tensor term can be expressed via the total spin $\vec{S}$, using
\begin{eqnarray}
(\vec{S}\cdot\vec{r}\,)^2&=&\left[(\vec{s}_1+\vec{s}_2)\cdot\vec{r}\,\right]^2=\left[(\vec{s}_1\cdot\vec{r}\,)+(\vec{s}_2\cdot\vec{r}\,)\right]^2\nonumber\\
&=&(\vec{s}_1\cdot\vec{r}\,)^2+(\vec{s}_2\cdot\vec{r}\,)^2+2(\vec{s}_1\cdot\vec{r}\,)(\vec{s}_2\cdot\vec{r}\,)\nonumber\\
&=&\frac{1}{2}\vec{r}^{\,2}+2(\vec{s}_1\cdot\vec{r}\,)(\vec{s}_2\cdot\vec{r}\,)\:.
\end{eqnarray}
It follows that
\begin{eqnarray}
(\vec{s}_1\cdot\vec{r}\,)(\vec{s}_2\cdot\vec{r}\,)&=&\frac{1}{2}(\vec{S}\cdot\vec{r}\,)^2-\frac{1}{4}\vec{r}^{\,2}\:.
\end{eqnarray}
With
\begin{eqnarray}
\vec{s}_1\cdot\vec{s}_2&=&\frac{1}{2}\vec{S}^{\,2}-\frac{3}{4}\:,
\end{eqnarray}
this leads to
\begin{eqnarray}
S_{12}&=&2\left[3\frac{(\vec{S}\cdot\vec{r}\,)^2}{r^2}-\vec{S}^{\,2}\right]\:.
\end{eqnarray}
It is evident that the tensor term vanishes for $S=0$. Furthermore
\begin{eqnarray}
S_{12}&=&0\qquad \textrm{for }\qquad l=0\:,
\end{eqnarray}
since in this case
\begin{eqnarray}
\bigg\langle\frac{r_ir_j}{r^2}\bigg\rangle&=&\frac{1}{3}\delta_{ij}\:.
\end{eqnarray}
A lengthy calculation gives
\begin{eqnarray}
S_{12}=\frac{4}{(2l-1)(2l+3)}\left[\vec{S}^{\,2}\vec{L}^2-\frac{3}{2}\vec{L}\cdot\vec{S}-3(\vec{L}\cdot\vec{S})^2\right]\:,
\end{eqnarray}
for the diagonal elements of $S_{12}$ which are listed in Tab. \ref{Tab:S12-Coeff}.

\begin{table}[h!]
\begin{center}
\begin{displaymath}
\begin{array}{|c|c|c|c|}
\hline
&&&\\
 j & l+1 & l & l-1 \\
&&&\\
\hline
&&&\\
\quad S_{12} \quad& \quad-\frac{2l}{2l+3}\quad & \quad 2 \quad& \quad-\frac{2(l+1)}{2l-1}\quad\\
&&&\\
\hline
\end{array}
\end{displaymath}
\caption{\label{Tab:S12-Coeff} Non-vanishing diagonal elements of $S_{12}$.}
\end{center}
\end{table}

%% file: 2-Basics/2-Beyond-Basic-Model.tex
The basic model is usually considered to be built of a linear confinement potential plus a potential arising from one-gluon exchange. In the attempt to describe the quarkonium spectra in greater detail different modifications have been applied to the model, some of them purely phenomenological and others of more fundamental nature. We give a brief survey over the most prominent ones. 
\paragraph{Phenomenological $\alpha_s(r)$\\}
The basic model does not include a running coupling although the short range part is calculated using perturbative QCD. As this is a basic feature of QCD and because there is no exact expression for the running coupling constant $\alpha_s(r)$ in coordinate space available, a first idea is to model $\alpha_s(Q^2)$ in the relevant region and transform it to coordinate space \cite{Godfrey:1985xj, Zeng:1994vj}.
\paragraph{Relativistic corrections for the confinement potential\\}
Most of the more recent Coulomb-plus-Linear models introduce relativistic corrections to the confining part of their potential. These corrections are due to an assumed Lorentz-structure of the linear potential. Usually corrections analogous to a Lorentz-scalar \cite{Gupta:1982kp} or a mixture of Lorentz-scalar and Lorentz-vector exchange \cite{Ebert:2002pp, Radford:2007vd} are considered. These terms are implemented as they improve spectral properties of the $P$-states.
\paragraph{One-Gluon exchange and higher order effects\\}
In the basic model the one-gluon exchange with or without retardation is considered as origin of the short distance part of the potential. As the short distance part is due to a perturbative QCD calculation, higher order diagrams have been considered in some models \cite{Gupta:1982kp, Ebert:2002pp, Radford:2007vd}.
\vspace{0.8cm}\\
In this setting we want to study the following questions:
\begin{itemize}
\item How much phenomenological input is needed to obtain a satisfactory description of the $q\overline{q}$ spectra?
\item How important are higher order effects for the short range part of the potential?
\item Is it possible to build a model, with minimal phenomenological input, with parameters close to the commonly expected values
\begin{eqnarray}\label{Eq:Parameter-Intervals}
\alpha_s(m_c^2)\approx0.3\:,\qquad\qquad\!&\quad&\alpha_s(m_b^2)\approx0.2\:,\nonumber\\
m_c=1.15-1.35\:\textrm{GeV}\:,&\quad& m_b=4.6-4.9\:\textrm{GeV}\:,\nonumber\\
\sigma\approx1\:\textrm{GeV/fm}\:,\qquad\qquad\!\!\!\!\!&&
\end{eqnarray}
given by QCD determinations (see \cite{PDBook}) or lattice simulations (e.g. \cite{Bali:2000vr})?
\end{itemize}
To gain some insight into these issues, we will study Coulomb-plus-Linear models with perturbatively treated corrections from various origins. To keep the phenomenological input at a minimum, we neither consider a phenomenological running coupling $\alpha_s(r)$ nor assume any Lorentz-structure, taking the confining potential to be purely static, in these models. Having said this, the origins of the perturbatively treated corrections in our models are either perturbative QCD (through expansion in powers of $\alpha_s$) or relativistic corrections (through expansion in powers of $1/m_q$).

%% file: 3-Breit/3-Intro.tex
This chapter motivates and discusses a potential model with perturbative corrections based on the Breit interaction.

%% file: 3-Breit/3-Breit-Quark-Antiquark.tex
The origin of the short range part of the potential is perturbative QCD which is appropriate for the description of high energy scattering processes. Thus we actually consider a scattering process, from which we deduce the potential. Potentials including relativistic corrections have first been deduced for electron-electron and electron-positron scattering by Gregory Breit \cite{Breit29, Breit30, Breit32}. These potentials are usually referred to as Breit interaction. In analogy to these potentials we have rederived the Breit interaction for quark-antiquark in Appendix \ref{Sec:BreitInteraction}, which is equivalent to an one-gluon exchange potential including retardation corrections. After transformation into the center of mass frame, which effectively results in the substitution
\begin{eqnarray}\label{Eq:Effective-CM-Transformation}
m_q&=&m_1=m_2\:,\quad(q=\:c,\:b)\:,\nonumber\\
\vec{p}&=&\vec{p}_1=-\vec{p}_2\:,
\end{eqnarray}
the Breit interaction for the quark-antiquark system (Eq. (\ref{Eq:Breit-Interaction})) reads
\begin{eqnarray}\label{Eq:Potential-Breit-CM}
V^{\textrm{Breit}}(\vec{r};\vec{p}\,)&\!\!\!\!\!=\!\!\!\!\!&-\frac{4\alpha_s}{3r}+\frac{4\pi\alpha_s}{3m_q^2}\delta^{(3)}(\vec{r}\,)-\frac{2\alpha_s}{3m_q^2}\left[\frac{\vec{p}\cdot\vec{p}}{r}+\frac{(\vec{r}\cdot\vec{p}\,)(\vec{r}\cdot\vec{p}\,)}{r^3}\right]\\
&&\!\!\!\!\!\!\!\!\!\!\!\!\!\!\!\!\!\!\!+\frac{4\alpha_s}{3m_q^2}\left[\frac{8\pi}{3}\delta^{(3)}(\vec{r}\,)(\vec{s}_1\cdot\vec{s}_2)+\frac{3(\vec{s}_1\cdot\hat{r})(\vec{s}_2\cdot \hat{r})-\vec{s}_1\cdot\vec{s}_2}{r^3}\right]+\frac{2\alpha_s}{m_q^2}\frac{(\vec{r}\times\vec{p}\,)\cdot(\vec{s}_1+\vec{s}_2)}{r^3}\:.\nonumber
\end{eqnarray}
This is taken to be the perturbative QCD part of the potential, and thus the origin of the perturbatively treated corrections, used in this model.

%% file: 3-Breit/3-Potential-Model.tex
The potential in this model is composed of a linear confining part and the Breit interaction for quark-antiquark (Eq. (\ref{Eq:Potential-Breit-CM})):
\begin{eqnarray}\label{Eq:Potential-CpL-B}
V^{\textrm{[CpL-B]}}(\vec{r};\vec{p}\,)&\!\!\!\!\!=\!\!\!\!\!&-\frac{4\alpha_s}{3r}+\sigma r+\frac{4\pi\alpha_s}{3m_q^2}\delta^{(3)}(\vec{r}\,)-\frac{2\alpha_s}{3m_q^2}\left[\frac{\vec{p}\cdot\vec{p}}{r}+\frac{(\vec{r}\cdot\vec{p}\,)(\vec{r}\cdot\vec{p}\,)}{r^3}\right]\\
&&\!\!\!\!\!\!\!\!\!\!\!\!\!\!\!\!\!\!\!+\frac{4\alpha_s}{3m_q^2}\left[\frac{8\pi}{3}\delta^{(3)}(\vec{r}\,)(\vec{s}_1\cdot\vec{s}_2)+\frac{3(\vec{s}_1\cdot\hat{r})(\vec{s}_2\cdot \hat{r})-\vec{s}_1\cdot\vec{s}_2}{r^3}\right]+\frac{2\alpha_s}{m_q^2}\frac{(\vec{r}\times\vec{p}\,)\cdot(\vec{s}_1+\vec{s}_2)}{r^3}\:.\nonumber
\end{eqnarray}
As for all Coulomb-plus-Linear models, the potential used in the Schr\"odinger equation to find bound states is
\begin{eqnarray}
V_0(r)&=&-\frac{4\alpha_s}{3r}+\sigma r\:,
\end{eqnarray}
while the remaining parts of Eq. (\ref{Eq:Potential-CpL-B}) are treated within perturbation theory. The resulting mass formula is given by
\begin{eqnarray}\label{Eq:MassFormula-CpL-B}
M^{\textrm{[CpL-B]}}(k^{2S+1}l_j)&=&2m_q+E_{kl}+\frac{4\pi\alpha_s}{3m_q^2}|\psi(0)|^2+\frac{32\pi\alpha_s}{9m_q^2}\left(\frac{1}{2}S(S+1)-\frac{3}{4}\right)|\psi(0)|^2\nonumber\\
&&\!\!\!\!+\alpha_s\frac{j(j+1)-l(l+1)-S(S+1)}{m_q^2}\bigg\langle\frac{1}{r^3}\bigg\rangle+\alpha_s\frac{S_{12}}{3m_q^2}\bigg\langle\frac{1}{r^3}\bigg\rangle\nonumber\\
&&\!\!\!\!+\frac{2\alpha_s}{3m_q^2}\int\mathrm{d}^3r\:\psi^*(\vec{r}\,)\left(\frac{1}{r}\vec{\nabla}^2+\frac{1}{r}\frac{\partial^2}{\partial r^2}\right)\psi(\vec{r}\,)\:.
\end{eqnarray}
The last term represents the expectation value of the purely momentum dependent terms (see App. \ref{Chapt:Expect-values-of-MD}). Though written according to option 1 in this Appendix, the expectation value for the momentum dependent terms in the case of the Breit interaction is unique (see  App. \ref{Sec:MD-Breit-unique}). To mark figures, tables and quantities belonging to this model they are denoted, as already indicated for the potential and the mass formula, with [CpL-B].

%% file: 3-Breit/3-Charmonium.tex
\begin{center}
\begin{figure}[p]
\begin{center}
\includegraphics[width=0.9\textwidth]{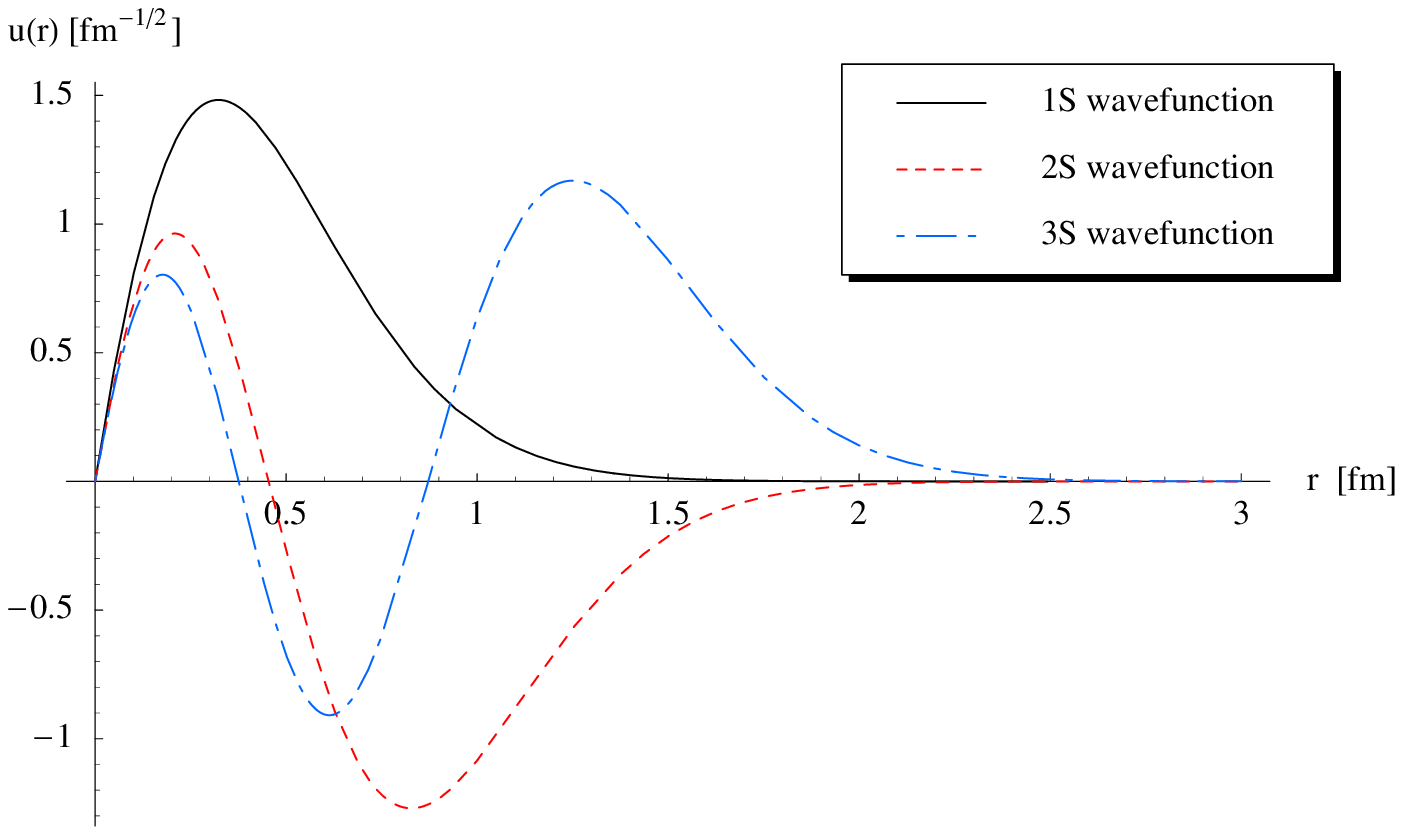}
\end{center}
\caption{\label{Fig:Wavefunctions-S-Charmonium-CpL-B} Charmonium $S$-state reduced radial wavefunctions calculated within model [CpL-B] ($\alpha_s=0.29$, $\sigma=1.306\:\textrm{GeV/fm}$, $m_c=1.2185\:\textrm{GeV}$).}
\end{figure} 
\end{center}
\begin{center}
\begin{figure}[p]\begin{center}
\includegraphics[width=0.9\textwidth]{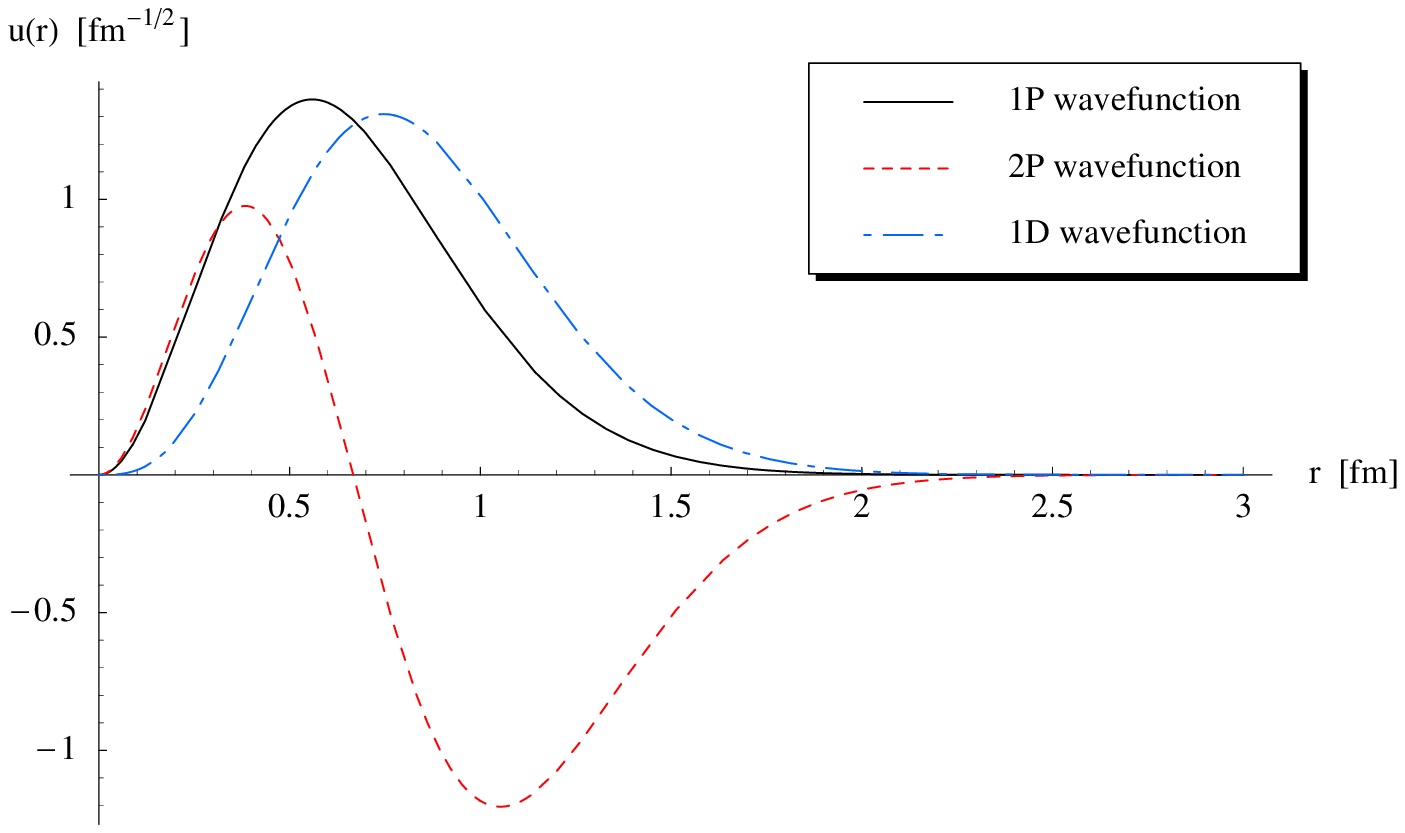}\end{center}
\caption{\label{Fig:Wavefunctions-PD-Charmonium-CpL-B} Charmonium $P$- and $D$-state reduced radial wavefunctions calculated within model [CpL-B] ($\alpha_s=0.29$, $\sigma=1.306\:\textrm{GeV/fm}$, $m_c=1.2185\:\textrm{GeV}$).}
\end{figure} 
\end{center}
\begin{center}
\begin{figure}[p]\begin{center}
\includegraphics[width=0.9\textwidth]{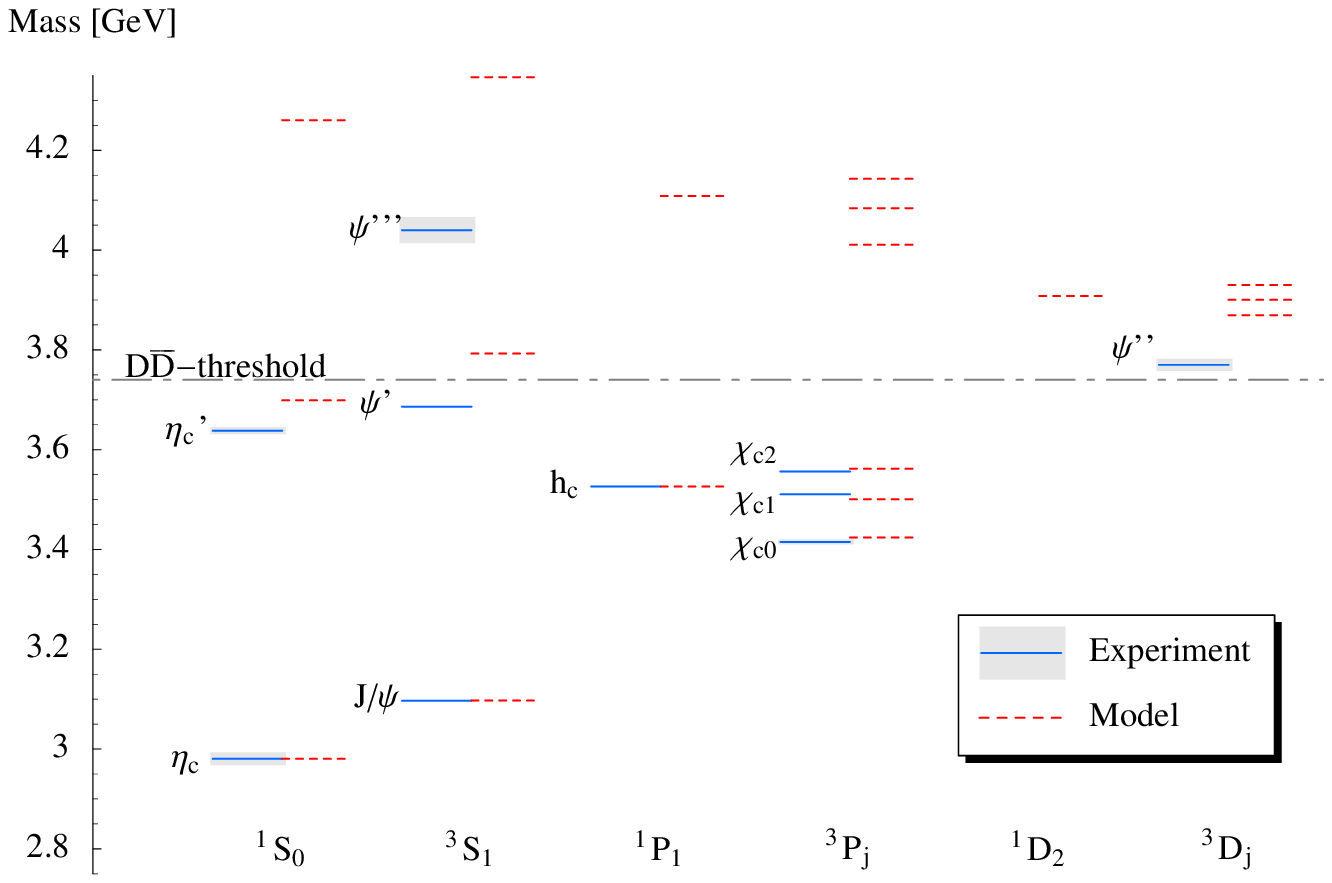}\end{center}
\caption{\label{Fig:SpectrumCharmoniumCpL-B} Charmonium spectra of experiment and model [CpL-B] ($\alpha_s=0.29$, $\sigma=1.306\:\textrm{GeV/fm}$, $m_c=1.2185\:\textrm{GeV}$); for experimental data masses and widths are shown.}
\end{figure} 
\end{center}
\begin{center}
\begin{figure}[p]
\begin{minipage}{7.45cm}\begin{center}
\includegraphics[width=\textwidth]{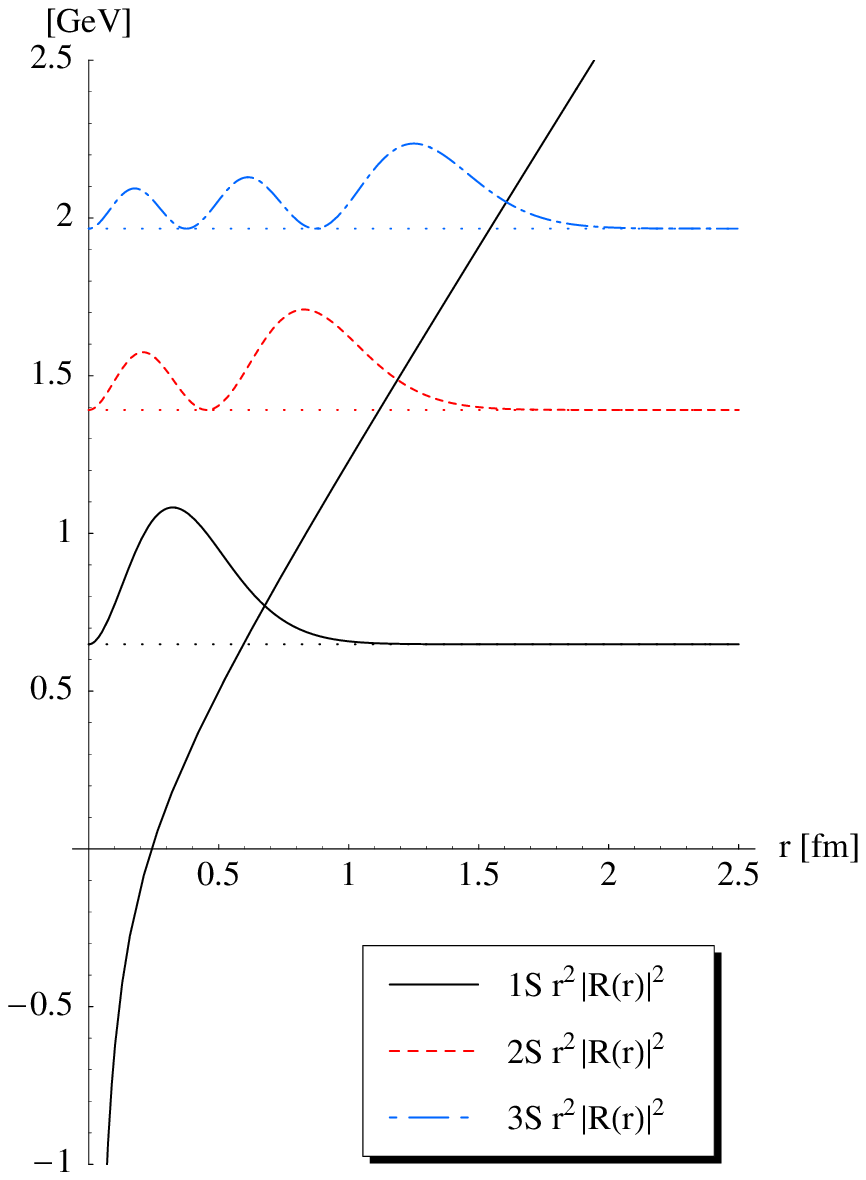}\end{center}
\end{minipage}\hfill
\begin{minipage}{7.45cm}\begin{center}
\includegraphics[width=\textwidth]{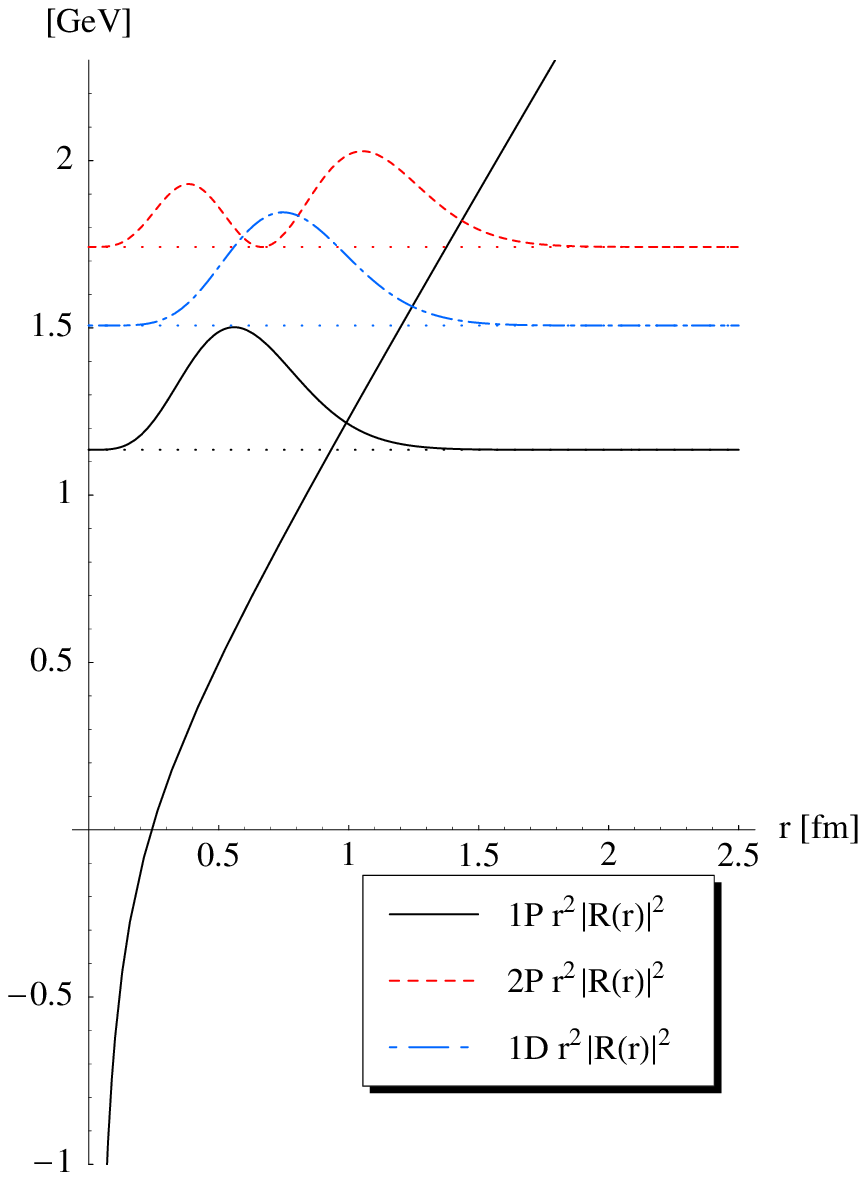}\end{center}
\end{minipage}\vspace{-0.5cm}
\caption{\label{Fig:RadialDensityCharmoniumCpL-B} Radial densities for charmonium states of model [CpL-B] plotted together with used potential ($\alpha_s=0.29$, $\sigma=1.306\:\textrm{GeV/fm}$, $m_c=1.2185\:\textrm{GeV}$); base lines of radial densities have been shifted by according $E_{kl}$.}
\end{figure} 
\end{center}\vspace{-1.5cm}
To determine the model parameters we use those states for which we have reliable experimental data and maximum confidence in the potential model. Therefore the parameters for the model [CpL-B] are fixed by a fit to the masses of the states $\eta_c$, $J/\psi$ and $h_c$, resulting in the set of parameters
\begin{eqnarray}
\alpha_s&=&0.29\:,\nonumber\\
m_c&=&1.2185\:\textrm{GeV}\:,\\
\sigma&=&1.306\:\textrm{GeV/fm}\:.\nonumber
\end{eqnarray}
One recognizes that $\alpha_s$ is in a region we can consider as reasonable (Eq. (\ref{Eq:Parameter-Intervals})) and that the charm quark mass $m_c$ is within the interval given in \cite{PDBook}. The value for the string tension $\sigma$ does not match with the usual value found in lattice simulations but is about 30\% larger than expected.

The resulting masses are compared with experimental data in Tab. \ref{Tab:MassesCharmoniumCpL-B}, while in Fig. \ref{Fig:SpectrumCharmoniumCpL-B} the emerging spectrum is confronted with the empirical spectrum. Taking a closer look at the spectrum one realizes that, for states below the $D\overline{D}$-threshold, model [CpL-B] is able to reproduce the general structure, though there are a few issues concerning details of the spectrum. Firstly, the spin-spin splitting for the $2S$-states is too large in magnitude. This is easy to understand by inspecting the radial wavefunctions (Fig. \ref{Fig:Wavefunctions-S-Charmonium-CpL-B} \& \ref{Fig:Wavefunctions-PD-Charmonium-CpL-B}), as the spin-spin splitting in model [CpL-B] is solely depending on the wavefunction at the origin $|\psi_{kl}(0)|$. This quantity is related via a power series ansatz to the slope of the reduced radial wavefunction $u_{kl}(r)$ at the origin. Investigating the reduced radial wavefunctions for the $S$-states (Fig. \ref{Fig:Wavefunctions-S-Charmonium-CpL-B}), we realize that the slopes near the origin are almost equal for all of these states. Considering, after a quick glance at the radial densities in Fig. \ref{Fig:RadialDensityCharmoniumCpL-B}, the linear part of the potential to be dominant for the charmonium bound states this is not surprising as one knows that for a purely linear potential
\begin{eqnarray}
|\psi_{k_1S}(0)|&=&|\psi_{k_2S}(0)|\:,\quad k_1,k_2\in \mathds{N}\:,
\end{eqnarray}
is valid. To reproduce the experimental spin-spin splitting for the $2S$-states one would need
\begin{eqnarray}
\frac{|\psi_{2S}(0)|}{|\psi_{1S}(0)|}\approx0.64\:,
\end{eqnarray}
instead. As other models \cite{Ebert:2002pp, Gupta:1995ps, Fulcher:1991dm, Godfrey:1985xj} face the same problems, we assume that the vicinity of the $D\overline{D}$-threshold leads to non-negligible corrections for these states.
\begin{table}[t]
\begin{center}
\begin{displaymath}
\begin{array}{||c|c|c|c|c||}
\hline\hline
&&\multicolumn{2}{|c|}{\qquad\textrm{Experiment \cite{PDBook}}\qquad} &\textrm{Theory [CpL-B]} \\
 \;\textrm{State}\; & \;\textrm{Candidate}\;&\multicolumn{2}{|c|}{\hrulefill} &  \\
&&\textrm{Mass [MeV]}&\textrm{Width [MeV]}&\textrm{Mass [MeV]}\\
\hline
1^1S_0 & \eta_c     & 2980.4   \pm 1.2   & 25.5   \pm 3.4    & 2980  \\
1^3S_1 & J/\psi     & 3096.916 \pm 0.011 & 0.0910 \pm 0.0032 & 3097  \\
\hline
1^1P_1 & h_c       & 3526.21  \pm 0.25  & <1.1              & 3526  \\
1^3P_0 & \chi_{c0}  & 3415.16  \pm 0.35  & 10.2   \pm 0.8      & 3424  \\
1^3P_1 & \chi_{c1}  & 3510.59  \pm 0.10  & 0.96   \pm 0.12     & 3501  \\
1^3P_2 & \chi_{c2}  & 3556.26  \pm 0.11  & 2.25   \pm 0.15     & 3561  \\
\hline
2^1S_0 & \eta_c'    & 3638     \pm 5     & 14     \pm 7        & 3699  \\
2^3S_1 & \psi'      & 3686.093 \pm 0.034 & 0.283  \pm 0.017    & 3793  \\
\hline
1^1D_2 &            &                    &                     & 3908  \\
1^3D_1 & \psi''     & 3770     \pm 2.4   & 23.6   \pm 2.7      & 3870  \\
1^3D_2 &            &                    &                     & 3900  \\
1^3D_3 &            &                    &                     & 3930  \\
\hline
2^1P_1 &            &                    &                     & 4109  \\
2^3P_0 &            &                    &                     & 4011  \\
2^3P_1 &            &                    &                     & 4084  \\
2^3P_2 &            &                    &                     & 4143  \\
\hline
3^1S_0 &            &                    &                     & 4260 \\
3^3S_1 & \psi'''    & 4040     \pm 10    & 52     \pm 10       & 4347  \\
\hline\hline
\end{array}
\end{displaymath}
\caption{\label{Tab:MassesCharmoniumCpL-B} $c\overline{c}$-state masses from experiment and model [CpL-B] ($\alpha_s=0.29$, $\sigma=1.306\:\textrm{GeV/fm}$, $m_c=1.2185\:\textrm{GeV}$); masses and widths are displayed for the experiment.}\vspace{-0.2cm}
\end{center}
\end{table}
Secondly, one realizes that, even though the magnitude of $M(1^3P_2)-M(1^3P_0)$ differs only slightly for model [CpL-B] and experiment, the model does not reproduce the triplet $P$-state splitting satisfactorily. This splitting is generated by the tensor and spin-orbit parts of the Hamiltonian, which are given for model [CpL-B] by
\begin{eqnarray}
H^{\textrm{LS}}+H^{\textrm{T}}&=&\left(6\vec{L}\cdot\vec{S}+S_{12}\right)\frac{\alpha_s}{3m_q^2}\bigg\langle\frac{1}{r^3}\bigg\rangle\:.
\end{eqnarray}
Thus the proportion
\begin{eqnarray}
\Phi_c^{\textrm{[CpL-B]}}(1P)&=&\left(\frac{M(\chi_{c2})-M(\chi_{c1})}{M(\chi_{c1})-M(\chi_{c0})}\right)_{\textrm{[CpL-B]}}=\frac{4}{5}\:,
\end{eqnarray}
does not depend on parameters and will never reproduce the experimental value
\begin{eqnarray}\label{Eq:P-ProportionCharmoniumExperiment}
\Phi_c^{\textrm{Exp}}(1P)&=&\left(\frac{M(\chi_{c2})-M(\chi_{c1})}{M(\chi_{c1})-M(\chi_{c0})}\right)_{\textrm{Exp}}=0.482\pm0.005\:,
\end{eqnarray}
leading to the conjecture that the Coulomb-plus-Linear model [CpL-B] is not sufficient to describe the spectrum in detail.\newpage

%% file: 3-Breit/3-Bottomonium.tex
In contrast to charmonium there are, apart from first hints for $\eta_b$ measured at ALEPH \cite{Heister:2002if}, no masses for bottomonium singlet states known. Thus we cannot fix the parameters of the model in the same manner as for charmonium. The states we use as input to determine parameters are $\Upsilon(1S)$, $\Upsilon(2S)$ and $C(1P)$, the center of gravity for the $1P$ triplet states defined as
\begin{eqnarray}\label{Eq:CoG-1P-Bottomonium}
C(1P)&=&\frac{1}{9}\left(5M(\chi_{b2})+3M(\chi_{b1})+M(\chi_{b0})\right)\approx9900\:\textrm{MeV}\:.
\end{eqnarray}
Identifying the $C(1P)$ with the $1^1P_1$ state of the model the resulting parameter set is
\begin{eqnarray}
\alpha_s&=&0.388\:,\nonumber\\
m_b&=&4.7645\:\textrm{GeV}\:,\\
\sigma&=&1.02\:\textrm{GeV/fm}\:.\nonumber
\end{eqnarray}
The string tension $\sigma$ and the bottom quark mass $m_b$ are in the expected regions, but starting from the value of $\alpha_s$ found for charmonium, one would expect a much smaller value for $\alpha_s$. As we will see this is due to the special properties of the $1S$ states.

\begin{center}
\begin{figure}[p]\begin{center}
\includegraphics[width=0.87\textwidth]{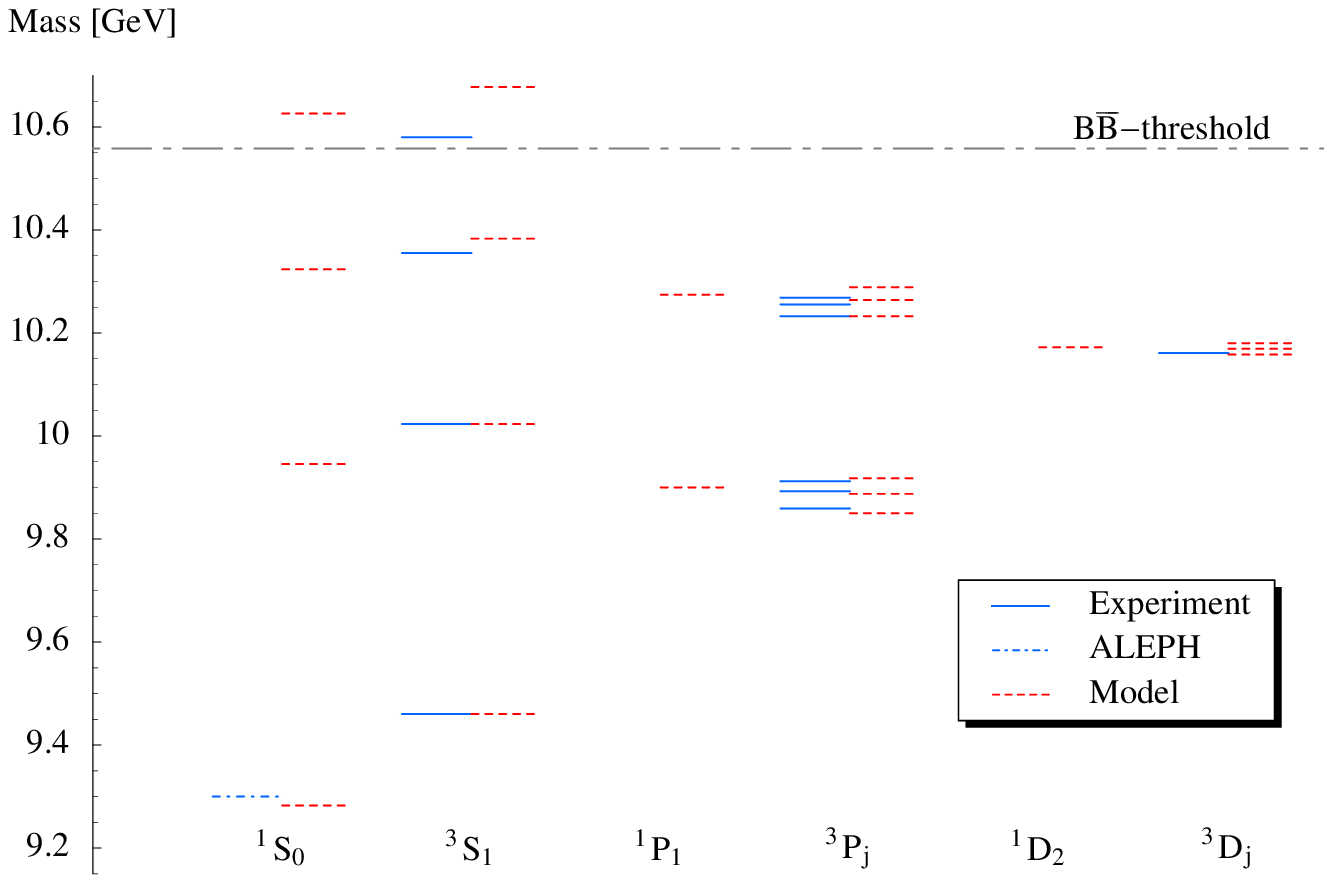}\end{center}
\caption{\label{Fig:SpectrumBottomoniumCpL-B} Bottomonium spectra of experiment and model [CpL-B] ($\alpha_s=0.388$, $\sigma=1.02\:\textrm{GeV/fm}$, $m_b=4.7645\:\textrm{GeV}$); $\eta_b$ mass measurement by ALEPH collaboration is included separately in the experimental spectrum.}
\end{figure} 
\end{center}

The masses computed within model [CpL-B] are compared to experimental data in Tab. \ref{Tab:MassesBottomoniumCpL-B}, while in Fig. \ref{Fig:SpectrumBottomoniumCpL-B} the resulting spectrum is displayed together with the experimental one.
\begin{table}[h!]
\begin{center}
\begin{displaymath}
\begin{array}{||c|c|c|c|c||}
\hline\hline
&&\multicolumn{2}{|c|}{\qquad\textrm{Experiment \cite{PDBook}}\qquad} & \textrm{Theory [CpL-B]} \\
 \;\textrm{State}\; & \;\textrm{Candidate}\;&\multicolumn{2}{|c|}{\hrulefill} &  \\
&&\textrm{Mass [MeV]}&\textrm{Width [MeV]}&\textrm{Mass [MeV]}\\
\hline
1^1S_0 & (\eta_b)       & (9300\pm20\pm20)       &                                 &  9283  \\
1^3S_1 & \Upsilon(1S)   & 9460\pm0.26            & (53.0\pm1.5)\textrm{ keV}       &  9460  \\
\hline
1^1P_1 &                &                        &                                 &  9900  \\
1^3P_0 & \chi_{b0}(1P)  & 9859.44\pm0.42\pm0.31  &                                 &  9850  \\
1^3P_1 & \chi_{b1}(1P)  & 9892.78\pm0.26\pm0.31  &                                 &  9888  \\
1^3P_2 & \chi_{b2}(1P)  & 9912.21\pm0.26\pm0.31  &                                 &  9918  \\
\hline
2^1S_0 &                &                        &                                 &  9946  \\
2^3S_1 & \Upsilon(2S)   & 10023.26\pm0.31        & (30.6\pm2.3)\textrm{ keV}       & 10023  \\
\hline
1^1D_2 &                &                        &                                 & 10172  \\
1^3D_1 &                &                        &                                 & 10158  \\
1^3D_2 & \Upsilon(1D)   & 10161.1\pm0.6\pm1.6    &                                 & 10169  \\
1^3D_3 &                &                        &                                 & 10180  \\
\hline
2^1P_1 &                &                        &                                 & 10274  \\
2^3P_0 & \chi_{b0}(2P)  & 10232.5\pm0.4\pm0.5    &                                 & 10233  \\
2^3P_1 & \chi_{b1}(2P)  & 10255.46\pm0.22\pm0.50 &                                 & 10264  \\
2^3P_2 & \chi_{b2}(2P)  & 10268.65\pm0.22\pm0.50 &                                 & 10289  \\
\hline
3^1S_0 &                &                        &                                 & 10324  \\
3^3S_1 & \Upsilon(3S)   & 10355.2\pm0.5          & (22.1\pm2.7)\textrm{ keV}       & 10383  \\
\hline
4^1S_0 &                &                        &                                 & 10626 \\
4^3S_1 & \Upsilon(4S)   & 10580.0\pm3.5          & 20\pm2\pm4                      & 10678  \\
\hline\hline
\end{array}
\end{displaymath}
\caption{\label{Tab:MassesBottomoniumCpL-B} $b\overline{b}$-state masses from experiment and model [CpL-B] ($\alpha_s=0.388$, $\sigma=1.02\:\textrm{GeV/fm}$, $m_b=4.7645\:\textrm{GeV}$); masses and widths are displayed for the experiment.}
\end{center}
\end{table}
\begin{center}
\begin{figure}[p]\begin{center}
\includegraphics[width=0.85\textwidth]{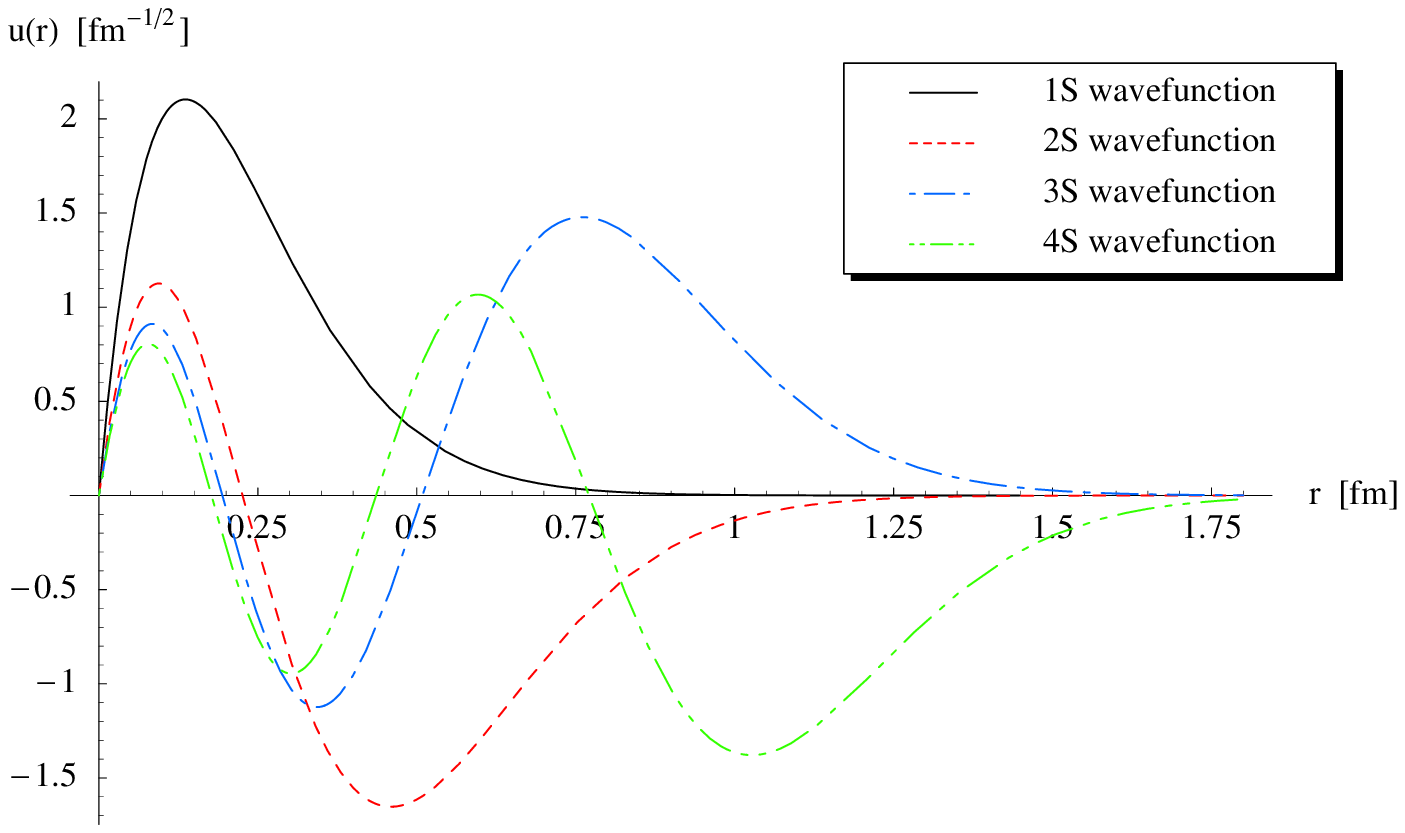}\end{center}
\caption{\label{Fig:Wavefunctions-S-Bottomonium-CpL-B} Bottomonium $S$-state reduced radial wavefunctions calculated within model [CpL-B] ($\alpha_s=0.388$, $\sigma=1.02\:\textrm{GeV/fm}$, $m_b=4.7645\:\textrm{GeV}$).}
\end{figure} 
\end{center}
\begin{center}
\begin{figure}[p]\begin{center}
\includegraphics[width=0.85\textwidth]{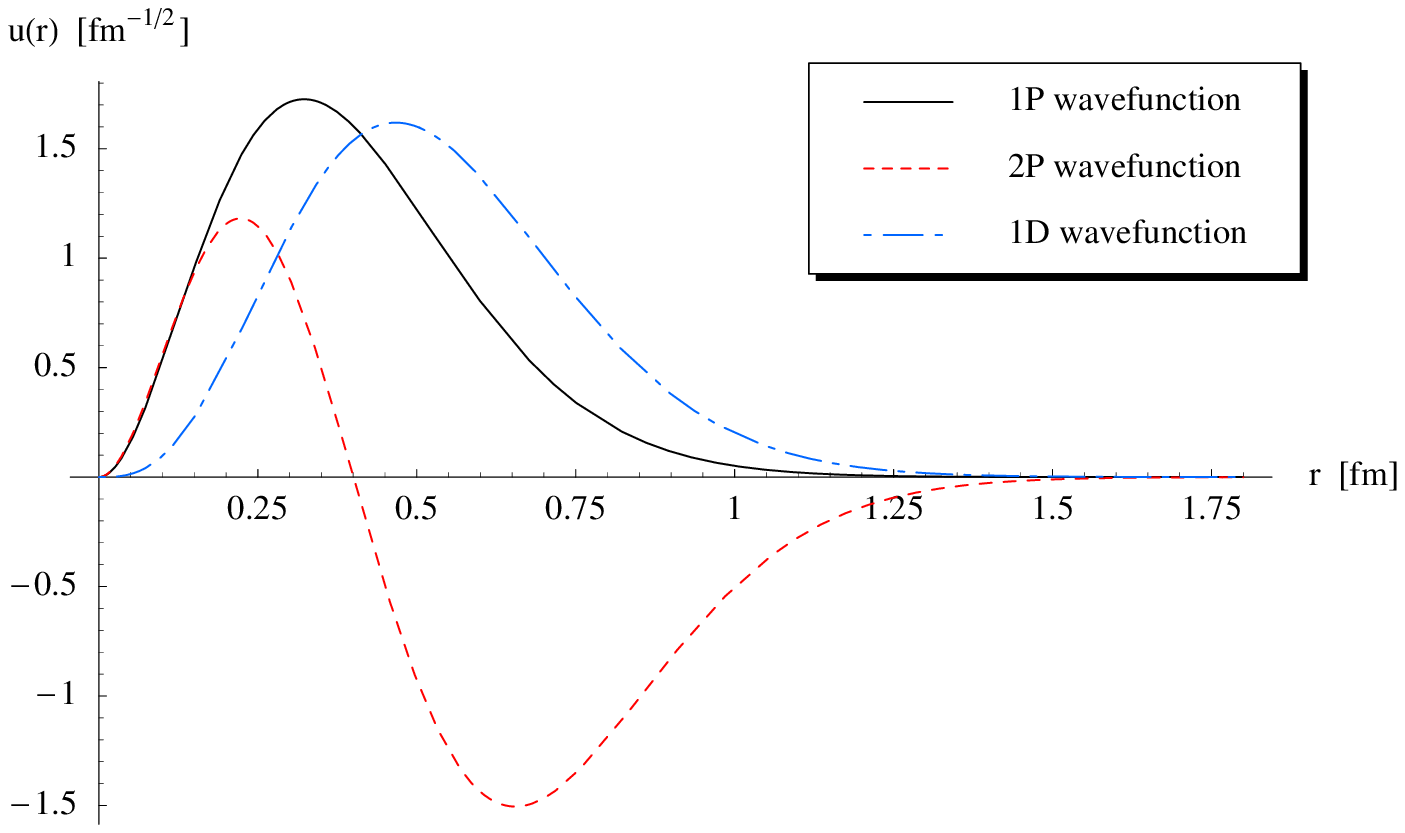}\end{center}
\caption{\label{Fig:Wavefunctions-PD-Bottomonium-CpL-B} Bottomonium $P$- and $D$-state reduced radial wavefunctions calculated within model [CpL-B] ($\alpha_s=0.388$, $\sigma=1.02\:\textrm{GeV/fm}$, $m_b=4.7645\:\textrm{GeV}$).}
\end{figure} 
\end{center}\vspace{-2cm}
\begin{center}
\begin{figure}[p]
\begin{minipage}{7.45cm}\begin{center}
\includegraphics[width=\textwidth]{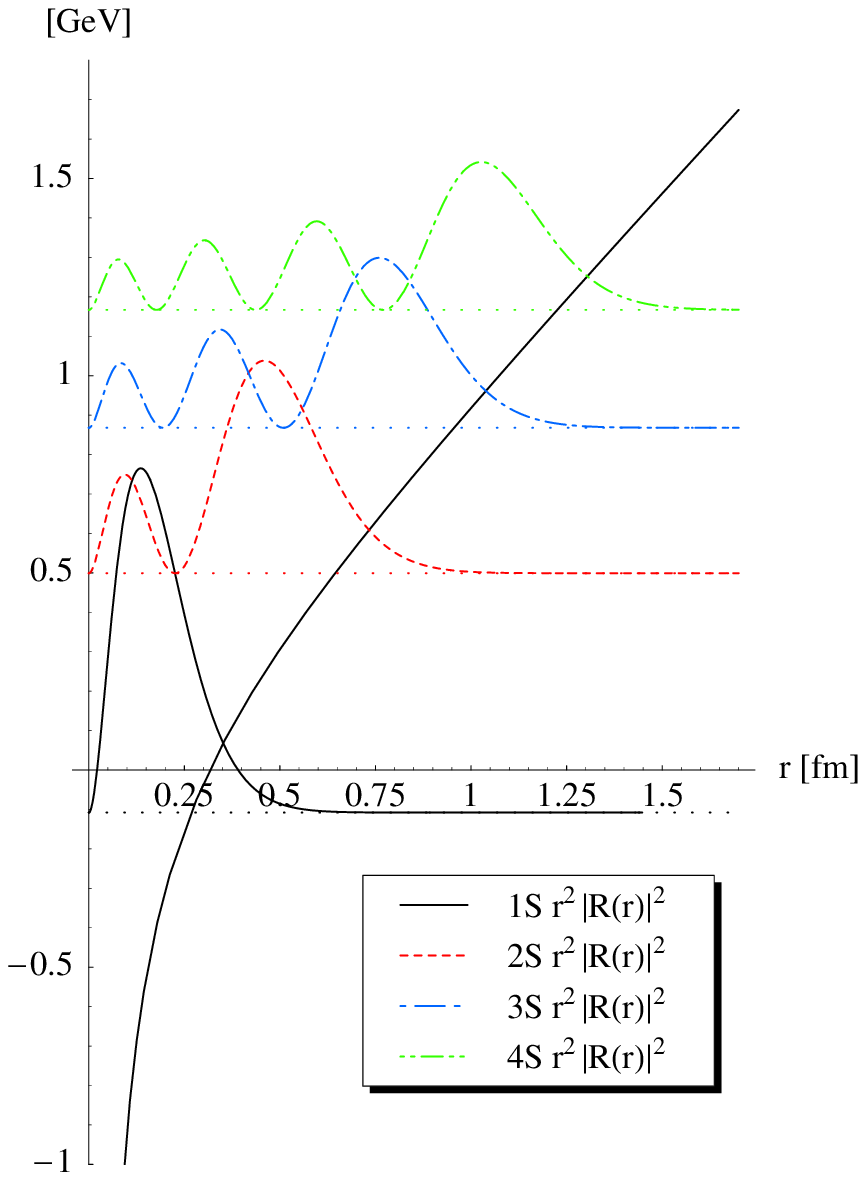}\end{center}
\end{minipage}\hfill
\begin{minipage}{7.45cm}\begin{center}
\includegraphics[width=\textwidth]{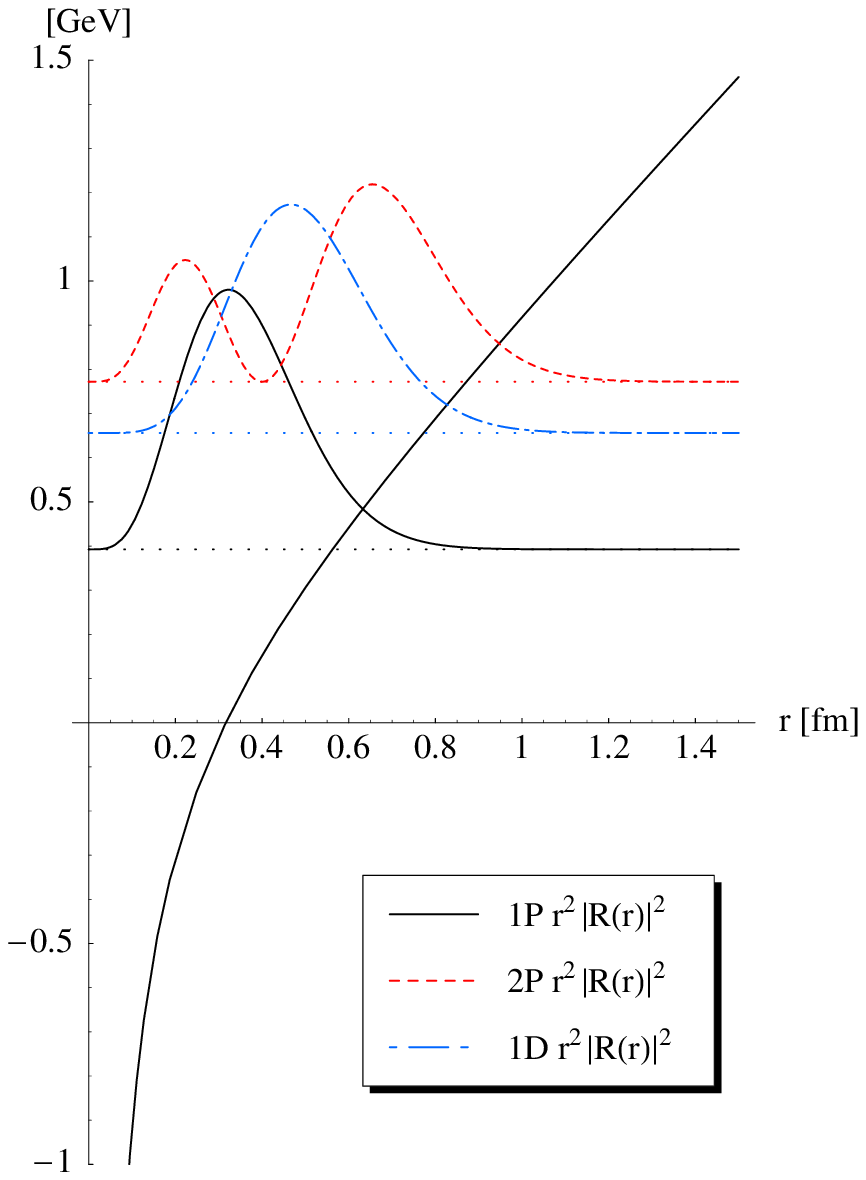}\end{center}
\end{minipage}\vspace{-0.5cm}
\caption{\label{Fig:RadialDensityBottomoniumCpL-B} Radial densities for bottomonium states of model [CpL-B] plotted together with used potential ($\alpha_s=0.388$, $\sigma=1.02\:\textrm{GeV/fm}$, $m_b=4.7645\:\textrm{GeV}$); base lines of radial densities have been shifted by according $E_{kl}$.}
\end{figure} 
\end{center}
The reduced radial wavefunctions calculated within model [CpL-B] are plotted in Fig. \ref{Fig:Wavefunctions-S-Bottomonium-CpL-B} and Fig. \ref{Fig:Wavefunctions-PD-Bottomonium-CpL-B}. Examining Fig. \ref{Fig:Wavefunctions-S-Bottomonium-CpL-B}, one realizes that the slope at the origin of the $1S$-state differs significantely from the slopes of the rest of the $S$-states, while their slopes again are all nearly equal. A quick glance at the radial densities (Fig. \ref{Fig:RadialDensityBottomoniumCpL-B}) shows that the $1S$-state is lying way deeper down in the funnel than all the other states. Both observations emphasize the special role of the $1S$-state for bottomonium, as it is, unlike the other states, strongly influenced by the Coulomb part of the potential.

From the spectrum (Fig. \ref{Fig:SpectrumBottomoniumCpL-B}) one realizes that the model matches experimental data quite well, with the exception of the triplet $P$-state splittings. The ratios
\begin{eqnarray}
\Phi_b^{\textrm{[CpL-B]}}(1P)&=&\left(\frac{M(\chi_{b2})-M(\chi_{b1})}{M(\chi_{b1})-M(\chi_{b0})}\right)_{\textrm{[CpL-B]}}=\frac{4}{5}\:,\nonumber\\
\Phi_b^{\textrm{[CpL-B]}}(2P)&=&\left(\frac{M(\chi'_{b2})-M(\chi'_{b1})}{M(\chi'_{b1})-M(\chi'_{b0})}\right)_{\textrm{[CpL-B]}}=\frac{4}{5}\:,
\end{eqnarray}
are independent of the model parameters and do not coincide with the experimental values
\begin{eqnarray}\label{Eq:P-ProportionBottomoniumExperiment}
\Phi_b^{\textrm{Exp}}(1P)&=&\left(\frac{M(\chi_{b2})-M(\chi_{b1})}{M(\chi_{b1})-M(\chi_{b0})}\right)_{\textrm{Exp}}=0.58\pm0.03\:,\nonumber\\
\Phi_b^{\textrm{Exp}}(2P)&=&\left(\frac{M(\chi'_{b2})-M(\chi'_{b1})}{M(\chi'_{b1})-M(\chi'_{b0})}\right)_{\textrm{Exp}}=0.57\pm0.05\:.
\end{eqnarray}
This reenforces the presumption that the Coulomb-plus-Linear model [CpL-B] is not sufficient to describe quarkonium spectra below threshold in detail. On the other hand the triplet $P$-state splittings are enlarged. These deviations in the overall size of the $P$-state splittings are understandable, since the magnitude of the relativistic corrections responsible for those splittings are governed by the strong coupling constant $\alpha_s$ and we already argued that the value one would expect for $\alpha_s$ is significantly smaller than the one used in this model. The agreement between the $1^1S_0$-state mass of model [CpL-B] and the (tentative) mass found by the ALEPH collaboration \cite{Heister:2002if} is a coincidence.

%% file: 3-Breit/3-Radius-Mass.tex
After studying the spectra generated by model [CpL-B], we examine whether the model leads to an approximately linear mass-radius relation\footnote{Of course, in the present context, the term "radius" means average distance between quark and antiquark.}. This is of interest in the study of string breaking \cite{Prkacin:2005dc, Bali:2005fu}.

Using the virial theorem for each given eigenstate $|\psi\rangle$,
\begin{eqnarray}
\langle\psi|T|\psi\rangle=\frac{1}{2}\langle\psi|\vec{r}\cdot\vec{\nabla}V(\vec{r}\,)|\psi\rangle\:,
\end{eqnarray}
with $T$ being the kinetic energy and $V(\vec{r}\,)$ the potential used in the Hamiltonian
\begin{eqnarray}
H=\frac{\vec{p}^{\,2}}{2m_\textrm{red}}+V(\vec{r}\,)\:,
\end{eqnarray}
one finds for a radially symmetric potential:
\begin{eqnarray}
\langle\psi|T|\psi\rangle=\frac{1}{2}\langle\psi|rV'(r)|\psi\rangle\:.
\end{eqnarray}
In the case of $V(r)=-\frac{4\alpha_s}{3r}+\sigma r$ we have
\begin{eqnarray}
E&=&\langle\psi|H|\psi\rangle=\langle\psi|T|\psi\rangle+\langle\psi|V|\psi\rangle=\frac{1}{2}\langle\psi|rV'|\psi\rangle+\langle\psi|V|\psi\rangle\nonumber\\
&=&\bigg\langle\frac{2\alpha_s}{3r}+\frac{\sigma r}{2}\bigg\rangle+\bigg\langle-\frac{4\alpha_s}{3r}+\sigma r\bigg\rangle=-\frac{2\alpha_s}{3}\bigg\langle\frac{1}{r}\bigg\rangle+\frac{3\sigma}{2}\langle r\rangle\:.
\end{eqnarray}
Adding the quark masses this leads to
\begin{eqnarray}\label{Eq:LinearGuess}
M=2m_q+E=2m_q+\frac{3\sigma}{2}\langle r\rangle-\frac{2\alpha_s}{3}\bigg\langle\frac{1}{r}\bigg\rangle\:,
\end{eqnarray}
with the leading term proportional to $\langle r\rangle$ accompanied by a small offset due to the Coulomb part.

When probing the relation of mass and radius one has to decide which mass to take into account as the perturbatively implemented corrections do not change the radius. One possibility is to consider only the bare masses, defined by
\begin{eqnarray}
M_{kl}^{\textrm{bare}}=2m_q+E_{kl}\:,
\end{eqnarray}
not including any of the perturbatively treated corrections.

%% file: 3-Breit/3-Radius-Mass-Charmonium.tex
It is instructive to have a look at the radial densities in Fig. \ref{Fig:RadialDensityCharmoniumCpL-B} before investigating the mass-radius dependence in detail. It occurs that all the states we consider are dominated by the linear part of the potential and thus the linear radius-mass relation (Eq. (\ref{Eq:LinearGuess})) is realized neglecting the small correction from the Coulomb term.

\begin{center}
\begin{figure}[p]\begin{center}
\includegraphics[width=\textwidth]{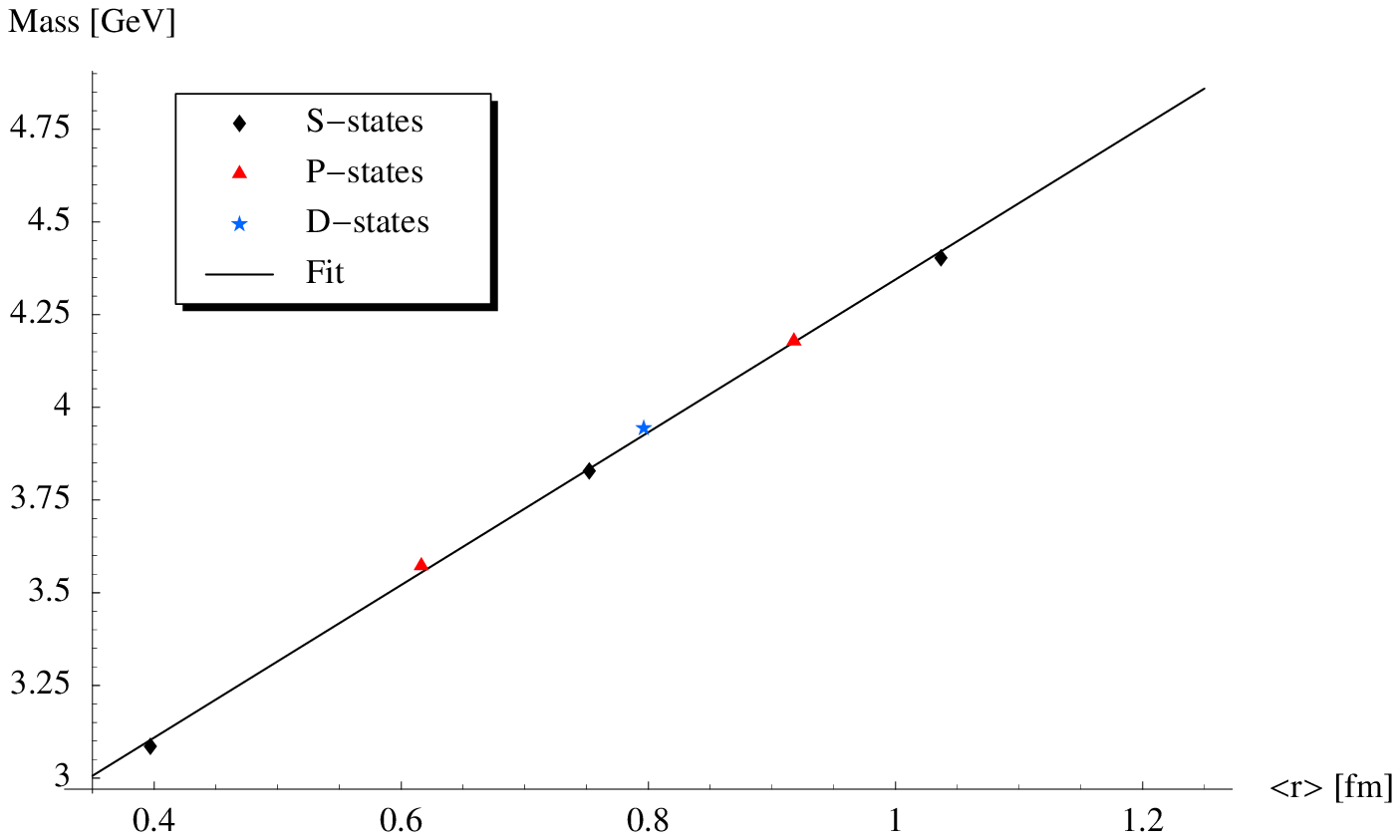}\end{center}
\caption{\label{Fig:CharmoniumRadiusMass-CpL-B} $M_{kl}^{\textrm{bare}}$ versus $\langle r\rangle$ for $c\overline{c}$-states in model [Cpl-B] ($\alpha_s=0.29$, $\sigma=1.306\:\textrm{GeV/fm}$, $m_c=1.2185\:\textrm{GeV}$).}
\end{figure} 
\end{center}

In Tab. \ref{Tab:CharmoniumRadiusMass-CpL-B} bare masses and radii for states in model [CpL-B] are presented.
\begin{table}[h!]
\begin{center}
\begin{displaymath}
\begin{array}{|c|c|c|}
\hline
&&\\
\quad\textrm{State}\quad &\quad M_{kl}^{\textrm{bare}}\textrm{ [MeV]} \quad &\quad \langle r\rangle \textrm{ [fm]}\quad\\
&&\\
\hline
1S & 3086 & 0.397\\
1P & 3573 & 0.616\\
2S & 3828 & 0.752\\
1D & 3944 & 0.796\\
2P & 4179 & 0.918\\
3S & 4403 & 1.037\\
\hline
\end{array}
\end{displaymath}
\caption{\label{Tab:CharmoniumRadiusMass-CpL-B} Charmonium radii and bare masses in model [CpL-B] ($\alpha_s=0.29$, $\sigma=1.306\:\textrm{GeV/fm}$, $m_c=1.2185\:\textrm{GeV}$).}
\end{center}
\end{table}\\
Matching a straight line to the data in Tab. \ref{Tab:CharmoniumRadiusMass-CpL-B} yields for the parameters of the best fit straight line
\begin{eqnarray}
M&=&a+b\langle r\rangle\:,\nonumber\\
a&=&2.285\: \textrm{GeV}\:,\\
b&=&2.060\: \textrm{GeV/fm}\:.\nonumber
\end{eqnarray}
The resulting straight line is displayed together with input data in Fig. \ref{Fig:CharmoniumRadiusMass-CpL-B}. As expected the fitted straight line reflects the dependence of mass and radius almost perfectly, even the virial theorem approximation for the slope of the straight line
\begin{eqnarray}
\frac{\partial M}{\partial \langle r\rangle}\approx\frac{3}{2}\sigma\:,
\end{eqnarray}
holds with high accuracy.

%% file: 3-Breit/3-Radius-Mass-Bottomonium.tex
\vspace{-0.5cm}
\begin{center}
\begin{figure}[p]\begin{center}
\includegraphics[width=\textwidth]{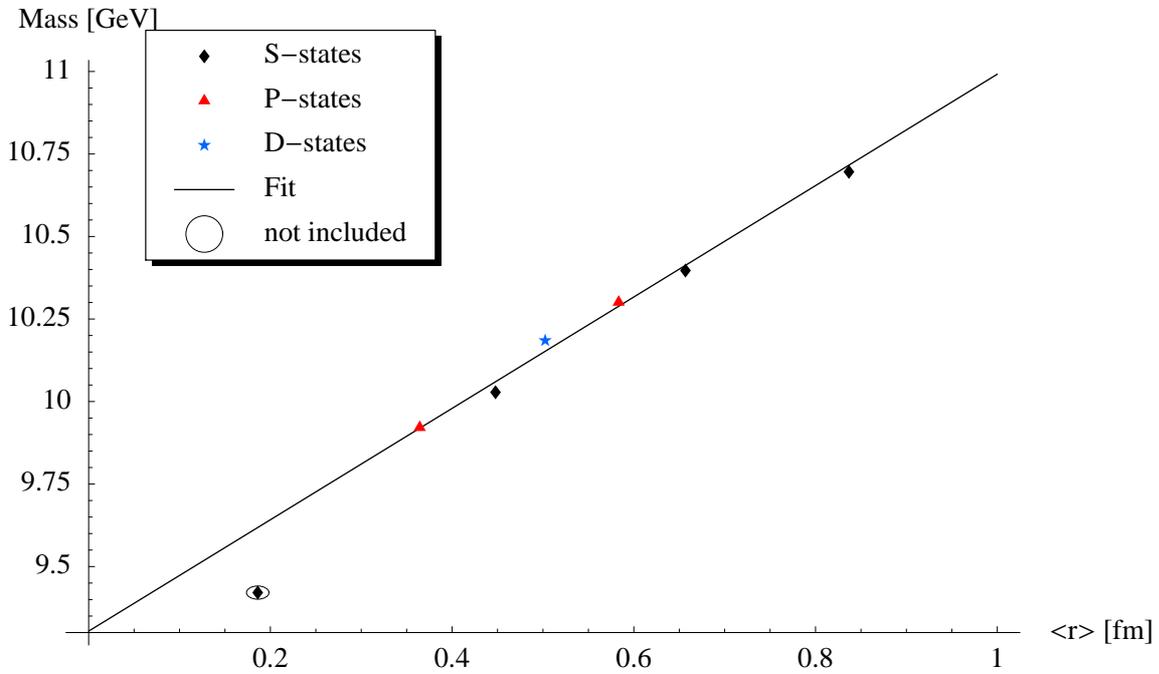}\end{center}
\caption{\label{Fig:BottomoniumRadiusMass-CpL-B} $M_{kl}^{\textrm{bare}}$ versus $\langle r\rangle$ for $b\overline{b}$-states in model [Cpl-B]  ($\alpha_s=0.388$, $\sigma=1.02\:\textrm{GeV/fm}$, $m_b=4.7645\:\textrm{GeV}$); 1S-state is not included in straight line fit.}
\end{figure} 
\end{center}
As for charmonium it is instructive to inspect the radial densities (Fig. \ref{Fig:RadialDensityBottomoniumCpL-B}) before studying the mass-radius relation in detail. One realizes that for bottomonium the $1S$-state is dominated by the Coulomb-part of the potential, while the remaining states seem to be mainly dominated by the linear part.

In Tab. \ref{Tab:BottomoniumRadiusMass-CpL-B} the radii and bare masses calculated within model [CpL-B] are presented.
\begin{table}[h!]
\begin{center}
\begin{displaymath}
\begin{array}{|c|c|c|}
\hline
&&\\
\quad\textrm{State}\quad &\quad M_{kl}^{\textrm{bare}}\textrm{ [MeV]} \quad &\quad \langle r\rangle \textrm{ [fm]}\quad\\
&&\\
\hline
1S &  9421 & 0.186\\
1P &  9921 & 0.364\\
2S & 10028 & 0.448\\
1D & 10185 & 0.502\\
2P & 10301 & 0.583\\
3S & 10397 & 0.657\\
4S & 10696 & 0.837\\
\hline
\end{array}
\end{displaymath}
\caption{\label{Tab:BottomoniumRadiusMass-CpL-B} Bottomonium radii and bare masses in model [CpL-B] ($\alpha_s=0.388$, $\sigma=1.02\:\textrm{GeV/fm}$, $m_b=4.7645\:\textrm{GeV}$).}
\end{center}
\end{table}
\\In Fig. \ref{Fig:BottomoniumRadiusMass-CpL-B} the bare masses are plotted over the radii of states together with the best fit straight line, which parameters are given by
\begin{eqnarray}
M&=&a+b\langle r\rangle\:,\nonumber\\
a&=&9.336\: \textrm{GeV}\:,\\
b&=&1.515\: \textrm{GeV/fm}\:,\nonumber
\end{eqnarray}
where the $1S$-state is not included into the fit. Though the linear behaviour for bottomonium is not as perfect as in the case of charmonium, the deviations of the considered states from the straight line are relatively small. Further one finds that the virial theorem approximation for the slope of the straight line
\begin{eqnarray}
\frac{\partial M}{\partial \langle r\rangle}\approx\frac{3}{2}\sigma\:,
\end{eqnarray}
is again approximately fulfilled.

%% file: 3-Breit/3-Summary.tex
In the study of model [CpL-B] we have seen that the spectra arising from the model already bear resemblance to the corresponding experimental spectra, though we encountered some problems.

It occurs that the string tension for charmonium in model [CpL-B] is about 30\% larger than the empirical value. Furthermore we find that the values for the strong coupling constant $\alpha_s$ for bottomonium and charmonium behaves in an unusual way, giving a larger value for bottomonium than for charmonium. As a consequence the spin-dependent splittings in bottomonium are, with respect to the experimental spectrum, generally enlarged.

For charmonium it is not possible to reproduce the spin splitting for the $2S$-states. The near $D\overline{D}$-threshold may have a non-negligible influence on these states. The most severe problem of model [CpL-B] is that the experiment provides a different structure for the triplet $P$-state splitting. The model yields the built-in value $\Phi_q^{\textrm{[CpL-B]}}(nP)=4/5$ independently of parameters or radial excitations, while the experiment significantly differs from this value both for bottomonium and charmonium. Based on this observation we conclude that the discussed model may be a first approximation for a quarkonium potential model, but has to be improved in order to reproduce details of the quarkonium spectra.

In addition we investigated the mass-radius dependence for states of charmonium and bottomonium. We confirmed a leading linear relation between mass and radius for charmonium and found that, with the exception of the 1S-state, the linear relation is also a good approximation for bottomonium.


%% file: 4-OGE/4-Intro.tex
This chapter motivates and discusses a potential model with perturbative corrections based on One-Gluon exchange without retardation.

%% file: 4-OGE/4-OGE.tex
In Chapter \ref{Chapt:CpL-Breit} we have investigated a Coulomb-plus-Linear model with perturbative corrections based on Breit interaction, model [CpL-B]. In the calculation of the Breit interaction (see App. \ref{Sec:BreitInteraction}) we explicitly used the on-shell condition for the energy transfer,
\begin{eqnarray}
q_0&=&E'_A-E_A=E_B-E'_B\:,
\end{eqnarray}
in the expansion of the propagator (Eq. (\ref{Eq:Propagator-Expanded})), leading to retardation. While suitable for the case of scattering, using this condition in the case of bound states is not valid. For a bound state $q_0=0$ should be the more natural choice. The corresponding potential (see App. \ref{Sec:OneGluonExchangeW/ORetardation}) is the one arising from one-gluon exchange without retardation corrections. Transformed into the center of mass frame, and using Eq. (\ref{Eq:Effective-CM-Transformation}), the potential  Eq. (\ref{Eq:OGE-Potential-noCM}) reads
\begin{eqnarray}\label{Eq:Potential-OGE-CM}
V^{(2)}(\vec{r};\vec{p}\,)&=&-\frac{4\alpha_s}{3r}+\frac{4\pi\alpha_s}{3m_q^2}\delta^{(3)}(\vec{r}\,)-\frac{4\alpha_s}{3m_q^2}\frac{\vec{p}\cdot\vec{p}}{r}+\frac{2\alpha_s}{m_q^2}\frac{(\vec{r}\times\vec{p}\,)\cdot(\vec{s}_1+\vec{s}_2)}{r^3}\nonumber\\
&&\quad+\frac{4\alpha_s}{3m_q^2}\left[\frac{8\pi}{3}\delta^{(3)}(\vec{r}\,)(\vec{s}_1\cdot\vec{s}_2)+\frac{3(\vec{s}_1\cdot\hat{r})(\vec{s}_2\cdot \hat{r})-\vec{s}_1\cdot\vec{s}_2}{r^3}\right]\:.
\end{eqnarray}

%% file: 4-OGE/4-Potential-Model.tex
The quark-antiquark potential in this model is composed of a linear confining part and the one-gluon exchange potential without retardation (Eq. (\ref{Eq:Potential-OGE-CM})):
\begin{eqnarray}\label{Eq:Potential-CpL-3}
V^{\textrm{[CpL-3]}}(\vec{r};\vec{p}\,)&\!\!\!\!\!=\!\!\!\!\!&-\frac{4\alpha_s}{3r}+\sigma r+\frac{4\pi\alpha_s}{3m_q^2}\delta^{(3)}(\vec{r}\,)-\frac{4\alpha_s}{3m_q^2}\frac{\vec{p}\cdot\vec{p}}{r}+\frac{2\alpha_s}{m_q^2}\frac{(\vec{r}\times\vec{p}\,)\cdot(\vec{s}_1+\vec{s}_2)}{r^3}\nonumber\\
&&\qquad\quad\!+\frac{4\alpha_s}{3m_q^2}\left[\frac{8\pi}{3}\delta^{(3)}(\vec{r}\,)(\vec{s}_1\cdot\vec{s}_2)+\frac{3(\vec{s}_1\cdot\hat{r})(\vec{s}_2\cdot \hat{r})-\vec{s}_1\cdot\vec{s}_2}{r^3}\right]\:.
\end{eqnarray}
As for all Coulomb-plus-Linear models, the potential used in the Schr\"odinger equation to find bound states is
\begin{eqnarray}
V_0(r)&=&-\frac{4\alpha_s}{3r}+\sigma r\:,
\end{eqnarray}
while the remaining parts of Eq. (\ref{Eq:Potential-CpL-3}) are treated perturbatively. The resulting mass formula is given by
\begin{eqnarray}\label{Eq:MassFormula-CpL-3}
M^{\textrm{[CpL-3]}}(k^{2S+1}l_j)&=&2m_q+E_{kl}+\frac{4\pi\alpha_s}{3m_q^2}|\psi(0)|^2+\frac{16\pi\alpha_s}{9m_q^2}\left(S(S+1)-\frac{3}{2}\right)|\psi(0)|^2\nonumber\\
&&\!\!\!\!+\alpha_s\frac{j(j+1)-l(l+1)-S(S+1)}{m_q^2}\bigg\langle\frac{1}{r^3}\bigg\rangle+\alpha_s\frac{S_{12}}{3m_q^2}\bigg\langle\frac{1}{r^3}\bigg\rangle\nonumber\\
&&\!\!\!\!+\frac{2\alpha_s}{3m_q^2}\left(\int\mathrm{d}^3r\:\psi^*(\vec{r}\,)\frac{1}{r}\left(\vec{\nabla}^{2}\psi(\vec{r}\,)\right)-\int\mathrm{d}^3r\:\frac{1}{r}\:|\:\vec{\nabla}\:\psi(\vec{r}\,)\:|^2\right)\:.
\end{eqnarray}
The last term represents the expectation value of the purely momentum dependent terms using option 3 in Appendix \ref{Chapt:Expect-values-of-MD}. Figures, tables and quantities related to this model are denoted as [CpL-3].

%% file: 4-OGE/4-Charmonium.tex
\begin{table}[b!]
\begin{center}
\begin{displaymath}
\begin{array}{||c|c|c|c|c||}
\hline\hline
&&\multicolumn{2}{|c|}{\qquad\textrm{Experiment \cite{PDBook}}\qquad} & \textrm{Theory [CpL-3]}\\
 \;\textrm{State}\; & \;\textrm{Candidate}\;&\multicolumn{2}{|c|}{\hrulefill} &  \\
&&\textrm{Mass [MeV]}&\textrm{Width [MeV]}&\textrm{Mass [MeV]}\\
\hline
1^1S_0 & \eta_c     & 2980.4   \pm 1.2   & 25.5   \pm 3.4      & 2980  \\
1^3S_1 & J/\psi     & 3096.916 \pm 0.011 & 0.0910 \pm 0.0032   & 3097  \\
\hline
1^1P_1 & h_c        & 3526.21  \pm 0.25  & <1.1                & 3527  \\
1^3P_0 & \chi_{c0}  & 3415.16  \pm 0.35  & 10.2   \pm 0.8      & 3430  \\
1^3P_1 & \chi_{c1}  & 3510.59  \pm 0.10  & 0.96   \pm 0.12     & 3503  \\
1^3P_2 & \chi_{c2}  & 3556.26  \pm 0.11  & 2.25   \pm 0.15     & 3560  \\
\hline
2^1S_0 & \eta_c'    & 3638     \pm 5     & 14     \pm 7        & 3674  \\
2^3S_1 & \psi'      & 3686.093 \pm 0.034 & 0.283  \pm 0.017    & 3765  \\
\hline
1^1D_2 &            &                    &                     & 3891  \\
1^3D_1 & \psi''     & 3770     \pm 2.4   & 23.6   \pm 2.7      & 3855  \\
1^3D_2 &            &                    &                     & 3884  \\
1^3D_3 &            &                    &                     & 3911  \\
\hline
2^1P_1 &            &                    &                     & 4077  \\
2^3P_0 &            &                    &                     & 3985  \\
2^3P_1 &            &                    &                     & 4054  \\
2^3P_2 &            &                    &                     & 4109  \\
\hline
3^1S_0 &            &                    &                     & 4207 \\
3^3S_1 & \psi'''    & 4040     \pm 10    & 52     \pm 10       & 4291  \\
\hline\hline
\end{array}
\end{displaymath}
\caption{\label{Tab:MassesCharmoniumCpL-3} $c\overline{c}$-state masses from experiment and model [CpL-3] ($\alpha_s=0.302$, $\sigma=1.21\:\textrm{GeV/fm}$, $m_c=1.272\:\textrm{GeV}$); masses and widths are displayed for the experiment.}
\end{center}
\end{table}
The parameters in model [CpL-3] are determined as in model [CpL-B], namely by a fit to the masses of states $\eta_c$, $J/\psi$ and $h_c$, providing the parameter set
\begin{eqnarray}
\alpha_s&=&0.302\:,\nonumber\\
m_c&=&1.272\:\textrm{GeV}\:,\\
\sigma&=&1.21\:\textrm{GeV/fm}\:.\nonumber
\end{eqnarray}
The resulting parameter values for quark mass $m_c$ and strong coupling constant $\alpha_s$ are similar to those in model [CpL-B]. The string tension $\sigma$ decreased by about 10\% with respect to model [CpL-B] and is somewhat closer to the value suggested by lattice simulations.

\begin{center}
\begin{figure}[p]\begin{center}
\includegraphics[width=0.9\textwidth]{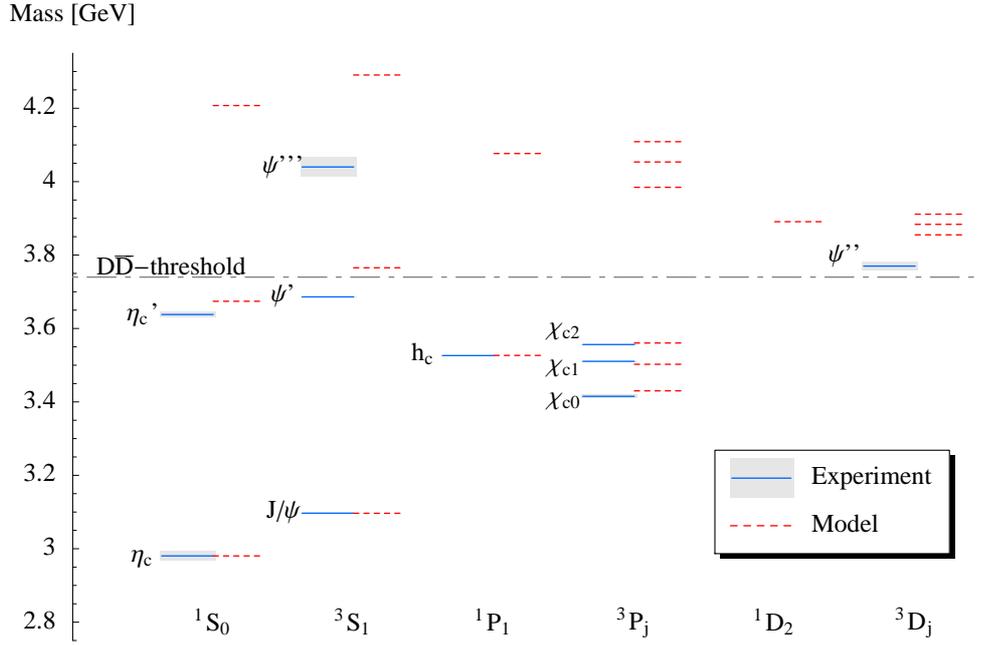}\end{center}
\caption{\label{Fig:SpectrumCharmoniumCpL-3} Charmonium spectra for experiment and model [CpL-3] ($\alpha_s=0.302$, $\sigma=1.21\:\textrm{GeV/fm}$, $m_c=1.272\:\textrm{GeV}$); for experimental data masses and widths are shown.}
\end{figure} 
\end{center}

The results of the model calculations are presented in comparison with experimental data in Tab. \ref{Tab:MassesCharmoniumCpL-3}. In Fig. \ref{Fig:SpectrumCharmoniumCpL-3} the spectrum is displayed together with the $c\overline{c}$-spectrum from experiments. As model [CpL-3] differs form model [CpL-B] only in the purely momentum dependent term, model [CpL-3] faces the same problems for the spin-dependent splittings as model [CpL-B], namely the built-in ratio
\begin{eqnarray}
\Phi_c^{\textrm{[CpL-3]}}&=&\left(\frac{M(\chi_{c2})-M(\chi_{c1})}{M(\chi_{c1})-M(\chi_{c0})}\right)_{\textrm{[CpL-3]}}=\frac{4}{5}\:,
\end{eqnarray}
which is not consistent with experimental data (Eq. (\ref{Eq:P-ProportionCharmoniumExperiment})). The decrease of the string tension $\sigma$ compared to model [CpL-B] leads to a slight reduction in size of the triplet $P$-state splitting.

%% file: 4-OGE/4-Bottomonium.tex
\begin{center}
\begin{figure}[p]\begin{center}
\includegraphics[width=0.9\textwidth]{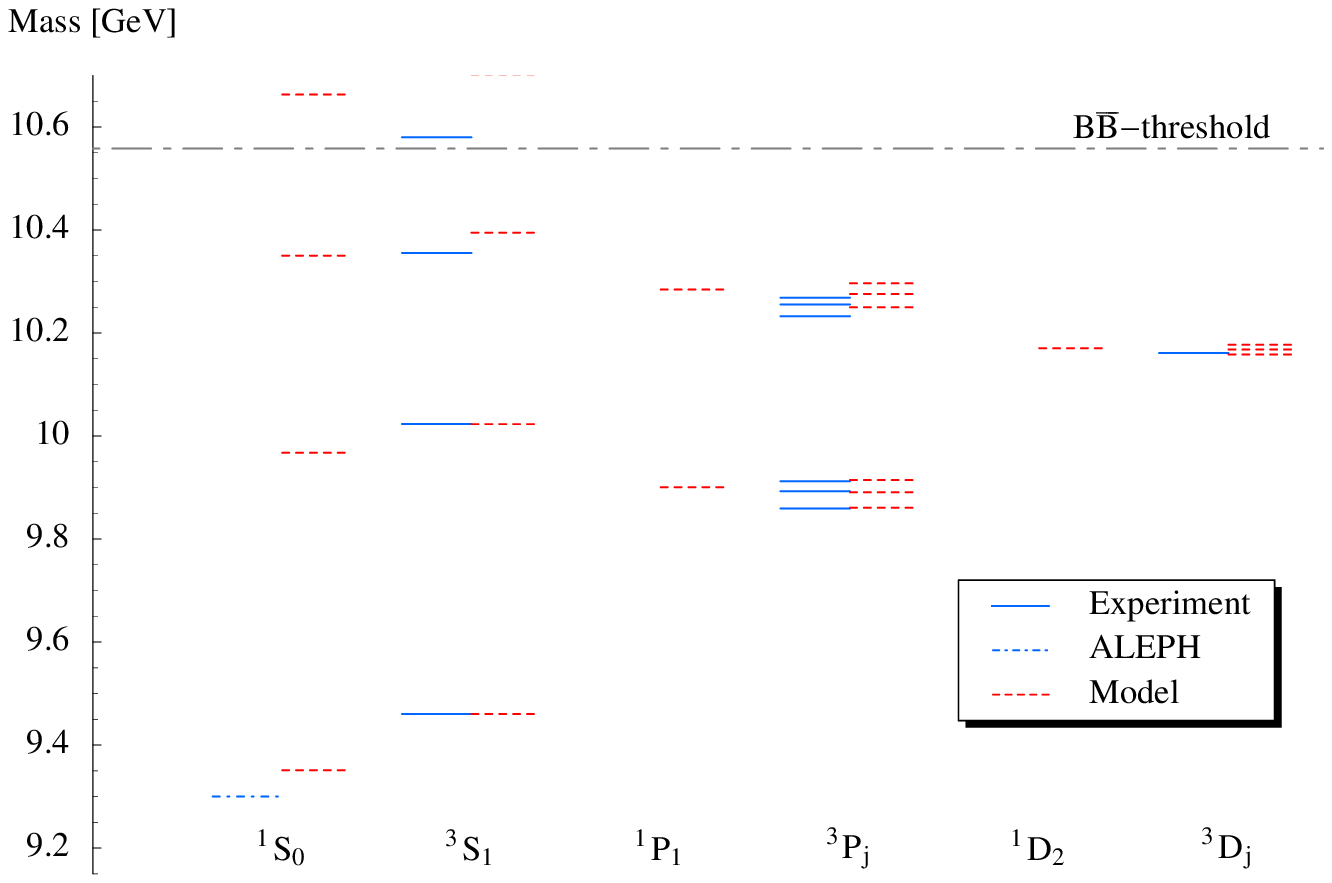}\end{center}
\caption{\label{Fig:SpectrumBottomoniumCpL-3} Bottomonium spectra from experiment and model [CpL-3] ($\alpha_s=0.326$, $\sigma=1.12\:\textrm{GeV/fm}$, $m_b=4.7215\:\textrm{GeV}$); $\eta_b$ mass measurement by ALEPH collaboration is included separately in the experimental spectrum.}
\end{figure} 
\end{center}
\begin{table}[b!]\vspace{-0.5cm}
\begin{center}
\begin{displaymath}
\begin{array}{||c|c|c|c|c||}
\hline\hline
&&\multicolumn{2}{|c|}{\qquad\textrm{Experiment \cite{PDBook}}\qquad} &  \textrm{Theory [CpL-3]}\\
 \;\textrm{State}\; & \;\textrm{Candidate}\;&\multicolumn{2}{|c|}{\hrulefill} &  \\
&&\textrm{Mass [MeV]}&\textrm{Width [MeV]}& \textrm{Mass [MeV]}\\
\hline
1^1S_0 & (\eta_b)       & (9300\pm20\pm20)       &                                 &  9351  \\
1^3S_1 & \Upsilon(1S)   & 9460\pm0.26            & (53.0\pm1.5)\textrm{ keV}       &  9460  \\
\hline
1^1P_1 &                &                        &                                 &  9901  \\
1^3P_0 & \chi_{b0}(1P)  & 9859.44\pm0.42\pm0.31  &                                 &  9861  \\
1^3P_1 & \chi_{b1}(1P)  & 9892.78\pm0.26\pm0.31  &                                 &  9890  \\
1^3P_2 & \chi_{b2}(1P)  & 9912.21\pm0.26\pm0.31  &                                 &  9915  \\
\hline
2^1S_0 &                &                        &                                 &  9967  \\
2^3S_1 & \Upsilon(2S)   & 10023.26\pm0.31        & (30.6\pm2.3)\textrm{ keV}       & 10023  \\
\hline
1^1D_2 &                &                        &                                 & 10170  \\
1^3D_1 &                &                        &                                 & 10158  \\
1^3D_2 & \Upsilon(1D)   & 10161.1\pm0.6\pm1.6    &                                 & 10168  \\
1^3D_3 &                &                        &                                 & 10177  \\
\hline
2^1P_1 &                &                        &                                 & 10284  \\
2^3P_0 & \chi_{b0}(2P)  & 10232.5\pm0.4\pm0.5    &                                 & 10250  \\
2^3P_1 & \chi_{b1}(2P)  & 10255.46\pm0.22\pm0.50 &                                 & 10276  \\
2^3P_2 & \chi_{b2}(2P)  & 10268.65\pm0.22\pm0.50 &                                 & 10296  \\
\hline
3^1S_0 &                &                        &                                 & 10350  \\
3^3S_1 & \Upsilon(3S)   & 10355.2\pm0.5          & (22.1\pm2.7)\textrm{ keV}       & 10395  \\
\hline
4^1S_0 &                &                        &                                 & 10663 \\
4^3S_1 & \Upsilon(4S)   & 10580.0\pm3.5          & 20\pm2\pm4                      & 10703  \\
\hline\hline
\end{array}
\end{displaymath}\vspace{-0.3cm}
\caption{\label{Tab:MassesBottomoniumCpL-3} $b\overline{b}$-state masses from experiment and model [CpL-3] ($\alpha_s=0.326$, $\sigma=1.12\:\textrm{GeV/fm}$, $m_b=4.7215\:\textrm{GeV}$); masses and widths are displayed for the experiment.}
\end{center}\vspace{-0.3cm}
\end{table}\vspace{-0.3cm}
The parameters of model [CpL-3]
\begin{eqnarray}
\alpha_s&=&0.326\:,\nonumber\\
m_b&=&4.7215\:\textrm{GeV}\:,\\
\sigma&=&1.12\:\textrm{GeV/fm}\:,\nonumber
\end{eqnarray}
are determined in the same manner as in model [CpL-B], by a fit to the masses of states $\Upsilon(1S)$, $\Upsilon(2S)$ and $C(1P)$ (Eq. (\ref{Eq:CoG-1P-Bottomonium})), identifying $C(1P)$ with the $1^1P_1$ state of the model. The strong coupling constant in model [CpL-3] is now significantely smaller than in model [CpL-B] but still definitely too large with respect to the strong coupling constant of model [CpL-3] for charmonium. Furthermore the reduction in $\alpha_s$ comes at the cost of a slightly increased string tension, while the quark mass stays about the same.

The resulting masses are compared to experimental data in Tab. \ref{Tab:MassesBottomoniumCpL-3}. In Fig. \ref{Fig:SpectrumBottomoniumCpL-3} the emerging spectrum is compared to the experimental one. The spin-dependent parts of [CpL-3] and [CpL-B] are identical. The problem with the triplet $P$-state splitting prevails, resulting once more in the ratios
\begin{eqnarray}
\Phi_b^{\textrm{[CpL-3]}}(1P)&=&\left(\frac{M(\chi_{b2})-M(\chi_{b1})}{M(\chi_{b1})-M(\chi_{b0})}\right)_{\textrm{[CpL-3]}}=\frac{4}{5}\:,\nonumber\\
\Phi_b^{\textrm{[CpL-3]}}(2P)&=&\left(\frac{M(\chi'_{b2})-M(\chi'_{b1})}{M(\chi'_{b1})-M(\chi'_{b0})}\right)_{\textrm{[CpL-3]}}=\frac{4}{5}\:,
\end{eqnarray}
which disagree with the experimental values (Eq. (\ref{Eq:P-ProportionBottomoniumExperiment})). Due to the smaller $\alpha_s$ the total size of the triplet $P$-state splitting is now close to the empirical one.

%% file: 4-OGE/4-Summary.tex
The study of model [CpL-3] is mainly inspired by the fact that the on-shell condition used in the computation of the Breit interaction is not valid for the investigated system.

As model [CpL-3] differs only in the purely momentum dependent terms from model [CpL-B], it provides similar spectra for the spin-dependent splittings as model [CpL-B]. Problems remain for the triplet $P$-states and to a lesser degree for the $2S$-state spin-spin splitting in charmonium. 

The problem with the $2S$-states for charmonium seems to be due to the near $D\overline{D}$-threshold, which we do not account for in any of the models.

The momentum dependent terms in [CpL-3] are not uniquely determined. The spectra depend on the ordering of gradient operators.

Model [CpL-3] yields, compared to model [CpL-B], a slightly decreased string tension for charmonium, and a significantly decreased but still definitely too large strong coupling constant $\alpha_s$ for bottomonium.

%% file: 5-OGE-induced/5-Intro.tex
The first section of this chapter is dedicated to the motivation and computation of the induced interaction, which is used to improve the triplet $P$-state structure of the Coulomb-plus-Linear model. The remaining sections present the improved model and discuss its results.

%% file: 5-OGE-induced/5-Induced-interaction.tex
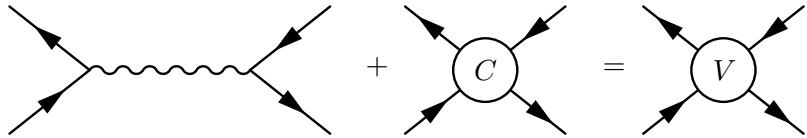
\begin{figure}[h!]
\vspace{0.7cm}
\begin{eqnarray}
	\vtop{\vskip-11mm\hbox{
\begin{fmfgraph*}(120,60)
\fmftop{t1,t2}
\fmfbottom{b1,b2}
\fmf{fermion}{b1,v1,t1}
\fmf{fermion}{t2,v2,b2}
\fmf{wiggly}{v1,v2}
\end{fmfgraph*}}}
\quad+
	\vtop{\vskip-11mm\hbox{
\begin{fmfgraph*}(60,60)
\fmftop{t3,t4}
\fmfbottom{b3,b4}
\fmf{fermion}{b3,v3,t3}
\fmf{fermion}{t4,v3,b4}
\fmfv{decor.shape=circle,decor.filled=empty,label=$C$,decor.size=.4w,label.dist=-4}{v3}
\end{fmfgraph*}}}
\quad=
	\vtop{\vskip-11mm\hbox{
\begin{fmfgraph*}(60,60)
\fmftop{t5,t6}
\fmfbottom{b5,b6}
\fmf{fermion}{b5,v5,t5}
\fmf{fermion}{t6,v5,b6}
\fmfv{decor.shape=circle,decor.filled=empty,label=$V$,decor.size=.4w,label.dist=-4}{v5}
\end{fmfgraph*}}}
\nonumber
\end{eqnarray}
\caption{\label{Fig:PotentialSE} Perturbative QCD short range part and phenomenological confining part composing the potential used in the Schr\"odinger equation.}
\end{figure}

In Chapters \ref{Chapt:CpL-Breit} \& \ref{Chapt:CpL-OGE} we observed that the Coulomb-plus-Linear models, with perturbative corrections arising from perturbative processes to Order $\mathcal{O}(\alpha_s)$, are not sufficient to describe the triplet $P$-state structure correctly. We now take a closer look at the Schr\"odinger equation, which we used to determine eigensolutions and eigenvalues. The potential used in the Schr\"odinger equation is composed of a phenomenological confinement part and a short range part derived from perturbative QCD
. The corresponding Feynman diagrams are shown in Fig. \ref{Fig:PotentialSE}, where the one-gluon exchange diagram is introduced as a symbol to represent the perturbative QCD part.

Solving the Schr\"odinger equation is equivalent to solving the Lippmann-Schwinger integral equation \cite{LandauQM290}
\begin{eqnarray}\label{Eq:LippmannSchwingerEquation}
|\psi\rangle&=&|\phi\rangle+GV|\psi\rangle=\frac{1}{1-GV}|\phi\rangle\\
&=&|\phi\rangle+GV|\phi\rangle+GVGV|\phi\rangle+GVGVGV|\phi\rangle+\dots\nonumber\:.
\end{eqnarray}
With the definition of the T-matrix
\begin{eqnarray}\label{Eq:DefT-Matrix}
T_\textrm{E}(\vec{k}^{\,'},\vec{k})&=&\langle\phi_{\vec{k}^{\,'}}|T|\phi_{\vec{k}}\rangle=\langle\phi_{\vec{k}^{\,'}}|V|\psi_{\vec{k}}\rangle\:,
\end{eqnarray}
the Lippman-Schwinger equation for the T matrix is
\begin{eqnarray}
\langle\phi_{\vec{k}^{\,'}}|V|\psi_{\vec{k}}\rangle&=&\langle\phi_{\vec{k}^{\,'}}|V|\phi_{\vec{k}}\rangle+\langle\phi_{\vec{k}^{\,'}}|VGV|\psi_{\vec{k}}\rangle\:,
\end{eqnarray}
or using Eq. (\ref{Eq:LippmannSchwingerEquation}) and rewritten as an operator equation:
\begin{eqnarray}\label{Eq:OperatorEquationT-Matrix}
T&=&V+VGV+VGVGV+\dots\:.
\end{eqnarray}
The Lippmann-Schwinger is commonly used for scattering problems. Bound states are identified as Poles in the T matrix $T_\textrm{E}$ at energies $E<0$.
\begin{figure}[h!]
\vspace{1.3cm}
\begin{eqnarray}
	\vtop{\vskip-11mm\hbox{
\begin{fmfgraph*}(60,60)
\fmftop{t5,t6}
\fmfbottom{b5,b6}
\fmf{fermion}{b5,v5,t5}
\fmf{fermion}{t6,v5,b6}
\fmfv{decor.shape=circle,decor.filled=empty,label=$V$,decor.size=.4w,label.dist=-4}{v5}
\end{fmfgraph*}}}
+
	\vtop{\vskip-21mm\hbox{
\begin{fmfgraph*}(60,120)
\fmftop{t7,t8}
\fmfbottom{b7,b8}
\fmf{fermion}{b7,v7,b8}
\fmf{fermion}{t8,v8,t7}
\fmf{fermion,left=.5,tension=.5}{v7,v8}
\fmf{fermion,left=.5,tension=.5}{v8,v7}
\fmfv{decor.shape=circle,decor.filled=empty,label=$V$,decor.size=.3w,label.dist=-4}{v7}
\fmfv{decor.shape=circle,decor.filled=empty,label=$V$,decor.size=.3w,label.dist=-4}{v8}
\end{fmfgraph*}}}
+
	\vtop{\vskip-21mm\hbox{
\begin{fmfgraph*}(60,120)
\fmftop{t7,t8}
\fmfbottom{b7,b8}
\fmf{fermion}{b7,v7,b8}
\fmf{fermion}{t8,v9,t7}
\fmf{fermion,left=.5,tension=.5}{v7,v8}
\fmf{fermion,left=.5,tension=.5}{v8,v7}
\fmf{fermion,left=.5,tension=.5}{v8,v9}
\fmf{fermion,left=.5,tension=.5}{v9,v8}
\fmfv{decor.shape=circle,decor.filled=empty,label=$V$,decor.size=.3w,label.dist=-4}{v7}
\fmfv{decor.shape=circle,decor.filled=empty,label=$V$,decor.size=.3w,label.dist=-4}{v8}
\fmfv{decor.shape=circle,decor.filled=empty,label=$V$,decor.size=.3w,label.dist=-4}{v9}
\end{fmfgraph*}}}
+\:\dots\:\nonumber
\end{eqnarray}
\vspace{-0.7cm}
\caption{\label{Fig:IterationSE}  Diagrams corresponding to the operator equation of the T-matrix.}
\end{figure}
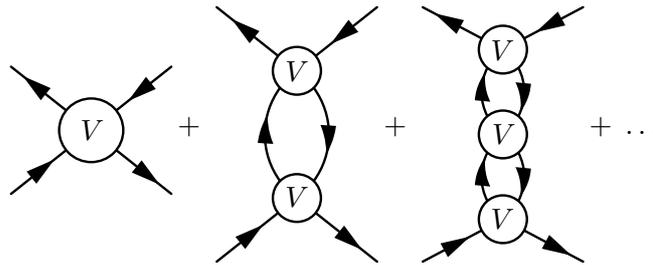\\
One realizes that the Schr\"odinger equation iterates the interaction to all orders and that finding an eigensolution corresponds to the formation of a bound state.
\begin{figure}[h!]
\begin{center}
\begin{fmfgraph*}(60,120)
\fmftop{t7,t8}
\fmfbottom{b7,b8}
\fmf{fermion}{b7,v7,b8}
\fmf{fermion}{t8,v8,t7}
\fmf{dbl_plain}{v7,v8}
\end{fmfgraph*}
\end{center}
\vspace{-0.5cm}
\caption{\label{Fig:S-Channel-Induced} s-channel diagram corresponding to the formation of a bound state.}
\end{figure}
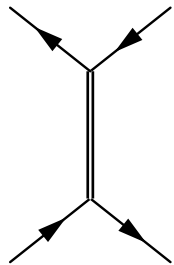
The formation of a bound state is equivalent to the s-channel diagram in Fig. \ref{Fig:S-Channel-Induced}.
\begin{figure}[b!]
\begin{center}
\begin{fmfgraph*}(120,60)
\fmftop{t10,t11}
\fmfbottom{b10,b11}
\fmf{fermion}{b10,v10,t10}
\fmf{fermion}{t11,v11,b11}
\fmf{dbl_plain}{v10,v11}
\end{fmfgraph*}
\end{center}
\vspace{-0.4cm}
\caption{\label{Fig:T-Channel-Induced} t-channel diagram corresponding to the exchange of a bound state.}
\end{figure}
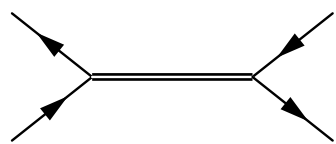
Crossing symmetry implies that, for any s-channel process forming a bound state of a particle-antiparticle system as in Fig. \ref{Fig:S-Channel-Induced}, there is also a t-channel process as shown in Fig. \ref{Fig:T-Channel-Induced}. This corresponds to the exchange of a bound state between quark and antiquark.

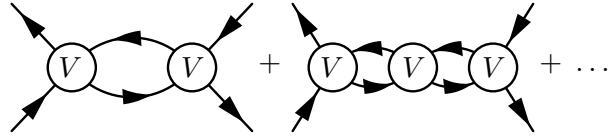
\begin{figure}[h!]
\vspace{0.7cm}
\begin{eqnarray}
	\vtop{\vskip-11mm\hbox{
\begin{fmfgraph*}(90,60)
\fmftop{t7,t8}
\fmfbottom{b7,b8}
\fmf{fermion}{b7,v7,t7}
\fmf{fermion}{t8,v8,b8}
\fmf{fermion,right=.5,tension=.5}{v7,v8}
\fmf{fermion,right=.5,tension=.5}{v8,v7}
\fmfv{decor.shape=circle,decor.filled=empty,label=$V$,decor.size=.2w,label.dist=-4}{v7}
\fmfv{decor.shape=circle,decor.filled=empty,label=$V$,decor.size=.2w,label.dist=-4}{v8}
\end{fmfgraph*}}}
+\!
	\vtop{\vskip-11mm\hbox{
\begin{fmfgraph*}(90,60)
\fmftop{t7,t8}
\fmfbottom{b7,b8}
\fmf{fermion}{b7,v7,t7}
\fmf{fermion}{t8,v9,b8}
\fmf{fermion,right=.5,tension=.5}{v7,v8}
\fmf{fermion,right=.5,tension=.5}{v8,v7}
\fmf{fermion,right=.5,tension=.5}{v8,v9}
\fmf{fermion,right=.5,tension=.5}{v9,v8}
\fmfv{decor.shape=circle,decor.filled=empty,label=$V$,decor.size=.2w,label.dist=-4}{v7}
\fmfv{decor.shape=circle,decor.filled=empty,label=$V$,decor.size=.2w,label.dist=-4}{v8}
\fmfv{decor.shape=circle,decor.filled=empty,label=$V$,decor.size=.2w,label.dist=-4}{v9}
\end{fmfgraph*}}}
+\:\dots
\nonumber
\end{eqnarray}
\caption{\label{Fig:InducedBubbles} Summation of bubbles corresponding to the exchange of a bound state.}
\end{figure}

In other words, by solving the Schr\"odinger equation we find bound states which correspond to s-channel processes. These s-channel processes induce t-channel interactions which have to be taken into account. While the s-channel diagram involves a sum over ladders (Fig. \ref{Fig:IterationSE}), the t-channel diagram represents the summation of bubbles (Fig. \ref{Fig:InducedBubbles}). Solving a Schr\"odinger or Bethe-Salpeter equation yields an approximation based on ladder diagrams, whereas taking into account the induced interaction generates a solution based on a more complete mix of bubble and ladder diagrams.

The Lorentz-structure of these additional interactions are given by the properties of the corresponding bound states. In principle one would have to take into account all bound states, but for our model we consider only the $1S$-states. These states should have the strongest influence on the spectrum, as they have the lowest mass and the effective range of the interaction decreases with increasing mass of the exchanged bound state. The $1S$-states for quarkonium are a pseudoscalar meson and a vector meson, hence we have to consider a pseudoscalar and a vector particle exchange interaction.

%% file: 5-OGE-induced/5-Pseudoscalar-exchange.tex
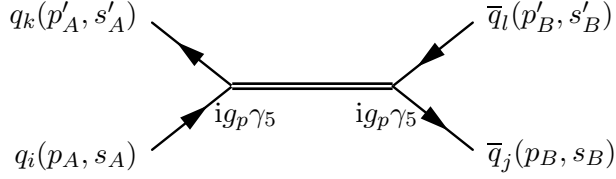
\begin{figure}[t!]
\vspace{0.7cm}
\begin{eqnarray}
	\vtop{\vskip-11mm\hbox{
\begin{fmfgraph*}(120,60)
\fmftop{t10,t11}
\fmfbottom{b10,b11}
\fmf{fermion}{b10,v10,t10}
\fmf{fermion}{t11,v11,b11}
\fmf{dbl_plain}{v10,v11}
\fmflabel{\vspace{-0.8cm}i$g_p\gamma_5$\hspace{-0.8cm}}{v10}
\fmflabel{\vspace{-0.8cm}\hspace{-0.7cm}i$g_p\gamma_5$}{v11}
\fmflabel{$q_k(p'_A,s'_A)$}{t10}
\fmflabel{$\overline{q}_l(p'_B,s'_B)$}{t11}
\fmflabel{$q_i(p_A,s_A)$}{b10}
\fmflabel{$\overline{q}_j(p_B,s_B)$}{b11}
\end{fmfgraph*}}}
\nonumber
\end{eqnarray}
\vspace{-0.2cm}
\caption{\label{Fig:PseudoscalarExchangeInducedInteraction} Quark-Antiquark pair exchanging a pseudoscalar meson.}
\end{figure}
{\noindent}The amplitude in Feynman gauge corresponding to the massive pseudoscalar exchange diagram (Fig. \ref{Fig:PseudoscalarExchangeInducedInteraction}) is, utilizing that initial and final quark-antiquark states are color neutral, given by
\begin{eqnarray}\label{Eq:AmplitudePseudoscalarRaw}
\mathcal{M}_{fi}&=&\left[\overline{u}(p'_A,s'_A)\left(\mathrm{i}g_\textrm{p}\gamma_5\right)u(p_A,s_a)\right]\frac{1}{q^2-M_\textrm{p}^2}\left[\overline{v}(p_B,s_B)\left(\mathrm{i}g_p\gamma_5\right)v(p'_B,s'_B)\right]\nonumber\\
&=&-\frac{g_\textrm{p}^2}{q^2-M_\textrm{p}^2}\left[\overline{u}(p'_A,s'_A)\gamma_5 u(p_A,s_A)\overline{v}(p_B,s_B)\gamma_5 v(p'_B,s'_B)\right]\:.
\end{eqnarray}
Applying the normalization conventions as in Appendix \ref{Sec:Conventions}, the Dirac spinors read
\begin{eqnarray}\label{Eq:Dirac-Spinors}
u(p_A,s_A)&=&\sqrt{\frac{E_A+m_A}{2m_A}}\left(\begin{array}{c}
\chi_{s_A}\\ 
\frac{\vec{\sigma}\cdot\vec{p}_A}{E_A+m_A}\chi_{s_A}
\end{array}\right)\:,\nonumber\\
\overline{u}(p'_A,s'_A)&=&\sqrt{\frac{E'_A+m_A}{2m_A}}\left(\begin{array}{c@{,}c}
\chi_{s'_A}^\dagger&\quad -\chi_{s'_A}^\dagger \frac{\vec{\sigma}\cdot\vec{p}\,'_A}{E'_A+m_A}
\end{array}\right)\:,\nonumber\\
v(p'_B,s'_B)&=&\sqrt{\frac{E'_B+m_B}{2m_B}}\left(\begin{array}{c}
\frac{\vec{\sigma}\cdot\vec{p}\,'_B}{E'_B+m_B}\chi^c_{s'_B}\\
\chi^c_{s'_B}
\end{array}\right)\:,\\
\overline{v}(p_B,s_B)&=&\sqrt{\frac{E_B+m_B}{2m_B}}\left(\begin{array}{c@{,}c}
\chi_{s_B}^{c\dagger} \frac{\vec{\sigma}\cdot\vec{p}_B\,}{E_B+m_B} & \quad -\chi_{s_B}^{c\dagger}
\end{array}\right)\:,\nonumber
\end{eqnarray}
with energies $E_X=\sqrt{\vec{p}_X^{\,2}+m_X^2}$ and quark masses $m_X$. Employing Eq. (\ref{Eq:Dirac-Spinors}), the pseudoscalar Dirac currents for particles and antiparticles are given by
\begin{eqnarray}
\overline{u}(p'_A,s'_A)\gamma_5 u(p_A,s_A)&=&\left(\begin{array}{c@{,}c}
 \chi^\dagger_{s'_A}&\quad -\chi^\dagger_{s'_A} \frac{\vec{\sigma}\cdot\vec{p}\,'_A}{E'_A+m_A}
\end{array}\right)\left(\begin{array}{cc} 0&1\\1&0\end{array}\right)\left(\begin{array}{c}\chi_{s_A}\\ \frac{\vec{\sigma}\cdot\vec{p}_A}{E_A+m_A}\chi_{s_A}\end{array}\right)\nonumber\\
&=&\chi^\dagger_{s'_A}\left(\frac{\vec{\sigma}\cdot\vec{p}_A}{E_A+m_A}-\frac{\vec{\sigma}\cdot\vec{p}\,'_A}{E'_A+m_A}\right)\chi_{s_A}\:,\\
\overline{v}(p_B,s_B)\gamma_5 v(p'_B,s'_B)&=&\left(\begin{array}{c@{,}c}\chi^{c\dagger}_{s_B}\frac{\vec{\sigma}\cdot\vec{p}_B}{E_B+m_B}&-\chi^{c\dagger}_{s_B}\end{array}\right)\left(\begin{array}{cc}0&1\\1&0\end{array}\right)\left(\begin{array}{c}\frac{\vec{\sigma}\cdot\vec{p}\,'_B}{E'_B+m_B}\chi^c_{s'_B}\\ \chi^c_{s'_B}\end{array}\right)\nonumber\\
&=&\chi^{c\dagger}_{s_B}\left(\frac{\vec{\sigma}\cdot\vec{p}_B}{E_B+m_B}-\frac{\vec{\sigma}\cdot\vec{p}\,'_B}{E'_B+m_B}\right)\chi^c_{s'_B}\:,
\end{eqnarray}
with Paulispinors $\chi_s$. Putting things together, remembering the definition of spin for particles (Eq. (\ref{Eq:Spin-Particles})) and antiparticles (Eq. (\ref{Eq:Spin-Antiparticles})), yields for the amplitude
\begin{eqnarray}\label{AmplitudePseudoscalarExpanded}
\mathcal{M}_{fi}\!\!\!\!\!&=&\!\!\!\!\!-\frac{g_\textrm{p}^2}{q^2-M_\textrm{p}^2}\chi^\dagger_{s'_A}\left[\frac{\vec{\sigma}\cdot\vec{p}_A}{E_A+m_A}-\frac{\vec{\sigma}\cdot\vec{p}\,'_A}{E'_A+m_A}\right]\chi_{s_A}\chi^{c\dagger}_{s_B}\left[\frac{\vec{\sigma}\cdot\vec{p}_B}{E_B+m_B}-\frac{\vec{\sigma}\cdot\vec{p}\,'_B}{E'_B+m_B}\right]\chi_{s'_B}^c\nonumber\\
\!\!\!\!\!&=&\!\!\!\!\!-\frac{g_\textrm{p}^2}{q^2-M_\textrm{p}^2}\frac{\left(\vec{s}_1\cdot\vec{q}\,\right)\left(\vec{s}_2\cdot\vec{q}\,\right)}{m_1m_2}+\mathcal{O}\left(m^{-3}\right)=\frac{g_\textrm{p}^2}{\vec{q}^{\,2}+M_\textrm{p}^2}\frac{\left(\vec{s}_1\cdot\vec{q}\,\right)\left(\vec{s}_2\cdot\vec{q}\,\right)}{m_1m_2}+\mathcal{O}\left(m^{-3}\right)\:,
\end{eqnarray}
where we have used
\begin{eqnarray}\label{Eq:PropagatorMassivePseudoscalar}
\frac{1}{q^2-M_\textrm{p}^2}\stackrel{q_0=0}{=}-\frac{1}{\vec{q}^{\,2}+M_\textrm{p}^2}\:.
\end{eqnarray}
The leading term in the amplitude (Eq. (\ref{AmplitudePseudoscalarExpanded})) is of order $O(m^{-2})$, hence the relation between potential and amplitude
\begin{eqnarray}
\frac{1}{(2\pi)^3}\tilde{V}_\textrm{p}(\vec{q}\,)=-\left(1+\mathcal{O}\left(\frac{1}{m^2}\right)\right)\mathcal{M}_{fi}
\end{eqnarray}
does not provide any further terms, as we are only expanding up to order $O(m^{-2})$. The Fourier transformation back into coordinate space
\begin{eqnarray}
V(\vec{r};\vec{p}_1,\vec{p}_2)=\int\mathrm{d}^3q\:\mathrm{e}^{-\mathrm{i}\vec{q}\cdot\vec{r}}\,\tilde{V}(\vec{q}\,)\:.
\end{eqnarray}
of the terms in Eq. (\ref{AmplitudePseudoscalarExpanded}) are given by
\begin{eqnarray}
W_\textrm{p}&=&\frac{1}{(2\pi)^3}\int\mathrm{d}^3q\:\mathrm{e}^{-\mathrm{i}\vec{q}\cdot\vec{r}}\frac{1}{\vec{q}^{\,2}+M_\textrm{p}^2}=\frac{\mathrm{e}^{-M_\textrm{p} r}}{4\pi r}\:,\\
W_{\textrm{p},1}&=&\frac{1}{(2\pi)^3}\int\mathrm{d}^3q\:q_iq_j\:\mathrm{e}^{-\mathrm{i}\vec{q}\cdot\vec{r}}\frac{1}{\vec{q}^{\,2}+M_\textrm{p}^2}=-\left[W''_\textrm{p}-\frac{1}{r}W'_\textrm{p}\right]\left(\frac{r_ir_j}{r^2}-\frac{1}{3}\delta_{ij}\right)-\frac{1}{3}\delta_{ij}\Delta W_\textrm{p}\:,\nonumber
\end{eqnarray}
with
\begin{eqnarray}
W'_\textrm{p}&=&-\frac{\mathrm{e}^{-M_\textrm{p} r}}{4\pi r}\left(M_\textrm{p}+\frac{1}{r}\right)\:,\nonumber\\
W''_\textrm{p}&=&\frac{\mathrm{e}^{-M_\textrm{p} r}}{4\pi r}\left(M_\textrm{p}^2+\frac{2M_\textrm{p}}{r}+\frac{2}{r^2}\right)\:,\nonumber\\
\Delta W_\textrm{p}&=&-\delta^{(3)}(\vec{r}\,)+\frac{M_\textrm{p}^2\,\mathrm{e}^{-M_\textrm{p} r}}{4\pi r}\:.
\end{eqnarray}
Collecting the pieces we eventually arrive at the potential for a massive pseudoscalar exchange 
\begin{eqnarray}\label{Eq:PotentialPseudoscalarInduced}
V_\textrm{p}(\vec{r}\,)&=&\frac{g_\textrm{p}^2}{4\pi m_1m_2}\left[\frac{\mathrm{e}^{-M_\textrm{p} r}}{3r}\left(M_\textrm{p}^2+\frac{3M_\textrm{p}}{r}+\frac{3}{r^2}\right)\left(3\frac{(\vec{s}_1\cdot\vec{r}\,)(\vec{s}_2\cdot\vec{r}\,)}{r^2}-(\vec{s}_1\cdot\vec{s}_2)\right)\right.\nonumber\\
&&\left.\qquad\qquad\qquad\qquad\qquad\qquad\quad-\frac{1}{3}\left(\delta^{(3)}(\vec{r}\,)4\pi-M_\textrm{p}^2\frac{\mathrm{e}^{-M_\textrm{p} r}}{r}\right)(\vec{s}_1\cdot\vec{s}_2)\right]\:.
\end{eqnarray}

%% file: 5-OGE-induced/5-Vector-exchange.tex
Employing Feynman rules one derives the amplitude in Feynman gauge for the massive vector exchange diagram (Fig. \ref{Fig:VectorExchangeInducedInteraction}):
\begin{eqnarray}\label{AmplitudeMassiveVectorExchange}
\mathcal{M}_{fi}=&=&-\left[\overline{u}(p'_A,s'_A)\left(\mathrm{i}g_\textrm{v}\gamma^\mu\right)u(p_A,s_a)\right]\frac{-g_{\mu\nu}}{q^2-M_\textrm{v}^2}\left[\overline{v}(p_B,s_B)\left(\mathrm{i}g_\textrm{v}\gamma^\nu\right)v(p'_B,s'_B)\right]\nonumber\\
&=&-\frac{g_\textrm{v}^2}{q^2-M_\textrm{v}^2}\left[\overline{u}(p'_A,s'_A)\gamma^\mu u(p_A,s_A)\overline{v}(p_B,s_B)\gamma_\mu v(p'_B,s'_B)\right]\:.
\end{eqnarray}
It is obvious that the amplitude for the massive vector exchange (Eq. (\ref{AmplitudeMassiveVectorExchange})) is similar to the amplitude for the one-gluon exchange (Eq. (\ref{Eq:Quark-Antiquark-Amplitude-Origin})), apart from the color factor and the propagator. One can follow the computation of the potential for the one-gluon exchange, with the replacement
\begin{eqnarray}
\frac{4}{3\:q^2}&\longrightarrow&\frac{1}{q^2-M_\textrm{v}^2}\:,
\end{eqnarray}
yielding
\begin{eqnarray}
\frac{1}{(2\pi)^3}\tilde{V}_\textrm{v}(\vec{q};\vec{p}_1,\vec{p}_2)\!\!\!\!\!&=&\!\!\!\!\!\frac{g_\textrm{v}^2}{q^2-M_\textrm{v}^2}\left[1+\frac{-\frac{1}{2}\vec{q}^{\,2}-2\mathrm{i}\vec{s}_1\cdot\left(\vec{q}\times\vec{p}_1\right)}{4m_1^2}+\frac{-\frac{1}{2}\vec{q}^{\,2}+2\mathrm{i}\vec{s}_2\cdot\left(\vec{q}\times\vec{p}_2\right)}{4m_2^2}\right.\\
&&\!\!\!\!\!\!\!\!\!\!\!\!\!\!\!\!\!\!\!\!\!\!\!\!\!\!\!\!\!\!\!\!\!\!\!\!\!\!\!\!\!\!\!\!\!\!\!\!+\!\left.\frac{1}{4m_1m_2}\Big(-4\vec{p}_1\vec{p}_2+4\mathrm{i}\vec{s}_1\cdot\left(\vec{q}\times\vec{p}_2\right)-4\mathrm{i}\vec{s}_2\cdot\left(\vec{q}\times\vec{p}_1\right)-4\vec{q}^{\,2}\left(\vec{s}_1\cdot\vec{s}_2\right)+4\left(\vec{q}\cdot\vec{s}_1\right)\left(\vec{q}\cdot\vec{s}_2\right)\Big)\right]\:.\nonumber
\end{eqnarray}
Using the equivalent to Eq. (\ref{Eq:PropagatorMassivePseudoscalar}) for the expansion of the propagator
\begin{eqnarray}
\frac{1}{q^2-M_\textrm{v}^2}\stackrel{q_0=0}{=}-\frac{1}{\vec{q}^{\,2}+M_\textrm{v}^2}\:,
\end{eqnarray}
the momentum space potential for the massive vector exchange is given by
\begin{eqnarray}\label{Eq:PotentialMomentumSpaceMassiveVector}
\frac{1}{(2\pi)^3}\tilde{V}_\textrm{v}(\vec{q};\vec{p}_1,\vec{p}_2)\!\!\!\!\!&=&\!\!\!\!\!-\frac{g_\textrm{v}^2}{\vec{q}^{\,2}+M_\textrm{v}^2}\left[1+\frac{-\frac{1}{2}\vec{q}^{\,2}-2\mathrm{i}\vec{s}_1\cdot\left(\vec{q}\times\vec{p}_1\right)}{4m_1^2}+\frac{-\frac{1}{2}\vec{q}^{\,2}+2\mathrm{i}\vec{s}_2\cdot\left(\vec{q}\times\vec{p}_2\right)}{4m_2^2}\right.\\
&&\!\!\!\!\!\!\!\!\!\!\!\!\!\!\!\!\!\!\!\!\!\!\!\!\!\!\!\!\!\!\!\!\!\!\!\!\!\!\!\!\!\!\!\!\!\!\!\!+\!\left.\frac{1}{4m_1m_2}\Big(-4\vec{p}_1\vec{p}_2+4\mathrm{i}\vec{s}_1\cdot\left(\vec{q}\times\vec{p}_2\right)-4\mathrm{i}\vec{s}_2\cdot\left(\vec{q}\times\vec{p}_1\right)-4\vec{q}^{\,2}\left(\vec{s}_1\cdot\vec{s}_2\right)+4\left(\vec{q}\cdot\vec{s}_1\right)\left(\vec{q}\cdot\vec{s}_2\right)\Big)\right]\:.\nonumber
\end{eqnarray}
The Fourier transformation back into coordinate space
\begin{eqnarray}
V(\vec{r};\vec{p}_1,\vec{p}_2)&=&\int\mathrm{d}^3q\:\mathrm{e}^{-\mathrm{i}\vec{q}\cdot\vec{r}}\,\tilde{V}(\vec{q};\vec{p}_1,\vec{p}_2)\:,
\end{eqnarray}
of the various terms in Eq. (\ref{Eq:PotentialMomentumSpaceMassiveVector}) is given by
\begin{eqnarray}
W_\textrm{v}&=&\frac{1}{(2\pi)^3}\int\mathrm{d}^3q\:\mathrm{e}^{-\mathrm{i}\vec{q}\cdot\vec{r}}\frac{1}{\vec{q}^{\,2}+M_\textrm{v}^2}=\frac{\mathrm{e}^{-M_\textrm{v} r}}{4\pi r}\:,\nonumber\\
W_{\textrm{v},1}&=&\frac{1}{(2\pi)^3}\int\mathrm{d}^3q\:\vec{q}^{\,2}\:\mathrm{e}^{-\mathrm{i}\vec{q}\cdot\vec{r}}\frac{1}{\vec{q}^{\,2}+M_\textrm{v}^2}=-\Delta W_\textrm{v}\:,\nonumber\\
W_{\textrm{v},2}&=&\frac{1}{(2\pi)^3}\int\mathrm{d}^3q\:q_j\:\mathrm{e}^{-\mathrm{i}\vec{q}\cdot\vec{r}}\frac{1}{\vec{q}^{\,2}+M_\textrm{v}^2}=\mathrm{i}\nabla_jW_\textrm{v}=\mathrm{i}\frac{r_j}{r}W'_\textrm{v}\:,\\
W_{\textrm{v},3}&=&\frac{1}{(2\pi)^3}\int\mathrm{d}^3q\:q_iq_j\:\mathrm{e}^{-\mathrm{i}\vec{q}\cdot\vec{r}}\frac{1}{\vec{q}^{\,2}+M_\textrm{v}^2}=-\left[W''_\textrm{v}-\frac{1}{r}W'_\textrm{v}\right]\left(\frac{r_ir_j}{r^2}-\frac{1}{3}\delta_{ij}\right)-\frac{1}{3}\delta_{ij}\Delta W_\textrm{v}\:,\nonumber
\end{eqnarray}
with
\begin{eqnarray}
W'_\textrm{v}&=&-\frac{\mathrm{e}^{-M_\textrm{v} r}}{4\pi r}\left(M_\textrm{v}+\frac{1}{r}\right)\:,\nonumber\\
W''_\textrm{v}&=&\frac{\mathrm{e}^{-M_\textrm{v} r}}{4\pi r}\left(M_\textrm{v}^2+\frac{2M_\textrm{v}}{r}+\frac{2}{r^2}\right)\:,\\
\Delta W_\textrm{v}&=&-\delta^{(3)}(\vec{r}\,)+\frac{M_\textrm{v}^2\,\mathrm{e}^{-M_\textrm{v} r}}{4\pi r}\:.\nonumber
\end{eqnarray}
Collecting all the pieces we eventually arrive at the potential in coordinate space
\begin{eqnarray}\label{Eq:PotentialVectorInduced}
V_\textrm{v}(\vec{r};\vec{p}_1,\vec{p}_2)\!\!\!\!&=&\!\!\!\!-\frac{g_\textrm{v}^2}{4\pi}\left[\frac{\mathrm{e}^{-M_\textrm{v}r}}{r}+\frac{1}{8}\left(M_\textrm{v}^2\frac{\mathrm{e}^{-M_\textrm{v}r}}{r}-4\pi\,\delta^{(3)}(\vec{r}\,)\right)\left(\frac{1}{m_1^2}+\frac{1}{m_2^2}\right)\right.\nonumber\\
&&\!\!\!\!\!+\frac{2(\vec{s}_1\cdot\vec{s}_2)}{3m_1m_2}\left(M_\textrm{v}^2\frac{\mathrm{e}^{-M_\textrm{v}r}}{r}-4\pi\delta^{(3)}(\vec{r}\,)\right)-\frac{\vec{p}_1\cdot\vec{p}_2}{m_1m_2}\frac{\mathrm{e}^{-M_\textrm{v}r}}{r}\nonumber\\
&&\!\!\!\!\!-\frac{\left(M_\textrm{v}^2+3M_\textrm{v}/r+3/r^2\right)}{3m_1m_2}\frac{\mathrm{e}^{-M_\textrm{v}r}}{r}\left(3\frac{(\vec{s}_1\cdot\vec{r}\,)(\vec{s}_2\cdot\vec{r}\,)}{r^2}-(\vec{s}_1\cdot\vec{s_2})\right)\\
&&\!\!\!\!\!+\frac{\mathrm{e}^{-M_\textrm{v}r}}{r^2}\left(M_\textrm{v}+\frac{1}{r}\right)\left(-\frac{(\vec{r}\times\vec{p}_1)\cdot\vec{s}_1}{2m_1^2}+\frac{(\vec{r}\times\vec{p}_2)\cdot\vec{s}_2}{2m_2^2}\right.\nonumber\\
&&\qquad\qquad\qquad\qquad\qquad\qquad\qquad\qquad\left.\left.-\frac{(\vec{r}\times\vec{p}_1)\cdot\vec{s}_2-(\vec{r}\times\vec{p}_2)\cdot\vec{s}_1}{m_1m_2}\right)\right]\:.\nonumber
\end{eqnarray}
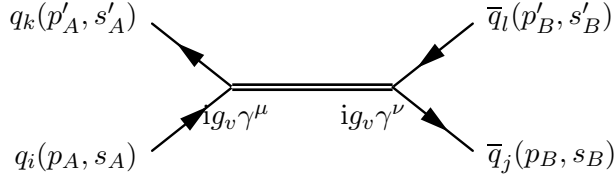
\begin{figure}[t!]
\vspace{0.7cm}
\begin{eqnarray}
	\vtop{\vskip-11mm\hbox{
\begin{fmfgraph*}(120,60)
\fmftop{t10,t11}
\fmfbottom{b10,b11}
\fmf{fermion}{b10,v10,t10}
\fmf{fermion}{t11,v11,b11}
\fmf{dbl_plain}{v10,v11}
\fmflabel{\vspace{-0.8cm}i$g_v\gamma^\mu$\hspace{-0.7cm}}{v10}
\fmflabel{\vspace{-0.8cm}\hspace{-0.9cm}i$g_v\gamma^\nu$}{v11}
\fmflabel{$q_k(p'_A,s'_A)$}{t10}
\fmflabel{$\overline{q}_l(p'_B,s'_B)$}{t11}
\fmflabel{$q_i(p_A,s_A)$}{b10}
\fmflabel{$\overline{q}_j(p_B,s_B)$}{b11}
\end{fmfgraph*}}}
\nonumber
\end{eqnarray}
\vspace{-0.4cm}
\caption{\label{Fig:VectorExchangeInducedInteraction} Quark-Antiquark pair exchanging a vector meson.}
\end{figure}

%% file: 5-OGE-induced/5-1S-state-Induced-interaction.tex
The potential for the $1S$-state t-channel exchange interaction, referred to as induced interaction from here onward, is composed of the pseudoscalar meson exchange potential (Eq. (\ref{Eq:PotentialPseudoscalarInduced})) and the vector meson exchange potential (Eq. (\ref{Eq:PotentialVectorInduced})). Merging these potentials yields, in principle, two different couplings $g_\textrm{p}$ and $g_\textrm{v}$, and two different masses for the exchanged particles $M_\textrm{p}$ and $M_\textrm{v}$. The couplings $g_\textrm{p}$, $g_\textrm{v}$ are related to the asymptotic normalization constant of the considered bound state (see App. 12 of \cite{EricsonWeise88}). In the Coulomb-plus-Linear model the wavefunctions for the pseudoscalar state $\eta_c$ and the vector state $J/\psi$ are identical as the potential $V_0$ is spin-independent. Therefore the couplings $g_\textrm{p}$ and $g_\textrm{v}$ are equal and we define, analogous to the strong coupling constant, the coupling constant for the induced interaction via
\begin{eqnarray}
\alpha_i&\equiv&\frac{g_\textrm{p}^2}{4\pi}=\frac{g_\textrm{v}^2}{4\pi}\:.
\end{eqnarray}
Based on the same argument, we use the degenerate $1S$ mass. In the case of charmonium this mass is taken to be the mass of a $1S$-state without spin-dependent splittings
\begin{eqnarray}
M&=&\frac{1}{4}\Big(M(\eta_c)+3M(J/\psi)\Big)\approx3.07\:\textrm{GeV}\:,
\end{eqnarray}
while for bottomonium, due to the lack of information about the singlet mass, we choose
\begin{eqnarray}
M&=&2\,m_b\:,
\end{eqnarray}
as this choice is usually close to the mass of $\Upsilon(1S)$. With these preparations, the potential of the induced interaction reads:
\begin{eqnarray}
V_\textrm{ind}(\vec{r};\vec{p}_1,\vec{p}_2)\!\!\!\!&=&\!\!\!\!-\alpha_i\left[\frac{\mathrm{e}^{-Mr}}{r}+\frac{1}{8}\left(M^2\frac{\mathrm{e}^{-Mr}}{r}-4\pi\,\delta^{(3)}(\vec{r}\,)\right)\left(\frac{1}{m_1^2}+\frac{1}{m_2^2}\right)\right.\nonumber\\
&&\!\!\!\!\!\qquad+\frac{(\vec{s}_1\cdot\vec{s}_2)}{3m_1m_2}\left(M^2\frac{\mathrm{e}^{-Mr}}{r}-4\pi\delta^{(3)}(\vec{r}\,)\right)-\frac{\vec{p}_1\cdot\vec{p}_2}{m_1m_2}\frac{\mathrm{e}^{-Mr}}{r}\nonumber\\
&&\!\!\!\!\!\qquad-2\frac{\left(M^2+3M/r+3/r^2\right)}{3m_1m_2}\frac{\mathrm{e}^{-Mr}}{r}\left(3\frac{(\vec{s}_1\cdot\vec{r}\,)(\vec{s}_2\cdot\vec{r}\,)}{r^2}-(\vec{s}_1\cdot\vec{s_2})\right)\\
&&\!\!\!\!\!\qquad+\frac{\mathrm{e}^{-Mr}}{r^2}\left(M+\frac{1}{r}\right)\left(-\frac{(\vec{r}\times\vec{p}_1)\cdot\vec{s}_1}{2m_1^2}+\frac{(\vec{r}\times\vec{p}_2)\cdot\vec{s}_2}{2m_2^2}\right.\nonumber\\
&&\qquad\qquad\qquad\qquad\qquad\qquad\qquad\qquad\left.\left.-\frac{(\vec{r}\times\vec{p}_1)\cdot\vec{s}_2-(\vec{r}\times\vec{p}_2)\cdot\vec{s}_1}{m_1m_2}\right)\right]\:.\nonumber
\end{eqnarray}

%% file: 5-OGE-induced/5-OGE-Induced.tex
The origin of the perturbative corrections in this model is on the one hand the one-gluon exchange potential without retardation (Eq. (\ref{Eq:Potential-OGE-CM})) and on the other hand the induced interaction discussed in the first part of this chapter. Transformed into the center of mass frame by utilizing the substitution Eq. (\ref{Eq:Effective-CM-Transformation}) the potential for the induced interaction reads
\begin{eqnarray}\label{Eq:PotentialInducedInteraction-CM}
V_{\textrm{ind}}(\vec{r};\vec{p}\,)\!\!\!\!&=&\!\!\!\!-\frac{\alpha_i\mathrm{e}^{-Mr}}{r}-\frac{1}{4m^2_q}\left[\alpha_i M^2\frac{\mathrm{e}^{-Mr}}{r}-\alpha_i4\pi\delta^{(3)}(\vec{r}\,)\right]-\frac{\alpha_i\mathrm{e}^{-Mr}}{r}\frac{\vec{p}^{\,2}}{m^2_q}\nonumber\\
&&\!\!\!\!\!-\alpha_i\frac{(\vec{s}_1\cdot\vec{s}_2)}{3m^2_q}\left[ M^2\frac{\mathrm{e}^{-Mr}}{r}-4\pi\delta^{(3)}(\vec{r}\,)\right]+\alpha_i\left(M+\frac{1}{r}\right)\frac{\mathrm{e}^{-Mr}}{r^2}\left(\frac{3(\vec{r}\times\vec{p}\,)\cdot(\vec{s}_1+\vec{s}_2)}{2m^2_q}\right)\nonumber\\
&&\!\!\!\!\!+\frac{2\alpha_i}{3m^2_q}\left(M^2+\frac{3M}{r}+\frac{3}{r^2}\right)\frac{\mathrm{e}^{-Mr}}{r}\left(3\frac{(\vec{s}_1\cdot\vec{r}\,)(\vec{s}_2\cdot\vec{r}\,)}{r^2}-(\vec{s}_1\cdot\vec{s}_2)\right)\:.
\end{eqnarray}
By analysing this potential one notices that the perturbative corrections provided by the induced interaction differ in structure from the ones obtained by the one-gluon exchange (Eq. (\ref{Eq:Potential-OGE-CM})). In particular, the relative strength of the tensor and spin-orbit terms and their radius dependence differ from the ones in one-gluon exchange. The induced interaction might therefore be able to provide, together with the perturbative QCD part, a satisfactorily description of the triplet $P$-state splittings.

We have not yet succeeded in establishing a rigorous relation between the asymptotic normalization constant and the coupling constant of the induced interaction. This is mainly due to the linearly rising potential. There exists no upper boundary and therefore no binding energy. In practice we treat $\alpha_i$ as free parameter in this model.

%% file: 5-OGE-induced/5-Potential-Model.tex
The potential in this model is composed of a linear confining part, the one-gluon exchange potential (Eq. (\ref{Eq:Potential-OGE-CM})) and the potential for the induced interaction (Eq. (\ref{Eq:PotentialInducedInteraction-CM})):
\begin{eqnarray}\label{Eq:Potential-CpLpI-3}
V^\textrm{[CpLpI-3]}(\vec{r},\vec{p}\,)\!\!\!\!\!&=&\!\!\!\!\!-\frac{4\alpha_s}{3r}+\sigma r-\frac{\alpha_i\mathrm{e}^{-Mr}}{r}-\frac{1}{4m^2_q}\left[\alpha_i M^2\frac{\mathrm{e}^{-Mr}}{r}-\left(\frac{4}{3}\alpha_s+\alpha_i\right)4\pi\delta^{(3)}(\vec{r}\,)\right]\nonumber\\
&&\!\!\!\!\!-\left(\frac{4\alpha_s}{3r}+\frac{\alpha_i\mathrm{e}^{-Mr}}{r}\right)\frac{\vec{p}\cdot\vec{p}}{m^2_q}-\frac{(\vec{s}_1\cdot\vec{s}_2)}{3m^2_q}\left[\alpha_i M^2\frac{\mathrm{e}^{-Mr}}{r}-\left(\frac{8}{3}\alpha_s+\alpha_i\right)4\pi\delta^{(3)}(\vec{r}\,)\right]\nonumber\\
&&\!\!\!\!\!+\left(\frac{4\alpha_s}{3r^3}+\alpha_i M\frac{\mathrm{e}^{-Mr}}{r^2}+\alpha_i\frac{\mathrm{e}^{-Mr}}{r^3}\right)\left(\frac{3(\vec{r}\times\vec{p}\,)\cdot(\vec{s}_1+\vec{s}_2)}{2m^2_q}\right)\\
&&\!\!\!\!\!+\frac{2}{3m^2_q}\left[\frac{2\alpha_s}{r^3}+\alpha_i\frac{\mathrm{e}^{-Mr}}{r}\left(M^2+\frac{3M}{r}+\frac{3}{r^2}\right)\right]\left(3\frac{(\vec{s}_1\cdot\vec{r}\,)(\vec{s}_2\cdot\vec{r}\,)}{r^2}-(\vec{s}_1\cdot\vec{s}_2)\right)\:.\nonumber
\end{eqnarray}
As for all Coulomb-plus-Linear models the potential used in the Schr\"odinger equation to find bound states is
\begin{eqnarray}
V_0&=&-\frac{4\alpha_s}{3r}+\sigma r\:,
\end{eqnarray}
while the remaining parts of Eq. (\ref{Eq:Potential-CpLpI-3}) are treated within perturbation theory. The resulting formula for the mass of quarkonium states is given by
\begin{eqnarray}
M^\textrm{[CpLpI-3]}(k^{2S+1}l_j)\!\!\!\!\!&=&\!\!\!\!\!2m_q+E_{kl}-\alpha_i\left(1+\frac{M^2}{4m_q^2}\right)\bigg\langle\frac{\mathrm{e}^{-Mr}}{r}\bigg\rangle+\frac{\pi}{m_q^2}\left(\frac{4\alpha_s}{3}+\alpha_i\right)|\psi(0)|^2\nonumber\\
&&\!\!\!\!\!\!\!\!\!\!\!\!\!\!\!\!\!\!\!\!\!\!\!\!\!+\frac{1}{3m_q^2}\left(\frac{1}{2}S(S+1)-\frac{3}{4}\right)\left[\left(\frac{32\pi}{3}\alpha_s+4\pi\alpha_i\right)|\psi(0)|^2-\alpha_i M^2\bigg\langle\frac{\mathrm{e}^{-Mr}}{r}\bigg\rangle\right]\nonumber\\
&&\!\!\!\!\!\!\!\!\!\!\!\!\!\!\!\!\!\!\!\!\!\!\!\!\!+\frac{3(j(j+1)-l(l+1)-S(S+1))}{4m_q^2}\left[\alpha_s\bigg\langle\frac{4}{3r^3}\bigg\rangle+\alpha_i\left(\bigg\langle M\frac{\mathrm{e}^{-Mr}}{r^2}\bigg\rangle+\bigg\langle\frac{\mathrm{e}^{-Mr}}{r^3}\bigg\rangle\right)\right]\nonumber\\
&&\!\!\!\!\!\!\!\!\!\!\!\!\!\!\!\!\!\!\!\!\!\!\!\!\!+\frac{S_{12}}{4m_q^2}\left[\alpha_s\bigg\langle\frac{4}{3r^3}\bigg\rangle+2\alpha_i\left(\bigg\langle\frac{\mathrm{e}^{-Mr}}{r^3}\bigg\rangle+\bigg\langle M\frac{\mathrm{e}^{-Mr}}{r^2}\bigg\rangle+\bigg\langle M^2\frac{\mathrm{e}^{-Mr}}{3r}\bigg\rangle\right)\right]\\
&&\!\!\!\!\!\!\!\!\!\!\!\!\!\!\!\!\!\!\!\!\!\!\!\!\!+\frac{1}{m_q^2}\frac{1}{2}\int\mathrm{d}^3r\left(\psi^*\left[\frac{4\alpha_s}{3r}+\frac{\alpha_i\mathrm{e}^{-Mr}}{r}\right](\vec{\nabla}^{\,2}\psi)-\left[\frac{4\alpha_s}{3r}+\frac{\alpha_i\mathrm{e}^{-Mr}}{r}\right]\big|\:\vec{\nabla}\:\psi\:\big|^2\right)\:,\nonumber
\end{eqnarray}
The last term represents the expectation value of the purely momentum dependent terms with option 3 (see App. \ref{Chapt:Expect-values-of-MD}). Figures, tables and quantities related to this model are marked as [CpLpI-3].

%% file: 5-OGE-induced/5-Charmonium.tex
The determination of parameters in model [CpLpI-3] is accomplished by a fit to the masses of states $\eta_c$, $J/\psi$ and $h_c$, and the ratio $\Phi_c^\textrm{Exp}(1P)$ (Eq. (\ref{Eq:P-ProportionCharmoniumExperiment})), with the additional constraint that the parameters $\alpha_s$, $m_c$ and $\sigma$ should be as close as possible to their empirical values. This leads to the parameter set
\begin{eqnarray}
\alpha_s&=&0.25\:,\nonumber\\
\alpha_i&=&1.1\:,\nonumber\\
m_c&=&1.2877\:\textrm{GeV}\:,\\
\sigma&=&1.11\:\textrm{GeV/fm}\:.\nonumber
\end{eqnarray}
The charm quark mass of model [CpLpI-3] is similar to the ones obtain in the previously discussed models. The strong coupling constant takes a small value, while the string tension in this model is closer to the empirical lattice value than in the model without induced interaction.

\begin{center}
\begin{figure}[t!]\begin{center}
\includegraphics[width=0.9\textwidth]{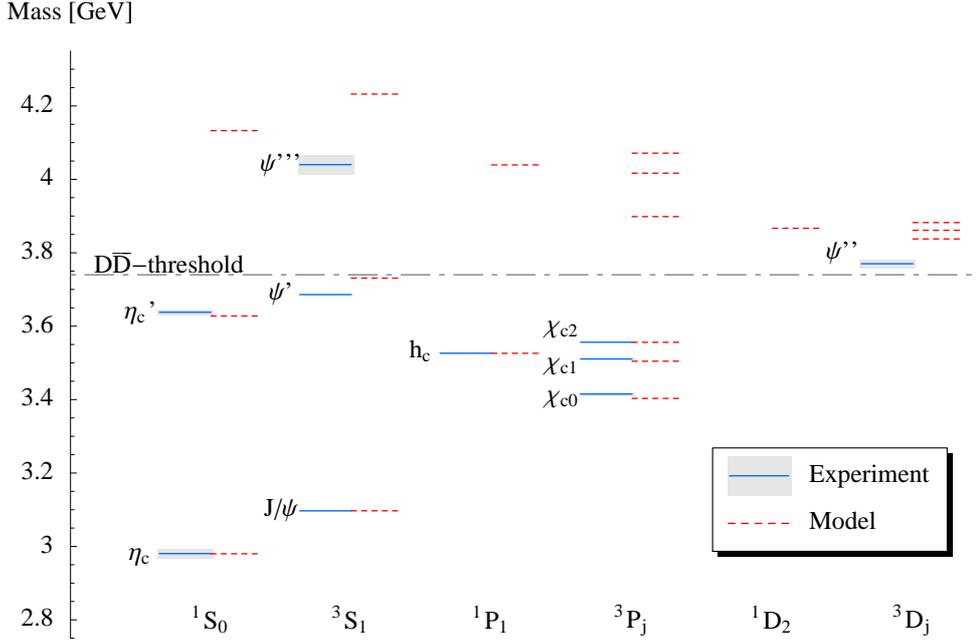}\end{center}
\caption{\label{Fig:SpectrumCharmoniumCpLpi-3} Charmonium spectra for experiment and model [CpLpI-3] ($\alpha_s=0.25$, $\alpha_i=1.1$, $\sigma=1.11\:\textrm{GeV/fm}$, $m_c=1.2877\:\textrm{GeV}$,  $M=3.07\:\textrm{GeV}$); for experimental data masses and widths are shown.}
\end{figure} 
\end{center}

The masses calculated in model [CpLpI-3] are compared with experimental data in Tab. \ref{Tab:MassesCharmoniumCpLpI-3}. The emerging spectrum is displayed together with the spectrum deduced from experiment in Fig. \ref{Fig:SpectrumCharmoniumCpLpi-3}. 
\begin{table}[b!]
\begin{center}
\begin{displaymath}
\begin{array}{||c|c|c|c|c||}
\hline\hline
&&\multicolumn{2}{|c|}{\qquad\textrm{Experiment \cite{PDBook}}\qquad} & \textrm{Theory [CpLpI-3]} \\
 \;\textrm{State}\; & \;\textrm{Candidate}\;&\multicolumn{2}{|c|}{\hrulefill} &  \\
&&\textrm{Mass [MeV]}&\textrm{Width [MeV]}&\textrm{Mass [MeV]}\\
\hline
1^1S_0 & \eta_c     & 2980.4   \pm 1.2   & 25.5   \pm 3.4      &  2980 \\
1^3S_1 & J/\psi     & 3096.916 \pm 0.011 & 0.0910 \pm 0.0032   &  3097 \\
\hline
1^1P_1 & h_c        & 3526.21  \pm 0.25  & <1.1                &  3526 \\
1^3P_0 & \chi_{c0}  & 3415.16  \pm 0.35  & 10.2   \pm 0.8      &  3403 \\
1^3P_1 & \chi_{c1}  & 3510.59  \pm 0.10  & 0.96   \pm 0.12     &  3505 \\
1^3P_2 & \chi_{c2}  & 3556.26  \pm 0.11  & 2.25   \pm 0.15     &  3556 \\
\hline
2^1S_0 & \eta_c'    & 3638     \pm 5     & 14     \pm 7        &  3628 \\
2^3S_1 & \psi'      & 3686.093 \pm 0.034 & 0.283  \pm 0.017    &  3731 \\
\hline
1^1D_2 &            &                    &                     &  3867 \\
1^3D_1 & \psi''     & 3770     \pm 2.4   & 23.6   \pm 2.7      &  3838 \\
1^3D_2 &            &                    &                     &  3861 \\
1^3D_3 &            &                    &                     &  3882 \\
\hline
2^1P_1 &            &                    &                     &  4039 \\
2^3P_0 &            &                    &                     &  3899 \\
2^3P_1 &            &                    &                     &  4017 \\
2^3P_2 &            &                    &                     &  4071 \\
\hline
3^1S_0 &            &                    &                     &  4133 \\
3^3S_1 & \psi'''    & 4040     \pm 10    & 52     \pm 10       &  4232 \\
\hline\hline
\end{array}
\end{displaymath}\vspace{-0.4cm}
\caption{\label{Tab:MassesCharmoniumCpLpI-3} $c\overline{c}$-state masses from experiment and model [CpLpI-3] ($\alpha_s=0.25$, $\alpha_i=1.1$, $\sigma=1.11\:\textrm{GeV/fm}$, $m_c=1.2877\:\textrm{GeV}$,  $M=3.07\:\textrm{GeV}$); experimental widths are also displayed.}
\end{center}
\end{table}

The main motivation to implement the induced interaction has been to reproduce the triplet $P$-state structure in greater detail. The resulting model ratio
\begin{eqnarray}\label{ProportionCpLpI-3}
\Phi_c^\textrm{[CpLpI-3]}(1P)=\left(\frac{M(\chi_{c,2})-M(\chi_{c,1})}{M(\chi_{c,1})-M(\chi_{c,0})}\right)_{\textrm{[CpLpI-3]}}\approx0.5\:,
\end{eqnarray}
is now very close to the experimental value
\begin{eqnarray}
\Phi_c^\textrm{Exp}(1P)=0.482\pm0.05\:.
\end{eqnarray}
Although the magnitude of the triplet $P$-state splitting is increased by about 10\%, the consideration of the induced interacton within model [CpLpI-3] leads to an enormous improvement compared to model [CpL-3], with its built-in ratio
\begin{eqnarray}
\Phi_q^{\textrm{[CpL-3]}}(nP)=\frac{4}{5}\:.
\end{eqnarray}

%% file: 5-OGE-induced/5-Bottomonium.tex
The input used to determine the parameters of model [CpLpI-3] is given by the states $\Upsilon(1S)$, $\Upsilon(2S)$ and $C(1P)$, and the ratio $\Phi_b^\textrm{Exp}(1P)$, with the additional constraint that the parameters $\alpha_s$, $m_b$ and $\sigma$ should be as close as possible to their empirical values. In the analysis of model [CpLpI-3] we found that we are able to improve the structure the triplet $P$-state splitting, but this improvement comes at the cost of parameters which are even further apart the empirical values as the ones in the model without induced interaction, or the incapability to match the states $\Upsilon(1S)$, $\Upsilon(2S)$ and $C(1P)$. It occurs that this is again a problem due to the special role of the $1S$-state.

%% file: 5-OGE-induced/5-Summary.tex
In this chapter we exploited crossing symmetry to obtain an induced interaction which is equivalent to the t-channel exchange of bound states. By considering only the lowest lying states in the spectra, we computed a potential for the t-channel exchange of $1S$-states which arises from pseudoscalar and vector meson exchange processes. 

On the basis of these concepts we introduced the induced interaction to the Coulomb-plus-Linear model [CpL-3], leading to additional perturbative corrections which allowed us to reproduce the triplet $P$-state structure. 

The model works very well for charmonium, yielding a value for the string tension close to the empirical one for the first time. For bottomonium we face problems due to the distinct influence  of the Coulomb-part of the potential on the $1S$-states as compared to the rest of the bound states. 
The situation for the strong coupling constant does not improve up to this point, still taking a larger value for bottomonium than for charmonium.

%% file: 6-Fogtpp/6-Intro.tex
In this chapter we adopt a gluon exchange potential derived by Gupta \textit{et al.}, which includes two-gluon exchange diagrams. The potential is used, with slight modifications, as short range part in a Coulomb-plus-linear model. After the dicussion and motivation of the model in the first two sections, we examine the effect of additional corrections found by Titard and Yndur\`ain.

%% file: 6-Fogtpp/6-Fogtpp.tex
The Coulomb-plus-Linear models we have discussed until now have employed an one-gluon exchange process to construct the short range part of the potential. The natural next step for a further improvement of the model is to consider not only an one-gluon exchange process but processes of higher order in perturbative QCD. Such a potential has already been calculated by Gupta \textit{et al.} in the early 80's \cite{Gupta:1981pd,Gupta:1982qc}. They derived a fourth-order gluonic quark-antiquark potential, which has later been slightly corrected by Titard and Yndur\`ain \cite{Titard:1993nn}. Gupta \textit{et al.} used this potential as short range part of a quarkonium potential model \cite{Gupta:1982kp}. The perturbative part of this potential model is, with a slight modification, adopted by us and taken as the short range part of the model potential.

The modification we apply to this potential is given by a reduction of the spin-independent contact term and its radiative correction by a factor of two. We think that this adjustment is necessary as we already know the strength of the leading term of the contact structure from one-gluon exchange potentials. The consideration of higher order processes in perturbative QCD should not change the strength of the leading term. We can compare the potential in \cite{Gupta:1982kp} with Eq. (\ref{Eq:Potential-OGE-CM}) and realize that the strength has to be reduced by a factor of two. The reason for this discrepancy seems to be the choice of kinematics. As we point out in Appendix \ref{Sec:Kinematics}, the choice of kinematics is important, meaning that different choices may lead to differences in the momentum dependent parts of these potentials. The kinematics choice we use to obtain a potential is dictated by invariance under time reversal, while the kinematics employed in \cite{Gupta:1981pd,Titard:1993nn} lead to a potential which is not invariant under the time reversal operation (see App. \ref{Sec:Kinematics}). In the center of mass frame the additional non-time reversal invariant terms of these potentials can be transformed into an additional contact term, by exploiting certain identities in the momentum space representation of the potential. Hence we modify the strength of the contact term in \cite{Gupta:1982kp} to half of its value and obtain
\begin{eqnarray}\label{Eq:PotentialGRm}
V^{(4)}(\vec{r};\vec{p}\,)\!\!\!\!\!&=&\!\!\!\!\!-\frac{4\alpha_s}{3r}\left[1-\frac{3\alpha_s}{2\pi}+\frac{\alpha_s}{6\pi}(33-2n_f)[\ln(\mu_{\textrm{GR}} r)+\gamma_\textrm{E}]\right]\nonumber\\
&&\!\!\!\!\!-\frac{4\alpha_s}{3m_q^2r}\left[1-\frac{3\alpha_s}{2\pi}+\frac{\alpha_s}{6\pi}(33-2n_f)[\ln(\mu_{\textrm{GR}} r)+\gamma_\textrm{E}]\right]\vec{p}^{\,2}\nonumber\\
&&\!\!\!\!\!+\frac{4\pi\alpha_s}{3m_q^2}\left[\left(1-\frac{3\alpha_s}{2\pi}\right)\delta^{(3)}\left(\vec{r}\,\right)-\frac{\alpha_s}{24\pi^2}(33-2n_f)\vec{\nabla}^{\,2}\left(\frac{\ln(\mu_{\textrm{GR}} r)+\gamma_\textrm{E}}{r}\right)\right]-\frac{14\alpha_s^2}{9m_qr^2}\nonumber\\
&&\!\!\!\!\!+\frac{32\pi\alpha_s}{9m_q^2}(\vec{s}_1\cdot\vec{s}_2)\left[\left(1-\frac{\alpha_s}{12\pi}(26+9\ln2)\right)\delta^{(3)}(\vec{r}\,)\right.\nonumber\\
&&\qquad\qquad\left.-\frac{\alpha_s}{24\pi^2}(33-2n_f)\vec{\nabla}^{\,2}\left(\frac{\ln(\mu_{\textrm{GR}} r)+\gamma_\textrm{E}}{r}\right)+\frac{21\alpha_s}{16\pi^2}\vec{\nabla}^{\,2}\left(\frac{\ln(m_qr)+\gamma_\textrm{E}}{r}\right)\right]\nonumber\\
&&\!\!\!\!\!+\frac{4\alpha_s}{3m_q^2}\frac{3(\vec{s}_1\cdot\hat{r})(\vec{s}_2\cdot\hat{r})-\vec{s_1}\cdot\vec{s}_2}{r^3}\left[1+\frac{4\alpha_s}{3\pi}\right.\\
&&\qquad\qquad\qquad\left.+\frac{\alpha_s}{6\pi}(33-2n_f)[\ln(\mu_{\textrm{GR}} r)+\gamma_\textrm{E}-\frac{4}{3}]-\frac{3\alpha_s}{\pi}[\ln(m_q r)+\gamma_\textrm{E}-\frac{4}{3}]\right]\nonumber\\
&&\!\!\!\!\!+\frac{2\alpha_s}{m_q^2}\frac{\vec{L}\cdot\vec{S}}{r^3}\!\left[1\!-\!\frac{11\alpha_s}{18\pi}+\frac{\alpha_s}{6\pi}(33-2n_f)[\ln(\mu_{\textrm{GR}} r)+\!\gamma_\textrm{E} -1]\!-\!\!\frac{2\alpha_s}{\pi}[\ln(m_qr)+\gamma_\textrm{E}-1]\right]\:.\nonumber
\end{eqnarray}
This perturbative QCD part of the potential, to order $\alpha_s^2$, is then treated as a perturbative correction to the confining potential.

%% file: 6-Fogtpp/6-Potential-Model.tex
The potential in this model is composed of the potential given in Eq. (\ref{Eq:PotentialGRm}) and a linear confining part:
\begin{eqnarray}\label{Eq:PotentialCpL*}
V^\textrm{[CpL*-2]}(\vec{r};\vec{p}\,)=V^{(4)}(\vec{r};\vec{p}\,)+\sigma r\:.
\end{eqnarray}

When trying to implement the radiative corrections to the leading Coulomb term into the potential used in the Schr\"odinger equation, this would lead to problems. One realizes that
\begin{eqnarray}
\lim_{r\rightarrow 0}\left(-\frac{4\alpha_s}{3r}\left[1-\frac{3\alpha_s}{2\pi}+\frac{\alpha_s}{6\pi}(33-2n_f)[\ln(\mu_{\textrm{GR}} r)+\gamma_\textrm{E}]\right]\right)&\longrightarrow&+\infty\:,
\end{eqnarray}
due to the logarithmic correction. This would imply that all wavefunctions should vanish at the origin
\begin{eqnarray}\label{Eq:WavefunctionOriginLog}
|\psi_{kl}(0)|&=&0\:,
\end{eqnarray}
leading to an almost vanishing splitting for the ${}^3S_1$ and ${}^1S_0$-states. A meaningful procedure is to start from solving the Schr\"odinger equation with the potential
\begin{eqnarray}
V_0(r)&=&-\frac{4\alpha_s}{3r}+\sigma r\:,
\end{eqnarray}
while the remaining parts of Eq. (\ref{Eq:PotentialCpL*}) are treated as a perturbation. The resulting mass formula is given by
\begin{eqnarray}\label{Eq:MassformularCpL*-2}
M^\textrm{[CpL*-2]}(k^{2S+1}l_j)\!\!\!\!\!&=&\!\!\!\!\!2m_q+E_{kl}+\frac{2\alpha_s^2}{\pi}\left(\bigg\langle\frac{1}{r}\bigg\rangle-\frac{33-2n_f}{9}\bigg\langle\frac{\ln\left(\mu_{\textrm{GR}}r\right)+\gamma_\textrm{E}}{r}\bigg\rangle\right)-\bigg\langle\frac{14\alpha_s^2}{9m_qr^2}\bigg\rangle\nonumber\\
&&\!\!\!\!\!\!\!\!\!\!\!\!\!\!\!\!\!\!\!\!+\frac{4\pi\alpha_s}{3m_q^2}\left(1-\frac{3\alpha_s}{2\pi}\right)\big|\:\psi(0)\:\big|^2-\frac{\alpha_s^2(33-2n_f)}{18\pi m_q^2}\int\mathrm{d}^3r\:\frac{\ln(\mu_{\textrm{GR}}r)+\gamma_\textrm{E}}{r}\Delta\big|\:\psi(\vec{r}\,)\:\big|^2\nonumber\\
&&\!\!\!\!\!\!\!\!\!\!\!\!\!\!\!\!\!\!\!\!+\frac{32\pi\alpha_s}{9m_q^2}\left(\frac{1}{2}S(S+1)-\frac{3}{4}\right)\Bigg\{\left[1-\frac{\alpha_s}{12\pi}\left(26+9\ln2\right)\right]\:\big|\:\psi(0)\:\big|^2\nonumber\\
&&\!\!\!\!\!\!\!\!\!\!\!\!\!\!\!\quad+\int\mathrm{d}^3r\left(\frac{21\alpha_s}{16\pi^2}\frac{\ln(m_q r)+\gamma_\textrm{E}}{r}-\frac{\alpha_s(33-2n_f)}{24\pi^2}\frac{\ln(\mu_{\textrm{GR}}r)+\gamma_\textrm{E}}{r}\right)\Delta\big|\:\psi(\vec{r}\,)\:\big|^2\Bigg\}\nonumber\\
&&\!\!\!\!\!\!\!\!\!\!\!\!\!\!\!\!\!\!\!\!+\frac{\alpha_sS_{12}}{3m_q^2}\Bigg\{\left(1+\frac{4\alpha_s}{3\pi}\right)\bigg\langle\frac{1}{r^3}\bigg\rangle+\frac{\alpha_s}{6\pi}(33-2n_f)\bigg\langle\frac{\ln(\mu_{\textrm{GR}} r)+\gamma_\textrm{E}-\frac{4}{3}}{r^3}\bigg\rangle\nonumber\\
&&\!\!\!\!\!\!\!\!\!\!\!\!\!\!\!\!\!\!\!\!\qquad\qquad\qquad\qquad\qquad\qquad\qquad\qquad\qquad\qquad\qquad-\frac{3\alpha_s}{\pi}\bigg\langle\frac{\ln(m_q r)+\gamma_\textrm{E}-\frac{4}{3}}{r^3}\bigg\rangle\Bigg\}\nonumber\\
&&\!\!\!\!\!\!\!\!\!\!\!\!\!\!\!\!\!\!\!\!+\frac{2\alpha_s}{m_q^2}\frac{j(j+1)-l(l+1)-S(S+1)}{2}\Bigg\{\left(1-\frac{11\alpha_s}{18\pi}\right)\bigg\langle\frac{1}{r^3}\bigg\rangle\\
&&\!\!\!\!\!\!\!\!\!\!\!\!\!\!\!\!\!\!\!\!\qquad\qquad\quad+\frac{\alpha_s(33-2n_f)}{6\pi}\bigg\langle\frac{\ln(\mu_{\textrm{GR}} r)+\gamma_\textrm{E} -1}{r^3}\bigg\rangle-\frac{2\alpha_s}{\pi}\bigg\langle\frac{\ln(m_q r)+\gamma_\textrm{E}-1}{r^3}\bigg\rangle\Bigg\}\nonumber\\
&&\!\!\!\!\!\!\!\!\!\!\!\!\!\!\!\!\!\!\!\!-\frac{4\alpha_s}{3m_q^2}\int\mathrm{d}^3r\:\frac{1}{r}\left[1-\frac{3\alpha_s}{2\pi}+\frac{\alpha_s}{6\pi}(33-2n_f)[\ln(\mu r)+\gamma_\textrm{E}]\right]\big|\:\vec{\nabla}\:\psi(\vec{r}\,)\:\big|^2\:.\nonumber
\end{eqnarray}
The last line represents the expectation value of the purely momentum dependent term using option 2 (see App. \ref{Chapt:Expect-values-of-MD}), as this scheme provides parameters closest to the empirical values.

One recognizes that Eq. (\ref{Eq:MassformularCpL*-2}) includes the number of light flavours and a renormalization scale. These quantities will not be treated as free parameters. Moreover the renormalization scale is given for a scheme by Gupta and Radford \cite{Gupta:1982im}, which differs from the common $\overline{\mathrm{MS}}$-scheme. Gupta and Radford give the relation 
\begin{eqnarray}\label{Eq:GRMS}
\mu_{\overline{\textrm{MS}}}= \mu_{\textrm{GR}}\exp\left(-\frac{49-\frac{10}{3}n_f}{2(33-2n_f)}\right)\:,
\end{eqnarray}
which connects their subtraction scheme with the renormalization scale of the $\overline{\textrm{MS}}$ scheme. We have checked and confirmed this relation by comparing \cite{Titard:1993nn} and \cite{Gupta:1982kp}. The number of light flavours is set to
\begin{eqnarray}
n_f&=&3\quad \textrm{for Charmonium,}\nonumber\\
n_f&=&4\quad \textrm{for Bottomonium.}
\end{eqnarray}
Usually the first guess for the renormalization scale in the modified minimal subtraction scheme $\overline{\mathrm{MS}}$ would be $\mu_{\overline{\textrm{MS}}}=m_q$.
Titard and Yndur\`ain maintain in \cite{Titard:1993nn} that the natural scale is not $\mu_{\overline{\textrm{MS}}}=m_q$, but determined by the average momentum of the quarks. Typically the average momentum for the ground state in our model is
\begin{eqnarray}\label{Eq:TypicalMomenta}
\langle \vec{p}^{\,2}\rangle^{1/2}&\approx&0.4m_c\quad\textrm{for Charmonium}\:,\nonumber\\
\langle \vec{p}^{\,2}\rangle^{1/2}&\approx&0.25m_b\quad\textrm{for Bottomonium}\:.
\end{eqnarray}
We take
\begin{eqnarray}
\mu_{\overline{\textrm{MS}}}&\approx&\frac{1}{2}m_c\quad\textrm{for Charmonium}\:,\nonumber\\
\mu_{\overline{\textrm{MS}}}&\approx&\frac{1}{3}m_b\quad\textrm{for Bottomonium}\:.
\end{eqnarray}
The analysis in \cite{Titard:1993nn} showed that this should be a good choice for bottomonium, while it might be unreliable for charmonium. The coupling constant evaluated at this scale is $\alpha_s(m_c^2/4)>0.5$, and therefore the calculation of the short range part via perturbative QCD is not justified. Nevertheless we consider this value for the scale as we do not know how to fix the scale more reliably. The model calculations are performed in the renormalization scheme by Gupta and Radford \cite{Gupta:1982im}. The relation Eq. (\ref{Eq:GRMS})
results for the considered cases of $n_f=3$ or $n_f=4$ in
\begin{eqnarray}
\mu_{\overline{\textrm{MS}}}\approx\frac{\mu_{\textrm{GR}}}{2}\:.
\end{eqnarray}
This implies
\begin{eqnarray}\label{Eq:ScaleGR}
\textrm{Charmonium:}&&\mu_{\textrm{GR}}=m_c\:,\nonumber\\
\textrm{Bottomonium:}&&\mu_{\textrm{GR}}=\frac{2}{3}m_b\:.
\end{eqnarray}
Figures, tables, and quantities related to this model are marked as [CpL*-2].

%% file: 6-Fogtpp/6-Charmonium.tex
We tried to fix the parameters of model [CpL*-2] analogously to models [CpL-B] and [CpL-3], by matching the masses of states $\eta_c$, $J/\psi$, and $h_c$, with the additional constraint that the parameters should be as close as possible to their empirical values. Problems were encountered in reproducing the $1S$ spin-spin splitting, which is, for parameters within reasonable intervals, significantly to small in this model. The parameter set
\begin{eqnarray}\label{Eq:ParameterFogtpp}
\alpha_s&=&0.27\:,\nonumber\\
m_c&=&1.326\:\textrm{GeV}\:,\\
\sigma&=&1.2\:\textrm{GeV/fm}\:,\nonumber
\end{eqnarray}
is constrained by reproduction of $h_c $, $\chi_1$ and the center of gravity $1S$-state mass
\begin{eqnarray}
C(1S)=\frac{1}{4}(M(\eta_c)+3M(J/\psi))\:,
\end{eqnarray}
with the spin-spin splitting as close as possible to the experimental one.

\begin{center}
\begin{figure}[p]\begin{center}
\includegraphics[width=0.9\textwidth]{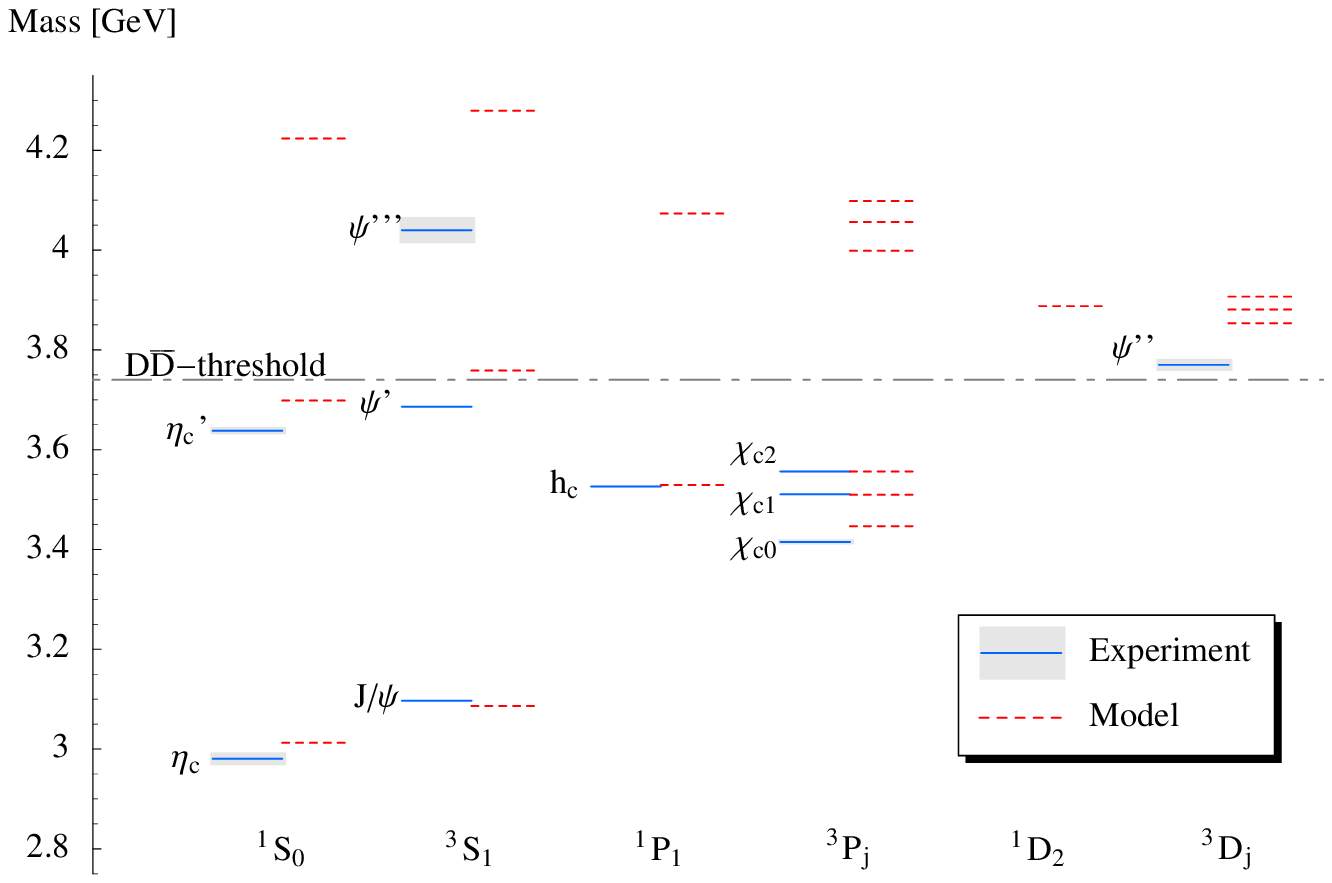}\end{center}
\caption{\label{Fig:SpectrumCharmoniumCpL*-2} Charmonium spectra for experiment and model [CpL*-2] ($\alpha_s=0.27$, $\sigma=1.2\:\textrm{GeV/fm}$, $m_c=1.326\:\textrm{GeV}$, $\mu_{\textrm{GR}}=m_c$, $n_f=3$); for experimental data masses and widths are shown.}
\end{figure} 
\end{center}

The resulting masses are shown together with experimental values for the masses and widths in Tab. \ref{Tab:MassesCharmoniumCpL*-2}. In Fig \ref{Fig:SpectrumCharmoniumCpL*-2} the arising spectrum is opposed to the experimental one.
\begin{table}[ht!]
\begin{center}
\begin{displaymath}
\begin{array}{||c|c|c|c|c||}
\hline\hline
&&\multicolumn{2}{|c|}{\qquad\textrm{Experiment \cite{PDBook}}\qquad} & \textrm{Theory [CpL*-2]}\\
 \;\textrm{State}\; & \;\textrm{Candidate}\;&\multicolumn{2}{|c|}{\hrulefill} &  \\
&&\textrm{Mass [MeV]}&\textrm{Width [MeV]}&\textrm{Mass [MeV]}\\
\hline
1^1S_0 & \eta_c     & 2980.4   \pm 1.2   & 25.5   \pm 3.4      &  3012 \\
1^3S_1 & J/\psi     & 3096.916 \pm 0.011 & 0.0910 \pm 0.0032   &  3086 \\
\hline
1^1P_1 & h_c        & 3526.21  \pm 0.25  & <1.1                &  3529 \\
1^3P_0 & \chi_{c0}  & 3415.16  \pm 0.35  & 10.2   \pm 0.8      &  3447 \\
1^3P_1 & \chi_{c1}  & 3510.59  \pm 0.10  & 0.96   \pm 0.12     &  3510 \\
1^3P_2 & \chi_{c2}  & 3556.26  \pm 0.11  & 2.25   \pm 0.15     &  3556 \\
\hline
2^1S_0 & \eta_c'    & 3638     \pm 5     & 14     \pm 7        &  3698 \\
2^3S_1 & \psi'      & 3686.093 \pm 0.034 & 0.283  \pm 0.017    &  3759 \\
\hline
1^1D_2 &            &                    &                     &  3888 \\
1^3D_1 & \psi''     & 3770     \pm 2.4   & 23.6   \pm 2.7      &  3853 \\
1^3D_2 &            &                    &                     &  3881 \\
1^3D_3 &            &                    &                     &  3907 \\
\hline
2^1P_1 &            &                    &                     &  4073 \\
2^3P_0 &            &                    &                     &  3999 \\
2^3P_1 &            &                    &                     &  4056 \\
2^3P_2 &            &                    &                     &  4098 \\
\hline
3^1S_0 &            &                    &                     &  4224 \\
3^3S_1 & \psi'''    & 4040     \pm 10    & 52     \pm 10       &  4279 \\
\hline\hline
\end{array}
\end{displaymath}
\caption{\label{Tab:MassesCharmoniumCpL*-2} $c\overline{c}$-state masses from experiment and model [CpL*-2] ($\alpha_s=0.27$, $\sigma=1.2\:\textrm{GeV/fm}$, $m_c=1.326\:\textrm{GeV}$, $\mu_{\textrm{GR}}=m_c$, $n_f=3$); masses and widths are displayed for the experiment.}
\end{center}
\end{table}
One recognizes that this model yields only about 60\% of the experimental $1^3S_1-1^1S_0$ splitting. Examining Eq. (\ref{Eq:MassformularCpL*-2}) one finds that there is no built-in value
\begin{eqnarray}\label{Eq:PWaveProportionCpL}
\left(\frac{M(\chi_{q,2})-M(\chi_{q,1})}{M(\chi_{q,1})-M(\chi_{q,0})}\right)=\frac{4}{5}\:,
\end{eqnarray}
as in the models [CpL-B] and [CpL-3] anymore. In model [CpL*-2] this ratio is now parameter dependent. With the parameter set Eq. (\ref{Eq:ParameterFogtpp}) one finds
\begin{eqnarray}
\Phi_c^\textrm{[CpL*-2]}(1P)=\left(\frac{M(\chi_{c,2})-M(\chi_{c,1})}{M(\chi_{c,1})-M(\chi_{c,0})}\right)_{\textrm{[CpL*-2]}}\approx0.74\:.
\end{eqnarray}
This illustrates, that while this proportion differs only slightly from its value for the basic Coulomb-plus-Linear model, it changes at least in the right direction.

%% file: 6-Fogtpp/6-Bottomonium.tex
\begin{center}
\begin{figure}[p]\begin{center}
\includegraphics[width=0.9\textwidth]{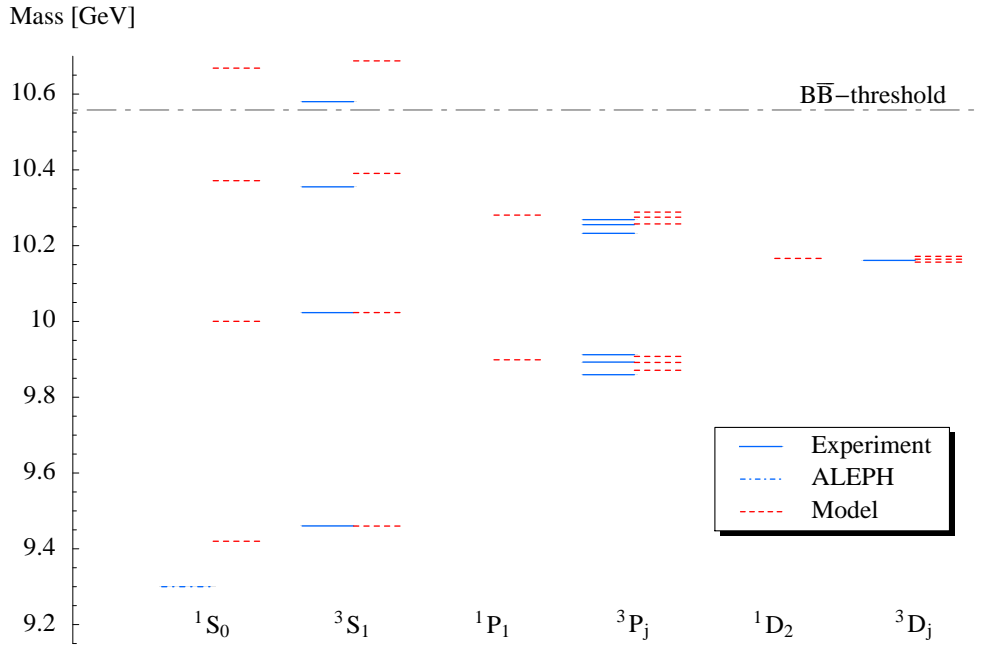}\end{center}
\caption{\label{Fig:SpectrumBottomoniumCpL*-2} Bottomonium spectra from experiment and model [CpL*-2] ($\alpha_s=0.27$, $\sigma=1.05\:\textrm{GeV/fm}$, $m_b=4.7877\:\textrm{GeV}$, $\mu_{\textrm{GR}}=2/3m_b$, $n_f=4$); $\eta_b$ mass measurement by ALEPH collaboration is included separately in the experimental spectrum.}
\end{figure} 
\end{center}
The parameters of model [CpL*-2] for bottomonium
\begin{eqnarray}
\alpha_s&=&0.27\:,\nonumber\\
m_b&=&4.7877\:\textrm{GeV}\:,\\
\sigma&=&1.05\:\textrm{GeV/fm}\:,\nonumber
\end{eqnarray}
are determined by a fit to the masses of states $\Upsilon(1S)$, $\Upsilon(2S)$ and $C(1P)$ (Eq. (\ref{Eq:CoG-1P-Bottomonium})), by identifying $C(1P)$ with the $1^1P_1$ state of model [CpL*-2]. Remarkably the model yields for the first time a coupling strength which is smaller than $\alpha_s=0.3$, while the string tension and the bottom quark mass coincide with empirical values.

The results for Eq. (\ref{Eq:MassformularCpL*-2}) are presented together with experimental data for masses and widths in Tab. \ref{Tab:MassesBottomCpL*-2}, while the according spectra are shown in Fig. \ref{Fig:SpectrumBottomoniumCpL*-2}.
\begin{table}[t!]
\begin{center}
\begin{displaymath}
\begin{array}{||c|c|c|c|c||}
\hline\hline
&&\multicolumn{2}{|c|}{\qquad\textrm{Experiment \cite{PDBook}}\qquad} & \textrm{Theory [CpL*-2]} \\
 \;\textrm{State}\; & \;\textrm{Candidate}\;&\multicolumn{2}{|c|}{\hrulefill} &  \\
&&\textrm{Mass [MeV]}&\textrm{Width [MeV]}& \textrm{Mass [MeV]}\\
\hline
1^1S_0 & (\eta_b)       & (9300\pm20\pm20)       &                                 &   9420 \\
1^3S_1 & \Upsilon(1S)   & 9460\pm0.26            & (53.0\pm1.5)\textrm{ keV}       &   9460 \\
\hline
1^1P_1 &                &                        &                                 &   9899 \\
1^3P_0 & \chi_{b0}(1P)  & 9859.44\pm0.42\pm0.31  &                                 &   9871 \\
1^3P_1 & \chi_{b1}(1P)  & 9892.78\pm0.26\pm0.31  &                                 &   9892 \\
1^3P_2 & \chi_{b2}(1P)  & 9912.21\pm0.26\pm0.31  &                                 &   9908 \\
\hline
2^1S_0 &                &                        &                                 &  10003 \\
2^3S_1 & \Upsilon(2S)   & 10023.26\pm0.31        & (30.6\pm2.3)\textrm{ keV}       &  10023 \\
\hline
1^1D_2 &                &                        &                                 &  10166 \\
1^3D_1 &                &                        &                                 &  10157 \\
1^3D_2 & \Upsilon(1D)   & 10161.1\pm0.6\pm1.6    &                                 &  10164 \\
1^3D_3 &                &                        &                                 &  10172 \\
\hline
2^1P_1 &                &                        &                                 &  10281 \\
2^3P_0 & \chi_{b0}(2P)  & 10232.5\pm0.4\pm0.5    &                                 &  10257 \\
2^3P_1 & \chi_{b1}(2P)  & 10255.46\pm0.22\pm0.50 &                                 &  10275 \\
2^3P_2 & \chi_{b2}(2P)  & 10268.65\pm0.22\pm0.50 &                                 &  10289 \\
\hline
3^1S_0 &                &                        &                                 &  10371 \\
3^3S_1 & \Upsilon(3S)   & 10355.2\pm0.5          & (22.1\pm2.7)\textrm{ keV}       &  10391 \\
\hline
4^1S_0 &                &                        &                                 &  10669 \\
4^3S_1 & \Upsilon(4S)   & 10580.0\pm3.5          & 20\pm2\pm4                      &  10688 \\
\hline\hline
\end{array}
\end{displaymath}\vspace{-0.3cm}
\caption{\label{Tab:MassesBottomCpL*-2} $b\overline{b}$-state masses from experiment and model [CpL*-2] ($\alpha_s=0.27$, $\sigma=1.05\:\textrm{GeV/fm}$, $m_b=4.7877\:\textrm{GeV}$, $\mu_{\textrm{GR}}=2/3m_b$, $n_f=4$); masses and widths are displayed for the experiment.}
\end{center}\vspace{-0.5cm}
\end{table}
Examining the spectrum one immediately notes that the triplet $P$-state splitting is to small by about 25\%. The proportions
\begin{eqnarray}
\Phi_b^\textrm{[CpL*-2]}(1P)=\left(\frac{M(\chi_{b2})-M(\chi_{b1})}{M(\chi_{b1})-M(\chi_{b0})}\right)_{\textrm{[CpL*-2]}}&\approx&0.77\:,\\
\Phi_b^\textrm{[CpL*-2]}(2P)=\left(\frac{M(\chi'_{b2})-M(\chi_{b1})}{M(\chi'_{b1})-M(\chi'_{b0})}\right)_{\textrm{[CpL*-2]}}&\approx&0.76\:.
\end{eqnarray}
show the same slight decrease compared to Eq. (\ref{Eq:PWaveProportionCpL}), which one has already observed for charmonium.

%% file: 6-Fogtpp/6-Titard-Yndurain.tex
In \cite{Titard:1993nn} Titard and Yndur\`ain show additional terms to the potential by Gupta \textit{et al.} \cite{Gupta:1982kp}. They maintain that the order $\mathcal{O}(m^{-2})$ for the spin-independent terms is not complete, plus they show a recipe how they calculated the part involving a fictive gluon mass. Having implemented perturbative corrections according to the potential computed by Gupta \textit{et al.} we have to check the effect of these additional corrections.

The additional terms shown in \cite{Titard:1993nn} are
\begin{eqnarray}\label{Eq:AdditionalTermsYndurain}
\delta V_\textrm{SI}^\textrm{Y}(\vec{q}\,)=\frac{4\alpha_s^2}{3m_q^2}\left(-\frac{133}{9}\ln\frac{|\vec{q}\,|}{m_q}+\frac{28}{9}\ln\frac{\lambda}{m_q}+\frac{128}{15}+2\ln2\right)\:,
\end{eqnarray}
with $\lambda$ the fictive gluon mass. Titard and Yndur\`ain define the spin-spin part in a way different from Gupta \textit{et al.}, leading to different terms in spin-independent $\mathcal{O}(m^{-2})$. We first have to relate these terms to the ones we use. We subtract the spin-independent part of
\begin{eqnarray}
\delta V_\textrm{SS}^\textrm{GR}(\vec{q}\,)&=&-\frac{2\alpha_s^2}{3m_q^2}\left(\frac{8}{9}+\ln2-7\ln\frac{|\vec{q}\,|}{m_q}\right)\left(\vec{\sigma}_1\cdot\vec{\sigma}_2\right)\nonumber\\
&=&-\frac{2\alpha_s^2}{3m_q^2}\left(\frac{8}{9}+\ln2-7\ln\frac{|\vec{q}\,|}{m_q}\right)\left(2\vec{S}^{\,2}-3\right)\:,
\end{eqnarray}
from Eq. (\ref{Eq:AdditionalTermsYndurain}) to obtain the additional terms in our definition:
\begin{eqnarray}
Y(\vec{q}\,)=\frac{4\alpha_s^2}{3m_q^2}\left(-\frac{77}{18}\ln\frac{|\vec{q}\,|}{m_q}+\frac{28}{9}\ln\frac{\lambda}{m_q}+\frac{36}{5}+\frac{1}{2}\ln2\right)\:.
\end{eqnarray}
Transforming to $r$-space this gives
\begin{eqnarray}\label{Eq:AdditionalTermsYndurainGRDefCoordinate}
Y(\vec{r}\,)=\frac{4\alpha_s^2}{3m_q^2}\left[-\frac{77}{72\pi}\vec{\nabla}^{2}\left(\frac{\ln(m_qr)+\gamma_\textrm{E}}{r}\right)+\delta^{(3)}(\vec{r}\,)\left(\frac{28}{9}\ln\frac{\lambda}{m_q}+\frac{36}{5}+\frac{1}{2}\ln2\right)\right]\:.
\end{eqnarray}
\begin{table}[t!]
\begin{center}
\begin{displaymath}
\begin{array}{||c|c|c||}
\hline\hline
&  \qquad   \langle Y(\vec{r}\,)\rangle\qquad    \textrm{[MeV]}  \qquad  & \qquad \langle Y(\vec{r}\,)\rangle \qquad\textrm{[MeV]}\qquad \\
\qquad\textrm{State}\qquad &    &\\
               & \textrm{with Eq. (\ref{Eq:GluonZero})}  & \textrm{with Eq. (\ref{Eq:GluonLamb})}  \\
\hline
1S             & 17.4 &  1.1\\
2S             & 11.6 & -1.5\\
3S             &  9.5 & -2.4\\
1P             &  1.9 &  1.9\\
2P             &  1.8 &  1.8\\
1D             &  0.5 &  0.5\\
\hline\hline
\end{array}
\end{displaymath}
\caption{\label{Tab:CharmoniumYndurainterme} Charmonium expectation values for the additional terms presented in \cite{Titard:1993nn} calculated within model [CpL*-2] ($\alpha_s=0.27$, $\sigma=1.2\:\textrm{GeV/fm}$, $m_c=1.326\:\textrm{GeV}$, $\mu_{\textrm{GR}}=m_c$, $n_f=3$).}
\end{center}\vspace{-0.5cm}
\end{table}
\begin{table}[t!]
\begin{center}
\begin{displaymath}
\begin{array}{||c|c|c||}
\hline\hline
&  \qquad   \langle Y(\vec{r}\,)\rangle\qquad    \textrm{[MeV]}  \qquad  & \qquad \langle Y(\vec{r}\,)\rangle \qquad\textrm{[MeV]}\qquad \\
\qquad\textrm{State}\qquad &    &\\
               & \textrm{with Eq. (\ref{Eq:GluonZero})}  & \textrm{with Eq. (\ref{Eq:GluonLamb})}  \\
\hline
1S             & 14.7 & 3.5\\
2S             &  7.6 & 1.3\\
3S             &  6.0 & 0.8\\
4S             &  5.2 & 0.5\\
1P             &  0.6 & 0.6\\
2P             &  0.6 & 0.6\\
1D             &  0.2 & 0.2\\
\hline\hline
\end{array}
\end{displaymath}
\caption{\label{Tab:BottomoniumYndurainterme} Bottomonium expectation values for the additional terms presented in \cite{Titard:1993nn} calculated within model [CpL*-2] ($\alpha_s=0.27$, $\sigma=1.05\:\textrm{GeV/fm}$, $m_b=4.7877\:\textrm{GeV}$, $\mu_{\textrm{GR}}=2/3m_b$, $n_f=4$).}
\end{center}
\end{table}
It is obvious that the fictive gluon mass $\lambda$ poses a problem for the calculation of an expectation value of Eq. (\ref{Eq:AdditionalTermsYndurainGRDefCoordinate}). To calculate the expectation values of $Y(\vec{r}\,)$ we try two differnet approaches, one where we set
\begin{eqnarray}\label{Eq:GluonZero}
\ln\frac{\lambda}{m_q}=0\:,
\end{eqnarray}
and simply get rid of this term, and a second one where we use the recipe
\begin{eqnarray}\label{Eq:GluonLamb}
\ln\frac{\lambda}{m_q} \rightarrow -\frac{5}{6}-\ln\overline{n}\:,
\end{eqnarray}
with $\ln\overline{n}\approx0.9$ as given in \cite{Titard:1993nn}.

The resulting expectation values are shown in Tab. \ref{Tab:CharmoniumYndurainterme} for charmonium and in Tab. \ref{Tab:BottomoniumYndurainterme} for bottomonium. As one can see the size of the corrections is relatively small, but the discrepancies between the results with different assumptions are significant. As we do not know a unique way to treat the term $\ln(\lambda/m_q)$ and for the reason that the corrections $\langle Y(\vec{r} \,)\rangle$ are very small with assumption Eq. (\ref{Eq:GluonLamb}), we consider them to be negligible in our model.

%% file: 6-Fogtpp/6-Summary.tex
In this chapter we adopted a fourth-order gluonic quark-antiquark potential by Gupta \textit{et al.}, which has been slightly modified by us to guarantee time reversal invariance of the potential. This potential is then treated as a perturbative correction. We have predetermined the renormalization scale appearing in the potential based on an argument by Titard and Yndur\`ain.

In the study of model [CpL*-2] it occurred that the spin-dependent splittings are in general too small. For bottomonium the triplet $P$-state splittings are too small. Remarkably, however, this model yields the first time a strong coupling constant $\alpha_s<0.3$ for bottomonium.

In the last section of this chapter we investigated additional terms to the potential by Gupta \textit{et al.}, which have been found by Titard and Yndur\`ain. These terms include a fictive gluon mass. In the analysis we showed that these additional terms yield only small contributions, but are sensitive to assumptions about the fictive gluon mass. Because of this behaviour we decided not to take these contributions into account in our models.

%% file: 7-Fogtpp-induced/7-Perturbative-Corrections.tex
In the previous chapters it became clear that on the one hand the induced interaction (Eq. (\ref{Eq:PotentialInducedInteraction-CM})) provides a suitable framework for the triplet $P$-state splittings. On the other hand corrections at 1-loop order to the one-gluon exchange potential (Eq. (\ref{Eq:PotentialGRm})) seem to remedy some of the parametrical oddities of the basic model. The next natural step is to merge these two models.

\begin{figure}[b!]\vspace{-0.4cm}
\begin{center}
\begin{fmfgraph*}(120,120)
\fmftop{t71,t101}
\fmfbottom{b71,b101}
\fmf{fermion,tension=-.5}{b71,v71,v81,b101}
\fmf{fermion,tension=-.5}{t101,v101,v91,t71}
\fmf{wiggly,tension=-.5}{v71,v91}
\fmf{wiggly,tension=-.5}{v81,v101}
\end{fmfgraph*}
\end{center}
\vspace{-0.5cm}
\caption{\label{Fig:Box-Diagram} Box diagram included in the compution of Eq. (\ref{Eq:PotentialGRm}).}\vspace{-0.4cm}
\end{figure}
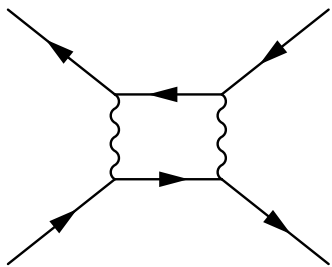

The combination of induced interaction and the fourth-order gluonic potential has an inherent double counting problem. The calculation of Eq. (\ref{Eq:PotentialGRm}) \cite{Gupta:1981pd,Titard:1993nn} includes the box diagram in Fig. \ref{Fig:Box-Diagram}.
One realizes that this boxed diagram is also included in the induced interaction. In principle one would have to subtract this diagram (Fig. \ref{Fig:Box-Diagram}) from the combination of Eq. (\ref{Eq:PotentialInducedInteraction-CM}) and Eq. (\ref{Eq:PotentialGRm}). This is not done in the present work, but we correct for this deficiency by treating the coupling constant $\alpha_i$ as free parameter.

%% file: 7-Fogtpp-induced/7-Potential-Model.tex
The potential in this model is composed of the perturbative QCD potential Eq. (\ref{Eq:PotentialGRm}), the induced interaction potential Eq. (\ref{Eq:PotentialInducedInteraction-CM}) and a linear confining part:
\begin{eqnarray}\label{Eq:PotentialCpL*pI-2}
V^\textrm{[CpL*pI-2]}(\vec{r};\vec{p}\,)=V^{(4)}(\vec{r};\vec{p}\,)+V_{\textrm{ind}}(\vec{r};\vec{p}\,)+\sigma r\:.
\end{eqnarray}
As for all Coulomb-plus-Linear models the potential used in the Schr\"odinger equation to calculate bound states is
\begin{eqnarray}
V_0(r)&=&-\frac{4\alpha_s}{3r}+\sigma r\:,
\end{eqnarray}
while the remaining parts of Eq. (\ref{Eq:PotentialCpL*pI-2}) are treated perturbatively. The mass formula is written as:
\begin{eqnarray}
M(k^{2S+1}l_j)\!\!\!\!\!&=&\!\!\!\!\!2m_q+E_{kl}+M_\textrm{g}^\textrm{Cc}(k^{2S+1}l_j)+M_\textrm{g}^\textrm{rsq}(k^{2S+1}l_j)+M_\textrm{g}^\textrm{con}(k^{2S+1}l_j)+M_\textrm{g}^\textrm{SS}(k^{2S+1}l_j)\nonumber\\
&+&\!\!\!\!\!M_\textrm{g}^\textrm{T}(k^{2S+1}l_j)+M_\textrm{g}^\textrm{LS}(k^{2S+1}l_j)+M_\textrm{g}^\textrm{MD}(k^{2S+1}l_j)+M_\textrm{i}^\textrm{C}(k^{2S+1}l_j)+M_\textrm{i}^\textrm{con}(k^{2S+1}l_j)\nonumber\\
&+&\!\!\!\!\!M_\textrm{i}^\textrm{SS}(k^{2S+1}l_j)+M_\textrm{i}^\textrm{T}(k^{2S+1}l_j)+M_\textrm{i}^\textrm{LS}(k^{2S+1}l_j)+M_\textrm{i}^\textrm{MD}(k^{2S+1}l_j)\:,
\end{eqnarray}
with the following contributions:
\begin{eqnarray}
M_\textrm{g}^\textrm{Cc}(k^{2S+1}l_j)\!\!\!\!\!&=&\!\!\!\!\!\frac{2\alpha_s^2}{\pi}\left(\bigg\langle\frac{1}{r}\bigg\rangle-\frac{33-2n_f}{9}\bigg\langle\frac{\ln\left(\mu_{\textrm{GR}}r\right)+\gamma_\textrm{E}}{r}\bigg\rangle\right)\:,\nonumber\\
M_\textrm{g}^\textrm{rsq}(k^{2S+1}l_j)\!\!\!\!\!&=&\!\!\!\!\!-\bigg\langle\frac{14\alpha_s^2}{9m_qr^2}\bigg\rangle\:,\nonumber\\
M_\textrm{g}^\textrm{con}(k^{2S+1}l_j)\!\!\!\!\!&=&\!\!\!\!\!\frac{4\pi\alpha_s}{3m_q^2}\left(1-\frac{3\alpha_s}{2\pi}\right)\big|\:\psi(0)\:\big|^2-\frac{\alpha_s^2(33-2n_f)}{18\pi m_q^2}\int\mathrm{d}^3r\:\frac{\ln(\mu_{\textrm{GR}}r)+\gamma_\textrm{E}}{r}\Delta\big|\:\psi(\vec{r}\,)\:\big|^2\:,\nonumber\\
M_\textrm{g}^\textrm{SS}(k^{2S+1}l_j)\!\!\!\!\!&=&\!\!\!\!\!\frac{32\pi\alpha_s}{9m_q^2}\left(\frac{1}{2}S(S+1)-\frac{3}{4}\right)\Bigg\{\left[1-\frac{\alpha_s}{12\pi}\left(26+9\ln2\right)\right]\:\big|\:\psi(0)\:\big|^2\nonumber\\
&&\!+\int\mathrm{d}^3r\left(\frac{21\alpha_s}{16\pi^2}\frac{\ln(m_q r)+\gamma_\textrm{E}}{r}-\frac{\alpha_s(33-2n_f)}{24\pi^2}\frac{\ln(\mu_{\textrm{GR}}r)+\gamma_\textrm{E}}{r}\right)\Delta\big|\:\psi(\vec{r}\,)\:\big|^2\Bigg\}\nonumber\:,\\
M_\textrm{g}^\textrm{T}(k^{2S+1}l_j)\!\!\!\!\!&=&\!\!\!\!\!\frac{\alpha_sS_{12}}{3m_q^2}\Bigg\{\left(1+\frac{4\alpha_s}{3\pi}\right)\bigg\langle\frac{1}{r^3}\bigg\rangle+\frac{\alpha_s}{6\pi}(33-2n_f)\bigg\langle\frac{\ln(\mu_{\textrm{GR}} r)+\gamma_\textrm{E}-\frac{4}{3}}{r^3}\bigg\rangle\\
&&\!\!\!\!\!\qquad\qquad\qquad\qquad\qquad\qquad\qquad\qquad\qquad\qquad-\frac{3\alpha_s}{\pi}\bigg\langle\frac{\ln(m_q r)+\gamma_\textrm{E}-\frac{4}{3}}{r^3}\bigg\rangle\Bigg\}\:,\nonumber\\
M_\textrm{g}^\textrm{LS}(k^{2S+1}l_j)\!\!\!\!\!&=&\!\!\!\!\!\frac{2\alpha_s}{m_q^2}\frac{j(j+1)-l(l+1)-S(S+1)}{2}\Bigg\{\left(1-\frac{11\alpha_s}{18\pi}\right)\bigg\langle\frac{1}{r^3}\bigg\rangle\nonumber\\
&&\!\!\!\!\!\qquad\quad+\frac{\alpha_s(33-2n_f)}{6\pi}\bigg\langle\frac{\ln(\mu_{\textrm{GR}} r)+\gamma_\textrm{E} -1}{r^3}\bigg\rangle-\frac{2\alpha_s}{\pi}\bigg\langle\frac{\ln(m_q r)+\gamma_\textrm{E}-1}{r^3}\bigg\rangle\Bigg\}\:,\nonumber\\
M_\textrm{g}^\textrm{MD}(k^{2S+1}l_j)\!\!\!\!\!&=&\!\!\!\!\!-\frac{4\alpha_s}{3m_q^2}\int\mathrm{d}^3r\:\frac{1}{r}\left[1-\frac{3\alpha_s}{2\pi}+\frac{\alpha_s}{6\pi}(33-2n_f)[\ln(\mu r)+\gamma_\textrm{E}]\right]\big|\:\vec{\nabla}\:\psi(\vec{r}\,)\:\big|^2\:,\nonumber
\end{eqnarray}
\begin{eqnarray}
M_\textrm{i}^\textrm{C}(k^{2S+1}l_j)\!\!\!\!\!&=&\!\!\!\!\!-\alpha_i\bigg\langle\frac{\mathrm{e}^{-Mr}}{r}\bigg\rangle\:,\nonumber\\
M_\textrm{i}^\textrm{con}(k^{2S+1}l_j)\!\!\!\!\!&=&\!\!\!\!\!\frac{\alpha_i}{4m_q^2}\left(4\pi\big|\:\psi(0)\:\big|^2-M^2\bigg\langle\frac{\mathrm{e}^{-Mr}}{r}\bigg\rangle\right)\:,\\
M_\textrm{i}^\textrm{SS}(k^{2S+1}l_j)\!\!\!\!\!&=&\!\!\!\!\!\frac{\alpha_i}{3m_q^2}\left(\frac{1}{2}S(S+1)-\frac{3}{4}\right)\left[4\pi\big|\:\psi(0)\:\big|^2-M^2\bigg\langle\frac{\mathrm{e}^{-Mr}}{r}\bigg\rangle\right]\:,\nonumber\\
M_\textrm{i}^\textrm{T}(k^{2S+1}l_j)\!\!\!\!\!&=&\!\!\!\!\!\alpha_i\frac{S_{12}}{2m_q^2}\left(\bigg\langle\frac{\mathrm{e}^{-Mr}}{r^3}\bigg\rangle+\bigg\langle M\frac{\mathrm{e}^{-Mr}}{r^2}\bigg\rangle+\bigg\langle M^2\frac{\mathrm{e}^{-Mr}}{3r}\bigg\rangle\right)\:,\nonumber\\
M_\textrm{i}^\textrm{LS}(k^{2S+1}l_j)\!\!\!\!\!&=&\!\!\!\!\!\alpha_i\frac{3\left(j(j+1)-l(l+1)-S(S+1)\right)}{4m_q^2}\left(\bigg\langle\frac{\mathrm{e}^{-Mr}}{r^3}\bigg\rangle+\bigg\langle M\frac{\mathrm{e}^{-Mr}}{r^2}\bigg\rangle\right)\:,\nonumber\\
M_\textrm{i}^\textrm{MD}(k^{2S+1}l_j)\!\!\!\!\!&=&\!\!\!\!\!-\frac{\alpha_i}{m_q^2}\int\mathrm{d}^3r\:\frac{\mathrm{e}^{-Mr}}{r}\:\big|\:\vec{\nabla}\:\psi(\vec{r}\,)\:\big|^2\:.\nonumber
\end{eqnarray}
The purely momentum dependent terms of Eq. (\ref{Eq:PotentialCpL*pI-2}) are treated according to option 2 of Appendix \ref{Chapt:Expect-values-of-MD}. Employing any other options results in much lower string tensions and higher values for the quark masses, outside the range of empirical values. Figures, tables and quantities related to this model are marked as [CpL*pI-2].

%% file: 7-Fogtpp-induced/7-Charmonium.tex
\vspace{-1.5cm}
\begin{center}
\begin{figure}[p]\begin{center}
\includegraphics[width=0.9\textwidth]{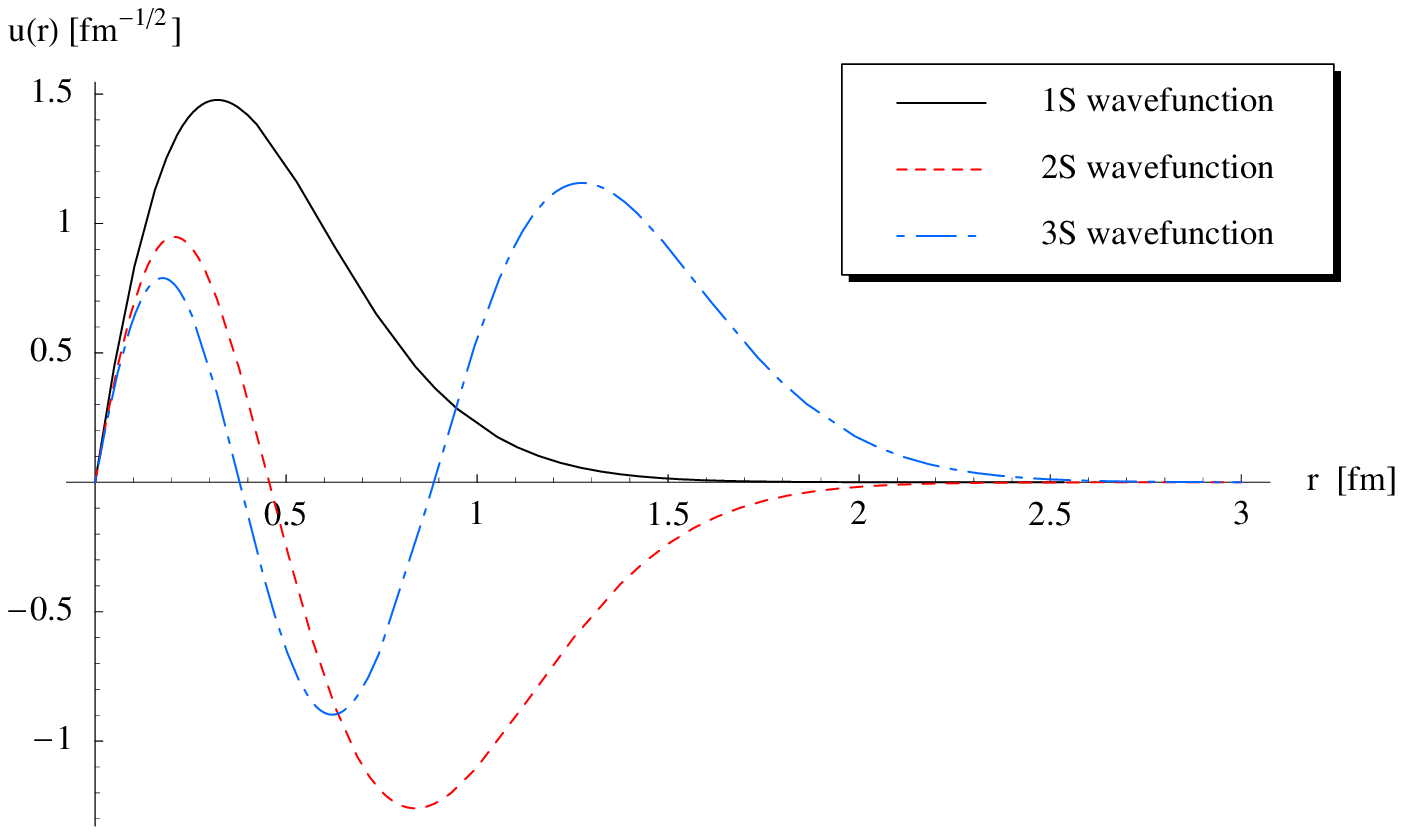}\end{center}
\caption{\label{Fig:Wavefunctions-S-Charmonium-CpL*pI-2} Charmonium $S$-state reduced radial wavefunctions calculated within model [CpL*pI-2] ($\alpha_s=0.295$, $\alpha_i=1.0$, $\sigma=1.04\:\textrm{GeV/fm}$, $m_c=1.4056\:\textrm{GeV}$, $M=3.07\:\textrm{GeV}$, $\mu_{\textrm{GR}}=m_c$, $n_f=3$).}
\end{figure} 
\end{center}
\begin{center}
\begin{figure}[p]\begin{center}
\includegraphics[width=0.9\textwidth]{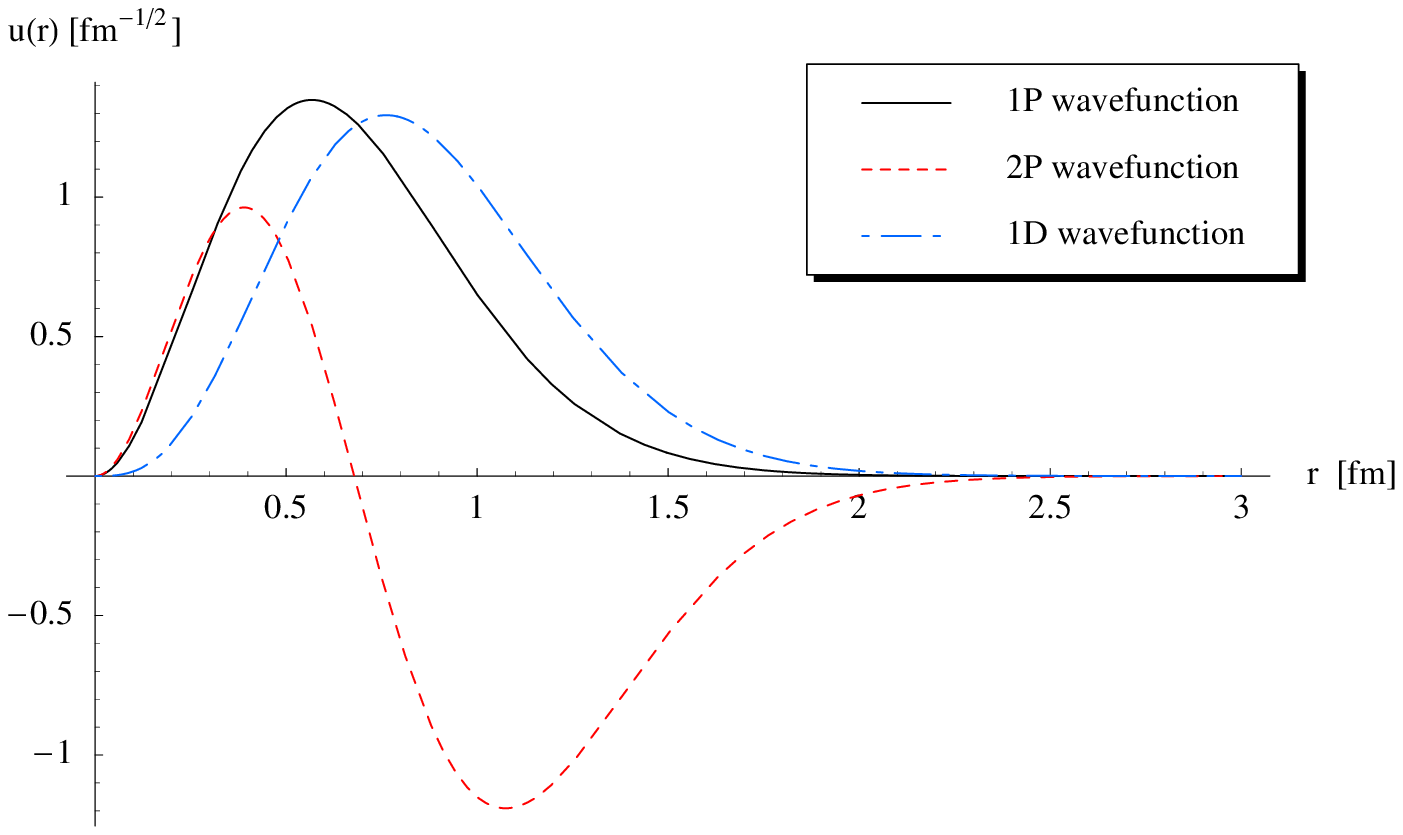}\end{center}
\caption{\label{Fig:Wavefunctions-PD-Charmonium-CpL*pI-2} Charmonium $P$- and $D$-state reduced radial wavefunctions calculated within model [CpL*pI-2] ($\alpha_s=0.295$, $\alpha_i=1.0$, $\sigma=1.04\:\textrm{GeV/fm}$, $m_c=1.4056\:\textrm{GeV}$, $M=3.07\:\textrm{GeV}$, $\mu_{\textrm{GR}}=m_c$, $n_f=3$).}
\end{figure} 
\end{center}
\begin{center}
\begin{figure}[p]\begin{center}
\begin{minipage}{7.45cm}\begin{center}
\includegraphics[width=\textwidth]{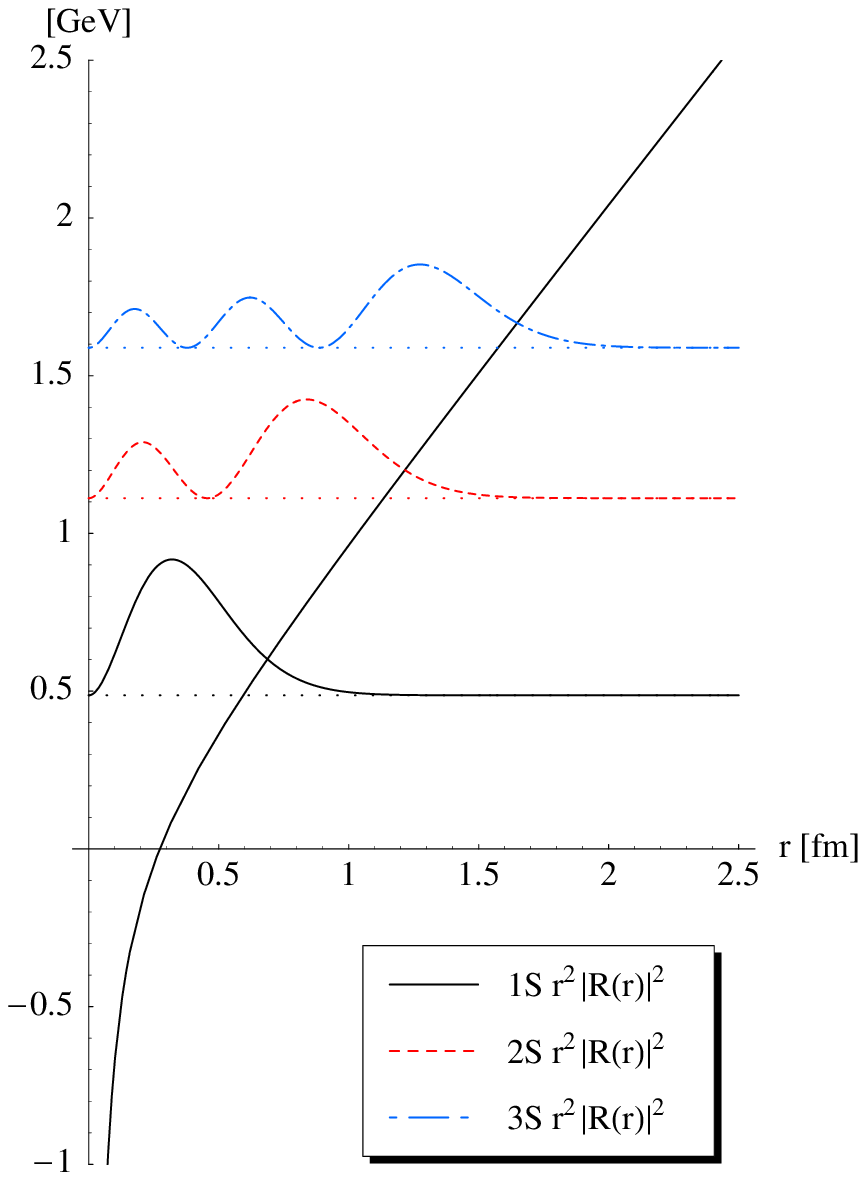}\end{center}
\end{minipage}\hfill
\begin{minipage}{7.45cm}\begin{center}
\includegraphics[width=\textwidth]{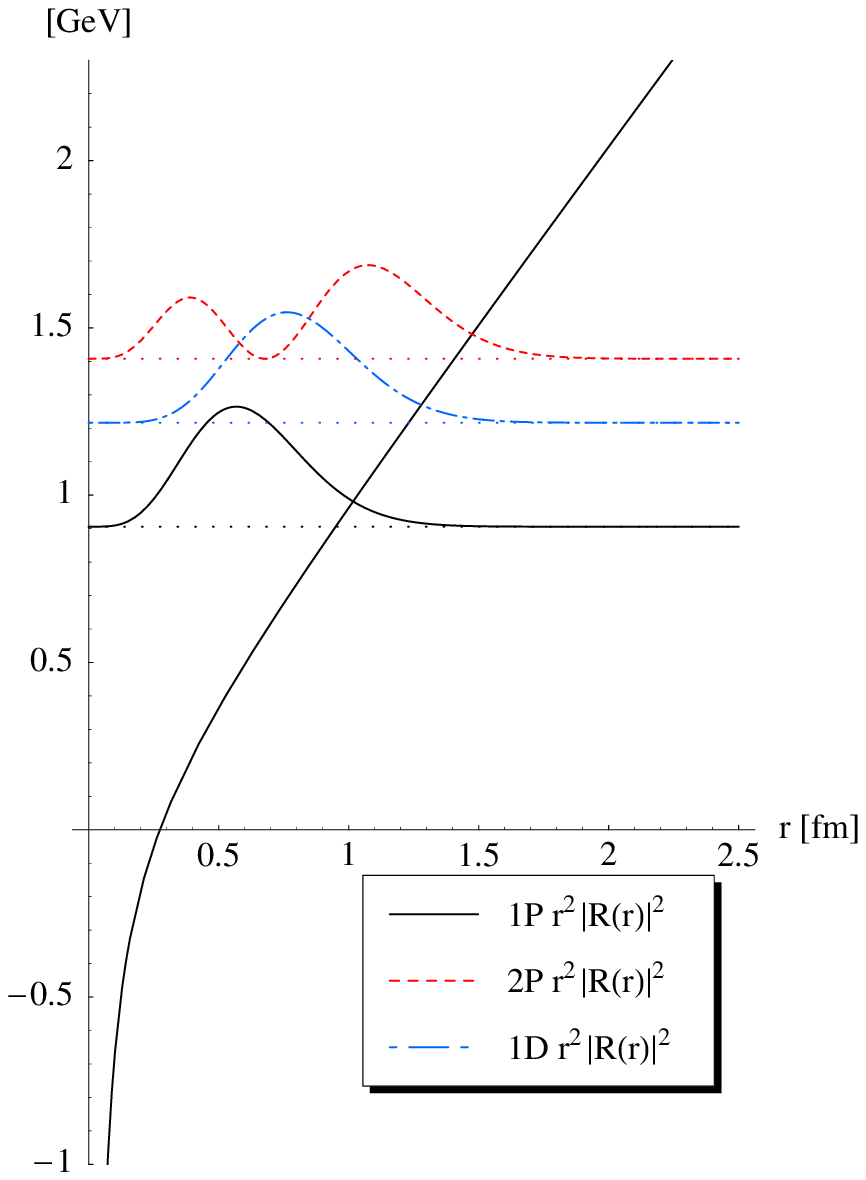}\end{center}
\end{minipage}\end{center}\vspace{-1.2cm}
\caption{\label{Fig:RadialDensityCharmoniumCpL*pI-2} Radial densities for charmonium states of model [CpL*pI-2] plotted together with used potential ($\alpha_s=0.295$, $\alpha_i=1.0$, $\sigma=1.04\:\textrm{GeV/fm}$, $m_c=1.4056\:\textrm{GeV}$, $M=3.07\:\textrm{GeV}$, $\mu_{\textrm{GR}}=m_c$, $n_f=3$); base lines of radial densities have been shifted by according $E_{kl}$.}
\end{figure} 
\end{center}
\begin{center}
\begin{figure}[p]\begin{center}
\includegraphics[width=0.9\textwidth]{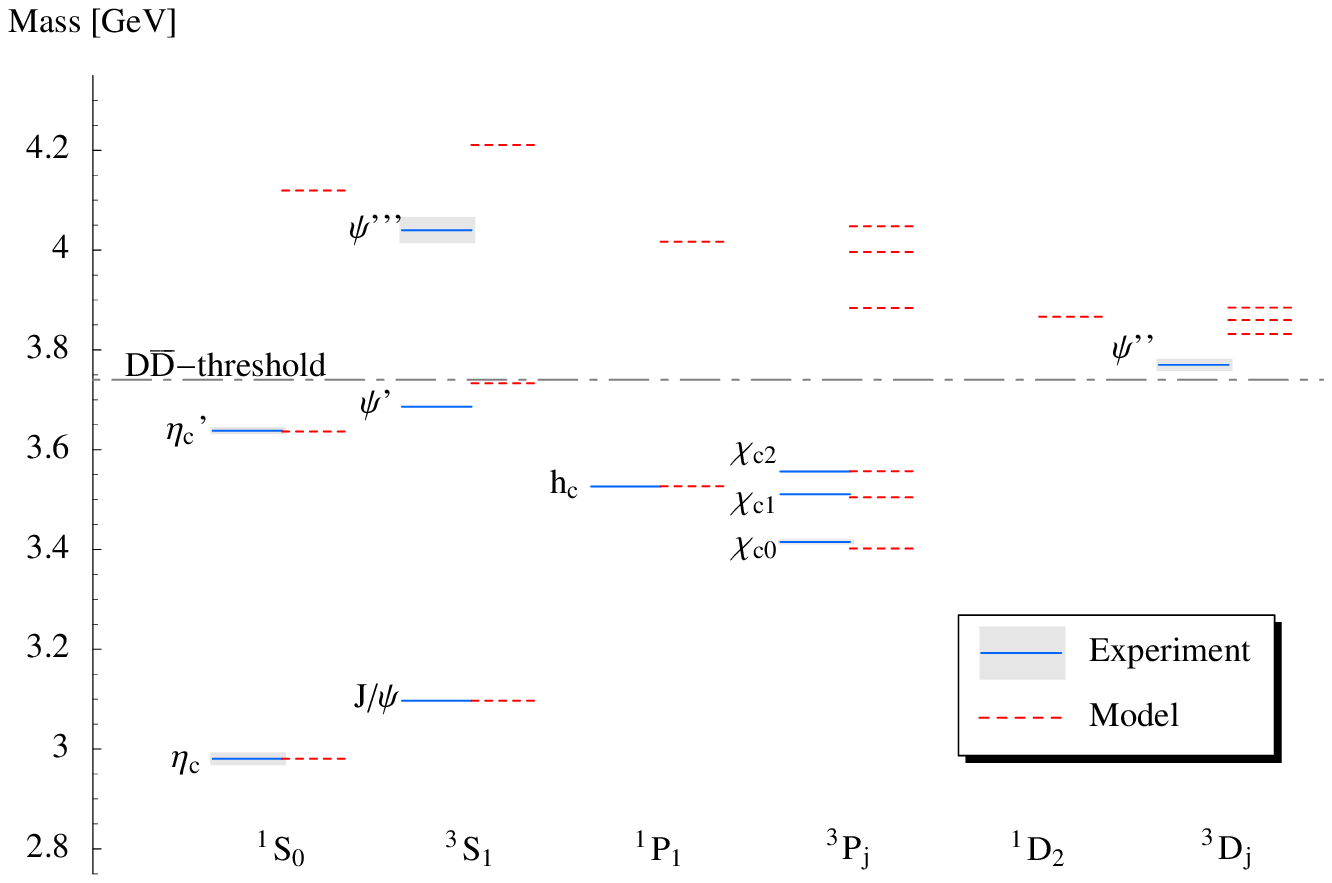}\end{center}
\caption{\label{Fig:SpectrumCharmoniumCpL*pI-2} Charmonium spectra from experiment and model [CpL*pI-2] ($\alpha_s=0.295$, $\alpha_i=1.0$, $\sigma=1.04\:\textrm{GeV/fm}$, $m_c=1.4056\:\textrm{GeV}$, $M=3.07\:\textrm{GeV}$, $\mu_{\textrm{GR}}=m_c$, $n_f=3$); for experimental data masses and widths are shown.}
\end{figure} 
\end{center}
The parameters in model [CpL*pI-2] are determined by a fit to the masses of states $\eta_c$, $J/\psi$ and $h_c$, and the ratio $\Phi_c^\textrm{Exp}(1P)$ (Eq. (\ref{Eq:P-ProportionCharmoniumExperiment})), with the additional constraint that the parameters $\alpha_s$, $m_c$ and $\sigma$ are as close as possible to their empirical values. The resulting parameter set is:
\begin{eqnarray}
\alpha_s&=&0.295\:,\nonumber\\
\alpha_i&=&1.0\:,\nonumber\\
m_c&=&1.4056\:\textrm{GeV}\:,\\
\sigma&=&1.04\:\textrm{GeV/fm}\:.\nonumber
\end{eqnarray}
The model gives values for the string tension $\sigma$, the strong coupling constant $\alpha_s$ and the charm quark mass $m_c$ which are very close to empirical values.

\begin{table}[b!]
\begin{center}
\begin{displaymath}
\begin{array}{||c|c|c|c|c||}
\hline\hline
&&\multicolumn{2}{|c|}{\qquad\textrm{Experiment \cite{PDBook}}\qquad} & \textrm{Theory [CpL*pI-2]}\\
 \;\textrm{State}\; & \;\textrm{Candidate}\;&\multicolumn{2}{|c|}{\hrulefill} &  \\
&&\textrm{Mass [MeV]}&\textrm{Width [MeV]}& \textrm{Mass [MeV]}\\
\hline
1^1S_0 & \eta_c     & 2980.4   \pm 1.2   & 25.5   \pm 3.4      &  2980 \\
1^3S_1 & J/\psi     & 3096.916 \pm 0.011 & 0.0910 \pm 0.0032   &  3097 \\
\hline
1^1P_1 & h_c        & 3526.21  \pm 0.25  & <1.1                &  3527 \\
1^3P_0 & \chi_{c0}  & 3415.16  \pm 0.35  & 10.2   \pm 0.8      &  3402 \\
1^3P_1 & \chi_{c1}  & 3510.59  \pm 0.10  & 0.96   \pm 0.12     &  3504 \\
1^3P_2 & \chi_{c2}  & 3556.26  \pm 0.11  & 2.25   \pm 0.15     &  3557 \\
\hline
2^1S_0 & \eta_c'    & 3638     \pm 5     & 14     \pm 7        &  3636 \\
2^3S_1 & \psi'      & 3686.093 \pm 0.034 & 0.283  \pm 0.017    &  3733 \\
\hline
1^1D_2 &            &                    &                     &  3866 \\
1^3D_1 & \psi''     & 3770     \pm 2.4   & 23.6   \pm 2.7      &  3832 \\
1^3D_2 &            &                    &                     &  3860 \\
1^3D_3 &            &                    &                     &  3885 \\
\hline
2^1P_1 &            &                    &                     &  4017 \\
2^3P_0 &            &                    &                     &  3884 \\
2^3P_1 &            &                    &                     &  3996 \\
2^3P_2 &            &                    &                     &  4048 \\
\hline
3^1S_0 &            &                    &                     &  4119 \\
3^3S_1 & \psi'''    & 4040     \pm 10    & 52     \pm 10       &  4211 \\
\hline\hline
\end{array}
\end{displaymath}
\caption{\label{Tab:MassesCharmoniumCpL*pI-2} $c\overline{c}$-state masses from experiment and model [CpL*pI-2] ($\alpha_s=0.295$, $\alpha_i=1.0$, $\sigma=1.04\:\textrm{GeV/fm}$, $m_c=1.4056\:\textrm{GeV}$, $M=3.07\:\textrm{GeV}$, $\mu_{\textrm{GR}}=m_c$, $n_f=3$); masses and widths are displayed for the experiment.}
\end{center}
\end{table}
\begin{table}[ht!]
\begin{center}\vspace{-0.4cm}
\begin{displaymath}
\begin{array}{||c|c|c|c|c|c|c|c||}
\hline\hline
       \multirow{2}{*}{ \textrm{State} }& E_{kl} & M_\textrm{g}^\textrm{Cc} & M_\textrm{g}^\textrm{rsq} & M_\textrm{g}^\textrm{con} & M_\textrm{i}^\textrm{con} & M_\textrm{g}^\textrm{SS} & M_\textrm{i}^\textrm{SS}\\
&\textrm{[MeV]}&\textrm{[MeV]}&\textrm{[MeV]}&\textrm{[MeV]}&\textrm{[MeV]}&\textrm{[MeV]}&\textrm{[MeV]}\\
\hline
1^1S_0 &  \multirow{2}{*}{486.8} & \multirow{2}{*}{-86.6} & \multirow{2}{*}{-75.2} & \multirow{2}{*}{24.6} &  \multirow{2}{*}{31.7} & -55.6& -31.7\\
1^3S_1 &    & & & && 18.5 & 10.6\\
\hline
1^1P_1 &  \multirow{4}{*}{905.8} & \multirow{4}{*}{-91.3} & \multirow{4}{*}{-15.5} & \multirow{4}{*}{2.0} &\multirow{4}{*}{-2.7}&0.7&2.7\\\cline{7-8}
1^3P_0 &  & & &  &&\multirow{3}{*}{-0.2}&\multirow{3}{*}{-0.9} \\
1^3P_1 &  & & & &&& \\
1^3P_2 &  &  &  & &&& \\
\hline
2^1S_0 & \multirow{2}{*}{1111.7} & \multirow{2}{*}{-72.3} & \multirow{2}{*}{-45.8} & \multirow{2}{*}{16.3} &\multirow{2}{*}{29.2}& -43.4&-29.2 \\
2^3S_1 & &  &  &  && 14.5 &9.7\\
\hline
1^1D_2 & \multirow{4}{*}{1216.4} & \multirow{4}{*}{-82.2} & \multirow{4}{*}{-7.7} & \multirow{4}{*}{0.6} &\multirow{4}{*}{-0.2}&0.2&0.2\\\cline{7-8}
1^3D_1 & & & &  &&\multirow{3}{*}{-0.1}&\multirow{3}{*}{-0.1}\\
1^3D_2 & & & &  &&&\\
1^3D_3 & & & &  &&&\\
\hline
2^1P_1 & \multirow{4}{*}{1407.8} & \multirow{4}{*}{-77.3} & \multirow{4}{*}{-12.1} & \multirow{4}{*}{0.4} &\multirow{4}{*}{-3.3}&0.6&3.3\\\cline{7-8}
2^3P_0 & &  & &  &&\multirow{3}{*}{-0.2}&\multirow{3}{*}{-1.1}\\
2^3P_1 & &  &  &  &&&\\
2^3P_2 & &  &  &  &&&\\
\hline
3^1S_0 & \multirow{2}{*}{1588.5} & \multirow{2}{*}{-63.5} & \multirow{2}{*}{-36.6} & \multirow{2}{*}{13.7} &\multirow{2}{*}{29.0}&-39.7&-29.0\\
3^3S_1 & & & & && 13.2 & 9.7 \\
\hline\hline
\multicolumn{8}{c}{}\\
\hline\hline
       \multirow{2}{*}{ \textrm{State} }& M_\textrm{g}^\textrm{T} & M_\textrm{i}^\textrm{T}& M_\textrm{g}^\textrm{LS} & M_\textrm{i}^\textrm{LS}& M_\textrm{g}^\textrm{MD} & M_\textrm{i}^\textrm{MD}& M^\textrm{C}_\textrm{i} \\
&\textrm{[MeV]}&\textrm{[MeV]}&\textrm{[MeV]}&\textrm{[MeV]}&\textrm{[MeV]}&\textrm{[MeV]}&\textrm{[MeV]}\\
\hline
1^1S_0 &  \multirow{2}{*}{0} & \multirow{2}{*}{0} & \multirow{2}{*}{0} & \multirow{2}{*}{0} & \multirow{2}{*}{-70.9} &\multirow{2}{*}{-4.0} &\multirow{2}{*}{-49.8}\\
1^3S_1 &    & & & &&& \\
\hline
1^1P_1 & 0 & 0 & 0 & 0 &\multirow{4}{*}{-76.6}&\multirow{4}{*}{-7.3}&\multirow{4}{*}{-2.2}\\
1^3P_0 & -21.0 & -21.4 & -56.6 & -21.3 & & & \\
1^3P_1 & 10.4 & 10.7 & -28.3 & -10.7 &&& \\
1^3P_2 & -2.1 & -2.1 & 28.3 & 10.7 &&& \\
\hline
2^1S_0 & \multirow{2}{*}{0} & \multirow{2}{*}{0} & \multirow{2}{*}{0} & \multirow{2}{*}{0} &\multirow{2}{*}{-101.7}&\multirow{2}{*}{-5.9} &\multirow{2}{*}{-34.0} \\
2^3S_1 & &  &  &  &&&\\
\hline
1^1D_2 & 0 & 0 & 0 & 0 &\multirow{4}{*}{-71.4}&\multirow{4}{*}{-0.6}&\multirow{4}{*}{-0.2}\\
1^3D_1 & -3.3 & -0.6 & -28.7 & -1.3 &&&\\
1^3D_2 & 3.3 & 0.6 & -9.6 & -0.4 &&&\\
1^3D_3 & -1.0 & -0.2 & 19.1 & 0.9 &&&\\
\hline
2^1P_1 & 0 & 0 & 0 & 0 &\multirow{4}{*}{-99.9}&\multirow{4}{*}{-10.3}&\multirow{4}{*}{-2.8}\\
2^3P_0 & -19.3 & -28.8 & -50.6 & -29.6 &&&\\
2^3P_1 & 9.6 & 14.4 & -25.3 & -14.8 &&&\\
2^3P_2 & -1.9 & -2.9 & 25.3 & 14.8 &&&\\
\hline
3^1S_0 & \multirow{2}{*}{0} & \multirow{2}{*}{0} & \multirow{2}{*}{0} & \multirow{2}{*}{0} &\multirow{2}{*}{-118.4}&\multirow{2}{*}{-7.4} &\multirow{2}{*}{-28.6}\\
3^3S_1 & & & & &&& \\
\hline\hline
\end{array}
\end{displaymath}\vspace{-0.6cm}
\caption{\label{Tab:MassesContributionsCharmoniumCpL*pI-2} Contributions to the masses of $c\overline{c}$-states in model [CpL*pI-2] ($\alpha_s=0.295$, $\alpha_i=1.0$, $\sigma=1.04\:\textrm{GeV/fm}$, $m_c=1.4056\:\textrm{GeV}$, $M=3.07\:\textrm{GeV}$, $\mu_{\textrm{GR}}=m_c$, $n_f=3$).}\vspace{-0.7cm}
\end{center}
\end{table}

The reduced radial wavefunctions for the lowest lying bound states are presented in Fig. \ref{Fig:Wavefunctions-S-Charmonium-CpL*pI-2} and Fig. \ref{Fig:Wavefunctions-PD-Charmonium-CpL*pI-2}, while in Fig. \ref{Fig:RadialDensityCharmoniumCpL*pI-2} the radial densities for charmonium states are shown. Comparing the radial densities of model [CpL*pI-2] with the ones of model [CpL-B], one realizes that they are positioned lower in the [CpL*pI-2] plot and therefore the eigenvalues are smaller. This is mainly due to the decrease of the string tension by approximately 25 percent.

In Tab. \ref{Tab:MassesCharmoniumCpL*pI-2} masses and widths from the experiment are displayed together with masses calculated in model [CpL*pI-2]. The experimental spectrum is compared to the one arising from model [CpL*pI-2] in Fig. \ref{Fig:SpectrumCharmoniumCpL*pI-2}. The data presented in Tab. \ref{Tab:MassesCharmoniumCpL*pI-2} shows, that the three masses of our input states $\eta_c$, $J/\psi$ and $h_c$ are matched with 1 MeV accuracy. The ratio
\begin{eqnarray}
\Phi_c^\textrm{[CpL*pI-2]}(1P)=\left(\frac{M(\chi_{c,2})-M(\chi_{c,1})}{M(\chi_{c,1})-M(\chi_{c,0})}\right)_{\textrm{[CpL*pI-2]}}&\approx&0.51\:,
\end{eqnarray}
which is used as a further criterion to determine the parameters, agrees quite well with the experimental value
\begin{eqnarray}
\Phi_c^\textrm{Exp}(1P)=\left(\frac{M(\chi_{c,2})-M(\chi_{c,1})}{M(\chi_{c,1})-M(\chi_{c,0})}\right)_{\textrm{Exp}}=0.482\pm0.005\:.
\end{eqnarray}
We consider this adequate to constrain the model parameters, as we restricted ourselves to certain intervals for three of the parameters and thus the fit is not free. The size of splittings of the induced interaction and fourth-order potential did not compensate each other exactly, resulting in an enlargement of the triplet $P$-state splitting by about seven percent. The splitting between the singlet and the triplet $2S$-states is enlarged, but we match the singlet state $\eta'_c$ within the uncertainties of the experiment and realize that the triplet state $\psi'$ is positioned slightly below the $D\overline{D}$-threshold. 
Therefore the position of $\psi'$ could change and the $2S$-splitting might get a value close to the experimental one. Not surprisingly, the potential model is not sufficient to describe the spectrum above threshold.

In Tab. \ref{Tab:MassesContributionsCharmoniumCpL*pI-2} the various contributions to the masses of $c\overline{c}$-states are displayed. One notices that
\begin{eqnarray}
M^\textrm{bare}_{kl}=2m_c+E_{kl}
\end{eqnarray}
leads to values significantly larger than the masses of the respective charmonium states. Hence large negative shifts are required. The purely momentum dependent $M_\textrm{g}^\textrm{MD}$ term yields a large negative contribution to the masses. An additional large contribution $M_\textrm{g}^\textrm{Cc}$ for all states comes from the radiative correction to the Coulomb part of the potential. The  $M_\textrm{g}^\textrm{rsq}$ term of order $\mathcal{O}(m^{-1})$ in the interaction also provides large negative mass shifts, but acts mainly on the $S$-states. Together with the contact term $M_\textrm{g}^\textrm{con}$ this influences the string tension parameter in the present model as the $1S$-states are further separated from the $1P$-states.

Tab. \ref{Tab:MassesContributionsCharmoniumCpL*pI-2} shows the importance of the induced interaction. The combination $M_\textrm{i}^\textrm{C}+M_\textrm{i}^\textrm{con}+M_\textrm{i}^\textrm{MD}$ yields an additional separation between the $S$-states and the angular excited states, leading to a slightly lower string tension and thus to a value close to the estimate from lattice QCD simulations. The induced interaction provides about 40 percent of the singlet-triplet-splitting of the $S$-states. It also leads to a triplet $P$-state splitting with the correct structure. This is due to the enhancement of the tensor part over the spin-orbit part in the induced interaction as compared to the gluon exchange interaction.

%% file: 7-Fogtpp-induced/7-Bottomonium.tex
\vspace{-0.5cm}
\begin{center}
\begin{figure}[p]\begin{center}
\includegraphics[width=0.9\textwidth]{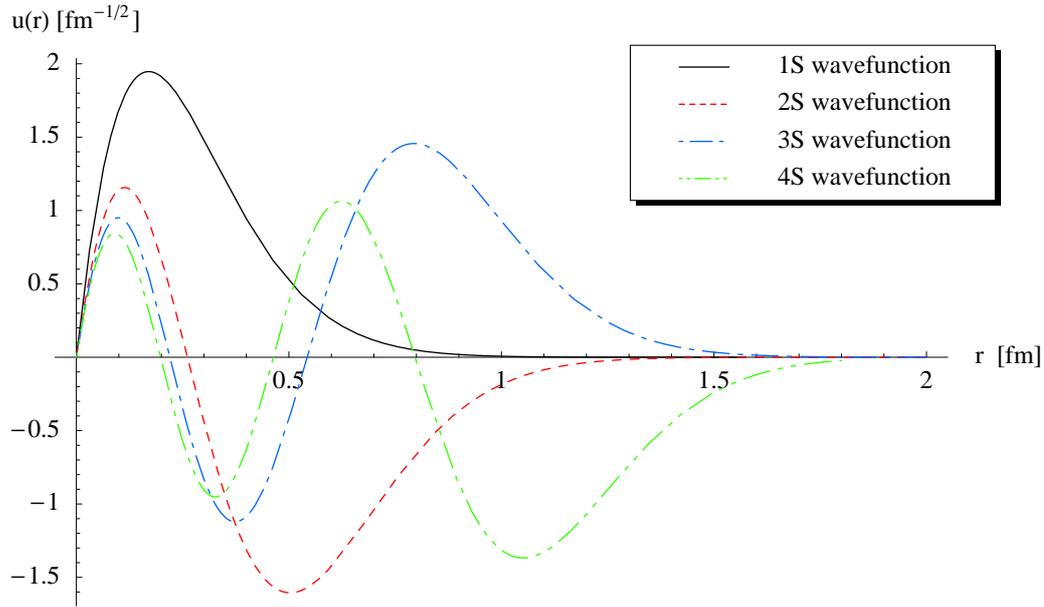}\end{center}
\caption{\label{Fig:Wavefunctions-S-Bottomonium-CpL*pI-2} Bottomonium $S$-state reduced radial wavefunctions calculated within model [CpL*pI-2] ($\alpha_s=0.255$, $\alpha_i=1.0$, $\sigma=1.08\:\textrm{GeV/fm}$, $m_b=4.768\:\textrm{GeV}$, $M=2m_b$, $\mu_{\textrm{GR}}=2/3m_b$, $n_f=4$).}
\end{figure} 
\end{center}
\vspace{-1cm}
\begin{center}
\begin{figure}[p]\begin{center}
\includegraphics[width=0.9\textwidth]{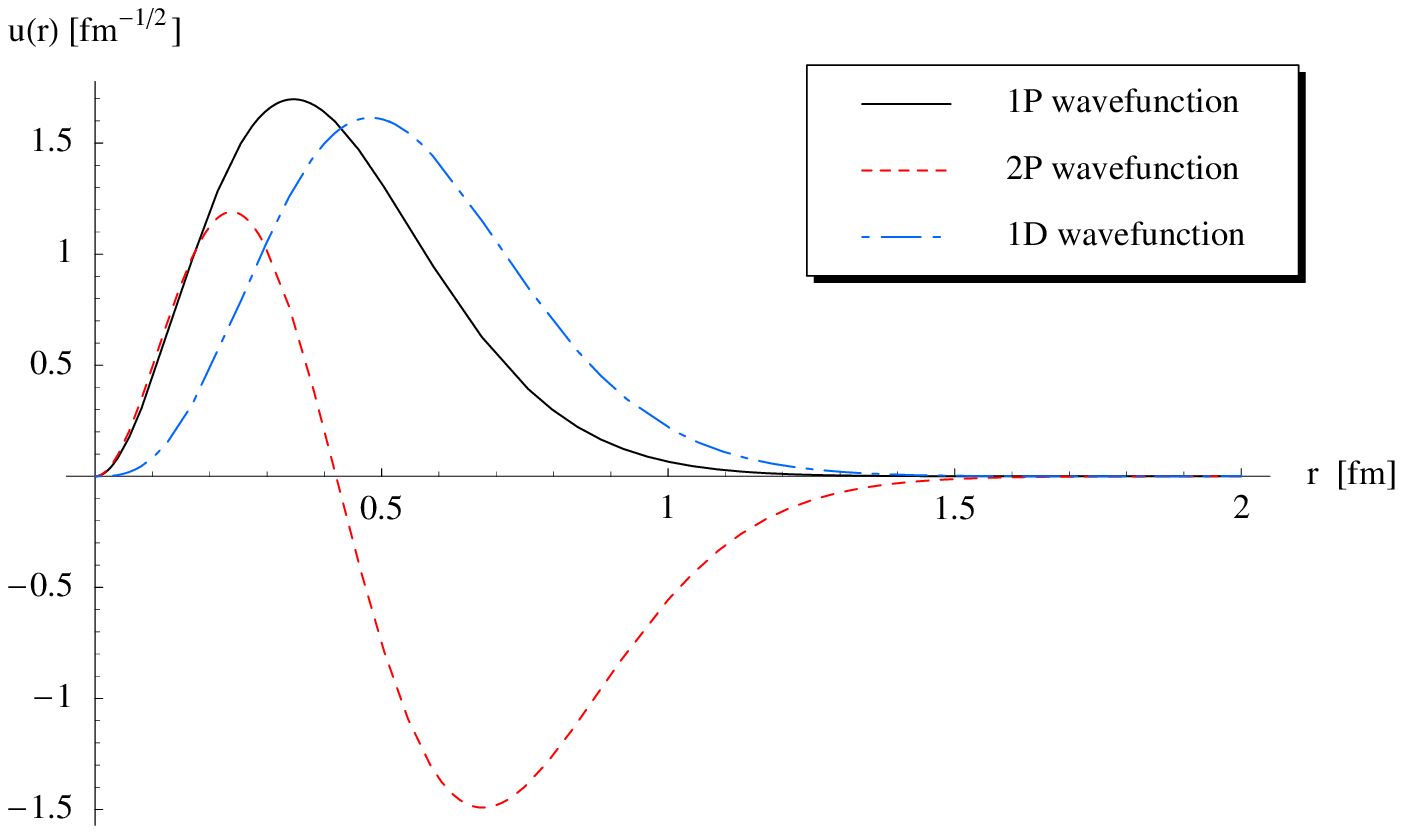}\end{center}
\caption{\label{Fig:Wavefunctions-PD-Bottomonium-CpL*pI-2} Bottomonium $P$- and $D$-state reduced radial wavefunctions calculated within model [CpL*pI-2] ($\alpha_s=0.255$, $\alpha_i=1.0$, $\sigma=1.08\:\textrm{GeV/fm}$, $m_b=4.768\:\textrm{GeV}$, $M=2m_b$, $\mu_{\textrm{GR}}=2/3m_b$, $n_f=4$).}
\end{figure} 
\end{center}
\begin{center}
\begin{figure}[p]\begin{center}
\begin{minipage}{7.45cm}\begin{center}
\includegraphics[width=\textwidth]{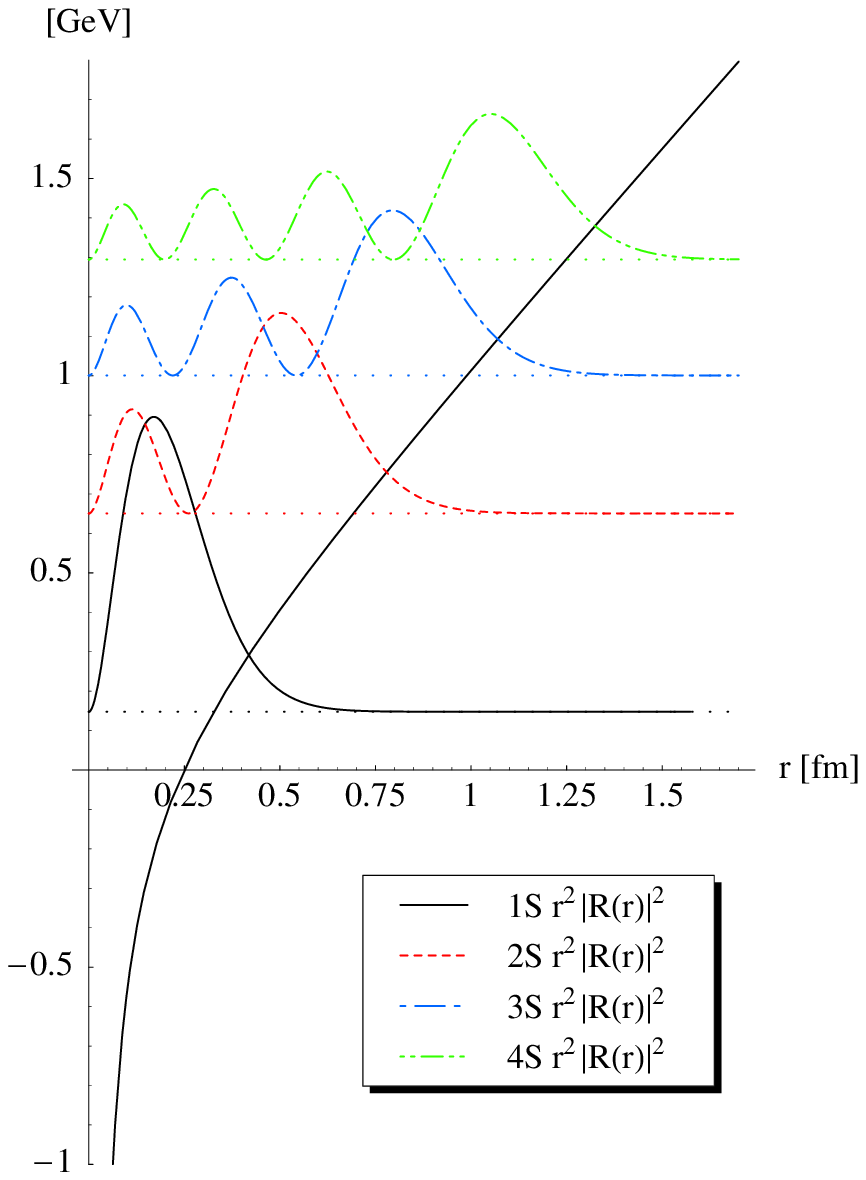}\end{center}
\end{minipage}
\begin{minipage}{7.45cm}\begin{center}
\includegraphics[width=\textwidth]{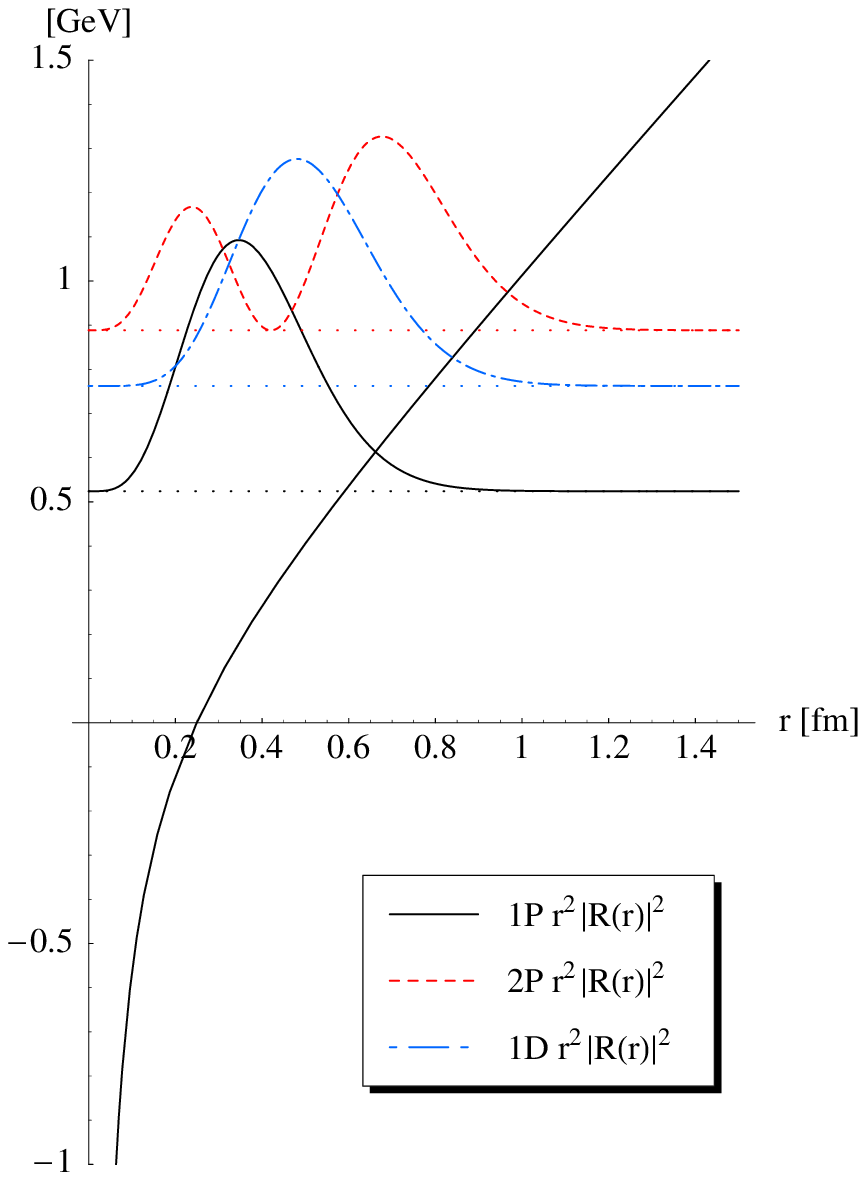}\end{center}
\end{minipage}\end{center}\vspace{-1cm}
\caption{\label{Fig:RadialDensityBottomoniumCpL*pI-2} Radial densities for bottomonium states of model [CpL*pI-2] plotted together with used potential ($\alpha_s=0.255$, $\alpha_i=1.0$, $\sigma=1.08\:\textrm{GeV/fm}$, $m_b=4.768\:\textrm{GeV}$, $M=2m_b$, $\mu_{\textrm{GR}}=2/3m_b$, $n_f=4$); base lines of radial densities have been shifted by according $E_{kl}$.}
\end{figure} 
\end{center}
\begin{center}
\begin{figure}[p]\begin{center}
\includegraphics[width=0.9\textwidth]{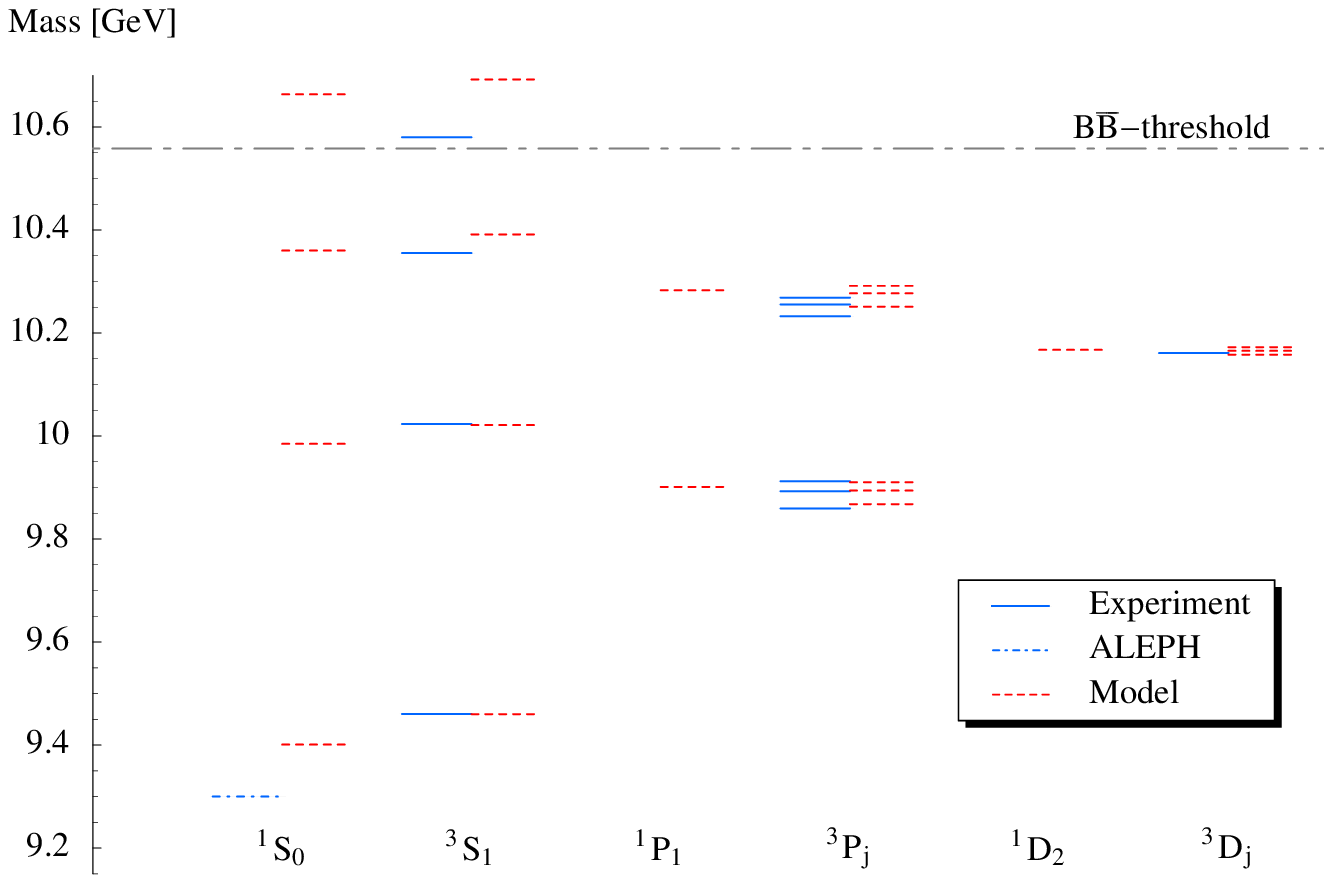}\end{center}
\caption{\label{Fig:SpectrumBottomoniumCpL*pI-2} Bottomonium spectra from experiment and model [CpL*pI-2] ($\alpha_s=0.255$, $\alpha_i=1.0$, $\sigma=1.08\:\textrm{GeV/fm}$, $m_b=4.768\:\textrm{GeV}$, $M=2m_b$, $\mu_{\textrm{GR}}=2/3m_b$, $n_f=4$); $\eta_b$ mass measurement by ALEPH collaboration is included separately in the experimental spectrum.}
\end{figure} 
\end{center}
The input used to determine the parameters of model [CpL*pI-2] for bottomonium is given by the masses of states $\Upsilon(1S)$ and $\Upsilon(2S)$, the center-of-gravity mass $C(1P)$, and the ratio
\begin{eqnarray}
\Phi_b^\textrm{Exp}(1P)=\frac{M(\chi_{b2})-M(\chi_{b1})}{M(\chi_{b1})-M(\chi_{b0})}\:,
\end{eqnarray}
identifying $C(1P)$ with the $1^1P_1$-state of the model. The parameters fullfill the constraint to be as close as possible to their empirical values. The resulting parameter set is given by
\begin{eqnarray}\label{Eq:ParameterSetFinalBottom}
\alpha_s&=&0.255\:,\nonumber\\
\alpha_i&=&1.0\:,\nonumber\\
m_b&=&4.768\:\textrm{GeV}\:,\\
\sigma&=&1.08\:\textrm{GeV/fm}\:.\nonumber
\end{eqnarray}
Remarkably, model [CpL*pI-2] implies a value of $\alpha_s$ for bottomonium which is significantly smaller than for charmonium, while the bottom quark mass and the string tension are close to their empirical determinations.

The resulting reduced radial wavefunctions are presented in Fig. \ref{Fig:Wavefunctions-S-Bottomonium-CpL*pI-2} and Fig. \ref{Fig:Wavefunctions-PD-Bottomonium-CpL*pI-2}, while in Fig. \ref{Fig:RadialDensityBottomoniumCpL*pI-2} the radial densities of the lowest lying $b\overline{b}$-states are displayed. The radial densities are drawn at their respective energy eigenvalues to indicate their position in the potential. The comparison between Fig. \ref{Fig:RadialDensityBottomoniumCpL*pI-2} and the corresponding figure for model [CpL-B] (Fig. \ref{Fig:RadialDensityBottomoniumCpL-B}) shows that the eigenvalues in model [CpL*pI-2] are significantly larger. This is mainly a consequence of the smaller coupling $\alpha_s$ and to a lesser degree due to the slight increase of the string tension $\sigma$.

Experimental masses and widths for bottomonium are displayed together with the masses of model [CpL*pI-2] in Tab. \ref{Tab:MassesBottomoniumCpL*pI-2}, while Fig. \ref{Fig:SpectrumBottomoniumCpL*pI-2} presents the spectrum arising from experiments together with the one of model [CpL*pI-2].
\begin{table}[t!]
\begin{center}
\begin{displaymath}
\begin{array}{||c|c|c|c|c||}
\hline\hline
&&\multicolumn{2}{|c|}{\qquad\textrm{Experiment \cite{PDBook}}\qquad} &  \textrm{Theory [CpL*pI-2]}\\
 \;\textrm{State}\; & \;\textrm{Candidate}\;&\multicolumn{2}{|c|}{\hrulefill} &  \\
&&\textrm{Mass [MeV]}&\textrm{Width [MeV]}& \textrm{Mass [MeV]}\\
\hline
1^1S_0 & (\eta_b)       & (9300\pm20\pm20)       &                                 &   9401 \\
1^3S_1 & \Upsilon(1S)   & 9460\pm0.26            & (53.0\pm1.5)\textrm{ keV}       &   9460 \\
\hline
1^1P_1 &                &                        &                                 &   9901 \\
1^3P_0 & \chi_{b0}(1P)  & 9859.44\pm0.42\pm0.31  &                                 &   9868 \\
1^3P_1 & \chi_{b1}(1P)  & 9892.78\pm0.26\pm0.31  &                                 &   9894 \\
1^3P_2 & \chi_{b2}(1P)  & 9912.21\pm0.26\pm0.31  &                                 &   9910 \\
\hline
2^1S_0 &                &                        &                                 &   9985 \\
2^3S_1 & \Upsilon(2S)   & 10023.26\pm0.31        & (30.6\pm2.3)\textrm{ keV}       &  10021 \\
\hline
1^1D_2 &                &                        &                                 &  10167 \\
1^3D_1 &                &                        &                                 &  10158 \\
1^3D_2 & \Upsilon(1D)   & 10161.1\pm0.6\pm1.6    &                                 &  10165 \\
1^3D_3 &                &                        &                                 &  10173 \\
\hline
2^1P_1 &                &                        &                                 &  10283 \\
2^3P_0 & \chi_{b0}(2P)  & 10232.5\pm0.4\pm0.5    &                                 &  10251 \\
2^3P_1 & \chi_{b1}(2P)  & 10255.46\pm0.22\pm0.50 &                                 &  10277 \\
2^3P_2 & \chi_{b2}(2P)  & 10268.65\pm0.22\pm0.50 &                                 &  10292 \\
\hline
3^1S_0 &                &                        &                                 &  10360 \\
3^3S_1 & \Upsilon(3S)   & 10355.2\pm0.5          & (22.1\pm2.7)\textrm{ keV}       &  10391 \\
\hline
4^1S_0 &                &                        &                                 &  10663 \\
4^3S_1 & \Upsilon(4S)   & 10580.0\pm3.5          & 20\pm2\pm4                      &  10692 \\
\hline\hline
\end{array}
\end{displaymath}
\caption{\label{Tab:MassesBottomoniumCpL*pI-2} Comparison of experimental masses and widths of $b\overline{b}$-states with masses calculated with model [CpL*pI-2] ($\alpha_s=0.255$, $\alpha_i=1.0$, $\sigma=1.08\:\textrm{GeV/fm}$, $m_b=4.768\:\textrm{GeV}$, $M=2m_b$, $\mu_{\textrm{GR}}=2/3m_b$, $n_f=4$).}
\end{center}
\end{table}
\begin{table}[p]
\begin{center}
\begin{displaymath}
\begin{array}{||c|c|c|c|c|c|c|c||}
\hline\hline
       \multirow{2}{*}{ \textrm{State} }& E_{kl} & M_\textrm{g}^\textrm{Cc} & M_\textrm{g}^\textrm{rsq} & M_\textrm{g}^\textrm{con} & M_\textrm{i}^\textrm{con} & M_\textrm{g}^\textrm{SS} & M_\textrm{i}^\textrm{SS}\\
&\textrm{[MeV]}&\textrm{[MeV]}&\textrm{[MeV]}&\textrm{[MeV]}&\textrm{[MeV]}&\textrm{[MeV]}&\textrm{[MeV]}\\
\hline
1^1S_0 &  \multirow{2}{*}{147.8} & \multirow{2}{*}{-132.8} & \multirow{2}{*}{-59.4} & \multirow{2}{*}{15.1} &  \multirow{2}{*}{17.4} &-26.9 & -17.4 \\
1^3S_1 &    & & & && 9.0 & 5.8 \\
\hline
1^1P_1 &  \multirow{4}{*}{524.1} & \multirow{4}{*}{-123.4} & \multirow{4}{*}{-9.3} & \multirow{4}{*}{0.5} &\multirow{4}{*}{-0.5}&0.3&0.5\\\cline{7-8}
1^3P_0 &  & & &  &&\multirow{3}{*}{-0.1}&\multirow{3}{*}{-0.2} \\
1^3P_1 &  & & & &&& \\
1^3P_2 &  &  &  & &&& \\
\hline
2^1S_0 & \multirow{2}{*}{650.8} & \multirow{2}{*}{-102.7} & \multirow{2}{*}{-29.5} & \multirow{2}{*}{8.3} &\multirow{2}{*}{11.2}& -16.2 & -11.2 \\
2^3S_1 & &  &  &  && 5.4 & 3.7\\
\hline
1^1D_2 & \multirow{4}{*}{762.5} & \multirow{4}{*}{-106.0} & \multirow{4}{*}{-4.3} & \multirow{4}{*}{0.1} &\multirow{4}{*}{$\sim0$}&0.1&\sim0\\\cline{7-8}
1^3D_1 & & & &  &&\multirow{3}{*}{$\sim0$}&\multirow{3}{*}{$\sim0$}\\
1^3D_2 & & & &  &&&\\
1^3D_3 & & & &  &&&\\
\hline
2^1P_1 & \multirow{4}{*}{888.4} & \multirow{4}{*}{-101.4} & \multirow{4}{*}{-6.9} & \multirow{4}{*}{$\sim0$} &\multirow{4}{*}{-0.6}&0.3&0.6\\\cline{7-8}
2^3P_0 & &  & &  &&\multirow{3}{*}{-0.1}&\multirow{3}{*}{-0.2}\\
2^3P_1 & &  &  &  &&&\\
2^3P_2 & &  &  &  &&&\\
\hline
3^1S_0 & \multirow{2}{*}{1000.6} & \multirow{2}{*}{-87.4} & \multirow{2}{*}{-22.3} & \multirow{2}{*}{6.6} &\multirow{2}{*}{9.9}&-13.6&-9.9\\
3^3S_1 & & & & && 4.5 & 3.3 \\
\hline
4^1S_0 & \multirow{2}{*}{1294.5} & \multirow{2}{*}{-77.9} & \multirow{2}{*}{-18.9} & \multirow{2}{*}{1.8} &\multirow{2}{*}{9.3}&-12.4&-9.3\\
4^3S_1 & & & & && 4.1 & 3.1 \\
\hline\hline
\multicolumn{8}{c}{}\\
\hline\hline
       \multirow{2}{*}{ \textrm{State} }& M_\textrm{g}^\textrm{T} & M_\textrm{i}^\textrm{T}& M_\textrm{g}^\textrm{LS} & M_\textrm{i}^\textrm{LS}& M_\textrm{g}^\textrm{MD} & M_\textrm{i}^\textrm{MD}& M^\textrm{C}_\textrm{i} \\
&\textrm{[MeV]}&\textrm{[MeV]}&\textrm{[MeV]}&\textrm{[MeV]}&\textrm{[MeV]}&\textrm{[MeV]}&\textrm{[MeV]}\\
\hline
1^1S_0 &  \multirow{2}{*}{0} & \multirow{2}{*}{0} & \multirow{2}{*}{0} & \multirow{2}{*}{0} & \multirow{2}{*}{-33.7} &\multirow{2}{*}{-15.2} &\multirow{2}{*}{-43.7}\\
1^3S_1 &    & & & &&& \\
\hline
1^1P_1 & 0 & 0 & 0 & 0 &\multirow{4}{*}{-25.6}&\multirow{4}{*}{-11.2}&\multirow{4}{*}{-0.5}\\
1^3P_0 & -6.6 & -3.5 & -18.9 & -3.3 & & & \\
1^3P_1 & 3.3 & 1.7 & -9.5 & -1.7 &&& \\
1^3P_2 & -0.7 & 0.3 & 9.5 & 1.7 &&& \\
\hline
2^1S_0 & \multirow{2}{*}{0} & \multirow{2}{*}{0} & \multirow{2}{*}{0} & \multirow{2}{*}{0} &\multirow{2}{*}{-36.0}&\multirow{2}{*}{-12.1} &\multirow{2}{*}{-24.7} \\
2^3S_1 & &  &  &  &&&\\
\hline
1^1D_2 & 0 & 0 & 0 & 0 &\multirow{4}{*}{-21.1}&\multirow{4}{*}{$\sim0$}&\multirow{4}{*}{$\sim0$}\\
1^3D_1 & -0.9 & \sim0 & -8.3 & -0.1 &&&\\
1^3D_2 & 0.9 & \sim0 & -2.7 & \sim0  &&&\\
1^3D_3 & -0.2 & \sim0 & 5.5 & \sim0 &&&\\
\hline
2^1P_1 & 0 & 0 & 0 & 0 &\multirow{4}{*}{-31.0}&\multirow{4}{*}{-15.0}&\multirow{4}{*}{-0.6}\\
2^3P_0 & -5.8 & -4.6 & -16.3 & -4.4  &&&\\
2^3P_1 & 2.9 & 2.3 & -8.1 & -2.2  &&&\\
2^3P_2 & -0.6 & -0.5 & 8.1 & 2.2 &&&\\
\hline
3^1S_0 & \multirow{2}{*}{0} & \multirow{2}{*}{0} & \multirow{2}{*}{0} & \multirow{2}{*}{0} &\multirow{2}{*}{-38.6}&\multirow{2}{*}{-12.1} &\multirow{2}{*}{-20.0}\\
3^3S_1 & & & & &&& \\
\hline
4^1S_0 & \multirow{2}{*}{0} & \multirow{2}{*}{0} & \multirow{2}{*}{0} & \multirow{2}{*}{0} &\multirow{2}{*}{-40.8}&\multirow{2}{*}{-12.5} &\multirow{2}{*}{-17.7}\\
4^3S_1 & & & & &&& \\
\hline\hline
\end{array}
\end{displaymath}\vspace{-0,6cm}
\caption{\label{Tab:MassesContributionsBottomoniumCpL*pI-2} Contributions to the masses of $b\overline{b}$-states in model [CpL*pI-2] ($\alpha_s=0.255$, $\alpha_i=1.0$, $\sigma=1.08\:\textrm{GeV/fm}$, $m_b=4.768\:\textrm{GeV}$, $M=2m_b$, $\mu_{\textrm{GR}}=2/3m_b$, $n_f=4$).}
\end{center}
\end{table}
The value for the mean $1P$ mass is
\begin{eqnarray*}
C(1P)\approx9900\:\textrm{MeV}\:.
\end{eqnarray*}
The largest deviation for our input states is less than 2 MeV for the $\Upsilon(2S)$-state. The ratio
\begin{eqnarray*}
\Phi_b^\textrm{[CpL*pI-2]}(1P)=\left(\frac{M(\chi_{b2})-M(\chi_{b1})}{M(\chi_{b1})-M(\chi_{b0})}\right)_{\textrm{[CpL*pI-2]}}&\approx&0.62\:,\\
\end{eqnarray*}
agrees with the experimental value
\begin{eqnarray*}
\Phi_b^\textrm{Exp}(1P)=\left(\frac{M(\chi_{b2})-M(\chi_{b1})}{M(\chi_{b1})-M(\chi_{b0})}\right)_{\textrm{Exp}}&=&0.58\pm0.03\:,\\
\end{eqnarray*}
almost within the uncertainties. Matching $\Phi(1P)$ exactly to the experimental value implies that the parameters tend to be not as close to their natural values anymore. In particular, the string tension increases further after tuning the remaining parameters. Further, the ratio $\Phi_b(2P)$ for the $2P$-states will be reduced from its value
\begin{eqnarray*}
\Phi_b^\textrm{[CpL*pI-2]}(2P)\left(\frac{M(\chi'_{b2})-M(\chi'_{b1})}{M(\chi'_{b1})-M(\chi'_{b0})}\right)_{\textrm{[CpL*pI-2]}}&\approx&0.56\:,
\end{eqnarray*}
calculated with the set Eq. (\ref{Eq:ParameterSetFinalBottom}), which agrees perfectly with the empirical
\begin{eqnarray*}
\Phi_b^\textrm{Exp}(1P)\left(\frac{M(\chi'_{b2})-M(\chi'_{b1})}{M(\chi'_{b1})-M(\chi'_{b0})}\right)_{\textrm{Exp}}&=&0.57\pm0.05\:.
\end{eqnarray*}

The contributions to the masses of bottomonium states are displayed in detail in Tab. \ref{Tab:MassesContributionsBottomoniumCpL*pI-2}. It is obvious that
\begin{eqnarray}
M_{kl}^\textrm{bare}=2m_b+E_{kl}\:,
\end{eqnarray}
leads to significantly larger values than the experimental masses. As in the case of charmonium, there exist large downward shifts for the bottomonium masses. The purely momentum dependent term $M_\textrm{g}^\textrm{MD}$ yields large negative contributions. The radiative correction to the Coulomb term $M_\textrm{g}^\textrm{Cc}$ provides a major negative shift to all states. The remaining terms which provide spin-independent splittings, $M_\textrm{g}^\textrm{rsq}$, $M_\textrm{g}^\textrm{con}$, $M_\textrm{i}^\textrm{C}$ and $M_\textrm{i}^\textrm{con}$, act primarily on the $S$-states and separate these states and from the rest. The $M_\textrm{i}^\textrm{MD}$ contribution shows quite a different behaviour for the $S$-states in bottomonium as compared to charmonium. While in charmonium the magnitude of $M_\textrm{i}^\textrm{MD}$ increases with the radial excitation, bottomonium yields the largest contribution for the groundstate. This is due to the influence of the Coulomb part of the potential. The induced interaction gives rise to about 40 percent of the ${}^3S_1-{}^1S_0$ splitting. The importance of the induced interaction for our model manifests itself once more in the structure of the triplet $P$-state splittings. One recognizes that due to the increased weight of the tensor part in the induced interaction compared to the gluon exchange interaction, the model agrees with the experimental ratios $\Phi_b(1P)$ and $\Phi_b(2P)$.

%% file: 7-Fogtpp-induced/7-Remarks.tex
In the effort of describing the quarkonium spectra within potential models, the model [CpL*pI-2] represents a substantial improvement compared to the basic model [CpL-B]. It provides a detailed description of the spectra below the heavy-light meson thresholds with minimal phenomenological input and realistic parameter values. Nevertheless some details have to be commented on.

There remain questions about the renormalization scale which we predetermined inspired by \cite{Titard:1993nn}. At least for charmonium the link between $\alpha_s$ and predetermined scale $\mu_{\overline{\textrm{MS}}}$ is not obvious. The fact that, for charmonium, $\alpha_s(\mu_{\overline{\textrm{MS}}})>0.5$, might be a sign that a perturbative approach is not justified anymore.

\begin{table}[t!]
\begin{center}
\begin{displaymath}
\begin{array}{||c|c|c||}
\hline\hline
&&\\
\quad\textrm{State}\quad &\quad J^{PC} \quad &\quad \textrm{Structure} \quad \\
&&\\
\hline
k^1P_1& 1^{+-} & \overline{\psi}\left(\sigma^{\mu\nu}\right)\psi\\
k^3P_0& 0^{++} & \overline{\psi}\psi\\
k^3P_1& 1^{++} & \overline{\psi}\left(\gamma^{\mu}\gamma_5\right)\psi\\
k^3P_2& 2^{++} & ?\quad\frac{1}{2m}\overline{\psi}\left(\gamma^\mu\left(p+p'\right)^\nu\right)\psi\quad?\\
\hline\hline
\end{array}
\end{displaymath}\vspace{-0.4cm}
\caption{\label{Tab:P-waveInducedInteraction} Experimentally determined quantum numbers for the $P$-states and presumptive structures for the t-channel exchange of $P$-states.}\vspace{-0.6cm}
\end{center}
\end{table}
The induced interaction is only implemented for the $1S$-states. In principle one would have to implement the induced interaction for all remaining bound states, starting with the induced interaction for the $1P$-states (Tab. \ref{Tab:P-waveInducedInteraction}). The major argument against considering the exchange of other bound states is, that this leads to additional parameters.

The combination of induced interaction and gluon exchange potential to order $\mathcal{O}(\alpha_s^2)$ bears the double counting problem mentioned previously. One would in principle have to subtract the double counted diagram, which is not yet done in the present calculations.

%% file: 7-Fogtpp-induced/7-Radius-Mass.tex
After studying the spectrum gained via model [CpL*pI-2] we will again test the mass-radius relation. In Chapter \ref{Chapt:CpL-Breit-Radius-Mass} we have already seen that, according to the virial theorem, the energy eigenvalues should follow
\begin{eqnarray}
E&=&\bigg\langle-\frac{2\alpha_s}{3r}+\frac{3\sigma}{2} r\bigg\rangle\:,
\end{eqnarray}
resulting in the mass
\begin{eqnarray}
M=2m_q+E=2m_q+\bigg\langle-\frac{2\alpha_s}{3r}+\frac{3\sigma r}{2}\bigg\rangle\:.
\end{eqnarray}
Thus one expects that the quarkonium states show an almost linear mass-radius relation.

To probe the relation of radius and mass one has to decide which masses to take into account for the quarkonium states as the perturbatively implemented corrections do not change the radius. In Chapter \ref{Chapt:CpL-Breit-Radius-Mass} we argued that one should take the bare mass
\begin{eqnarray}
M_{kl}^\textrm{bare}=2m_q+E_{kl}\:,
\end{eqnarray}
which does not include any of the perturbatively treated corrections as none of them occur explicitely in the virial theorem. In the discussion of model [CpL*pI-2] we realized that the values of $M_{kl}^\textrm{bare}$ are about 200 MeV larger than the respective experimental quarkonium state masses. A possibility to test the mass-radius relation with masses close to the experimental ones is to include the spin-independent corrections of model [CpL*pI-2] into the considered mass 
\begin{eqnarray}
M_{kl}^\textrm{SI}=2m_q+E_{kl}+M_\textrm{g}^\textrm{Cc}+M_\textrm{g}^\textrm{rsq}+M_\textrm{g}^\textrm{con}+M_\textrm{g}^\textrm{MD}+M_\textrm{i}^\textrm{C}+M_\textrm{i}^\textrm{con}+M_\textrm{i}^\textrm{MD}\:.
\end{eqnarray}
Hence we will investigate the mass-radius relation in two ways, a first with $M_{kl}^\textrm{bare}$ as input and a second in which  $M_{kl}^\textrm{SI}$ is used.

%% file: 7-Fogtpp-induced/7-Radius-Mass-Charmonium.tex
\vspace{-0.8cm}
\begin{center}
\begin{figure}[p]\begin{center}
\includegraphics[width=0.9\textwidth]{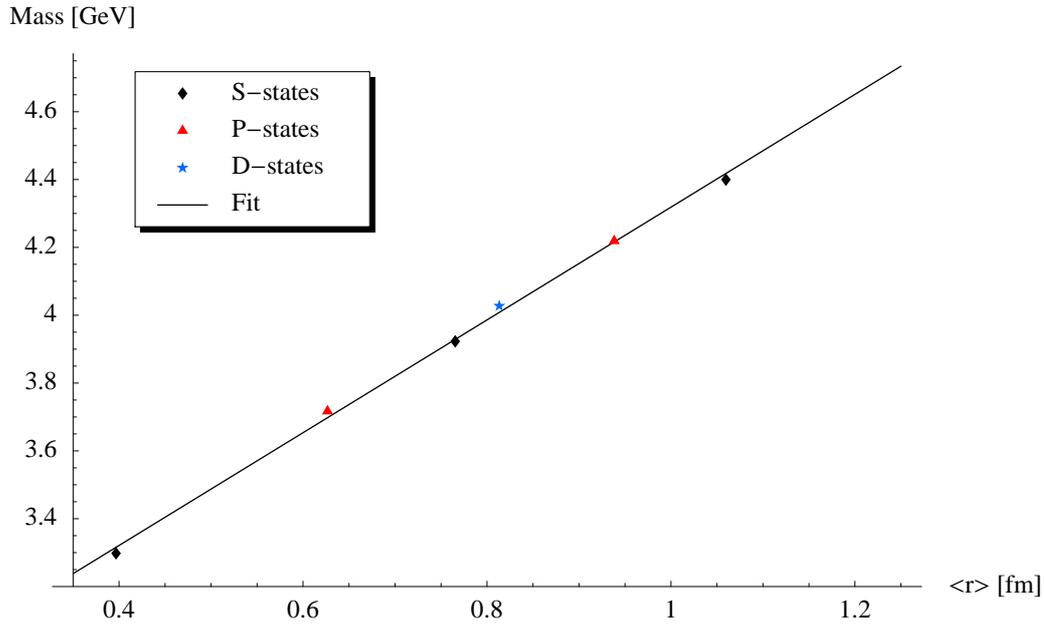}\end{center}
\caption{\label{Fig:CharmoniumRadiusMassBare-CpL*pI-2} $M_{kl}^{\textrm{bare}}$ versus $\langle r\rangle$ for $c\overline{c}$-states in model [CpL*pI-2] ($\alpha_s=0.295$, $\alpha_i=1.0$, $\sigma=1.04\:\textrm{GeV/fm}$, $m_c=1.4056\:\textrm{GeV}$, $M=3.07\:\textrm{GeV}$, $\mu_{\textrm{GR}}=m_c$, $n_f=3$).}
\end{figure} 
\end{center}
\begin{center}
\begin{figure}[p]\begin{center}
\includegraphics[width=0.9\textwidth]{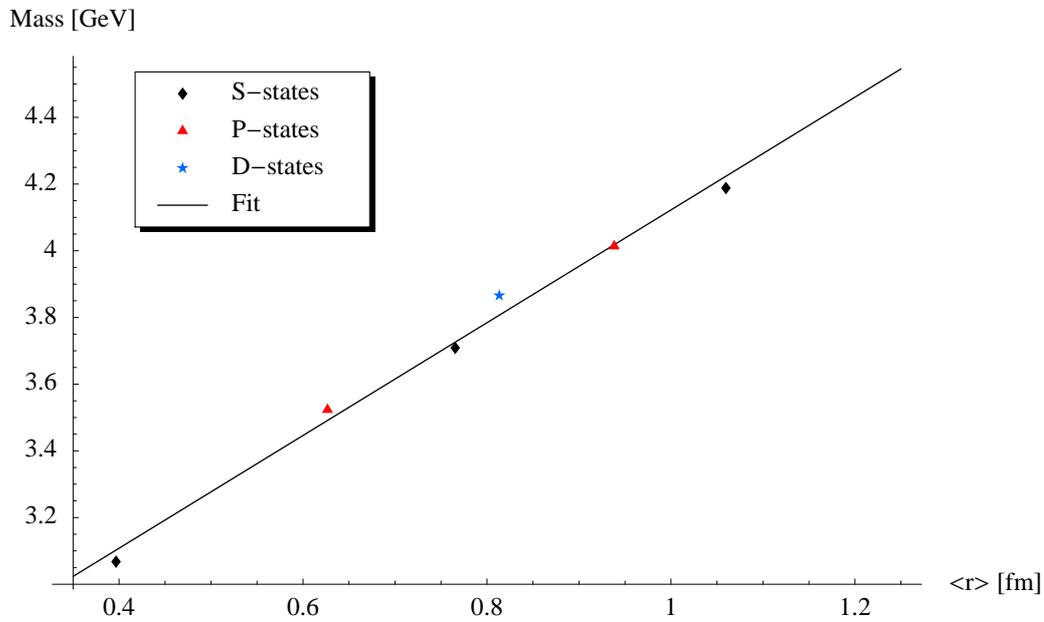}\end{center}
\caption{\label{Fig:CharmoniumRadiusMassSI-CpL*pI-2} $M_{kl}^{\textrm{SI}}$ versus $\langle r\rangle$ for $c\overline{c}$-states in model [CpL*pI-2] ($\alpha_s=0.295$, $\alpha_i=1.0$, $\sigma=1.04\:\textrm{GeV/fm}$, $m_c=1.4056\:\textrm{GeV}$, $M=3.07\:\textrm{GeV}$, $\mu_{\textrm{GR}}=m_c$, $n_f=3$).}
\end{figure} 
\end{center}
It is instructive to examine the radial densities in Fig. \ref{Fig:RadialDensityCharmoniumCpL*pI-2}, which demonstrate that the charmonium states are dominated by the linear part of the potential and thus a linear mass-radius relation emerges.

In Tab. \ref{Tab:CharmoniumRadiusMass-CpL*pI-2} the considered $c\overline{c}$-masses and radii within model [CpL*pI-2] are presented.
\begin{table}[ht!]
\begin{center}\vspace{-0.2cm}
\begin{displaymath}
\begin{array}{|c|c|c|c|}
\hline
&&&\\
\quad\textrm{State}\quad &\quad M_{kl}^\textrm{bare}\textrm{ [MeV]} \quad &\quad M_{kl}^\textrm{SI}\textrm{ [MeV]} \quad &\quad \langle r\rangle \textrm{ [fm]}\quad\\
&&&\\
\hline
1S & 3298 & 3068 & 0.397\\
1P & 3717 & 3523 & 0.627\\
2S & 3922 & 3709 & 0.765\\
1D & 4028 & 3866 & 0.813\\
2P & 4219 & 4014 & 0.938\\
3S & 4399 & 4188 & 1.060\\
\hline
\end{array}
\end{displaymath}\vspace{-0.6cm}
\caption{\label{Tab:CharmoniumRadiusMass-CpL*pI-2} Charmonium radii and masses in model [CpL*pI-2] ($\alpha_s=0.295$, $\alpha_i=1.0$, $\sigma=1.04\:\textrm{GeV/fm}$, $m_c=1.4056\:\textrm{GeV}$, $M=3.07\:\textrm{GeV}$, $\mu_{\textrm{GR}}=m_c$, $n_f=3$).}
\end{center}
\end{table}
Matching straight lines to the data in Tab. \ref{Tab:CharmoniumRadiusMass-CpL*pI-2} yields, using $M_{kl}^\textrm{bare}$ as input, for the parameters of the best fit straight line 
\begin{eqnarray}\label{Eq:RadiusMassCharmoniumBare}
M_\textrm{fit}^\textrm{bare}&=&a^\textrm{bare}+b^\textrm{bare}\langle r\rangle\:,\nonumber\\
a^\textrm{bare}&=&2.66\: \textrm{GeV}\:,\\
b^\textrm{bare}&=&1.66\: \textrm{GeV/fm}\:,\nonumber
\end{eqnarray}
while using $M_{kl}^\textrm{SI}$ results in
\begin{eqnarray}
M_\textrm{fit}^\textrm{SI}&=&a^\textrm{SI}+b^\textrm{SI}\langle r\rangle\:,\nonumber\\
a^\textrm{SI}&=&2.43\: \textrm{GeV}\:,\\
b^\textrm{SI}&=&1.69\: \textrm{GeV/fm}\:.\nonumber
\end{eqnarray}

The straight line $M_\textrm{fit}^\textrm{bare}$ is displayed together with the input data in Fig. \ref{Fig:CharmoniumRadiusMassBare-CpL*pI-2}. The deviations from a linear behaviour are small, which seems natural as the viral theorem predicts a linear relation between mass and radius in this region. One also recognizes that the value for $b^\textrm{bare}$ is, as predicted by the virial theorem, very close to $3/2\:\sigma$.

In Fig. \ref{Fig:CharmoniumRadiusMassSI-CpL*pI-2} the resulting straight line $M_\textrm{fit}^\textrm{SI}$ and the input data for the second case are presented. One realizes that, while the parameter $a^\textrm{SI}$ is, as expected, about 200 MeV smaller than $a^\textrm{bare}$, the slope parameter $b^\textrm{SI}$ increased slightly comared to the bare mass fit. This is understandable if one remembers that some of the perturbatively treated corrections yield an additional division between the $S$-states and the angular excited states. This additional separation is similar to an increase in string tension and is therfore noticeable via a larger value for the slope parameter of the straight line. Further one realizes that the deviations to the linear behaviour are considerably larger in this case, but still small enough to justify the assumption of a linear mass-radius relation.

%% file: 7-Fogtpp-induced/7-Radius-Mass-Bottomonium.tex
\vspace{-0.8cm}
\begin{center}
\begin{figure}[p]\begin{center}
\includegraphics[width=0.9\textwidth]{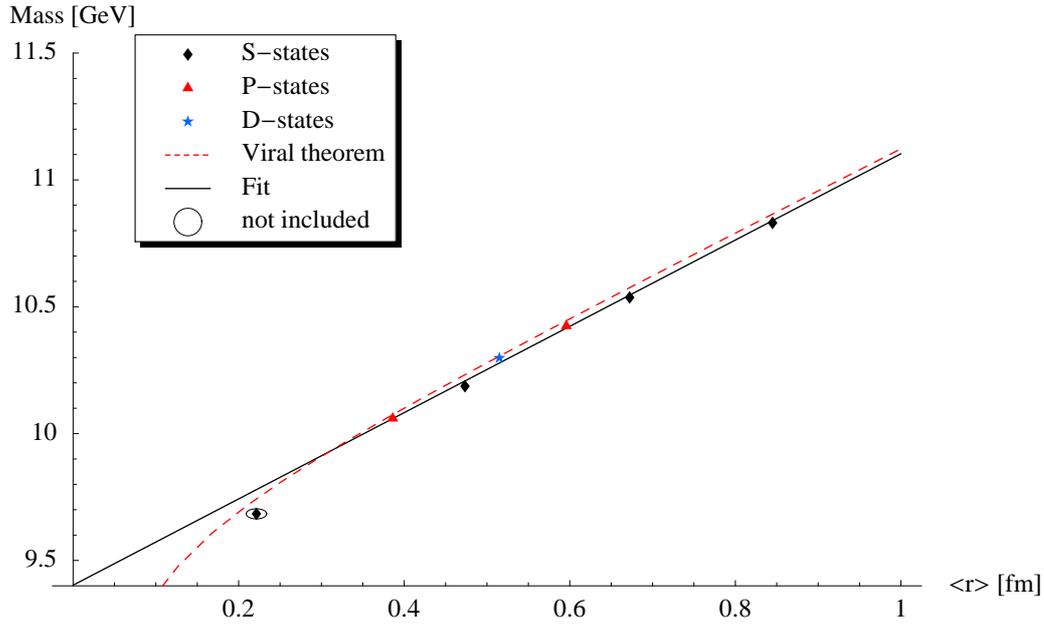}\end{center}
\caption{\label{Fig:BottomoniumRadiusMassBare-CpL*pI-2} $M_{kl}^{\textrm{bare}}$ versus $\langle r\rangle$ for $b\overline{b}$-states in model [CpL*pI-2] ($\alpha_s=0.255$, $\alpha_i=1.0$, $\sigma=1.08\:\textrm{GeV/fm}$, $m_b=4.768\:\textrm{GeV}$, $M=2m_b$, $\mu_{\textrm{GR}}=2/3m_b$, $n_f=4$) including the virial theorem curve (Eq. (\ref{Eq:VirialMass})); 1S-state is not included in straight line fit.}
\end{figure} 
\end{center}
\begin{center}
\begin{figure}[p]\begin{center}
\includegraphics[width=0.9\textwidth]{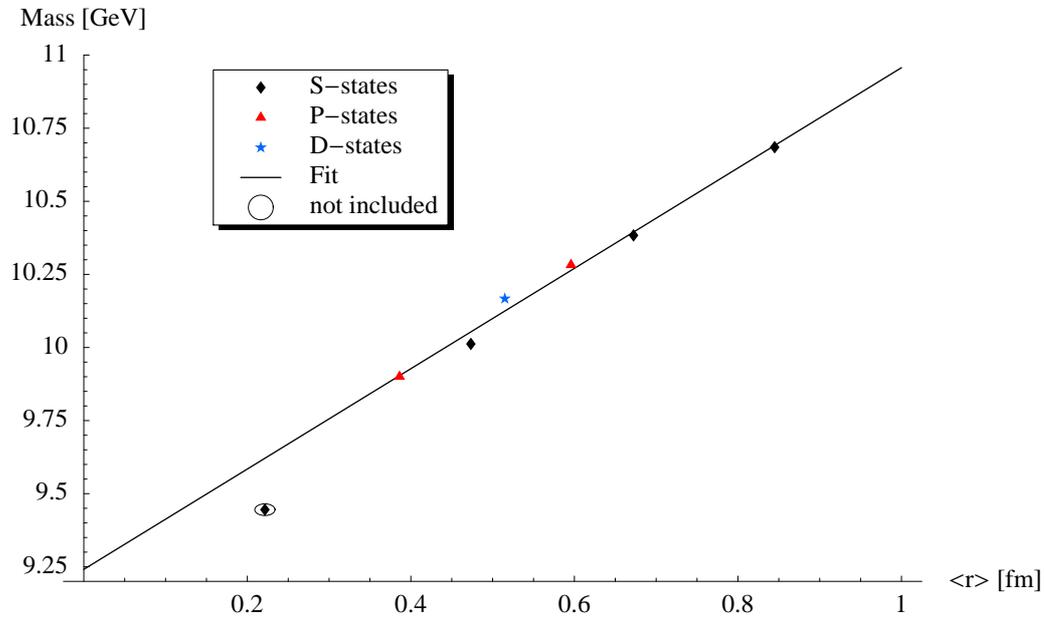}\end{center}
\caption{\label{Fig:BottomoniumRadiusMassSI-CpL*pI-2} $M_{kl}^{\textrm{SI}}$ versus $\langle r\rangle$ for $b\overline{b}$-states in model [CpL*pI-2] ($\alpha_s=0.255$, $\alpha_i=1.0$, $\sigma=1.08\:\textrm{GeV/fm}$, $m_b=4.768\:\textrm{GeV}$, $M=2m_b$, $\mu_{\textrm{GR}}=2/3m_b$, $n_f=4$); 1S-state is not included in straight line fit.}
\end{figure} 
\end{center}

For guidance one takes once more a quick look at the radial density plot (Fig. \ref{Fig:RadialDensityBottomoniumCpL*pI-2}), where one can easily behold that the $1S$-state seems to be special as it is, in contrast to the other states, strongly influenced by the Coulomb part of the potential. Consequently a linear dependence of mass and radius might only be a good guess for states other than the $1S$-state.

In Tab. \ref{Tab:BottomoniumRadiusMass-CpL*pI-2} the radii and masses calculated within model [CpL*pI-2] are presented.
\begin{table}[htb!]
\begin{center}
\begin{displaymath}
\begin{array}{|c|c|c|c|}
\hline
&&&\\
\quad\textrm{State}\quad &\quad M_{kl}^\textrm{bare}\textrm{ [MeV]} \quad &\quad M_{kl}^\textrm{SI}\textrm{ [MeV]} \quad &\quad \langle r\rangle \textrm{ [fm]}\quad\\
&&&\\
\hline
1S &  9684 &  9445 & 0.221 \\
1P & 10060 &  9900 & 0.386 \\
2S & 10187 & 10012 & 0.473 \\
1D & 10299 & 10167 & 0.515 \\
2P & 10424 & 10283 & 0.596 \\
3S & 10537 & 10384 & 0.672 \\
4S & 10831 & 10685 & 0.845 \\
\hline
\end{array}
\end{displaymath}
\caption{\label{Tab:BottomoniumRadiusMass-CpL*pI-2} Bottomonium radii and masses in model [CpL*pI-2] ($\alpha_s=0.255$, $\alpha_i=1.0$, $\sigma=1.08\:\textrm{GeV/fm}$, $m_b=4.768\:\textrm{GeV}$, $M=2m_b$, $\mu_{\textrm{GR}}=2/3m_b$, $n_f=4$).}
\end{center}
\end{table}
Matching straight lines to the data in Tab. \ref{Tab:BottomoniumRadiusMass-CpL*pI-2} yields, using $M_{kl}^\textrm{bare}$ as input, for the parameters of the best fit straight line 
\begin{eqnarray}
M_\textrm{fit}^\textrm{bare}&=&a^\textrm{bare}+b^\textrm{bare}\langle r\rangle\:,\nonumber\\
a^\textrm{bare}&=&9.40\: \textrm{GeV}\:,\\
b^\textrm{bare}&=&1.70\: \textrm{GeV/fm}\:,\nonumber
\end{eqnarray}
while using $M_{kl}^\textrm{SI}$ results in
\begin{eqnarray}
M_\textrm{fit}^\textrm{SI}&=&a^\textrm{SI}+b^\textrm{SI}\langle r\rangle\:,\nonumber\\
a^\textrm{SI}&=&9.24\: \textrm{GeV}\:,\\
b^\textrm{SI}&=&1.72\: \textrm{GeV/fm}\:,\nonumber
\end{eqnarray}
where in both fits the $1S$-state data has not been included.

The straight line $M_\textrm{fit}^\textrm{bare}$ is displayed together with the input data in Fig. \ref{Fig:BottomoniumRadiusMassBare-CpL*pI-2}. In addition the prediction of the virial theorem is included in Fig. \ref{Fig:BottomoniumRadiusMassBare-CpL*pI-2} by displaying 
\begin{eqnarray}\label{Eq:VirialMass}
M(\langle r\rangle)=2m_b-\frac{2\alpha_s}{3\langle r\rangle}+\frac{3\sigma \langle r\rangle}{2}\:.
\end{eqnarray}
Fig. \ref{Fig:BottomoniumRadiusMassBare-CpL*pI-2} illustrates that the deviations of the model state masses from the linear behaviour are tiny. Furthermore, even the slope parameter $b^\textrm{bare}$ is near the prediction of the virial theorem. The model data points are lying slightly below the curve inspired by the viral theorem, which is due to the fact that one is not able to calculate an expectation value for the hole curve. To test this statement we have checked the expectation values of Eq. (\ref{Eq:VirialMass}) for the $S$-states and found that the values agree with the bare mass values.

In Fig. \ref{Fig:BottomoniumRadiusMassSI-CpL*pI-2} the resulting straight line $M_\textrm{fit}^\textrm{SI}$ and the input data for the second case are presented. One realizes that the deviations from a linear behaviour are significantely larger than in the case of bare masses. Compared to the bare mass fit one observes again a slight increase of the slope parameter.

The approximation of linear mass-radius relation for bottomonium is justified for both of the discussed cases, as long as the $1S$-state is excluded.

%% file: 7-Fogtpp-induced/7-Summary.tex
In this chapter we have combined the previously discussed models [CpLpI-3] and [CpL*-2] into a new model with perturbative corrections based on induced interaction and a fourth-order gluonic quark-antiquark potential. The resulting model [CpL*pI-2] incorporates the benefits of both models, yielding a quantitatively good spectrum by using reasonable parameters. At present this success still comes at the cost of a double counting problem and issues about the link between the relevant renormalization scale and the running coupling.

We found that within model [CpL*pI-2] the strong coupling constants $\alpha_s$ for bottomonium and charmonium are, for the first time in this work, correlated in the proper way, namely taking a significantly smaller value for bottomonium than for charmonium. Furthermore the parameter values for the quark masses and the string tension are close to their empirical values.

The triplet $P$-state structure of model [CpL*pI-2] coincides, up to some percent, with the experimental one for charmonium as well as for bottomonium.

In a last step we investigated, analogously to the discussion of model [CpL-B], the relation between mass and radius. In contrast to the discussion in Chapter \ref{Chapt:CpL-Breit-Radius-Mass}, we used two distinct assumptions for the considered masses, as the bare masses used in Chapter \ref{Chapt:CpL-Breit-Radius-Mass} are significantly larger in model [CpL*pI-2] than the experimental masses. The analysis for both cases shows that a linear mass-radius relation is a good approximation for charmonium in general, and also for bottomonium if one excludes the $1S$-state.

%% file: 9-Summary/9-Discussion.tex
In this work, we have developed a potential model which gives satisfactory results for the quarkonium spectra with only four adjustable parameters and minimal phenomenological input.

Starting out from the conventional Breit interaction combined with a linear confinement potential (Chapt. \ref{Chapt:CpL-Breit}), we argued that the retardation corrections included in the Breit interaction are not justified, since the on-mass shell condition is not valid for a bound state. This led us to a $q\overline{q}$-potential consisting of linear confinement and the one-gluon exchange without retardation (Chapt. \ref{Chapt:CpL-OGE}). Such restricted potentials face the following problems: The relative splittings of the triplet $P$-states ($\chi_{qj}$, $j=0,1,2$) are grossly overestimated. The string tension for charmonium comes out too large. Moreover, the effective coupling $\alpha_s$ for bottomonium is significantly larger than for charmonium.

In Chapt. \ref{Chapt:CpL-induced} we have presented a possible solution to the problems concerning the triplet $P$-state splitting, by including in the $q\overline{q}$-potential contributions from t-channel bound state exchanges (induced interaction). Such an additional interaction is a natural consequence of crossing symmetry together with the fact that the Schr\"odinger equation resums ladder-diagrams only in the s-channel. However, the problems concerning the string tension and the coupling $\alpha_s$ remained.

In the effort to cure also these problems (Chapt. \ref{Chapt:Fogtpp}), we made use of a perturbative QCD potential by Gupta and Radford, including 1-loop corrections at order $\mathcal{O}(\alpha_s^2)$. We have taken special care of time reversal invariance by choosing a more symmetric kinematics in the center of mass frame (under time reversal: $\vec{p}\rightarrow-\vec{p}$, $\vec{q}\rightarrow\vec{q}$, with $\vec{q}$ Fourier conjugated to $\vec{r}$\,). This procedure led to a reduction of the strength of some contact term by a factor of two. We have demonstrated that the inclusion of the fourth-order gluonic potential results in a reduced strong coupling $\alpha_s$ for bottomonium.

Finally, in Chapt. \ref{Chapt:FogtppInduced} both types of corrections - induced interaction and two gluon exchange - have been combined. Altogether this led to a satisfactory description of the charmonium and bottomonium spectra below the heavy-light meson thresholds $D\overline{D}$ and $B\overline{B}$. In particular, such an improved potential model is able to reproduce correctly the relative splitting of triplet $P$-states. This comes about through additional interactions of spin-orbit and tensor type. At the same time the parameters, quark mass $m_q$, string tension $\sigma$ and strong coupling $\alpha_s$, of the model take on values close to those given by QCD determinations and lattice QCD simulations.\medskip\\
As an application we have investigated the relation between masses and mean radii of the $q\overline{q}$-states in the model (Chapt. \ref{Chapt:CpL-Breit-Radius-Mass} \& Chapt. \ref{Chapt:FogtppInducedRadiusMass}). In each case we find an approximately linear mass-radius relation for all states except the $1S$-state of bottomonium. This underlines that the majority of the $q\overline{q}$-bound states is primarily determined by the linear confining potential.

%% file: 9-Summary/9-Conclusion.tex
In this work, we have constructed an improved potential model for heavy quarkonium. Our aim has been to get a satisfactory description of the quarkonium spectra with minimal phenomenological input. Our corrections beyond linear confinement and one-gluon exchange are physically well motivated. They originate either as 1-loop corrections from perturbative QCD or can be seen as non-perturbative effects (bound state exchanges in the t-channel) required by crossing symmetry. In comparison to other works we did neither allow for an additive constant in the confining potential nor did we make any assumptions about its Lorentz-structure (scalar versus vector confinement). The conclusions to be drawn from the present work are the following:
\begin{itemize}
\item Potential models with the spin-dependent interaction terms from the one-gluon exchange alone are not sufficient to give a quantitative description of the quarkonium spectra (in particular fine and hyperfine splittings). Moreover, the parameters - quark mass $m_q$, string tension $\sigma$ and strong coupling $\alpha_s$ - in such models take on values incompatible with QCD determinations and lattice calculations.
\item Corrections from perturbative QCD such as the fourth-order gluonic potential proportional to $\alpha_s^2$ are essential for an improved description of quarkonium. Their inclusion allows to have $\alpha_s$ approximately equal for bottomonium and charmonium.
\item The induced interaction (t-channel exchange of $q\overline{q}$-bound states) provides a distinct way to improve the description of the triplet $P$-state splittings, different from assuming a Lorentz-structure for the phenomenological confinement potential. In the latter case new fit parameters are unavoidably introduced.
\item We have demonstrated that a satisfactory description of the quarkonium spectra is possible within an improved potential model with reasonable values for the quark mass $m_q$, string tension $\sigma$ and strong coupling $\alpha_s$, and without using additional, uncontrolled phenomenological input.
\end{itemize}

%% file: 9-Summary/9-Outlook.tex
Even though the potential model constructed in this work provides a satisfactory reproduction of the quarkonium spectra, it leaves some open questions.

The combination of induced interaction and fourth-oder potential potentially includes some double counting. However, subtracting the contribution of the particular box diagram (Fig. \ref{Fig:Box-Diagram}) cannot be straight forwardly done since it involves a regulating gluon mass.

The coupling $\alpha_i$ for the induced interaction ($1S$-state t-channel exchange) can be related to the asymptotic normalization constant of the wavefunction. This would fix the coupling $\alpha_i$. Furthermore, exchanges of bound states with different quantum numbers ($P$-states, ...)
can also contribute to the induced interaction. At present their effect is hidden in the adjusted value of $\alpha_i$.

A quantitative description of quarkonium near and above the heavy-light meson thresholds clearly requires a coupled channel analysis.\medskip\\
In this work the improved $q\overline{q}$-potential has been used exclusively for the description of the quarkonium spectra. In the future one should use it also to calculate other properties such as decay widths and branching ratios.




%% file: A-Appendices/A-Conventions.tex
The different conventions employed in quantum field theory only differ in the parameters $J_i$, $K_i$, $L_i$, $N_i$ ($i=\; B,\; F$ distinguishes bosons from fermions) \cite{DonoghueDotSM92}, occurring in the normalization of fields,
\begin{eqnarray}
\varphi&=&\int\frac{\mathrm{d}^3k}{J_B}(a(\vec{k})\mathrm{e}^{-ik\cdot x}+a^\dagger(\vec{k})\mathrm{e}^{ip\cdot x})\:,\nonumber\\
\psi(x)&=&\sum_s\int\frac{\mathrm{d}^3p}{J_F}(b(\vec{p},s)u(\vec{p},s)\mathrm{e}^{-ip\cdot x}+d^\dagger(\vec{p},s)v(\vec{p},s)\mathrm{e}^{ip\cdot x})\:,
\end{eqnarray}
in the normalization of single particle states, 
\begin{eqnarray}
|\vec{k}\rangle&=&L_Ba^\dagger(\vec{k})|0\rangle\:,\nonumber\\
|\vec{p},s\rangle&=&L_Fb^\dagger(\vec{p},s)|0\rangle\:,
\end{eqnarray}
in momentum space algebraic relations,
\begin{eqnarray}
[a(\vec{k}),a^\dagger(\vec{k}\,')]&=&K_B\,\delta^{(3)}(\vec{k}-\vec{k}\,')\:,\nonumber\\
\lbrace b(\vec{p},r),b^\dagger(\vec{p}\,',s) \rbrace&=&K_F\,\delta_{rs}\,\delta^{(3)}(\vec{p}-\vec{p}\,')\:,
\end{eqnarray}
and in the normalization of the fermion spinor
\begin{eqnarray}
u^\dagger(\vec{p},r)u(\vec{p},s)&=&N_F2E_p\delta_{rs}\:.
\end{eqnarray}
The parameters $J_i$, $K_i$, $N_i$ are constrained by the canonical commutation or anticommutation relations to obey
\begin{eqnarray}
\frac{K_iN_i}{J_i^2}=\frac{1}{(2\pi)^32E}\qquad(i=\:B,\:F)\:.
\end{eqnarray}
Using the above, the single-particle expectation value of the quantum mechanical probability density can be expressed as
\begin{eqnarray}
\rho_i&=&\frac{K_iL_i^2}{(2\pi)^3}\qquad(i=\:B,\:F)\:.
\end{eqnarray}
The parameters of choice used for the calculations in this work are
\begin{eqnarray}
J_{F,j}=(2\pi)^{3/2}\sqrt{\frac{E_j}{m_j}}\:,\qquad K_{F.j}=1\:,\qquad L_{F,j}=(2\pi)^{3/2}\sqrt{2E_j}\:,\qquad N_{F,j}=\frac{1}{2m_j}\:.
\end{eqnarray}

%% file: A-Appendices/A-Cross-sections.tex
We now want to gain the cross section in a general form, employing the parameters of choice defined in App. \ref{Sec:Conventions}. The cross section for two particles to many particles reactions is given by
\begin{eqnarray}
\mathrm{d}\sigma(a_1+a_2\rightarrow f)=\frac{P_{fi}}{F_{\textrm{inc}}}\mathrm{d}N_{\textrm{fin}}\:,
\end{eqnarray}
where $F_{\textrm{inc}}$ is the flux of incoming particles
\begin{eqnarray}
F_{\textrm{inc}}=\rho_1\rho_2|\vec{v}_1-\vec{v}_2|=\frac{K_1L_1^2}{(2\pi)^3}\frac{K_2L_2^2}{(2\pi)^3}|\vec{v}_1-\vec{v}_2|\:,
\end{eqnarray}
$P_{fi}$ 
the transition probability per unit time per unit volume for the process $i\rightarrow f$\,
\begin{eqnarray}
P_{fi}=(2\pi)^4\delta^4(\sum_f p_f-\sum_i p_i)\left(\prod_j\frac{L_jK_j}{J_j}\right)^2\sum_{\textrm{int}}|\mathcal{M}_{fi}|^2\:,
\end{eqnarray}
and $\mathrm{d}N_{\textrm{fin}}$ the density of the final state
\begin{eqnarray}
\mathrm{d}N_{\textrm{fin}}=\frac{1}{\mathcal{Z}}\prod_{f=1}^n\frac{\mathrm{d}^3p_f}{K_fL_f^2}\:,
\end{eqnarray}
with $\mathcal{Z}=\prod_k n_k!$ a statistical factor accounting for the presence of $n_k$ identical particles of type $k$ in the final state and $\sum_{\textrm{int}}$ a sum over internal degrees of freedom. Collecting the pieces, we obtain the cross section in the desired form:
\begin{eqnarray}
\!\!\mathrm{d}\sigma\!\!\!\!\!&=&\!\!\!\!\!\frac{1}{\mathcal{Z}}\!\!\left(\prod_{f}\frac{\mathrm{d}^3p_f}{K_fL_f^2}\right)\!\!\!\!\left(\prod_j\frac{L_jK_j}{J_j}\right)^{\!\!2}\!\!\!\left(\prod_i\frac{(2\pi)^3}{K_iL_i^2}\right)\!\!(2\pi)^4\delta^{(4)}(\sum_f p_f-\!\sum_i p_i)\sum_{\textrm{int}}\frac{|\mathcal{M}_{fi}|^2}{|\vec{v}_1-\vec{v}_2|}
\:,
\end{eqnarray}
where the indices of the sums or products are given by $i$ for particles in the initial state, $f$ for particles of the final state and $j$ for particles of initial and final states.

%% file: A-Appendices/B-Pert-QCD-Potentials.tex
The short range part of the quark-antiquark potentials in our models is calculated via perturbative QCD. Therefore, we explicitly calculate one-gluon exchange potentials and dicuss some basic problems in their derivation. 

%% file: A-Appendices/B-Breit.tex
The Breit interaction has first been deduced by Gregory Breit \cite{Breit29, Breit30, Breit32} for the electron-electron scattering process. The according potential includes, apart form a leading Coulomb term, relativistic corrections which have their origin in the one-photon exchange process and an expansion of the propagator, leading to retardation. Likewise we will rederive the Breit interaction for the quark-antiquark scattering process
\begin{eqnarray}\label{Eq:Qark-Antiquark-Scattering}
q_i(p_A,s_A)+\overline{q}_j(p_B,s_B)\longrightarrow q_k(p'_A,s'_A)+\overline{q}_l(p'_B,s'_B)\:,
\end{eqnarray}
with four-momenta $p_X$, $p'_X$ and spins $s_X$, $s'_X$ as indicated, and $i,j,k,l=1,2,3$ being color indices. In this case the Breit interaction is equivalent to the potential arising from one-gluon exchange, including retardation corrections. The scattering process Eq. (\ref{Eq:Qark-Antiquark-Scattering}) is described, at lowest order perturbative QCD, by two Feynman diagrams, a s-channel diagram (Fig. \ref{OGE-S-Channel}) and a t-channel diagram (Fig. \ref{OGE-T-Channel}). We have to keep in mind, that the purpose of this perturbative QCD calculation is to obtain a potential describing quarkonium. As quarkonium states are mesons which are color neutral objects, we possess already some knowledge about the color wavefunction
\begin{eqnarray}
|\textrm{Meson}\rangle\sim\frac{1}{\sqrt{3}}\sum_{m=1}^{3}|\overline{q}_iq_i\rangle\:,
\end{eqnarray}
resulting for Eq. (\ref{Eq:Qark-Antiquark-Scattering}) in the conditions
\begin{eqnarray}\label{Eq:Meson-Color-Singlet-Condition}
\frac{1}{\sqrt{3}}\delta_{ij} \quad \textrm{and}\quad \frac{1}{\sqrt{3}}\delta_{kl}\:.
\end{eqnarray}
Taking this condition into account the s-channel diagram does not contribute. Employing Feynman rules (see e.g. \cite{ThomasWeise01, PeskinQFT95}) the amplitude, corresponding to the s-channel diagram (Fig. \ref{OGE-S-Channel}), in Feynman gauge reads
\begin{eqnarray}
\mathcal{M}_{fi}\!\!\!\!\!&=&\!\!\!\!\!\left[\overline{u}(p'_A,s'_A)\left(\mathrm{i}g\gamma^\mu\frac{\lambda_{kl}^a}{2}\right) v(p'_B,s'_B)\right]\frac{-g_{\mu\nu}}{(p_A+p_B)^2+\mathrm{i}\epsilon}\left[\overline{v}(p_B,s_B)\left(\mathrm{i}g\gamma^\nu\frac{\lambda_{ij}^a}{2}\right) u(p_A,s_A)\right]\nonumber\\
\!\!\!\!\!&=&\!\!\!\!\!\frac{g^2}{(p_A+p_B)^2+\mathrm{i}\epsilon}\frac{\lambda_{ij}^a}{2}\frac{\lambda_{kl}^a}{2}\overline{u}(p'_A,s'_A)\gamma_\mu v(p'_B,s'_B)\overline{v}(p_B,s_B)\gamma^\mu u(p_A,s_A)\:.
\end{eqnarray}
\begin{figure}[ht!]
\begin{center}
\begin{fmfgraph*}(60,120)
\fmfset{arrow_len}{3mm}
\fmfset{arrow_ang}{15}
\fmftop{t1,t2}
\fmflabel{$q_k(p'_A,s'_A)$}{t1}
\fmflabel{$\overline{q}_l(p'_B,s'_B)$}{t2}
\fmflabel{$q_i(p_A,s_A)$}{b1}
\fmflabel{$\overline{q}_j(p_B,s_B)$}{b2}
\fmfbottom{b1,b2}
\fmf{fermion}{b1,v1,b2}
\fmf{fermion}{t2,v2,t1}
\fmf{wiggly}{v1,v2}
\end{fmfgraph*}
\end{center}
\caption{\label{OGE-S-Channel} One-gluon exchange s-channel diagram for quark-antiquark.}
\end{figure}
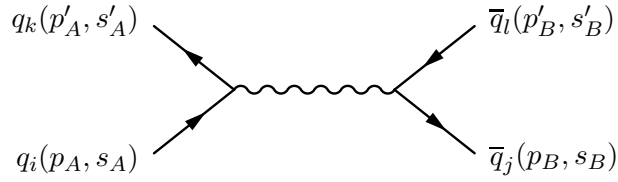
\begin{figure}[ht!]
\begin{center}
\begin{fmfgraph*}(120,60)
\fmfset{arrow_len}{3mm}
\fmfset{arrow_ang}{15}
\fmftop{t1,t2}
\fmflabel{$q_k(p'_A,s'_A)$}{t1}
\fmflabel{$\overline{q}_l(p'_B,s'_B)$}{t2}
\fmflabel{$q_i(p_A,s_A)$}{b1}
\fmflabel{$\overline{q}_j(p_B,s_B)$}{b2}
\fmfbottom{b1,b2}
\fmf{fermion}{b1,v1,t1}
\fmf{fermion}{t2,v2,b2}
\fmf{wiggly}{v1,v2}
\end{fmfgraph*}
\end{center}
\caption{\label{OGE-T-Channel} One-gluon exchange t-channel diagram for quark-antiquark.}
\end{figure}
Using Eq. (\ref{Eq:Meson-Color-Singlet-Condition}) this amplitude vanishes, as
\begin{eqnarray}
\frac{1}{\sqrt{3}}\delta_{ij}\frac{\lambda_{ij}^a}{2}=\frac{1}{2\sqrt{3}}\mathrm{tr}(\lambda^a)=0\:,
\end{eqnarray}
or in words, the gluon as color-octet does not couple to a color-singlet. This leaves the t-channel diagram (Fig. \ref{OGE-T-Channel}) to deal with. Employing Feynman rules the t-channel amplitude in Feynman gauge is given by
\begin{eqnarray}\label{Eq:Quark-Antiquark-Amplitude-Origin}
\mathcal{M}_{fi}&=&\left[\overline{u}(p'_A,s'_A)\left(\mathrm{i}g\gamma^\mu\frac{\lambda_{ki}^a}{2}\right) u(p_A,s_A)\right]\frac{g_{\mu\nu}}{q^2+\mathrm{i}\epsilon}\left[\overline{v}(p_B,s_B)\left(\mathrm{i}g\gamma^\nu\frac{\lambda_{jl}^a}{2}\right) v(p'_B,s'_B)\right]\nonumber\\
&=&-\frac{g^2}{q^2+\mathrm{i}\epsilon}\frac{\lambda_{ki}^a}{2}\frac{\lambda_{jl}^a}{2}\overline{u}(p'_A,s'_A)\gamma_\mu u(p_A,s_A)\overline{v}(p_B,s_B)\gamma^\mu v(p'_B,s'_B)\:,
\end{eqnarray}
with $q^2=(p_A-p_A')^2=(p_B'-p_B)^2$. Remembering Eq. (\ref{Eq:Meson-Color-Singlet-Condition}) yields the factor
\begin{eqnarray}
\frac{1}{\sqrt{3}}\sum_{i,j=1}^{3}\delta_{ij}\frac{1}{\sqrt{3}}\sum_{k,l=1}^{3}\delta_{kl}\sum_{a=1}^{8}\frac{\lambda_{ki}^a}{2}\frac{\lambda_{jl}^a}{2}=\frac{1}{12}\sum_{i,k=1}^{3}\sum_{a=1}^{8}\lambda_{ki}^a\lambda_{ik}^a=\frac{1}{12}\sum_{a=1}^{8}\mathrm{tr}[(\lambda^a)^2]=\frac{4}{3}\:.
\end{eqnarray}
Thus, for color neutral initial and final states the process Eq. (\ref{Eq:Qark-Antiquark-Scattering}) leads to the amplitude
\begin{eqnarray}\label{Eq:Quark-Antiquark_Amplitude}
\mathcal{M}_{fi}&=&-\frac{4}{3}\frac{g^2}{q^2+\mathrm{i}\epsilon}\overline{u}(p'_A,s'_A)\gamma_\mu u(p_A,s_A)\overline{v}(p_B,s_B)\gamma^\mu v(p'_B,s'_B)\:.
\end{eqnarray}
The computation of the amplitude Eq. (\ref{Eq:Quark-Antiquark_Amplitude}) is done partwise. To begin with, we evaluate the Dirac currents $\overline{u}(p'_A,s'_A)\gamma^\mu u(p_A,s_A)$ and $\overline{v}(p_B,s_B)\gamma^\mu v(p'_B,s'_B)$, after applying the normalization conventions for Dirac spinors as in Appendix \ref{Sec:Conventions}:
\begin{eqnarray}
u(p_A,s_A)&=&\sqrt{\frac{E_A+m_A}{2m_A}}\left(\begin{array}{c}
\chi_{s_A}\\ 
\frac{\vec{\sigma}\cdot\vec{p}_A}{E_A+m_A}\chi_{s_A}
\end{array}\right)\:,\nonumber\\
\overline{u}(p'_A,s'_A)&=&\sqrt{\frac{E'_A+m_A}{2m_A}}\left(\begin{array}{c@{,}c}
\chi_{s'_A}^\dagger&\quad -\chi_{s'_A}^\dagger \frac{\vec{\sigma}\cdot\vec{p}\,'_A}{E'_A+m_A}
\end{array}\right)\:,\nonumber\\
v(p'_B,s'_B)&=&\sqrt{\frac{E'_B+m_B}{2m_B}}\left(\begin{array}{c}
\frac{\vec{\sigma}\cdot\vec{p}\,'_B}{E'_B+m_B}\chi^c_{s'_B}\\
\chi^c_{s'_B}
\end{array}\right)\:,\\
\overline{v}(p_B,s_B)&=&\sqrt{\frac{E_B+m_B}{2m_B}}\left(\begin{array}{c@{,}c}
\chi_{s_B}^{c\dagger} \frac{\vec{\sigma}\cdot\vec{p}_B\,}{E_B+m_B} & \quad -\chi_{s_B}^{c\dagger}
\end{array}\right)\:,\nonumber
\end{eqnarray}
with energies $E_X=\sqrt{\vec{p}_X^{\,2}+m_X^2}$ and quark masses $m_X$. Using the auxiliary calculations
\begin{eqnarray}
&&\!\!\!\!\!\!\!\!\!\!\left(\begin{array}{c@{,}c}
 \chi^\dagger_{s'_A}&\quad \chi^\dagger_{s'_A} \frac{\vec{\sigma}\cdot\vec{p}_A\,\!\!\!\!'}{E'_A+m_A}
\end{array}\right)
\left(\begin{array}{cc}
0 & \sigma_i \\ 
\sigma_i & 0 
\end{array}\right)
\left(\begin{array}{c}
\chi_{s_A} \\ 
\frac{\vec{\sigma}\cdot\vec{p}_A}{E_A+m_A}\chi_{s_A}
\end{array}\right)\nonumber\\
&&=\left(\begin{array}{c@{,}c}
\chi^\dagger_{s'_A}\frac{\vec{\sigma}\cdot\vec{p}_A\,\!\!\!\!'}{E'_A+m_A}\sigma_i &\quad\chi^\dagger_{s'_A} \sigma_i
\end{array}\right)
\left(\begin{array}{c}
\chi_{s_A}\\ 
\frac{\vec{\sigma}\cdot\vec{p}_A}{E_A+m_A}\chi_{s_A}
\end{array}\right)=\chi^\dagger_{s'_A}\left(\frac{\vec{\sigma}\cdot\vec{p}_A\,\!\!\!\!'}{E'_A+m_A}\sigma_i+\sigma_i\frac{\vec{\sigma}\cdot\vec{p}_A}{E_A+m_A}\right)\chi_{s_A}\:,\nonumber\\
&&\!\!\!\!\!\!\!\!\!\!\left(\begin{array}{c@{,}c}
\chi^{c\dagger}_{s_B} \frac{\vec{\sigma}\cdot\vec{p}_B}{E_B+m_B} &\quad  \chi^{c\dagger}_{s_B}
\end{array}\right)
\left(\begin{array}{cc}
0 & \sigma_i \\ 
\sigma_i & 0 
\end{array}\right)
\left(\begin{array}{c}
\frac{\vec{\sigma}\cdot\vec{p}_B\,\!\!\!\!'}{E'_B+m_B}\chi_{s'_B}^c \\ 
\chi_{s'_B}^c
\end{array}\right)\\
&&=\left(\begin{array}{c@{,}c}
\chi^{c\dagger}_{s_B}\sigma_i &\quad\chi^{c\dagger}_{s_B} \frac{\vec{\sigma}\cdot\vec{p}_B}{E_B+m_B} \sigma_i
\end{array}\right)
\left(\begin{array}{c}
\frac{\vec{\sigma}\cdot\vec{p}_B\,\!\!\!\!'}{E'_B+m_B}\chi_{s'_B}^c\\ 
\chi_{s'_B}^c
\end{array}\right)=\chi^{c\dagger}_{s_B}\left(\frac{\vec{\sigma}\cdot\vec{p}_B}{E_B+m_B}\sigma_i+\sigma_i\frac{\vec{\sigma}\cdot\vec{p}_B\,\!\!\!\!'}{E'_B+m_B}\right)\chi^c_{s'_B}\:,\nonumber
\end{eqnarray}
\begin{eqnarray}
(\vec{\sigma}\cdot\vec{\epsilon}\,)(\vec{\sigma}\cdot\vec{x}\,)+(\vec{\sigma}\cdot\vec{y}\,)(\vec{\sigma}\cdot\vec{\epsilon}\,)&=&\vec{\epsilon}\cdot\vec{x}+\mathrm{i}\vec{\sigma}\cdot(\vec{\epsilon}\times\vec{x}\,)+\vec{y}\cdot\vec{\epsilon}+\mathrm{i}\vec{\sigma}\cdot(\vec{y}\times\vec{\epsilon}\,)\nonumber\\
&=&\vec{\epsilon}\cdot(\vec{y}+\vec{x}\,)+\mathrm{i}\vec{\epsilon}\cdot\vec{\sigma}\times(\vec{y}-\vec{x}\,)\:,
\end{eqnarray}
one can easily verify
\begin{eqnarray}\label{Eq:Dirac-currents-Old-variables}
\overline{u}(p'_A,s'_A)\gamma^{\mu}u(p_A,s_A)\!\!\!\!\!&=&\!\!\!\!\!\left(\begin{array}{c}
u^\dagger(p'_A,s'_A)u(p_A,s_A) \\ 
u^\dagger(p'_A,s'_A)\gamma_0\vec{\gamma}u(p_A,s_A)
\end{array}\right)\nonumber\\
\!\!\!\!\!&=&\!\!\!\!\!\sqrt{\frac{E'_A+m_A}{2m_A}}\sqrt{\frac{E_A+m_A}{2m_A}}\left(
\begin{array}{c}
\chi_{s'_A}^\dagger\left(1+\frac{(\vec{\sigma}\cdot\vec{p}_A\,\!\!\!\!')(\vec{\sigma}\cdot\vec{p}_A)}{(E'_A+m)_A(E_A+m_A)}\right)\chi_{s_A}\\ 
\chi_{s'_A}^\dagger\left(\frac{\vec{\sigma}\cdot\vec{p}_A\,\!\!\!\!'}{E'_A+m_A}\vec{\sigma}+\vec{\sigma}\frac{\vec{\sigma}\cdot\vec{p}_A}{E_A+m_A}\right)\chi_{s_A}
\end{array}\right)\nonumber\\
\!\!\!\!\!&=&\!\!\!\!\!\sqrt{\frac{E'_A+m_A}{2m_A}}\sqrt{\frac{E_A+m_A}{2m_A}}\nonumber\\
&&\left(\begin{array}{c}
\chi_{s'_A}^\dagger\left(1+\frac{(\vec{\sigma}\cdot\vec{p}_A\,\!\!\!\!')(\vec{\sigma}\cdot\vec{p}_A)}{(E'_A+m_A)(E_A+m_A)}\right)\chi_{s_A} \\ 
\chi_{s'_A}^\dagger\left[ \left(\frac{\vec{p}_A}{E_A+m_A}+\frac{\vec{p}_A\,\!\!\!\!'}{E'_A+m_A}\right)+\mathrm{i}\vec{\sigma}\times\left(\frac{\vec{p}_A\,\!\!\!\!'}{E'_A+m_A}-\frac{\vec{p}_A}{E_A+m_A}\right)\right]\chi_{s_A}
\end{array}\right)\:,\nonumber\\
\overline{v}(p_B,s_B)\gamma^{\mu}v(p'_B,s'_B)
\!\!\!\!\!&=&\!\!\!\!\!\sqrt{\frac{E'_B+m_B}{2m_B}}\sqrt{\frac{E_B+m_B}{2m_B}}\\
&&\left(\begin{array}{c}
\chi_{s_B}^{c\dagger}\left(1+\frac{(\vec{\sigma}\cdot\vec{p}_B)(\vec{\sigma}\cdot\vec{p}_B\,\!\!\!\!')}{(E'_B+m_B)(E_B+m_B)}\right)\chi_{s'_B}^c \\ 
\chi_{s_B}^{c\dagger}\left[ \left(\frac{\vec{p}_B\,\!\!\!\!'}{E'_B+m_B}+\frac{\vec{p}_B}{E_B+m_B}\right)+\mathrm{i}\vec{\sigma}\times\left(\frac{\vec{p}_B}{E_B+m_B}-\frac{\vec{p}_B\,\!\!\!\!'}{E'_B+m_B}\right)\right]\chi_{s'_B}^c
\end{array}\right)\:,\nonumber
\end{eqnarray}
for the Dirac currents. For the further calculation we substitute
\begin{eqnarray}
&&\vec{p}_A=\vec{p}_1+\frac{1}{2}\vec{q}\:, \qquad \vec{p}_A\,\!\!\!\!'=\vec{p}_1-\frac{1}{2}\vec{q}\:,\qquad\vec{s}_A=\vec{s}_1\:,\qquad m_A=m_1\:,\nonumber\\
&&\vec{p}_B=\vec{p}_2-\frac{1}{2}\vec{q}\:, \qquad \vec{p}_B\,\!\!\!\!'=\vec{p}_2+\frac{1}{2}\vec{q}\:,\qquad\vec{s}_B=\vec{s}_2\:,\qquad m_B=m_2\:.
\end{eqnarray}
We choose these kinematics to get a coordinate space potential which is invariant under the usual time reversal transformation
\begin{eqnarray}
t&\longrightarrow& -t\:,\nonumber\\
\vec{r}&\longrightarrow& \vec{r}\:,\nonumber\\
\vec{p}_i&\longrightarrow& -\vec{p}_i\:,\\
\vec{J}&\longrightarrow& -\vec{J}\:.\nonumber
\end{eqnarray}
The kinematics-choice is essential due to the fact that, while the amplitude is time reversal invariant as long as it is expressed in Lorentz-covariants, its nonrelativistic expansion and the gained potential are not (see Appendix \ref{Sec:Kinematics}). With new variables the timelike and spacelike parts of Eq. (\ref{Eq:Dirac-currents-Old-variables}) read
\begin{eqnarray}
u^\dagger(p'_A,s'_A) u(p_A,s_A)\!\!\!\!&=&\!\!\!\!\sqrt{\frac{E_A+m_1}{2m_1}}\sqrt{\frac{E'_A+m_1}{2m_1}}\chi_{s_1'}^\dagger\left(1+\frac{(\vec{\sigma}\cdot(\vec{p}_1-\frac{1}{2}\vec{q}\,))(\vec{\sigma}\cdot(\vec{p}_{1}+\frac{1}{2}\vec{q}\,))}{(E_A+m_1)(E'_A+m_1)}\right)\chi_{s_1}\nonumber\\
\!\!\!\!&=&\!\!\!\!\sqrt{\frac{E_A+m_1}{2m_1}}\sqrt{\frac{E'_A+m_1}{2m_1}}\chi_{s_1'}^\dagger\left(1+\frac{(\vec{p}_1-\frac{1}{2}\vec{q}\,)\cdot(\vec{p}_1+\frac{1}{2}\vec{q}\,)}{(E_A+m_1)(E'_A+m_1)}\right.\nonumber\\
&&\left.\qquad\qquad\qquad\qquad\qquad\qquad\quad\:\:\!\!\!\!+\frac{\mathrm{i}\vec{\sigma}\cdot\left((\vec{p}_1-\frac{1}{2}\vec{q}\,)\times(\vec{p}_1+\frac{1}{2}\vec{q}\,)\right)}{(E_A+m_1)(E'_A+m_1)}\right)\chi_{s_1}\nonumber\\
\!\!\!\!&=&\!\!\!\!\sqrt{\frac{E_A+m_1}{2m_1}}\sqrt{\frac{E'_A+m_1}{2m_1}}\chi_{s_1'}^\dagger\left(1+\frac{\vec{p}_1^{\,2}-\frac{1}{4}\vec{q}^{\,2}-\mathrm{i}\vec{\sigma}\cdot(\vec{q}\times\vec{p}_1)}{(E_A+m_1)(E'_A+m_1)}\right)\chi_{s_1}\:,\nonumber\\
v^\dagger(p_B,s_B) v(p'_B,s'_B)\!\!\!\!&=&\!\!\!\!\sqrt{\frac{E_B+m_2}{2m_2}}\sqrt{\frac{E'_B+m_2}{2m_2}}\chi_{s_2}^{c\dagger}\left(1+\frac{\vec{p}_2^{\,2}-\frac{1}{4}\vec{q}^{\,2}-\mathrm{i}\vec{\sigma}\cdot(\vec{q}\times\vec{p}_2)}{(E_B+m_2)(E'_B+m_2)}\right)\chi_{s'_2}^c\:,\nonumber
\end{eqnarray}\vspace{-0.5cm}
\begin{eqnarray}
\bar{u}(p'_A,s'_A)\vec{\gamma}u(p_A,s_A)\!\!\!\!&=&\!\!\!\!\sqrt{\frac{E_A+m_1}{2m_1}}\sqrt{\frac{E'_A+m_1}{2m_1}}\nonumber\\
&&\quad\chi_{s_1'}^\dagger\left[\left(\frac{\vec{p}_1+\frac{1}{2}\vec{q}}{E_A+m_1}+\frac{\vec{p}_1-\frac{1}{2}\vec{q}}{E'_A+m_1}\right)+\mathrm{i}\vec{\sigma}\times\left(\frac{\vec{p}_1-\frac{1}{2}\vec{q}}{E'_A+m_1}-\frac{\vec{p}_1+\frac{1}{2}\vec{q}}{E_A+m_1}\right)\right]\chi_{s_1}\nonumber\\
&=&\frac{1}{2m_1}\chi_{s'_1}^\dagger\left[2\vec{p}_1-\mathrm{i}\vec{\sigma}\times\vec{q}\,\right]\chi_{s_1}+\mathcal{O}\left(\frac{1}{m_1^2}\right)\:,\nonumber\\
\bar{v}(p_B,s_B)\vec{\gamma}v(p'_B,s'_B)\!\!\!\!&=&\!\!\!\!\frac{1}{2m_2}\chi_{s_2}^{c\dagger}\left[2\vec{p}_2-\mathrm{i}\vec{\sigma}\times\vec{q}\,\right]\chi_{s'_2}^c+\mathcal{O}\left(\frac{1}{m_2^2}\right)\:.
\end{eqnarray}
For the spacelike part of the currents we have neglected corrections of order $\mathcal{O}(m_i^{-2})$. These order $\mathcal{O}(m_i^{-2})$ terms in the currents give order $\mathcal{O}(m_i^{-2}m_j^{-1})$ and $\mathcal{O}(m_i^{-2}m_j^{-2})$ contributions in the amplitude. These terms are negligible as we expand the amplitude to order $\mathcal{O}(m_i^{-1}m_j^{-1})$. The spin of particles is given by the expectation values
\begin{eqnarray}\label{Eq:Spin-Particles}
\vec{s}_1\:\delta_{s_1s'_1}\equiv\chi_{s'_1}^\dagger\vec{\sigma}\chi_{s_1}\:,
\end{eqnarray}
of the Pauli matrix $\vec{\sigma}$. In the case of antiparticles, we have to deal with charge conjugated Pauli spinors
\begin{eqnarray}
\vec{s}_2\:\delta_{s_2s'_2}\equiv\chi_{s_2}^{c\dagger}\sigma\chi_{s'_2}^c\:.
\end{eqnarray}
Thus, we have to relate the spin to charge conjugated spinors. The charge conjugation of wavefunctions is given by
\begin{eqnarray}
\psi^c&=&C\gamma^0\psi^*\:,\nonumber\\
\psi^{c\dagger}&=&\left(C\gamma^0\psi^*\right)^\dagger=\psi^\mathrm{t}\gamma^0 C^\dagger\:,\nonumber\\
C&=&\mathrm{i}\gamma^2\gamma^0\:.
\end{eqnarray}
Employing this we can express $\chi^{c\dagger}\vec{\sigma}\chi^c$ with Pauli spinors
\begin{eqnarray}
\left(\chi^{c\dagger}\vec{\sigma}\chi^c\right)&=&\chi^\mathrm{t}\sigma_2\vec{\sigma}\sigma_2\chi^*=-\left(\chi^\dagger\vec{\sigma}\chi\right)^*
\end{eqnarray}
where
\begin{eqnarray}
\sigma_2\vec{\sigma}\sigma_2&=&\left(\begin{array}{c} -\sigma_1\\ \sigma_2 \\-\sigma_3\end{array}\right)=-\vec{\sigma}^*\:,
\end{eqnarray}
is utilized in the last step. Knowing that the spin is a real quantity, we recognize that
\begin{eqnarray}
\left(\chi^{c\dagger}\vec{\sigma}\chi^c\right)&=&-\left(\chi^\dagger\vec{\sigma}\chi\right)\:,
\end{eqnarray}
and therefore
\begin{eqnarray}\label{Eq:Spin-Antiparticles}
\vec{s}_2\:\delta_{s_2s'_2}\equiv-\left(\chi_{s_2}^\dagger\vec{\sigma}\chi_{s'_2}\right)\:.
\end{eqnarray}
Setting the pieces together, we obtain for the amplitude
\begin{eqnarray}\label{Eq:Amplitude-Quark-Antiquark-Expanded}
\mathcal{M}_{fi}\!\!\!\!&=&\!\!\!\!-\frac{4}{3}\frac{g^2}{q^2+\mathrm{i}\epsilon}(\mathcal{A}+\mathcal{B})\:,\nonumber\\
\mathcal{A}\!\!\!\!&=&\!\!\!\!\EuScript{N}\chi_{s'_1}^\dagger\left(1+\frac{\vec{p}_1^{\,2}-\frac{1}{4}\vec{q}^{\,2}-\mathrm{i}\vec{\sigma}\cdot(\vec{q}\times\vec{p}_1)}{(E_A+m_1)(E'_A+m_1)}\right)\chi_{s_1}\chi_{s_2}^{c\dagger}\left(1+\frac{\vec{p}_2^{\,2}-\frac{1}{4}\vec{q}^{\,2}-\mathrm{i}\vec{\sigma}\cdot(\vec{q}\times\vec{p}_2)}{(E_B+m_2)(E'_B+m_2)}\right)\chi_{s'_2}^c\nonumber\\
\!\!\!\!&=&\!\!\!\!\EuScript{N}\delta_{s_1s'_1}\delta_{s_2s'_2}\left(1+\frac{\vec{p}_1^{\,2}-\frac{1}{4}\vec{q}^{\,2}-2\mathrm{i}\vec{s}_{1}\cdot(\vec{q}\times\vec{p}_1)}{(E_A+m_1)(E'_A+m_1)}+\frac{\vec{p}_2^{\,2}-\frac{1}{4}\vec{q}^{\,2}+2\mathrm{i}\vec{s}_{2}\cdot(\vec{q}\times\vec{p}_2)}{(E_B+m_2)(E'_B+m_2)}\right)\nonumber\\
\!\!\!\!&=&\!\!\!\!\delta_{s_1s'_1}\delta_{s_2s'_2}\left(1+\frac{2\vec{p}_1^{\,2}-2\mathrm{i}\vec{s}_1\cdot\left(\vec{q}\times\vec{p}_1^{\,2}\right)}{4m_1^2}+\frac{2\vec{p}_2^{\,2}+2\mathrm{i}\vec{s}_2\cdot\left(\vec{q}\times\vec{p}_2^{\,2}\right)}{4m_2^2}\right)+\mathcal{O}\left(\frac{1}{m_i^3}\right)\:,\nonumber\\
\mathcal{B}\!\!\!\!&=&\!\!\!\!-\chi_{s'_1}^\dagger\left(\frac{2\vec{p}_1-\mathrm{i}\vec{\sigma}\times\vec{q}}{2m_1}\right)\chi_{s_1}\chi_{s_2}^{c\dagger}\left(\frac{2\vec{p}_2-\mathrm{i}\vec{\sigma}\times\vec{q}}{2m_2}\right)\chi_{s'_2}^c+\mathcal{O}\left(\frac{1}{m_i^3}\right)\\
\!\!\!\!&=&\!\!\!\!\frac{\delta_{s_1s'_1}\delta_{s_2s'_2}}{4m_1m_2}\left(-4\vec{p}_1\cdot\vec{p}_2+4\mathrm{i}\vec{s}_1\cdot\left(\vec{q}\times\vec{p}_2\right)-4\mathrm{i}\vec{s}_2\cdot\left(\vec{q}\times\vec{p}_1\right)\right.\nonumber\\
&&\qquad\qquad\qquad\qquad\qquad\left.-4\vec{q}^{\,2}\left(\vec{s}_1\cdot\vec{s}_2\right)+4\left(\vec{q}\cdot\vec{s}_1\right)\left(\vec{q}\cdot\vec{s}_2\right)\frac{}{}\right)+\mathcal{O}\left(\frac{1}{m_i^3}\right)\:,\nonumber\\
\EuScript{N}\!\!\!\!&=&\!\!\!\!\sqrt{\frac{E_A+m_1}{2m_1}}\sqrt{\frac{E'_A+m_1}{2m_1}}\sqrt{\frac{E_B+m_2}{2m_2}}\sqrt{\frac{E'_B+m_2}{2m_2}}\nonumber\\
\!\!\!\!&=&\!\!\!\!1+\frac{\vec{p}_1^{\,2}+\frac{1}{4}\vec{q}^{\,2}}{4m_1^2}+\frac{\vec{p}_2^{\,2}+\frac{1}{4}\vec{q}^{\,2}}{4m_2^2}+\mathcal{O}\left(\frac{1}{m_i^3}\right)\:,\nonumber
\end{eqnarray}
where we have expanded the energies in the respective last step. In the following the $\delta_{s_is'_i}$ are suppressed. To find the potential corresponding to $\mathcal{M}_{fi}$, we have to relate the non-relativistic cross section from Schr\"odinger theory with the cross section computed in Appendix \ref{Sec:Cross-sections}. We recall that for the non-relativistic case, the differential cross section is related with the Born scattering amplitude via
\begin{eqnarray}
\mathrm{d}\sigma_\textrm{NR}=\frac{\mathrm{d}^3p'_1\mathrm{d}^3p'_2}{(2\pi)^3(2\pi)^3}(2\pi)^4\delta^{(4)}\left(\sum_f p_f-\sum_i p_i\right)\frac{|f_B|^2}{|v_\textrm{rel}|}\:.
\end{eqnarray}
Applying our conventions to the cross section given in Appendix \ref{Sec:Cross-sections}, we end up with
\begin{eqnarray}
\mathrm{d}\sigma_\textrm{R}=\mathrm{d}^3p'_1\mathrm{d}^3p'_2\frac{m_1^2m_2^2}{E_A E'_A E_B E'_B}(2\pi)^4\delta^{(4)}\left(\sum_f p_f-\sum_i p_i\right)\frac{|\mathcal{M}_{fi}|^2}{|v_\textrm{rel}|}\:.
\end{eqnarray}
Knowing that the Born scattering amplitude is equal to the Fourier transform of the potential,
\begin{eqnarray}
f_\textrm{B}=\int\frac{\mathrm{d}^3r}{(2\pi)^3}\mathrm{e}^{-\mathrm{i}\vec{p}\,'\!\cdot\vec{r}}V(\vec{r}\,)\mathrm{e}^{\mathrm{i}\vec{p}\cdot\vec{r}}=\tilde{V}(\vec{q}\,)\:,
\end{eqnarray}
we gain the relation between potential and amplitude
\begin{eqnarray}
\frac{1}{(2\pi)^3}\tilde{V}(\vec{q};\vec{p}_1,\vec{p}_2)=\pm\frac{m_1 m_2}{\sqrt{E_A}\sqrt{E_B}\sqrt{E'_A}\sqrt{E'_B}}\mathcal{M}_{fi}\:.
\end{eqnarray}
Expanding the energies the relation reads
\begin{eqnarray}\label{Eq:Relation-Potential-Amplitude}
\frac{1}{(2\pi)^3}\tilde{V}(\vec{q};\vec{p}_1,\vec{p}_2)=\pm\left(1-\frac{\vec{p}_1^{\,2}}{2m_1^2}-\frac{\vec{q}^{\,2}}{8m_1^2}-\frac{\vec{p}_2^{\,2}}{2m_2^2}-\frac{\vec{q}^{\,2}}{8m_2^2}+\mathcal{O}\left(\frac{1}{m_i^3}\right)\right)\mathcal{M}_{fi}\:.
\end{eqnarray}
The sign has to be determined by the fact that the interaction is repulsive in the case of the electron-electron one-photon exchange, giving a minus sign. Combining Eq. (\ref{Eq:Amplitude-Quark-Antiquark-Expanded}) and Eq. (\ref{Eq:Relation-Potential-Amplitude}) leads up to order $\mathcal{O}(m_i^{-1}m_j^{-1})$ to
\begin{eqnarray}\label{Eq:Potential-Momentum-Space}
\frac{1}{(2\pi)^3}\tilde{V}(\vec{q};\vec{p}_1,\vec{p}_2)\!\!\!\!&=&\!\!\!\!\frac{4g^2}{3(q^2+\mathrm{i}\epsilon)}\Biggl[1+\frac{-\frac{1}{2}\vec{q}^{\,2}-2\mathrm{i}\vec{s}_1\cdot\left(\vec{q}\times\vec{p}_1\right)}{4m_1^2}+\frac{-\frac{1}{2}\vec{q}^{\,2}+2\mathrm{i}\vec{s}_2\cdot\left(\vec{q}\times\vec{p}_2\right)}{4m_2^2}\\
&&\!\!\!\!\!\!\!\!\!\!\!\!\!\!\!\!\!\!\!\!\!\!\!\!\!\!\!\!\!\!\!\!\!\!\!\!\!\!\!+\frac{1}{4m_1m_2}\biggl(-4\vec{p}_1\vec{p}_2+4\mathrm{i}\vec{s}_1\cdot\left(\vec{q}\times\vec{p}_2\right)-4\mathrm{i}\vec{s}_2\cdot\left(\vec{q}\times\vec{p}_1\right)-4\vec{q}^{\,2}\left(\vec{s}_1\cdot\vec{s}_2\right)+4\left(\vec{q}\cdot\vec{s}_1\right)\left(\vec{q}\cdot\vec{s}_2\right)\biggr)\Biggr]\:.\nonumber
\end{eqnarray}
Since we want to obtain a nonrelativistic potential we have to expand the 4-vector in the propagator,
\begin{eqnarray}\label{Eq:Propagator-Expanded}
\frac{1}{q^2}&\approx&-\frac{1}{\vec{q}^{\,2}}-\frac{q_0^2}{\vec{q}^{\,4}}\:,\nonumber\\
q_0^2&=&(E'_A-E_A)(E_B-E'_B)=  \frac{(\vec{p}_1\cdot\vec{q}\,)(\vec{p}_2\cdot\vec{q}\,)}{m_1m_2}     +\mathcal{O}\left(\frac{1}{m^3}\right)\:,\nonumber\\
\frac{1}{q^2}&=&-\frac{1}{\vec{q}^{\,2}}-\frac{1}{\vec{q}^{\,4}}\frac{(\vec{p}_1\cdot\vec{q}\,)(\vec{p}_2\cdot\vec{q}\,)}{m_1m_2} +\mathcal{O}\left(\frac{1}{m^3}\right)\:,
\end{eqnarray}
resulting in
\begin{eqnarray}\label{Eq:Potential-incl.Retardation-Momentum-Space}
\frac{1}{(2\pi)^3}\tilde{V}^{\textrm{Breit}}(\vec{q};\vec{p}_1,\vec{p}_2)\!\!\!\!\!&=&\!\!\!\!\!-\frac{4g^2}{3\vec{q}^{\,2}}\Biggl[1+\frac{-\frac{1}{2}\vec{q}^{\,2}-2\mathrm{i}\vec{s}_1\cdot\left(\vec{q}\times\vec{p}_1\right)}{4m_1^2}+\frac{-\frac{1}{2}\vec{q}^{\,2}+2\mathrm{i}\vec{s}_2\cdot\left(\vec{q}\times\vec{p}_2\right)}{4m_2^2}\nonumber\\
&&\qquad\quad+\frac{4}{\vec{q}^{\,2}}\frac{(\vec{p}_1\cdot\vec{q}\,)(\vec{p}_2\cdot\vec{q}\,)}{4m_1m_2}+\frac{1}{4m_1m_2}\biggl(-4\vec{p}_1\vec{p}_2+4\mathrm{i}\vec{s}_1\cdot\left(\vec{q}\times\vec{p}_2\right)\nonumber\\
&&\qquad\quad-4\mathrm{i}\vec{s}_2\cdot\left(\vec{q}\times\vec{p}_1\right)-4\vec{q}^{\,2}\left(\vec{s}_1\cdot\vec{s}_2\right)+4\left(\vec{q}\cdot\vec{s}_1\right)\left(\vec{q}\cdot\vec{s}_2\right)\biggr)\Biggr]\:.
\end{eqnarray}
The Fourier transformation back into coordinate space
\begin{eqnarray}
V(\vec{r};\vec{p}_1,\vec{p}_2)=\int\mathrm{d}^3q\mathrm{e}^{-\mathrm{i}\vec{q}\cdot\vec{r}}\tilde{V}(\vec{q};\vec{p}_1\vec{p}_2)\:,
\end{eqnarray}
of the various terms in Eq. (\ref{Eq:Potential-incl.Retardation-Momentum-Space}) is given by
\begin{eqnarray}\label{Eq:Fouriertransform-Gluon-Exchange}
W&=&\frac{1}{(2\pi)^3}\int\mathrm{d}^3q\:\mathrm{e}^{-\mathrm{i}\vec{q}\cdot\vec{r}}\frac{1}{\vec{q}^{\,2}}=\frac{1}{4\pi r}\:,\nonumber\\
W_1&=&\frac{1}{(2\pi)^3}\int\mathrm{d}^3q\:q_j\,\mathrm{e}^{-\mathrm{i}\vec{q}\cdot\vec{r}}\frac{1}{\vec{q}^{\,2}}=\mathrm{i}\nabla_j W=\mathrm{i}\frac{r_j}{r}W'\:,\\
W_2&=&\frac{1}{(2\pi)^3}\int\mathrm{d}^3q\:\mathrm{e}^{-\mathrm{i}\vec{q}\cdot\vec{r}}\vec{q}^{\,2}\frac{1}{\vec{q}^{\,2}}=-\Delta\int\frac{\mathrm{d}^3q}{(2\pi)^3}\mathrm{e}^{-\mathrm{i}\vec{q}\cdot\vec{r}}\frac{1}{\vec{q}^{\,2}}=-\Delta W=\delta^{(3)}(\vec{r}\,)\:,\nonumber\\
W_3&=&\frac{1}{(2\pi)^3}\int\mathrm{d}^3q\:q_i\,q_j\mathrm{e}^{-\mathrm{i}\vec{q}\cdot\vec{r}}\frac{1}{\vec{q}^{\,2}}=-\nabla_i\nabla_j W=-\left[W''-\frac{1}{r}W'\right]\left(\frac{r_ir_j}{r^2}-\frac{1}{3}\delta_{ij}\right)-\frac{1}{3}\delta_{ij}\Delta W\:,\nonumber\\
U&=&\frac{1}{(2\pi)^3}\int\mathrm{d}^3q\:\mathrm{e}^{-\mathrm{i}\vec{q}\cdot\vec{r}}\frac{1}{\vec{q}^{\,4}}=-\frac{r}{8\pi}\:,\nonumber\\
U_1&=&\frac{1}{(2\pi)^3}\int\mathrm{d}^3q\:q_i\,q_j\mathrm{e}^{-\mathrm{i}\vec{q}\cdot\vec{r}}\frac{1}{\vec{q}^{\,4}}=-\nabla_i\nabla_j U=-\left[U''-\frac{1}{r}U'\right]\left(\frac{r_ir_j}{r^2}-\frac{1}{3}\delta_{ij}\right)-\frac{1}{3}\delta_{ij}\Delta U\:,\nonumber
\end{eqnarray}
with
\begin{eqnarray}
\Delta\frac{1}{r}&=&-4\pi\delta^{(3)}(\vec{r}\,)\:,\nonumber\\
W'&=&-\frac{1}{4\pi r^2}\:,\quad W''=\frac{1}{2\pi r^3}\:,\quad\Delta W=-\delta^{(3)}(\vec{r}\,)\:,\nonumber\\
U'&=&-\frac{1}{8\pi}\:,\quad U''=0\:,\quad\Delta U=-\frac{1}{4\pi r}\:,
\end{eqnarray}
and the computation of $\nabla_i\nabla_j W$ being performed explicitly in Appendix \ref{Sec:NablaNablaW}. Collecting all the pieces we obtain the Breit interaction for quark-antiquark scattering
\begin{eqnarray}\label{Eq:Breit-Interaction}
V^{\textrm{Breit}}(\vec{r};\vec{p}_1,\vec{p}_2)&\!\!\!\!\!=\!\!\!\!\!&-\frac{4\alpha_s}{3r}+\frac{2\pi\alpha_s}{3}\delta^{(3)}(\vec{r}\,)\left(\frac{1}{m_1^2}+\frac{1}{m_2^2}\right)+\frac{2\alpha_s}{3m_1m_2}\left[\frac{\vec{p}_1\cdot\vec{p}_2}{r}+\frac{(\vec{r}\cdot\vec{p}_1)(\vec{r}\cdot\vec{p}_2)}{r^3}\right]\nonumber\\
&&+\frac{4\alpha_s}{3m_1m_2}\left[\frac{8\pi}{3}\delta^{(3)}(\vec{r}\,)(\vec{s}_1\cdot\vec{s}_2)+\frac{3(\vec{s}_1\cdot\hat{r})(\vec{s}_2\cdot \hat{r})-\vec{s}_1\cdot\vec{s}_2}{r^3}\right]\\
&&+\frac{2\alpha_s}{3r^3}\left[\frac{(\vec{r}\times\vec{p}_1)\cdot\vec{s}_1}{m_1^2}-\frac{(\vec{r}\times\vec{p}_2)\cdot\vec{s}_2}{m_2^2}+\frac{2}{m_1m_2}\left((\vec{r}\times\vec{p}_1)\cdot\vec{s}_2-(\vec{r}\times\vec{p}_2)\cdot\vec{s}_1\right)\right]\:.\nonumber
\end{eqnarray}

%% file: A-Appendices/B-OGE.tex
The calculation of the Breit interaction which is equivalent to the one-gluon exchange with retardation corrections, and the one-gluon exchange without retardation differ only in the treatment of the propagator. Therefore, we may as well start the calculation of the one-gluon exchange without retardation on basis of Eq. (\ref{Eq:Potential-Momentum-Space}). Instead of Eq. (\ref{Eq:Propagator-Expanded}) we use
\begin{eqnarray}\label{Eq:Propagator-OGE}
\frac{1}{q^2}&\approx&-\frac{1}{\vec{q}^{\,2}}-\frac{q_0^2}{\vec{q}^{\,4}}\:,\nonumber\\
q_0^2&=&0\:,\nonumber\\
\frac{1}{q^2}&=&-\frac{1}{\vec{q}^{\,2}}\:,
\end{eqnarray}
for the propagator, leading to the momentum space potential
\begin{eqnarray}\label{Eq:Potential-Momentum-Space-OGE}
\frac{1}{(2\pi)^3}\tilde{V}^{(2)}(\vec{q};\vec{p}_1,\vec{p}_2)\!\!\!\!\!&=&\!\!\!\!\!-\frac{4g^2}{3\vec{q}^{\,^2}}\left[1+\frac{-\frac{1}{2}\vec{q}^{\,2}-2\mathrm{i}\vec{s}_1\cdot\left(\vec{q}\times\vec{p}_1\right)}{4m_1^2}+\frac{-\frac{1}{2}\vec{q}^{\,2}+2\mathrm{i}\vec{s}_2\cdot\left(\vec{q}\times\vec{p}_2\right)}{4m_2^2}\right.\\
&&\!\!\!\!\!\!\!\!\!\!\!\!\!\!\!\!\!\!\!\!\!\!\!\!\!\!\!\!\!\!\!\!\!\!\!\!\!\!\!\!\!\!\!\!\!\!\!\!+\!\left.\frac{1}{4m_1m_2}\Big(-4\vec{p}_1\vec{p}_2+4\mathrm{i}\vec{s}_1\cdot\left(\vec{q}\times\vec{p}_2\right)-4\mathrm{i}\vec{s}_2\cdot\left(\vec{q}\times\vec{p}_1\right)-4\vec{q}^{\,2}\left(\vec{s}_1\cdot\vec{s}_2\right)+4\left(\vec{q}\cdot\vec{s}_1\right)\left(\vec{q}\cdot\vec{s}_2\right)\Big)\right]\:.\nonumber
\end{eqnarray}
Employing the Fourier transformation of the various terms of Eq. (\ref{Eq:Potential-Momentum-Space-OGE}) given by Eq. (\ref{Eq:Fouriertransform-Gluon-Exchange}), yields for the potential in coordinate space
\begin{eqnarray}\label{Eq:OGE-Potential-noCM}
V^{(2)}(\vec{r};\vec{p}_1,\vec{p}_2)&\!\!\!\!\!=\!\!\!\!\!&-\frac{4\alpha_s}{3r}+\frac{2\pi\alpha_s}{3}\delta^{(3)}(\vec{r}\,)\left(\frac{1}{m_1^2}+\frac{1}{m_2^2}\right)+\frac{4\alpha_s}{3m_1m_2}\frac{\vec{p}_1\cdot\vec{p}_2}{r}\nonumber\\
&&+\frac{4\alpha_s}{3m_1m_2}\left[\frac{8\pi}{3}\delta^{(3)}(\vec{r}\,)(\vec{s}_1\cdot\vec{s}_2)+\frac{3(\vec{s}_1\cdot\hat{r})(\vec{s}_2\cdot \hat{r})-\vec{s}_1\cdot\vec{s}_2}{r^3}\right]\\
&&+\frac{2\alpha_s}{3r^3}\left[\frac{(\vec{r}\times\vec{p}_1)\cdot\vec{s}_1}{m_1^2}-\frac{(\vec{r}\times\vec{p}_2)\cdot\vec{s}_2}{m_2^2}+\frac{2}{m_1m_2}\left((\vec{r}\times\vec{p}_1)\cdot\vec{s}_2-(\vec{r}\times\vec{p}_2)\cdot\vec{s}_1\right)\right]\:.\nonumber
\end{eqnarray}
The label $V^{(2)}$ is chosen as the diagrams taken into account to compute the potential possess two vetices.

%% file: A-Appendices/B-Kinematics.tex
The kinematics-choice is essential in the derivation of potentials from perturbative QCD due to the fact that, while the amplitude is invariant under the usual time reversal transformation as long as it is expressed in Lorentz-covariants, its nonrelativistic expansion and the gained potential are not. Starting from Eq. (\ref{Eq:Dirac-currents-Old-variables}) we want to show the importance of the kinematics-choice for the potential and the effect of retardation, by explicit recalculation of a quark-antiquark interaction with different kinematics.
\paragraph{Non time reversal invariant kinematics-choice\\}
For this case we choose the kinematics:
\begin{eqnarray}
&&\vec{p}_A=\vec{p}_1\:, \qquad \vec{p}_A\,\!\!'=\vec{p}_1-\vec{q}\:,\qquad\vec{s}_A=\vec{s}_1\:,\qquad m_A=m_1\:,\nonumber\\
&&\vec{p}_b=\vec{p}_2\:, \qquad \vec{p}_B\,\!\!'=\vec{p}_2+\vec{q}\:,\qquad\vec{s}_B=\vec{s}_2\:,\qquad m_B=m_2\:.
\end{eqnarray}
For the timelike and spacelike parts this leads to the expressions
\begin{eqnarray}
u^\dagger(p'_A,s'_A) u(p_A,s_A)\!\!\!\!&=&\!\!\!\!\sqrt{\frac{E_A+m_1}{2m_1}}\sqrt{\frac{E'_A+m_1}{2m_1}}\chi_{s_1'}^\dagger\left(1+\frac{(\vec{\sigma}\cdot(\vec{p}_1-\vec{q}\,))(\vec{\sigma}\cdot\vec{p}_{1})}{(E_A+m_1)(E'_A+m_1)}\right)\chi_{s_1}\nonumber\\
\!\!\!\!&=&\!\!\!\!\sqrt{\frac{E_A+m_1}{2m_1}}\sqrt{\frac{E'_A+m_1}{2m_1}}\chi_{s_1'}^\dagger\left(1+\frac{(\vec{p}_1-\vec{q}\,)\cdot\vec{p}_1}{(E_A+m_1)(E'_A+m_1)}\right.\nonumber\\
&&\left.\qquad\qquad\qquad\qquad\qquad\qquad\quad\:\:\!\!\!\!+\frac{\mathrm{i}\vec{\sigma}\cdot\left((\vec{p}_1-\vec{q}\,)\times\vec{p}_1\right)}{(E_A+m_1)(E'_A+m_1)}\right)\chi_{s_1}\nonumber\\
\!\!\!\!&=&\!\!\!\!\sqrt{\frac{E_A+m_1}{2m_1}}\sqrt{\frac{E'_A+m_1}{2m_1}}\chi_{s_1'}^\dagger\left(1+\frac{\vec{p}_1^{\,2}-\vec{p}_1\cdot\vec{q}-\mathrm{i}\vec{\sigma}\cdot(\vec{q}\times\vec{p}_1)}{(E_A+m_1)(E'_A+m_1)}\right)\chi_{s_1}\:,\nonumber\\
v^{\dagger}(p_B,s_B) v(p'_B,s'_B)\!\!\!\!&=&\!\!\!\!\sqrt{\frac{E_B+m_2}{2m_2}}\sqrt{\frac{E'_B+m_2}{2m_2}}\chi_{s_2}^{c\dagger}\left(1+\frac{\vec{p}_2^{\,2}+\vec{p}_2\cdot\vec{q}-\mathrm{i}\vec{\sigma}\cdot(\vec{q}\times\vec{p}_2)}{(E_B+m_2)(E'_B+m_2)}\right)\chi_{s'_2}^c\:,\\
\end{eqnarray}
\begin{eqnarray}
\bar{u}(p'_A,s'_A)\vec{\gamma}u(p_A,s_A)\!\!\!\!&=&\!\!\!\!\sqrt{\frac{E_A+m_1}{2m_1}}\sqrt{\frac{E'_A+m_1}{2m_1}}\nonumber\\
&&\quad\!\!\!\!\!\chi_{s_1'}^\dagger\left[\left(\frac{\vec{p}_1}{E_A+m_1}+\frac{\vec{p}_1-\vec{q}}{E'_A+m_1}\right)+\mathrm{i}\vec{\sigma}\times\left(\frac{\vec{p}_1-\vec{q}}{E'_A+m_1}-\frac{\vec{p}_1}{E_A+m_1}\right)\right]\chi_{s_1}\nonumber\\
&=&\frac{1}{2m_1}\chi_{s'_1}^\dagger\left[2\vec{p}_1-\vec{q}-\mathrm{i}\vec{\sigma}\times\vec{q}\,\right]\chi_{s_1}+\mathcal{O}\left(\frac{1}{m_1^2}\right)\:,\nonumber\\
\bar{v}(p_B,s_B)\vec{\gamma}v(p'_B,s'_B)\!\!\!\!&=&\!\!\!\!\frac{1}{2m_2}\chi_{s_2}^{c\dagger}\left[2\vec{p}_2+\vec{q}-\mathrm{i}\vec{\sigma}\times\vec{q}\,\right]\chi_{s'_2}^c+\mathcal{O}\left(\frac{1}{m_2^2}\right)\:.
\end{eqnarray}
Employing these expressions in the amplitude Eq. (\ref{Eq:Quark-Antiquark_Amplitude}), remembering Eq. (\ref{Eq:Spin-Particles}) and Eq. (\ref{Eq:Spin-Antiparticles}), gives
\begin{eqnarray}
\mathrm{i}\mathcal{M}_{fi}\!\!\!\!&=&\!\!\!\!-\frac{4}{3}\frac{\mathrm{i}g^2}{q^2+\mathrm{i}\epsilon}(\mathcal{A}+\mathcal{B})\:,\nonumber\\
\mathcal{A}\!\!\!\!&=&\!\!\!\!\EuScript{N}\chi_{s'_1}^\dagger\left(1+\frac{\vec{p}_1^{\,2}-\vec{p}_1\cdot\vec{q}-\mathrm{i}\vec{\sigma}\cdot(\vec{q}\times\vec{p}_1)}{(E_A+m_1)(E'_A+m_1)}\right)\chi_{s_1}\chi_{s_2}^{c\dagger}\left(1+\frac{\vec{p}_2^{\,2}+\vec{p}_2\cdot\vec{q}-\mathrm{i}\vec{\sigma}\cdot(\vec{q}\times\vec{p}_2)}{(E_B+m_2)(E'_B+m_2)}\right)\chi_{s'_2}^c\nonumber\\
\!\!\!\!&=&\!\!\!\!\EuScript{N}\delta_{s_1s'_1}\delta_{s_2s'_2}\left(1+\frac{\vec{p}_1^{\,2}-\vec{p}_1\cdot\vec{q}-2\mathrm{i}\vec{s}_{1}\cdot(\vec{q}\times\vec{p}_1)}{(E_A+m_1)(E'_A+m_1)}+\frac{\vec{p}_2^{\,2}+\vec{p}_2\cdot\vec{q}+2\mathrm{i}\vec{s}_{2}\cdot(\vec{q}\times\vec{p}_2)}{(E_B+m_2)(E'_B+m_2)}\right)\nonumber\\
\!\!\!\!&=&\!\!\!\!\delta_{s_1s'_1}\delta_{s_2s'_2}\left(1+\frac{2\vec{p}_1^{\,2}-2\vec{p}_1\cdot\vec{q}+\frac{1}{2}\vec{q}^{\,2}-2\mathrm{i}\vec{s}_1\cdot\left(\vec{q}\times\vec{p}_1^{\,2}\right)}{4m_1^2}\right.\nonumber\\
&&\quad\quad\quad\quad\quad\left.+\frac{2\vec{p}_2^{\,2}+2\vec{p}_2\cdot\vec{q}+\frac{1}{2}\vec{q}^{\,2}+2\mathrm{i}\vec{s}_2\cdot\left(\vec{q}\times\vec{p}_2^{\,2}\right)}{4m_2^2}\right)+\mathcal{O}\left(\frac{1}{m_i^3}\right)\:,\nonumber\\
\mathcal{B}\!\!\!\!&=&\!\!\!\!-\chi_{s'_1}^\dagger\left(\frac{2\vec{p}_1-\vec{q}-\mathrm{i}\vec{\sigma}\times\vec{q}}{2m_1}\right)\chi_{s_1}\chi_{s_2}^{c\dagger}\left(\frac{2\vec{p}_2+\vec{q}+\mathrm{i}\vec{\sigma}\times\vec{q}}{2m_2}\right)\chi_{s'_2}^c+\mathcal{O}\left(\frac{1}{m_i^2m_j}\right)\nonumber\\
\!\!\!\!&=&\!\!\!\!\frac{\delta_{s_1s'_1}\delta_{s_2s'_2}}{4m_1m_2}\left(-4\vec{p}_1\cdot\vec{p}_2-2\vec{p}_1\cdot\vec{q}+2\vec{p}_2\cdot\vec{q}+\vec{q}^{\,2}\frac{}{}\right.\nonumber\\
&&\quad\quad\!\!\!\!\left.+4\mathrm{i}\vec{s}_1\cdot\left(\vec{q}\times\vec{p}_2\right)-4\mathrm{i}\vec{s}_2\cdot\left(\vec{q}\times\vec{p}_1\right)
-4\vec{q}^{\,2}\left(\vec{s}_1\cdot\vec{s}_2\right)+4\left(\vec{q}\cdot\vec{s}_1\right)\left(\vec{q}\cdot\vec{s}_2\right)\frac{}{}\right)+\mathcal{O}\left(\frac{1}{m_i^2m_j}\right)\:,\nonumber\\
\EuScript{N}\!\!\!\!&=&\!\!\!\!\sqrt{\frac{E_A+m_1}{2m_1}}\sqrt{\frac{E'_A+m_1}{2m_1}}\sqrt{\frac{E_B+m_2}{2m_2}}\sqrt{\frac{E'_B+m_2}{2m_2}}\nonumber\\
\!\!\!\!&=&\!\!\!\!1+\frac{\vec{p}_1^{\,2}-\vec{p}_1\cdot\vec{q}+\frac{1}{2}\vec{q}^{\,2}}{4m_1^2}+\frac{\vec{p}_2^{\,2}+\vec{p}_2\cdot\vec{q}+\frac{1}{2}\vec{q}^{\,2}}{4m_2^2}+\mathcal{O}\left(\frac{1}{m_i^3}\right)\:.
\end{eqnarray}
The relation between potential and amplitude 
\begin{eqnarray}
\frac{1}{(2\pi)^3}\tilde{V}(\vec{q};\vec{p}_1,\vec{p}_2)=\pm\frac{m_1 m_2}{\sqrt{E_A}\sqrt{E_B}\sqrt{E'_A}\sqrt{E'_B}}\mathcal{M}_{fi}\:,
\end{eqnarray}
leads then to the factor
\begin{eqnarray}
\frac{m_1 m_2}{\sqrt{E_A}\sqrt{E_B}\sqrt{E'_A}\sqrt{E'_B}}=1\!-\!\frac{2\vec{p}_1^{\,2}}{4m_1^2}-\frac{2\vec{p}_2^{\,2}}{4m_2^2}+\frac{2\vec{p}_1\cdot\vec{q}}{4m_1^2}-\frac{2\vec{p}_2\cdot\vec{q}}{4m_2^2}-\frac{\vec{q}^{\,2}}{4m_1^2}-\frac{\vec{q}^{\,2}}{4m_2^2}+\!\mathcal{O}\!\left(\!\frac{1}{m_i^3}\!\right)\:,
\end{eqnarray}
and we obtain 
\begin{eqnarray}
\frac{1}{(2\pi)^3}\tilde{V}(\vec{q};\vec{p}_1,\vec{p}_2)\!\!\!\!&=&\!\!\!\!\frac{4g^2}{3q^2}\Biggl[1+\frac{-\frac{1}{2}\vec{q}^{\,2}-2\mathrm{i}\vec{s}_1\cdot\left(\vec{q}\times\vec{p}_1\right)}{4m_1^2}+\frac{-\frac{1}{2}\vec{q}^{\,2}+2\mathrm{i}\vec{s}_2\cdot\left(\vec{q}\times\vec{p}_2\right)}{4m_2^2}\nonumber\\
&&\quad+\frac{1}{4m_1m_2}\biggl(-4\vec{p}_1\vec{p}_2-2\vec{p}_1\cdot\vec{q}+2\vec{p}_2\cdot\vec{q}+\vec{q}^{\,2}+4\mathrm{i}\vec{s}_1\cdot\left(\vec{q}\times\vec{p}_2\right)\nonumber\\
&&\qquad\qquad-4\mathrm{i}\vec{s}_2\cdot\left(\vec{q}\times\vec{p}_1\right)-4\vec{q}^{\,2}\left(\vec{s}_1\cdot\vec{s}_2\right)+4\left(\vec{q}\cdot\vec{s}_1\right)\left(\vec{q}\cdot\vec{s}_2\right)\biggr)\Biggr]\:.
\end{eqnarray}
Using Eq. (\ref{Eq:Propagator-OGE}) for the expansion of the propagator, the potential in momentum space reads
\begin{eqnarray}
\frac{1}{(2\pi)^3}\tilde{V}(\vec{q};\vec{p}_1,\vec{p}_2)\!\!\!\!&=&\!\!\!\!-\frac{4g^2}{3\vec{q}^{\,2}}\Biggl[1+\frac{-\frac{1}{2}\vec{q}^{\,2}-2\mathrm{i}\vec{s}_1\cdot\left(\vec{q}\times\vec{p}_1\right)}{4m_1^2}+\frac{-\frac{1}{2}\vec{q}^{\,2}+2\mathrm{i}\vec{s}_2\cdot\left(\vec{q}\times\vec{p}_2\right)}{4m_2^2}\nonumber\\
&&\quad+\frac{1}{4m_1m_2}\biggl(-4\vec{p}_1\vec{p}_2-2\vec{p}_1\cdot\vec{q}+2\vec{p}_2\cdot\vec{q}+\vec{q}^{\,2}+4\mathrm{i}\vec{s}_1\cdot\left(\vec{q}\times\vec{p}_2\right)\nonumber\\
&&\qquad\qquad-4\mathrm{i}\vec{s}_2\cdot\left(\vec{q}\times\vec{p}_1\right)-4\vec{q}^{\,2}\left(\vec{s}_1\cdot\vec{s}_2\right)+4\left(\vec{q}\cdot\vec{s}_1\right)\left(\vec{q}\cdot\vec{s}_2\right)\biggr)\Biggr]\:.
\end{eqnarray}
As one can see there are spin-independent terms of structure $\vec{p}_i\cdot\vec{q}$. These structures will after Fourier transformation result in terms which are not invariant under the usual time reversal transformation. The parameterization yields an additional $\vec{q}^{\,2}$ term assigned with a factor $(m_1m_2)^{-1}$.
\paragraph{Time reversal invariant parameterization\\}
For the case, which leads to a time reversal invariant potential, we substitute the variables in the amplitude by
\begin{eqnarray}
&&\vec{p}_A=\vec{p}_1+\frac{1}{2}\vec{q}\:, \qquad \vec{p}_A\,\!\!'=\vec{p}_1-\frac{1}{2}\vec{q}\:,\qquad\vec{s}_A=\vec{s}_1\:,\qquad m_A=m_1\:,\nonumber\\
&&\vec{p}_B=\vec{p}_2-\frac{1}{2}\vec{q}\:, \qquad \vec{p}_B\,\!\!'=\vec{p}_2+\frac{1}{2}\vec{q}\:,\qquad\vec{s}_B=\vec{s}_2\:,\qquad m_B=m_2\:.
\end{eqnarray}
As this is exactly the choice used in in App. \ref{Sec:BreitInteraction} and \ref{Sec:OneGluonExchangeW/ORetardation}, we skip the details and perform only the last steps. Employing Eq. (\ref{Eq:Propagator-OGE}) for the propagator
\begin{eqnarray}
\frac{1}{(2\pi)^3}\tilde{V}(\vec{q};\vec{p}_1,\vec{p}_2)\!\!\!\!&=&\!\!\!\!\frac{4g^2}{3q^2}\Biggl[1+\frac{-\frac{1}{2}\vec{q}^{\,2}-2\mathrm{i}\vec{s}_1\cdot\left(\vec{q}\times\vec{p}_1\right)}{4m_1^2}+\frac{-\frac{1}{2}\vec{q}^{\,2}+2\mathrm{i}\vec{s}_2\cdot\left(\vec{q}\times\vec{p}_2\right)}{4m_2^2}\\
&&\!\!\!\!\!\!\!\!\!\!\!\!\!\!\!\!\!\!\!\!\!\!\!\!\!\!\!\!\!\!\!\!\!\!\!\!\!\!\!\!\!+\frac{1}{4m_1m_2}\biggl(-4\vec{p}_1\vec{p}_2+4\mathrm{i}\vec{s}_1\cdot\left(\vec{q}\times\vec{p}_2\right)-4\mathrm{i}\vec{s}_2\cdot\left(\vec{q}\times\vec{p}_1\right)-4\vec{q}^{\,2}\left(\vec{s}_1\cdot\vec{s}_2\right)+4\left(\vec{q}\cdot\vec{s}_1\right)\left(\vec{q}\cdot\vec{s}_2\right)\biggr)\Biggr]\:,\nonumber
\end{eqnarray}
transforms to
\begin{eqnarray}\label{Eq:Time-Reversal-Potential-Momentum-Space}
\frac{1}{(2\pi)^3}\tilde{V}(\vec{q};\vec{p}_1,\vec{p}_2)\!\!\!\!&=&\!\!\!\!-\frac{4g^2}{3\vec{q}^{\,2}}\Biggl[1+\frac{-\frac{1}{2}\vec{q}^{\,2}-2\mathrm{i}\vec{s}_1\cdot\left(\vec{q}\times\vec{p}_1\right)}{4m_1^2}+\frac{-\frac{1}{2}\vec{q}^{\,2}+2\mathrm{i}\vec{s}_2\cdot\left(\vec{q}\times\vec{p}_2\right)}{4m_2^2}\\
&&\!\!\!\!\!\!\!\!\!\!\!\!\!\!\!\!\!\!\!\!\!\!\!\!\!\!\!\!\!\!\!\!\!\!\!\!\!\!\!\!\!+\frac{1}{4m_1m_2}\biggl(-4\vec{p}_1\vec{p}_2+4\mathrm{i}\vec{s}_1\cdot\left(\vec{q}\times\vec{p}_2\right)-4\mathrm{i}\vec{s}_2\cdot\left(\vec{q}\times\vec{p}_1\right)-4\vec{q}^{\,2}\left(\vec{s}_1\cdot\vec{s}_2\right)+4\left(\vec{q}\cdot\vec{s}_1\right)\left(\vec{q}\cdot\vec{s}_2\right)\biggr)\Biggr]\:.\nonumber 
\end{eqnarray}
One recognizes that there are no terms of the structure described in the previous paragraph in Eq. (\ref{Eq:Time-Reversal-Potential-Momentum-Space}).
\paragraph{The effect of retardation\\}
The previous paragraphs discussed the influence of the kinematics-choice on potentials without retardation corrections. Taking into account retardation in the same way as for the Breit interaction (App. \ref{Sec:BreitInteraction}), the influence of the kinematics-choice vanishes as both parameterizations result in
\begin{eqnarray}
\frac{1}{(2\pi)^3}\tilde{V}(\vec{q};\vec{p}_1,\vec{p}_2)\!\!\!\!\!\!&=&\!\!\!\!\!\!-\frac{4g^2}{3\vec{q}^{\,2}}\Biggl[1+\frac{-\frac{1}{2}\vec{q}^{\,2}-2\mathrm{i}\vec{s}_1\cdot\left(\vec{q}\times\vec{p}_1\right)}{4m_1^2}+\frac{-\frac{1}{2}\vec{q}^{\,2}+2\mathrm{i}\vec{s}_2\cdot\left(\vec{q}\times\vec{p}_2\right)}{4m_2^2}\nonumber\\
&&\qquad\quad\!+\frac{4}{\vec{q}^{\,2}}\frac{(\vec{p}_1\cdot\vec{q})(\vec{p}_2\cdot\vec{q})}{4m_1m_2}+\frac{1}{4m_1m_2}\biggl(-4\vec{p}_1\vec{p}_2+4\mathrm{i}\vec{s}_1\cdot\left(\vec{q}\times\vec{p}_2\right)\nonumber\\
&&\qquad\qquad-4\mathrm{i}\vec{s}_2\cdot\left(\vec{q}\times\vec{p}_1\right)-4\vec{q}^{\,2}\left(\vec{s}_1\cdot\vec{s}_2\right)+4\left(\vec{q}\cdot\vec{s}_1\right)\left(\vec{q}\cdot\vec{s}_2\right)\biggr)\Biggr]\:.
\end{eqnarray}
This is a consequence of the distinct expansions for the energies in
\begin{eqnarray}
\frac{1}{q^2}&\approx&-\frac{1}{\vec{q}^{\,2}}-\frac{q_0^2}{\vec{q}^{\,4}}\:,\nonumber\\
q_0^2&=&(E'_A-E_A)(E_B-E'_B)\:.
\end{eqnarray}
For the non time reversal invariant case we obtain,
\begin{eqnarray}
\frac{1}{q^2}&=&-\frac{1}{\vec{q}^{\,2}}-\frac{1}{\vec{q}^{\,4}}\frac{(\vec{p}_1\cdot\vec{q}\,)(\vec{p}_2\cdot\vec{q}\,)}{m_1m_2}-\frac{1}{2}\frac{1}{\vec{q}^{\,2}}\frac{\vec{p}_1\cdot\vec{q}-\vec{p}_2\cdot\vec{q}}{m_1m_2}+\frac{1}{4}\frac{1}{m_1m_2} +\mathcal{O}\left(\frac{1}{m^3}\right)\:,
\end{eqnarray}
while for the time reversal invariant parameterization it eventually reads,
\begin{eqnarray}
\frac{1}{q^2}&=&-\frac{1}{\vec{q}^{\,2}}-\frac{1}{\vec{q}^{\,4}}\frac{(\vec{p}_1\cdot\vec{q}\,)(\vec{p}_2\cdot\vec{q}\,)}{m_1m_2} +\mathcal{O}\left(\frac{1}{m^3}\right)\:.
\end{eqnarray}
Though the statement, that retardation compensates the influence of the kinematics-choice, is true for the One-Gluon exchange, it does not hold for the Lorentz-vector exchange of massive particles.

%% file: A-Appendices/C-Computation-nabla-nabla-W.tex
As one has to proceed with caution in computing an expression for $\nabla_i\nabla_jW(r)$, depending reclusively on derivations of $W$ with respect to $r$, we will show the calculation in detail. First, we search for the most general ansatz for two derivatives with two indices, and eventually find
\begin{eqnarray}
\nabla_i\nabla_j W=\left(a\delta_{ij}+b\frac{r_ir_j}{r^2}\right)\frac{1}{r}W'+\left(c\delta_{ij}+d\frac{r_ir_j}{r^2}\right)W''+\left(e\delta_{ij}+f\frac{r_ir_j}{r^2}\right)\Delta W\:.
\end{eqnarray}
The powers of $r$ in the denominator are chosen in this way to make sure that the coefficients $a,\dots\:,\:f$ are dimensionless. Distinguishing the case of $i=j$ from $i\neq j$ we multiply our ansatz by $\delta_{ij}$, using that
\begin{eqnarray}
\delta_{ij}\delta_{ij}=\mathrm{tr}(\delta_{ij})=3\:,
\end{eqnarray}
to arrive at
\begin{eqnarray}
\Delta W=\left(3a+b\right)\frac{1}{r}W'+\left(3c+d\right)W''+\left(3e+f\right)\Delta W\:.
\end{eqnarray}
This yields the following conditions for the coefficients
\begin{eqnarray}
a=-\frac{b}{3},\quad c=-\frac{d}{3},\quad e=\frac{1}{3}-\frac{f}{3}\:,
\end{eqnarray}
which, introduced into our ansatz, result in
\begin{eqnarray}\label{Eq:Ansatz-i=j}
\nabla_i\nabla_j W=\left(b\frac{1}{r}W'+dW''+f\Delta W\right)\left(\frac{r_ir_j}{r^2}-\frac{1}{3}\delta_{ij}\right)+\frac{1}{3}\delta_{ij}\Delta W\:.
\end{eqnarray}
The remaining constraint needed to identify our coefficients are gained by investigation  of the case $i\neq j$. Remembering that
\begin{eqnarray}
\nabla_i W(r)=\frac{r_i}{r}W'(r)\:,
\end{eqnarray}
we find 
\begin{eqnarray}
\nabla_i\nabla_j W=\frac{r_ir_j}{r^2}\left(W''-\frac{1}{r}W'\right)\:,\qquad i\neq j\:.
\end{eqnarray}
Then again, using the ansatz Eq. (\ref{Eq:Ansatz-i=j}) yields
\begin{eqnarray}
\nabla_i\nabla_j W=\frac{r_ir_j}{r^2}\left(b\frac{1}{r}W'+dW''+f\Delta W\right)\:,\qquad i\neq j\:.
\end{eqnarray}
Through comparison of the last two equations we eliminate the remaining unknowns, giving:
\begin{eqnarray}\label{Eq:nablainablajW}
\nabla_i\nabla_j W(r)=\left[W''(r)-\frac{1}{r}W'(r)\right]\left(\frac{r_ir_j}{r^2}-\frac{1}{3}\delta_{ij}\right)+\frac{1}{3}\delta_{ij}\Delta W(r)\:.
\end{eqnarray}

%% file: A-Appendices/D-MD-Expectation-values.tex
The computation of quarkonium masses within the model includes perturbatively treated terms. One class of terms are purely momentum dependent terms which are given by a Hamiltonian of structure
\begin{eqnarray}\label{Eq:Hamiltonian-momentum-dependent}
H^{\textrm{MD}}&=&f(r)\bigg(a\vec{p}^{\,2}+b(\hat{r}\cdot\vec{p}\,)(\hat{r}\cdot\vec{p}\,)\bigg)\:,
\end{eqnarray}
with an arbitrary function $f(r)$ and $a$, $b$ being constants. For the evaluation we have to translate the Hamilton function into the corresponding Hamilton operator. As the operators for $\vec{p}$ and $\vec{r}$ do not commute the transition from Hamilton function to Hamilton operator is not unique, meaning that there is no unique ordering for the operators corresponding to $\vec{p}$ and $\vec{r}$. Hence we have to look at several orderings of operators, with the prerequisite that the resulting expression has to be hermitean.

%% file: A-Appendices/D-MD-Interpretations.tex
In this work we choose one of the three options presented in the following for the momentum dependent terms.
\paragraph{Option 1\\}
In the first option the ordering is as such that the operator corresponding to the function $f(r)$ in Eq. (\ref{Eq:Hamiltonian-momentum-dependent}) is either completely left or right of the $\vec{p}$ operators. Therefore, the expectation value of Eq. (\ref{Eq:Hamiltonian-momentum-dependent}) in option 1 is given by a superposition of
\begin{eqnarray}
\langle f(r)\vec{p}^{\,2}\rangle_1&=&-\frac{1}{2}\int\mathrm{d}^3r\:\left(\psi^*\vec{\nabla}^{\,2}f(r)\psi+\psi^*f(r)\vec{\nabla}^{\,2}\psi\right)\nonumber\\
&=&-\frac{1}{2}\int\mathrm{d}^3r\:\left((\vec{\nabla}^{\,2}\psi^*)f(r)\psi+\psi^*f(r)\vec{\nabla}^{\,2}\psi\right)\nonumber\\
&=&-\int\mathrm{d}^3r\:\psi^*f(r)\vec{\nabla}^{\,2}\psi\:,
\end{eqnarray}
and
\begin{eqnarray}
\langle f(r)(\hat{r}\cdot\vec{p}\,)(\hat{r}\cdot\vec{p}\,)\rangle_1&=&-\frac{1}{2}\int\mathrm{d}^3r\:\psi^*\bigg(f(r)(\hat{r}\cdot\vec{\nabla})(\hat{r}\cdot\vec{\nabla})+(\hat{r}\cdot\vec{\nabla})(\hat{r}\cdot\vec{\nabla})f(r)\bigg)\psi\nonumber\\
&=&-\frac{1}{2}\int\mathrm{d}^3r\:\left[\left(\frac{\partial^2}{\partial r^2}\psi^*\right)f(r)\:\psi+\psi^* f(r)\frac{\partial^2}{\partial r^2}\psi\right]\nonumber\\
&=&-\int\mathrm{d}^3r\:\psi^* f(r)\frac{\partial^2}{\partial r^2}\psi\:.
\end{eqnarray}
In terms of the reduced radial wavefunction $u(r)$ these expression transform to
\begin{eqnarray}\label{Eq:MD-Interpretations1}
\langle f(r)\vec{p}^{\,2}\rangle_1&=&-\int\mathrm{d}^3r\:f(r)\psi^*\vec{\nabla}^2\psi\nonumber\\
&=&-\int\mathrm{d}r\:f(r)\left(u(r)u''(r)-\frac{\big(u(r)\big)^2}{r^2}l(l+1)\right)\:,\\
\langle f(r)(\hat{r}\cdot\vec{p}\,)(\hat{r}\cdot\vec{p}\,)\rangle_1&=&-\int\mathrm{d}^3r\:f(r)\psi^*\frac{\partial^2}{\partial r^2}\psi\nonumber\\
&=&-\int\mathrm{d}r\:f(r)\left(u(r)u''(r)-2\frac{u(r)u'(r)}{r}+2\frac{\big(u(r)\big)^2}{r^2}\right)\:.
\end{eqnarray}
\paragraph{Option 2\\}
For the second option we write down a more symmetric form, where the operator corresponding to the function $f(r)$ in Eq. (\ref{Eq:Hamiltonian-momentum-dependent}) stands in between the $\vec{p}$ operators. The expectation value of Eq. (\ref{Eq:Hamiltonian-momentum-dependent}) is then a superposition of the expressions
\begin{eqnarray}
\langle f(r)\vec{p}^{\,2}\rangle_2&=&-\int\mathrm{d}^3r\:\psi^*\vec{\nabla}f(r)\vec{\nabla}\psi=\int\mathrm{d}^3r\:f(r)\:\big|\:\vec{\nabla}\:\psi\:\big|^2\:,\\
\langle f(r)(\hat{r}\cdot\vec{p}\,)(\hat{r}\cdot\vec{p}\,)\rangle_2&=&-\int\mathrm{d}^3r\:\psi^*\:\hat{r}\cdot\vec{\nabla}\:f(r)\:\hat{r}\cdot\vec{\nabla}\:\psi=\int\mathrm{d}^3r\:f(r)\:\big|\:\hat{r}\cdot\vec{\nabla}\:\psi\:\big|^2\:.
\end{eqnarray}
The equivalent expressions in terms of the reduced radial wavefunctions are given by
\begin{eqnarray}\label{Eq:MD-Interpretations2a}
\langle f(r)\vec{p}^{\,2}\rangle_2&=&\int\mathrm{d}^3r\:f(r)\:\big|\:\vec{\nabla}\:\psi\:\big|^2=-\int\mathrm{d}^3r\:f(r)\:\bigg|\vec{\nabla}\left[\frac{u(r)}{r}Y_{lm}(\theta,\phi)\right]\bigg|^2\nonumber\\
&=&\int\mathrm{d}^3r\:f(r)\left(\bigg|\left[\vec{\nabla}\frac{u(r)}{r}\:\right]Y_{lm}(\theta,\phi)\:\bigg|^2+\:\bigg|\frac{u(r)}{r}\left[\vec{\nabla}\:Y_{lm}(\theta,\phi)\right]\bigg|^2\right)\nonumber\\
&=&\int\mathrm{d}^3r\:f(r)\left(\bigg|\left[\vec{\nabla}\frac{u(r)}{r}\right]Y_{lm}(\theta,\phi)\:\bigg|^2+\:\bigg|\frac{u(r)}{r}\left[\hat{r}\times\vec{\nabla}\:Y_{lm}(\theta,\phi)\right]\bigg|^2\right)\nonumber\\
&=&\int\mathrm{d}r\:r^2f(r)\left[\left(\frac{u'(r)}{r}-\frac{u(r)}{r^2}\right)^2+\left(\frac{u(r)}{r^2}\right)^2\int\mathrm{d}\Omega \:\bigg|\:\vec{L}\:Y_{lm}(\theta,\phi)\:\bigg|^2\right]\nonumber\\
&=&\int\mathrm{d}r\:f(r)\left[\big(u'(r)\big)^2-2\frac{u'(r)u(r)}{r}+\frac{\big(u(r)\big)^2}{r^2}\Big(1+l(l+1)\Big)\right]\:,
\end{eqnarray}
\begin{eqnarray}\label{Eq:MD-Interpretations2b}
\langle f(r)(\hat{r}\cdot\vec{p}\,)(\hat{r}\cdot\vec{p}\,)\rangle_2&=&\int\mathrm{d}^3r\:f(r)\:\bigg|\:\hat{r}\cdot\vec{\nabla}\:\psi\:\bigg|^2=-\int\mathrm{d}^3r\:f(r)\:\bigg|\:\frac{\partial}{\partial r}\:\psi\:\bigg|^2\nonumber\\
&=&\int\mathrm{d}^3r\:f(r)\:\bigg|\:\frac{\partial}{\partial r}\left(\frac{u(r)}{r}\:Y_{lm}(\theta,\phi)\right)\:\bigg|^2\nonumber\\
&=&\int\mathrm{d}r\:f(r)\left(\big(u'(r)\big)^2-2\frac{u'(r)u(r)}{r}+\frac{\big(u(r)\big)^2}{r^2}\right)\:.
\end{eqnarray}
\paragraph{Option 3\\}
In addition to these two basic options we also employ an option motivated by the Weyl-prescription \cite{Brambilla:1993zw}
\begin{eqnarray}
\left\{f(r)p^ip^j\right\}_W=\frac{1}{4}\left\{\left\{f(r),p^i\right\},p^j\right\}\:.
\end{eqnarray}
This option is essentially the arithmetic mean of options 1 \& 2.
\paragraph{}Although in principle any other normed mixing of options 1 \& 2 would be possible, we pick, for the momentum dependent terms within our model, one of the three options above.

%% file: A-Appendices/D-MD-Interpretations-and-Retardation.tex
The momentum dependent term of the Breit interaction (Eq. \ref{Eq:Breit-Interaction}) is a unique case as option 1 and option 2 and therefore any superposition of them are equivalent. In the following this feature of the Breit interaction is shown.

The momentum dependent term of the Breit interaction is given by
\begin{eqnarray}\label{Eq:MD-Term-Breit}
H^{\textrm{MD}}_{\textrm{Breit}}&=&-\frac{2\alpha_s}{3m^2}\left[\frac{\vec{p}^{\,2}}{r}+\frac{(\vec{r}\cdot\vec{p}\,)(\vec{r}\cdot\vec{p}\,)}{r^3}\right]\:.
\end{eqnarray}
The expectation value of Eq. (\ref{Eq:MD-Term-Breit}) in options 1 and 2 in terms of the reduced radial wavefunction (see App. \ref{Sec:MD-interpretations}) read
\begin{eqnarray}
A&=&\bigg\langle- \frac{2\alpha_s}{3m^2r}\left(\vec{p}^{\,2}+(\hat{r}\cdot\vec{p}\,)(\hat{r}\cdot\vec{p}\,)\right)\bigg\rangle_1\nonumber\\
&=&\frac{2\alpha_s}{3m^2}\int\mathrm{d} r\:\frac{1}{r}\left(2u(r)u''(r)-\frac{2u(r)u'(r)}{r}+\frac{\big(u(r)\big)^2}{r^2}\Big(2-l(l+1)\Big) \right)\:,\nonumber\\
B&=&\bigg\langle- \frac{2\alpha_s}{3m^2r}\left(\vec{p}^{\,2}+(\hat{r}\cdot\vec{p}\,)(\hat{r}\cdot\vec{p}\,)\right)\bigg\rangle_2\nonumber\\
&=&\frac{2\alpha_s}{3m^2}\int\mathrm{d} r\:\frac{1}{r}\left(-2\big(u'(r)\big)^2+4\frac{u'(r)u(r)}{r}-\frac{\big(u(r)\big)^2}{r^2}\Big(2+l(l+1)\Big) \right)\:.
\end{eqnarray}
Comparing the two options one finds that they differ only by a total derivation:
\begin{eqnarray}
A-B&=&\frac{2\alpha_s}{3m^2}\int\mathrm{d}r\:\frac{1}{r}\left(4\frac{\big(u(r)\big)^2}{r^2}-6\frac{u(r)u'(r)}{r}+2\big(u'(r)\big)^2+2u(r)u''(r)\right)\nonumber\\
&=&\frac{4\alpha_s}{3m^2}\int\mathrm{d}r\left[\frac{\mathrm{d}}{\mathrm{d}r}\left(u(r)\frac{\mathrm{d}}{\mathrm{d}r}\frac{u(r)}{r}\right)\right]=-\frac{4\alpha_s}{3m^2}\left(\frac{\big(u(r)\big)^2}{r^2}-\frac{u(r)u'(r)}{r}\right)\bigg|_{r=0}^{\infty}\:.
\end{eqnarray}
The boundary condition which the radial wavefunctions must fulfill at $r=0$ and the utilization of a power series ansatz for $u(r)$ near the origin give (see e.g. \cite{Friedrich:TAP06})
\begin{eqnarray}
u(0)&=&0\:,\nonumber\\
\lim_{r\rightarrow0} u'(r)&=&\lim_{r\rightarrow0}\frac{u(r)}{r}\:.
\end{eqnarray}
Furthermore, we know that in our models the linear part of the potential is dominant for large $r$. Thus, the reduced radial wavefunctions features for $r\rightarrow\infty$ the asymptotic behaviour of solutions for a linear potential which are given by  Airy-functions, yielding the conditions
\begin{eqnarray}
\lim_{r\rightarrow\infty}u(r)&=&0\:,\nonumber\\
\lim_{r\rightarrow\infty}u'(r)&=&0\:.
\end{eqnarray}
Considering these properties of the reduced radial wavefunction we can verify
\begin{eqnarray}
\left(\frac{u^2(r)}{r^2}-\frac{u(r)u'(r)}{r}\right)\bigg|_{r=0}^{\infty}=0\:,
\end{eqnarray}
implying that options 1 and 2 are equivalent.
\newpage
\addcontentsline{toc}{chapter}{Bibliography}

%% file: 0-Titlepage/0-Acknowledgements.tex
There are many people who have - in one way or another - contributed to the success of this work. I would like to thank all of them especially...
\begin{itemize}
\item Prof. Dr. Wolfram Weise, for his support and interest, for many insightful and motivating discussions, and for the opportunity to write this thesis.
\item Prof. Dr. Norbert Kaiser, for countless helpful discussions and the advice he gave me.
\item all members of T39, for the nice working climate.
\item my family, for their support before and during my studies.
\item Christina Wendl, for her love, and for her lenience concerning the many hours tainted with physics.
\end{itemize}

%% file: 0-Titlepage/0-CorrectionHistory.tex
The following corrections have been made with respect to the version handed in on January 31, 2007
\begin{itemize}
\item added entry 'Bibliography' in table of contents
\item corrected index in Eq. (\ref{Eq:nablainablajW})
\item corrected $M_\textrm{i}^\textrm{T}$ for $1^3D_j$-states in Tab. \ref{Tab:MassesContributionsCharmoniumCpL*pI-2}
\item removed widths for $k^3P_j$-states in Tab. \ref{Tab:MassesBottomoniumCpL-B}, Tab. \ref{Tab:MassesBottomoniumCpL-3}, Tab. \ref{Tab:MassesBottomCpL*-2} and Tab. \ref{Tab:MassesBottomoniumCpL*pI-2}
\item corrected mistakes in Tab. \ref{Tab:CharmoniumRadiusMass-CpL*pI-2} and Eq. (\ref{Eq:RadiusMassCharmoniumBare}), altered text in Chapter \ref{Sec:FogtppInducedRadiusMassCharmonium} and Chapter \ref{Sec:FogtppInducedRadiusMassBottomonium} concerning Eq. (\ref{Eq:RadiusMassCharmoniumBare})
\end{itemize}
\begin{flushright}
Date: July 03, 2007
\end{flushright}